\pdfoutput=1

\documentclass[11pt]{article}

\usepackage[preprint]{acl}

\usepackage{times}
\usepackage{latexsym}

\usepackage[T1]{fontenc}

\usepackage[utf8]{inputenc}

\usepackage{microtype}

\usepackage{inconsolata}

\usepackage{graphicx}

\usepackage{amssymb}
\usepackage{bm}
\usepackage{mathtools}
\usepackage{amsmath}
\usepackage{amssymb}
\usepackage{amsfonts}
\usepackage{algorithmic}
\usepackage[linesnumbered, ruled, vlined]{algorithm2e}
\usepackage{etoolbox}
\makeatletter
\patchcmd\algocf@Vline{\vrule}{\vrule \kern-0.4pt}{}{}
\patchcmd\algocf@Vsline{\vrule}{\vrule \kern-0.4pt}{}{}
\patchcmd\@makecaption{\scshape}{}{}{}
\patchcmd\@makecaption{\\}{:~}{}{}
\patchcmd\@makecaption{\raggedleft}{}{}{}
\makeatother
\SetAlgoNlRelativeSize{-1}
\usepackage{graphicx}
\usepackage{adjustbox}
\usepackage{diagbox}
\usepackage{slashbox}
\usepackage{multirow}
\usepackage{booktabs}
\usepackage{caption}
\usepackage{makecell}
\usepackage{enumitem}
\setlist{nosep,leftmargin=2.5em}
\usepackage{wrapfig}
\usepackage{float}
\usepackage{tikz}
\usepackage{courier}
\usepackage{color}
\usepackage{xcolor}
\definecolor{customblue}{rgb}{0.1, 0.1, 0.8}
\hypersetup{
    colorlinks=true,
    citecolor=customblue,
    linkcolor=customblue,
    urlcolor=customblue
}

\usepackage[most]{tcolorbox}
\tcbuselibrary{listings}
\tcbuselibrary{breakable}
\tcbuselibrary{skins}
\tcbset{mybreakable/.style={
    breakable,
    enhanced,
}}
\lstdefinelanguage{sysprompt}
{
    moredelim=**[is][\color{blue}]{@}{@},
    moredelim=**[is][\bfseries]{//}{//},
    moredelim=**[is][\color{green}]{<<}{>>},
    moredelim=**[is][\color{red}]{!!}{!!},
    moredelim=**[is][\colorbox{pink!20}]{&&&}{&&&},
}

\lstdefinelanguage{dialogue}
{
    moredelim=**[is][\bfseries]{<<}{>>},
}

\newtcblisting[auto counter]{exampledata}[1][]{
    colback=white,
    listing only,
    title={\mbox{Example \thetcbcounter:} #1},
    listing options={
        language=dialogue,
        basicstyle=\scriptsize\ttfamily,
        breaklines=true,
        breakindent=0pt,
        showstringspaces=false,
        escapechar=|,
        literate={µ}{{$\mu$}}1
    },
    boxrule=0.3mm,
    coltitle=white,
    mybreakable,
    colbacktitle=blue,
    colframe=blue,
}

\newtcblisting[auto counter]{dynamictemplate}[2][]{
    colback=white,
    listing only,
    title={\mbox{Dynamic instruction \thetcbcounter:} #1},
    listing options={
        language=sysprompt,
        basicstyle=\scriptsize\ttfamily,
        breaklines=true,
        breakindent=0pt,
        showstringspaces=false
    },
    boxrule=0.3mm,
    coltitle=white,
    mybreakable,
    colbacktitle=red,
    colframe=red,
    label=#2
}

\newtcblisting{custominput}[1][]{
    colback=white,
    listing only,
    title={#1},
    listing options={
        language=dialogue,
        basicstyle=\scriptsize\ttfamily,
        breaklines=true,
        breakindent=0pt,
        showstringspaces=false,
    },
    boxrule=0.3mm,
    mybreakable,
    coltitle=black,
    colbacktitle=gray!20
}

\newtcblisting{customdialogue}[1][]{
    colback=white,
    listing only,
    title={#1},
    listing options={
        language=dialogue,
        basicstyle=\scriptsize\ttfamily,
        breaklines=true,
        breakatwhitespace,
        breakindent=0pt,
        showstringspaces=false,
        literate={
          {µ}{{$\mu$}}1
        }
    },
    boxrule=0.3mm,
    mybreakable,
    coltitle=white,
}

\tcbset{
  subtitle style={
    fontupper=\small\normalfont
  }
}
\usepackage{rotating}
\usepackage{cuted}

\title{\includegraphics[width=0.7cm]{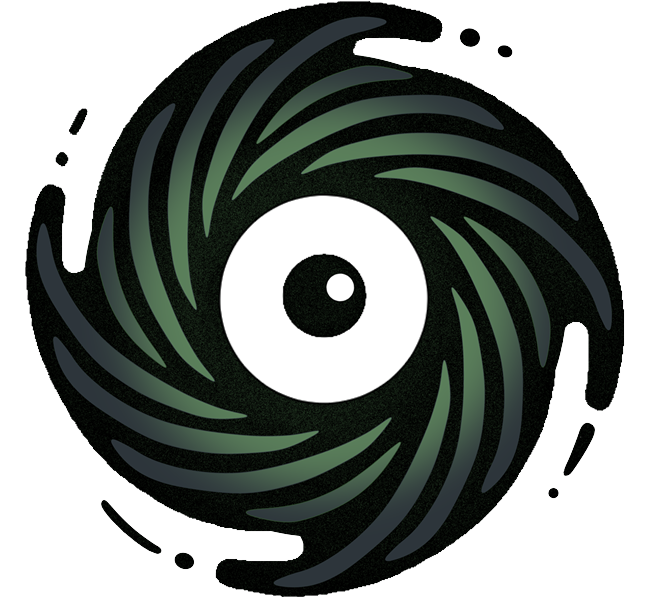} \raisebox{0.3\height}{\textsc{ChaosEater}:}\\ Fully Automating Chaos Engineering with Large Language Models}
\author{Daisuke Kikuta, Hiroki Ikeuchi, Kengo Tajiri \\
        NTT Corporation \\
        \texttt{\href{mailto:daisuke.kikuta@ntt.com}{daisuke.kikuta@ntt.com}}}

\begin{document}
\maketitle
\begin{strip}
    \begin{minipage}{\textwidth}\centering
        \vspace{-3\intextsep}
        \setlength{\tabcolsep}{0.3em}
        \begin{tabular}{rl}
            \raisebox{-0.15\height}{\includegraphics[width=0.4cm]{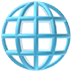}} & \textbf{Project page} : \url{https://ntt-dkiku.github.io/chaos-eater} \\
            \raisebox{-0.15\height}{\includegraphics[width=0.4cm]{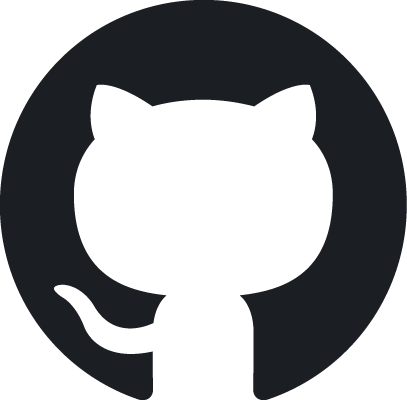}} & \textbf{Code} : \url{https://github.com/ntt-dkiku/chaos-eater} \\
            \raisebox{-0.15\height}{\includegraphics[width=0.4cm]{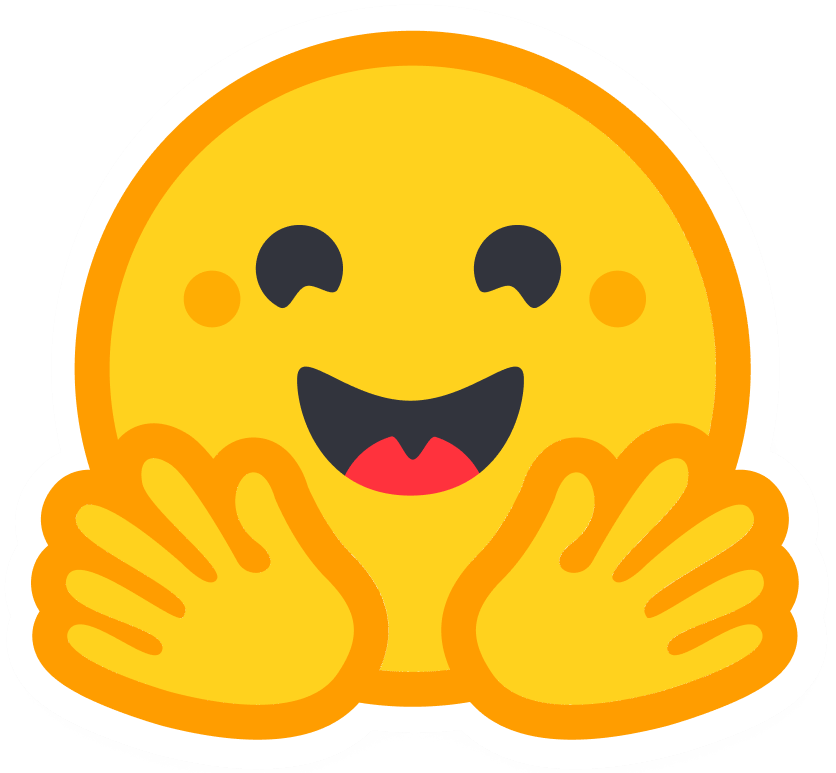}} & \textbf{Demo} : \url{https://huggingface.co/spaces/oookiku/chaos-eater}
        \end{tabular}
    \end{minipage}
\end{strip}

\begin{abstract}
Chaos Engineering (CE) is an engineering technique aimed at improving the resiliency of distributed systems.
It involves artificially injecting specific failures into a distributed system and observing its behavior in response. 
Based on the observation, the system can be proactively improved to handle those failures.
Recent CE tools implement the automated execution of predefined CE experiments.
However, defining these experiments and improving the system based on the experimental results still remain manual.
To reduce the costs of the manual operations, we propose \textsc{ChaosEater}, a system for automating the entire CE operations with Large Language Models (LLMs).
It predefines the agentic workflow according to a systematic CE cycle and assigns subdivided operations within the workflow to LLMs.
\textsc{ChaosEater} targets CE for Kubernetes systems, which are managed through code (i.e., Infrastructure as Code).
Therefore, the LLMs in \textsc{ChaosEater} perform software engineering tasks to complete CE cycles, including requirement definition, code generation, debugging, and testing.
We evaluate \textsc{ChaosEater} through case studies on both small and large Kubernetes systems.
The results demonstrate that it stably completes reasonable single CE cycles with significantly low time and monetary costs. The CE cycles are also qualitatively validated by human engineers and LLMs.
\end{abstract}

\section{Introduction}
Modern software applications are built on distributed systems, where the entire systems are composed of networks of subdivided component services.
This design, known as microservice architecture \citep{microservice}, enables scalable and continuous deployment while supporting the integration of heterogeneous technologies.
On the other hand, the complex dependencies among small services can lead to unexpected and chaotic behavior in the entire system, even from minor failures. 
However, proactively predicting and addressing such complex behavior is challenging.

To address this and improve the resiliency of distributed systems, numerous organizations, including Netflix, Amazon, and Microsoft, have recently adopted Chaos Engineering (CE) \citep{ce,ce2}.
Its concept is that \textit{rather than predicting the chaotic behavior, let's observe it directly by artificially injecting the failures into the system}. 
Based on the actual observation, we can proactively rebuild a new system that is resilient to the assumed failures.
Systematically, CE cycles through four phases for a target system:
\begin{enumerate}
    \item \textbf{\textit{Hypothesis}}: Define steady states (i.e., normal behavior) of the system and a failure scenario. Then, make a hypothesis that \textit{the steady states of the system are maintained even when the failures occur in the scenario}.
    \item \textbf{\textit{(Chaos) Experiment}}: Inject the failures into the system while logging the system's response behavior.
    \item \textbf{\textit{Analysis}}: Analyze the logged data and check if the hypothesis is satisfied. If so, this CE cycle is finished here. If not, move to (4).
    \item \textbf{\textit{Improvement}}: Reconfigure the system to satisfy the hypothesis. The reconfigured system is tested again in (2) and (3), i.e., repeat (2) to (4) until the hypothesis is satisfied.
\end{enumerate}

In recent years, several CE tools \citep{chaosmonkey,aws-fis,chaosmesh,azure-chaos} have advanced the automation of chaos-experiment execution. Moreover, monitoring tools \citep{prometheus,k6} enable automating metric collection, aggregation, and threshold-based testing during chaos experiments. 
Hence, the \textit{experiment} and \textit{analysis} phases have been mostly automated.
However, defining a hypothesis in the \textit{hypothesis} phase, planning a chaos experiment to test the hypothesis in the \textit{experiment} phase, and reconfiguring the system in the \textit{improvement} phase still remain manual.
These manual operations require a complex set of skills, including domain knowledge in networking and CE, the ability to interpret system configurations, logs, and error messages, as well as generative problem-solving for requirement definition, experiment planning, and system reconfiguration.
Consequently, while the costs of these operations remain high, their automation has not been achieved yet with existing algorithmic approaches.

We believe that Large Language Models (LLMs) are the key to overcoming this challenge.
LLMs have recently shown promising capabilities across a wide range of general tasks required for CE, including natural language processing, coding, and network operation \citep{general_llm_survey,llm_code,eval-llm4nw,eval-llm4nw2}.
Moreover, LLMs have demonstrated promising performance in software engineering (SE) tasks \citep{swe-agent,devin}, which is crucial for automating CE. 
Infrastructure as Code (IaC) enables software systems to be managed through code, and recent CE tools can manage chaos experiments in the same way.
Therefore, CE operations on software systems can be regarded as SE tasks, and the promising results in automating SE tasks suggest the feasibility of automating CE.
Given those general capabilities, domain knowledge in networking, and the compatibility between IaC and LLMs for SE, LLMs are a strong candidate for fully automating the CE cycle.

Here, we propose \textsc{ChaosEater}, a system for automating the entire CE cycle with LLMs.
It predefines the agentic workflow according to the systematic CE cycle and assigns subdivided CE operations within the workflow to each LLM with a specific role. 
The predefined workflow ensures that the multiple LLMs collaboratively perform CE operations as intended.
Considering the compatibility between IaC and LLMs for SE, \textsc{ChaosEater} especially targets CE on Kubernetes (K8s) \citep{k8s} systems, which are software systems managed through code.
Therefore, the LLMs perform SE tasks to complete CE cycles, including requirement definition, code generation, debugging, and testing.
In this paper, we present the workflow design, a set of system prompts for creating LLM agents for each CE operation, the interface between LLMs and existing CE tools, and a technique for conducting consistent and transparent testing.

We evaluate \textsc{ChaosEater} through case studies on both small and large K8s systems.
The results demonstrate that it stably completes reasonable single CE cycles with significantly low time and monetary costs (\$0.2--0.8 and 11--25m). The validity of these CE cycles is also confirmed by two human engineers and three different LLMs.

The main contributions of this paper are organized as follows:
\begin{itemize}
    \item We are the first to propose a system for fully automating the systematic CE cycle with LLMs, which significantly reduces time and monetary costs in CE cycles. This proposal would be a starting point towards the full automation of system resiliency improvement.
    \item We make all resources of \textsc{ChaosEater} public. This release provides a general development practice for constructing complex systems that combine LLMs and existing tools.
    \item We evaluate \textsc{ChaosEater} quantitatively in terms of cost and stability, and qualitatively in detail for each phase by both human engineers and LLMs. The results provide its fine-grained potential, limitations, and future directions (see Appendix \ref{adx:discussion}).
\end{itemize}

\section{Proposed System: \textsc{ChaosEater}}
\label{sec:chaoseater}

\begin{figure*}[tb] \centering
    \includegraphics[width=\textwidth]{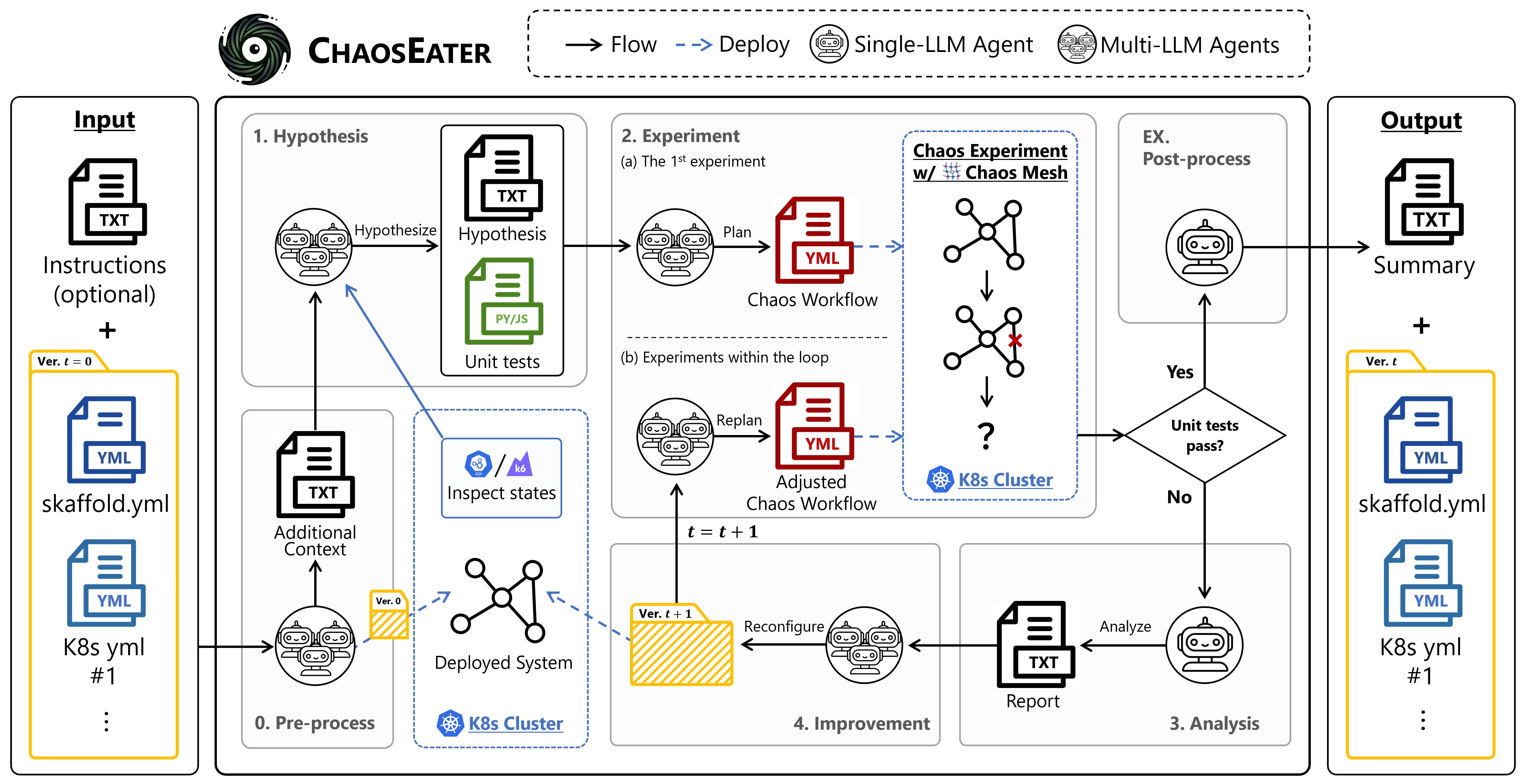}
    \caption{A simplified agentic workflow of \textsc{ChaosEater}. \textsc{ChaosEater} follows the workflow to autonomously complete the systematic CE cycle using LLM agents and existing tools. Note that only the representative inputs and outputs of agents are illustrated here. The two K8s clusters within the workflow refer to the same one.}
    \label{fig:arch}
\end{figure*}

In this section, we describe a technical overview of \textsc{ChaosEater}.
Figure \ref{fig:arch} shows its simplified agentic workflow.
It takes as input instructions for the CE cycle (optional) and a folder containing K8s manifests \citep{k8s} and a Skaffold configuration file \citep{skaffold}. In short, K8s manifests are system configuration files that define the resources (i.e., small services) that constitute a system, while a Skaffold configuration file defines the process to automatically deploy those resources in a K8s cluster. 
It then conducts a CE cycle for those inputs through five divided phases: \textit{pre-processing}, \textit{hypothesis}, \textit{experiment}, \textit{analysis}, \textit{improvement}, and \textit{post-processing} phases.
Finally, it outputs a summary of the completed CE cycle and a modified folder containing K8s manifests that have been reconfigured to satisfy the hypothesis defined in the \textit{hypothesis} phase, along with their corresponding Skaffold configuration file.

In the following, we describe \textsc{ChaosEater}'s workflow design from input to output, breaking it down into the five phases.
Note that, in this paper, we refer to each LLM assigned a specific role (i.e., a subdivided CE operation) as an LLM agent, and that the underlying LLMs do not require additional fine-tuning.
See Appendix \ref{adx:implementation} for implementation details, including the detailed agentic workflow, system prompt templates, graphical user interface, and system deployment.

\subsection{Phase 0: Pre-processing}
Given the user input, \textsc{ChaosEater} first deploys the user's system to the K8s cluster by running the Skaffold configuration file.
Then, each LLM agent sequentially fills in the implicit context of the user's input.
The filled context includes summary of the K8s manifests, their potential issues for resiliency and redundancy, and a possible application, which will be provided as auxiliary information in the subsequent phases.
The filtering of harmful prompts in user instructions is also performed here.

\subsection{Phase 1: Hypothesis}
The \textit{hypothesis} phase defines the system's resiliency for an assumed failure scenario, which corresponds to the requirements definition from a fault tolerance perspective.
Following the principles of CE \citep{ce}, \textsc{ChaosEater} first defines steady states and defines a failure scenario.

\paragraph{Steady-state definition}
Steady states are expected, normal behaviors of a system.
Each steady state is defined by a pair of a state value and a threshold, and a steady state is considered satisfied when the state value meets the threshold.
Therefore, the state values must be measurable outputs of the system, such as the number of active resources, error rates, and response time. 

Given the pre-processed user input, an LLM agent first defines a measurable state critical to maintaining the system's application.
Another agent then inspects the current value of the state using either K8s API or k6 \citep{k6}.
The inspection is conducted by a Python or JavaScript script written by the agent.
Based on the inspected value, an agent defines the threshold for the state, which, according to the definition of a steady state, must be satisfied under the current (normal) state.
Finally, an agent adds threshold-based assertions to the inspection script to generate a unit test script that validates whether the steady state is satisfied.
These processes are repeated to list multiple steady states without duplication until an agent determines that the number of steady states is sufficient.
The unit test script is used for mechanically validating the steady state during the \textit{experiment} phase;
we call this approach of having LLMs judge validity through unit test code \textit{Validation as Code} (VaC), which ensures consistency and transparency in the validation process.

\paragraph{Failure definition}
Given the pre-processed user inputs and the steady states, an LLM agent proposes a failure scenario that may occur in the system (e.g., a surge in access due to a promotional campaign, cyber attack, etc.), and defines a sequence of failures that simulate the scenario.
\textsc{ChaosEater} employs Chaos Mesh \citep{chaosmesh}, a CE tool that can manage chaos experiments on K8s systems through code; therefore, the failures are selected from ones supported by Chaos Mesh.
After drafting failures, an another agent refines the detailed parameters for each Chaos Mesh failure, such as the scope of the failure injection, the failure sub-type, the failure strength.
See Appendix \ref{adx:supported_failures} and \ref{adx:hypo} for supported failures and mechanism of defining failures, respectively.

At this point, the hypothesis can be reframed as \textit{all VaC scripts pass, even when the defined failure injections are performed}.

\subsection{Phase 2: (Chaos) Experiment}
The \textit{experiment} phase plans a chaos experiment to validate the hypothesis and executes it. 

\paragraph{Experiment planning}
To enable systematic planning, we propose dividing a chaos experiment into three stages: pre-validation, failure-injection, and post-validation.
In the pre-validation stage, VaC scripts are run to ensure that the steady states are satisfied under normal conditions.
In the failure-injection stage, failure injections are performed. If some steady states needs to be validated concurrently, the corresponding VaC scripts are also run here.
In the post-validation stage, VaC scripts are run to ensure that steady states have been properly recovered after the failure injections.

Given the pre-processed user inputs and the hypothesis, an LLM agent first determines the duration of each stage. Then, other agents determine the VaC scripts and failure injections to be executed in each stage, along with their execution timing and durations.
Finally, an agent writes a summary of the chaos experiment timeline, which is referenced during the \textit{analysis} phase to identify the causes of the hypothesis not being satisfied in the chaos experiment.
The three-stage chaos-experiment plan is then converted to a Chaos Mesh workflow manifest, which enables automated failure injection and hypothesis validation via VaC scripts according to the schedule defined in the manifest.
See Appendix \ref{adx:chaosmesh_workflow} for our proposed algorithm and LLMs' output format for this conversion.

\paragraph{Experiment replanning}
Resource types and metadata of K8s manifests may be reconfigured during the \textit{improvement} phase.
Therefore, inspection targets in VaC scripts and scopes of failure injections must be updated accordingly between the \textit{improvement} phase and the next experiment execution. 
Given the original and reconfigured K8s manifests, as well as their previous configurations, each LLM agent proposes new inspection targets and new scopes of failure injections.
Then, a new ChaosMesh workflow manifest is generated by updating the corresponding parts in the previous one.
Note that this update only makes minor adjustments to reflect the changes in K8s manifests, without altering the original intent of the chaos experiment.

\paragraph{Experiment execution}
After the Chaos Mesh workflow manifest is generated, \textsc{ChaosEater} applies it to the K8s cluster. 
Then, the scheduled failure injections and hypothesis validation are automatically executed by Chaos Mesh.
In the meantime, \textsc{ChaosEater} simply waits for the experiment to complete.

\subsection{Phase 3: Analysis}
After the chaos experiment is finished, \textsc{Chaos-\allowbreak Eater} mechanically checks whether the VaC scripts have passed.
If all of them have passed, that means the current system configurations (i.e., K8s manifests) already satisfy the hypothesis.
Therefore, \textsc{ChaosEater} finishes the current CE cycle at this point and moves to the \textit{post-processing} phase.
If at least one has failed, \textsc{ChaosEater} moves to the next \textit{improvement} phase after analyzing the experimental results.
In this analysis, given the K8s manifests, the timeline of the chaos experiments, and the list of failed VaC scripts with their logs, an LLM agent identifies the cause of the fails and then generates a report containing the causes and recommended countermeasures. 

\subsection{Phase 4: Improvement}
The \textit{improvement} phase reconfigures the K8s manifests to satisfy the hypothesis.
Given the K8s manifests, the hypothesis, the experiment plan, and the improvement loop history, an LLM agent reconfigures the K8s manifests so that all the VaC scripts pass in the chaos experiment.
There are three reconfiguration modes: \texttt{create}, \texttt{delete}, and \texttt{replace}.
The agent first selects modes while specifying file names, and then writes the reconfigured K8s manifests only when \texttt{create} or \texttt{replace} is specified.
The file management algorithm of \textsc{ChaosEater} then edits the folder from the previous improvement loop (in the first improvement, it corresponds to the user's input folder) according to the agent's output.

\paragraph{Improvement loop}
After the reconfiguration, \textsc{ChaosEater} applies the reconfigured K8s manifests to the K8s cluster.
Then, they will be validated again through the \textit{experiment} and \textit{analysis} phases.
That is, as in the systematic CE cycle, \textsc{Chaos-\allowbreak Eater} also repeats the \textit{experiment}, \textit{analysis}, \textit{improvement} phases until the hypothesis is satisfied.
We define this loop as the improvement loop.
The improvement loop history refers to the history of the experimental results, their analysis reports, and their reconfigurations within this improvement loop, which suppresses the repetition of the same reconfiguration.

\subsection{Extra Phase: Post-processing}
After the CE cycle is completed, \textsc{ChaosEater} finalizes its entire process by summarizing the completed CE cycle.
An LLM agent summarizes the user's input and the four completed phases.
Finally, \textsc{ChaosEater} provides the user with the summary of the completed CE cycle and the folder containing K8s manifests that have been reconfigured to satisfy the hypothesis defined in the \textit{hypothesis} phase, along with their Skaffold configuration file.

\begin{figure*}[!t]
    \centering
    \includegraphics[width=0.996\textwidth]{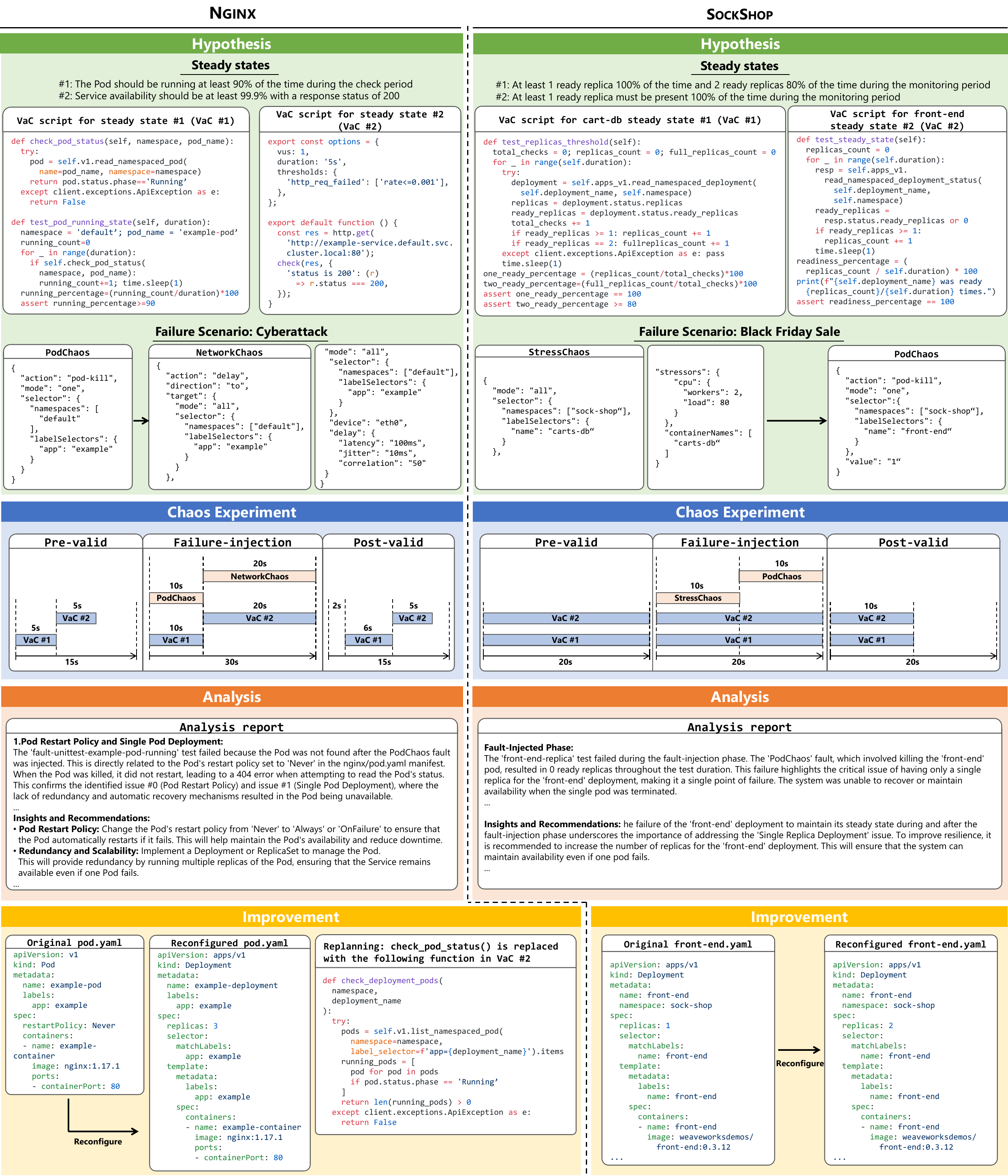}
    \caption{The highlighted outputs for \textsc{Nginx} and \textsc{SockShop}. See Appendix \ref{adx:nginx} and \ref{adx:sockshop} for their full versions.}
    \label{fig:highlight}
\end{figure*}

\section{Case Study}
CE cycles should not be evaluated solely based on whether appropriate reconfigurations are performed; it is equally important to evaluate whether each phase leading up to it is meaningful.
Therefore, rather than creating a benchmark that quantifies evaluation in the binary manner, we here evaluate \textsc{ChaosEater} through in-depth case studies focused on two critical cases.

The first case, \textsc{Nginx}, is a small-scale system consisting of two K8s manifests (i.e., two resources): \texttt{pod.yaml} and \texttt{service.yaml}.
The former defines a \texttt{Pod} resource including a Nginx server, and the latter defines \texttt{Service} resource routing TCP traffic to the \texttt{Pod}.
To verify whether \textsc{ChaosEater} can improve the system when there are resiliency issues, we intentionally configure the resource with a non-resilient setting; we set \texttt{restartPolicy} to \texttt{Never} in \texttt{Pod.yaml}.
With this configuration, once the \texttt{Pod} goes down, it will never restart, resulting in extended service outages.
The second case, \textsc{SockShop} \citep{sockshop}, is a practical and large-scale e-commerce system that consists of 29 manifests, which define the resources and databases for front-end pages, user information, order, payment, shipping, and so on. The number of replicas of all the \texttt{Deployment} resources is originally set to one. However, this setting could lead to downtime of the single replica when it goes down. 
To narrow down this original resiliency issue to a single point, we increase the replicas for \texttt{Deployment} resources other than \texttt{front-end-dep.yaml} to two, while keeping a single replica for \texttt{front-end-dep.yaml}.
This relatively reduces the redundancy/resiliency of the front-end resource.
In this case study, we validate whether \textsc{ChaosEater} correctly identifies and addresses these resiliency issues through a reasonable CE cycle.

To maintain the autonomy of \textsc{ChaosEater}, we input only the instruction to keep each chaos experiment within one minute (access methods are also input for \textsc{SockShop}).
We use gpt-4o-2024-08-06 \citep{openai2024gpt4ocard} as the underlying LLMs.
To improve the reproducibility of this case study, its temperature is set to 0 with the random seed fixed at 42.
We run a single CE cycle for each system five times under the same settings.
In the following, we first discuss quantitative metrics and then qualitatively validate the operations within single CE cycles by LLM-as-a-judge and human engineers.
See Appendix \ref{adx:evaluation} and \ref{adx:case study} for more details on the evaluation and results.

\begin{table*}[tb]
    \caption{Time and monetary costs of single CE cycles conducted by \textsc{ChaosEater}. The values for each phase are averaged across runs that did not skip that phase, while the values for overall are averaged across runs that involved system reconfiguration. API costs are calculated from the official OpenAI API pricing table in September 2024. Abbreviations of each phase name are as follows: `All' is the overall process; `Pre' is \textit{pre-processing}; `Hyp.' is \textit{hypothesis}; `Expt.' is \textit{experiment}; `Anlys.' is \textit{analysis}; `Imp.' is \textit{improvement}; `Post' is \textit{post-processing}.}
    \begin{center}
    \small
    \setlength{\tabcolsep}{4.2px}
    \begin{tabular}{l ccccccc ccccccc}
        \toprule
        &\multicolumn{7}{c}{\textsc{Ngnix}} &\multicolumn{7}{c}{\textsc{SockShop}}\\
        \cmidrule(lr){2-8} \cmidrule(lr){9-15}
        Metric &All &Pre &Hyp. &Expt. &Anlys. &Imp. &Post &All &Pre &Hyp. &Expt. &Anlys. &Imp. &Post\\
        \midrule
        Input tokens &59k &2.6k  &25k  &13k  &4.4k &5.5k &8.2k &284k &30k  &150k  &57k  &14k &15k &18k\\
        Output tokens &5.9k &0.5k  &2.5k  &1.7k  &0.6k &0.2k &0.4k &13k &5.7k  &3.8k  &1.8k  &0.7k &0.6k &0.5k\\
        API cost (\$) &0.21 &0.01  &0.09  &0.05  &0.02 &0.02 &0.02 &0.84 &0.13  &0.41  &0.16  &0.04 &0.04 &0.05\\
        Time &11m &21s  &2.6m  &4.4m  &50s &12s &21s &25m &4.6m  &4.3m  &3.3m  &36s &4.3m &21s\\
        \bottomrule
    \end{tabular}
    \label{tab:eval-costs}
    \end{center}
\end{table*}
\begin{figure*}[tb] \centering
    \includegraphics[width=\linewidth]{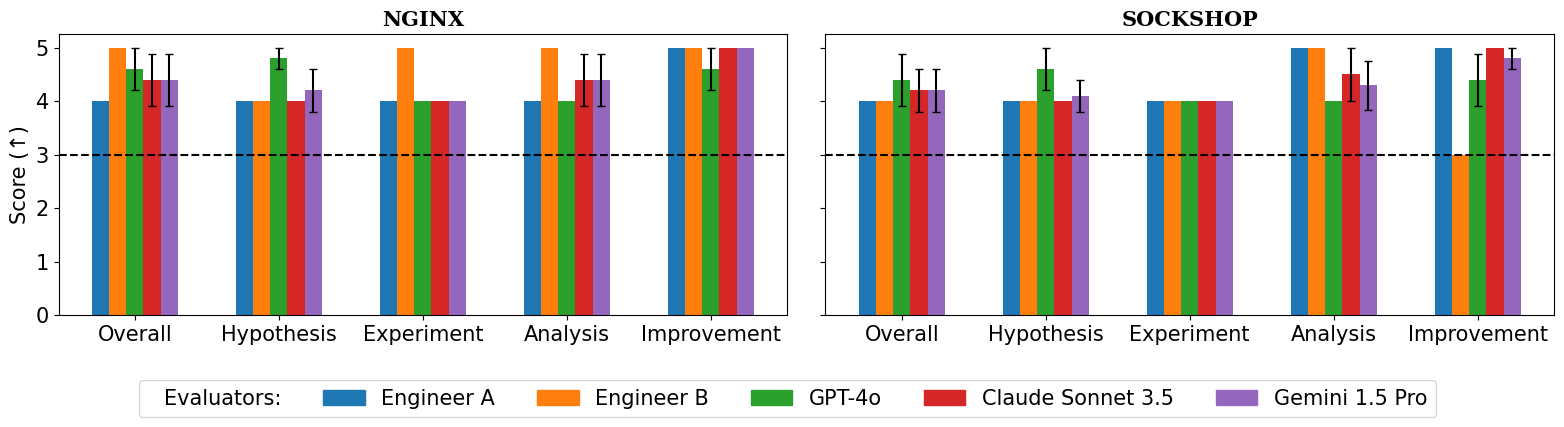}
    \caption{Qualitative evaluation results of CE cycles for each system. A score of 3 or higher is a positive rating.}
    \label{fig:qualitative_eval}
\end{figure*}

\paragraph{Costs and stability}
Table \ref{tab:eval-costs} shows the time and monetary costs of single CE cycles for each target system.
Although we do not have statistical data on the actual working time and labor costs for the same CE cycles performed by human engineers, these total operational time and monetary costs (\$0.21 and 11m) are obviously lower than that.
For \textsc{SockShop}, the monetary cost increases by approximately four times (\$0.84), and the time doubled (25m).
However, these values are still intuitively lower than those of human engineers.
Even with the number of resources increasing by more than ten times compared to \textsc{Nginx}, the cost increase remains minimal, demonstrating that \textsc{ChaosEater} maintains low costs even for large-scale systems.

In terms of stability, \textsc{ChaosEater} successfully completes the CE cycle for each system without runtime errors in all five runs. 
It also correctly reconfigures \textsc{Nginx} in all five runs and \textsc{SockShop} in four out of five runs. Even in the non-reconfigured case, we confirm that a valid CE cycle is completed without requiring reconfigurations. 

\paragraph{Qualitative validation}
To validate CE operations completed by \textsc{ChaosEater}, we select one of the five runs for each target system and qualitatively evaluate the four phases and the overall process using a five-point scale.
Here, the scale is designed so that a score of 3 or higher is a positive rating.
The evaluators are two external human engineers and three LLMs: GPT-4o, Claude Sonnet 3.5, and Gemini Pro 1.5.
The LLM evaluators evaluate each CE cycle five times with a temperature of 0, 
and the final score is calculated as the average of these five evaluations.
See Appendix \ref{adx:evaluation} for details on the evaluation scale and other settings.

Figure \ref{fig:highlight} shows the highlighted outputs from the evaluated cycles for \textsc{Nginx} and \textsc{SockShop}.
For both cases, \textsc{ChaosEater} defined pod availability as one of the steady states, and then hypothesized that they are maintained even in scenarios where \texttt{Pods} go down, such as cyberattacks or Black Friday sales. 
Through these experiments, it successfully identified the issues of \texttt{restartPolicy} and the number of replicas, and solved them by replacing \texttt{Pod} with a \texttt{Deployment} resource and increasing the number of replicas, respectively.
The authors have qualitatively confirmed that these operations are appropriate for addressing the issues intentionally introduced into the original systems.

Figure \ref{fig:qualitative_eval} shows the results of the qualitative evaluation of the two cycles conducted by the evaluators.
The results show that all evaluators rated every phase above the threshold for a positive rating for both systems, demonstrating that \textsc{ChaosEater} completed reasonable single CE cycles.
In their reviews, they pointed out the need for additional CE cycles to achieve broader resiliency improvement and suggested the inclusion of more complex failure sequences, such as cases where a single failure indirectly affects multiple resources through complex dependencies.
However, all of them agreed that, as a single CE cycle addressing the given issues, the operations are effective and appropriate.
In conclusion, the validity of the CE cycles is confirmed by the authors, the human engineers, and LLMs.
See Appendix \ref{adx:case study} for more detailed descriptions of the \textsc{ChaosEater}'s outputs and the full text of the reviews.

\section{Conclusion}
In this paper, we proposed \textsc{ChaosEater}, an LLM-based system for fully automating single CE cycles.
We presented its technical details and validated it through case studies on small and large K8s systems.
The results demonstrated that \textsc{ChaosEater} successfully completes reasonable single CE cycles with significantly low time and monetary costs. 

On the other hand, the current version has several limitations, such as restriction to development environments, support only for K8s manifest reconfiguration, and limited capability in vulnerability discovery (see Appendix \ref{adx:discussion} for comprehensive discussion on limitations and future directions).
As the automatic generation of software applications by LLMs has become widespread in recent years, the automation of improving their infrastructure resiliency is becoming increasingly important.
To support this, we will improve \textsc{ChaosEater} by addressing its current limitations.



\section*{Acknowledgements}
We would like to thank NTT TechnoCross members, especially Kensuke Yagura, Yoshitaka Sadohara, and Tamotsu Ohyama, for trying out \textsc{ChaosEater} and sharing their valuable feedback.
We also thank the anonymous reviewers for their constructive feedback and valuable suggestions.
Finally, the \textsc{ChaosEater} would not have been possible without LLM providers and relevant OSS projects.
We deeply appreciate their continuous efforts and dedication.

\bibliography{custom}

\appendix
\label{sec:appendix}
\section{Related Work}
\paragraph{Chaos engineering}
Since \citet{ce} introduced its name, CE has gained attention and is currently employed in various services \citep{ce2,ce_study}.
The major research question of CE is how we can efficiently test meaningful failures to identify system vulnerability.
To address this, various optimization methods for failure selection have been proposed \citep{failure_opt,security_ce,analysis_ce,failure_opt2,failure_opt3,failure_opt4}.
In application, several automation tools have been developed in both the open-source community \citep{chaosmesh, chaosmonkey} and the commercial sector \citep{aws-fis,azure-chaos}. 
While these technologies have advanced CE automation, its full automation has not yet been achieved due to the complexity of generative tasks, such as hypothesis formulation and system reconfiguration.
Our work is the first to fully automate CE using LLMs.

\paragraph{LLMs for software engineering}
LLMs for coding have been actively explored from various aspects: Pretraining models \citep{chen2021evaluating,codellama,deepseekcoder}, prompt engineering \citep{chen2022codet,jiang2023selfevolve}, and evaluation \citep{austin2021program,chen2021evaluating}.
For more general SE tasks, LLMs that solve issues in GitHub repositories have also emerged \citep{swe-agent,devin,agentless}.
Since CE for software systems are regarded as SE, our work can be considered a part of this trend.
Unlike existing SE benchmarks, CE requires autonomously defining and achieving its own goals.
Our work demonstrates the potential of LLMs to tackle such complex SE tasks that are new to them.

\paragraph{LLMs for networking (NW)}
LLMs for networking have also been explored from various aspects in recent years: 
Datasets \citep{finetuned-nw,dataset-5g}, 
benchmarks \citep{benchmark-nw,benchmark-nw2}, 
fine-tuned models \citep{finetuned-nw}, 
an agent framework for NW-related tasks \citep{nw-agent}, 
and comprehensive evaluation \citep{eval-llm4nw,eval-llm4nw2}.
These works empirically demonstrate the promise of applying LLMs to the NW domain.
In parallel with the research side, various NW applications have also been developed, especially for software systems.
They range from LLM-based IaC code generation \citep{aiac,kubectl-ai,pulumi-ai} to diagnostic tools \citep{k8sgpt,k8s-chatgpt-bot} and misconfiguration remediation \citep{genkubesec}.
Despite the advancements of LLMs in the NW domain, their application to CE remains unexplored.
Our work is the first to demonstrate the capabilities of LLMs in CE, which involves complex NW operations.

\paragraph{Concurrent works}
In parallel with our work, some other projects have shown promising results in applying LLMs to CE.
From a security perspective, \citet{llm4sce} leverage LLMs to construct attack-defense trees, which assist security analysts in designing security chaos experiments. 
The major differences from ours lie in their use of LLMs as a supplementary support tool and designing the LLM workflow specialized to the \textit{hypothesis} phase based on the methodology of security CE.
AIOpsLab \citep{aiopslab} is an evaluation framework for LLM agents that automates the operations of cloud (i.e., software) systems.
It includes failure injection as a feature and allows us to evaluate the capabilities of LLM agents in CE. The supported failures include both \textit{functional failures}, such as system misconfigurations, and independent \textit{symptomatic failures}, such as server downtime and network delays.
However, while it evaluates whether LLM agents can provide appropriate cause analyses and solutions for functional failures, the evaluation for symptomatic failures is limited to identifying the presence of failure and pinpointing their locations.
On the other hand, \textsc{ChaosEater} focuses on the symptomatic failures and is capable of performing more detailed and comprehensive CE by following the systematic CE cycle.
For example, it autonomously makes a appropriate hypothesis for a given system, designs complex failure scheduling to simulate symptomatic failure scenarios, analyzes system behavior during symptomatic failure injection, and enables reconfigurations to address unexpected behavior due to the failures.
While these concurrent works, including ours, share a common goal, each focuses on a different aspect of CE.

\section{Discussion}
\label{adx:discussion}
\paragraph{Broader impacts}
Numerous systems, including the increasing number of LLM applications in recent years, are built in the microservice architecture, and their number is expected to continue to grow in the future.
By fully automating CE, it will be possible for anyone to easily build resilient systems.
Moreover, it is also expected to combine \textsc{ChaosEater} with other LLM systems for creating software applications, such as improving the resiliency of applications created by other LLM systems through \textsc{ChaosEater}.
Although \textsc{ChaosEater} is not yet at a practical level, we believe that \textsc{ChaosEater} would be a good starting point toward such use cases.
Even at its current level, \textsc{ChaosEater} can be sufficiently used as training materials (including both good and bad practices) for the Chaos Game Day, which is a training exercise for CE engineers.

\paragraph{Limitations}
\textsc{ChaosEater} currently has the following limitations: 
\begin{enumerate}

\item \textbf{Limited deployment environment}; Although CE should ideally be conducted in actual production environments, \textsc{ChaosEater} is currently only supported in development environments.

\item \textbf{Limited to GPT-4o}; \textsc{ChaosEater}'s prompt templates are highly tuned only for GPT-4o. Therefore, other LLMs can not currently be used for \textsc{ChaosEater}.

\item \textbf{Limited to K8s manifest reconfiguration}; Software systems consist not only of K8s manifests but also of other types of codebases, such as HTML/CSS/JS and Python. Although K8s manifest reconfiguration can handle a majority of system resiliency issues, reconfiguration of all types of codebases is necessary to optimally improve system resiliency. However, \textsc{ChaosEater} currently supports reconfiguring only K8s manifests.

\item \textbf{Vulnerability discovery}; In the case study, \textsc{ChaosEater} improved systems with relatively simple resiliency issues.
However, for systems that already possess a certain level of resiliency, \textsc{ChaosEater} fails to find new hidden issues through a CE cycle.
Given that this is a challenging task even for human engineers, \textsc{ChaosEater} is currently considered to perform at a level comparable to, or lower than, that of engineers.
To find such issues, it is necessary to conduct multiple CE cycles over extended operational periods.

\end{enumerate}

\paragraph{Future directions}
Given the current limitations above, we share several future directions for \textsc{ChaosEater} and the full automation of system resiliency improvement:
\begin{enumerate}
    \item \textbf{Production deployment and security}; If \textsc{ChaosEater} is deployed in production environments, further research on security will be necessary. This includes controlling more carefully the impact range of failures (i.e., blast radius), preventing \textsc{ChaosEater} from becoming a proxy for attacking production services, and proposing emergency response measures, such as a higher-level monitoring system that always monitors \textsc{ChaosEater}.
    
    \item \textbf{Support for various LLMs}; As \textsc{ChaosEater}'s prompt templates are tuned manually, supporting various LLMs significantly increases their management costs. To address this, automatic prompt tuning is considered an effective solution. Our current prompt templates may be used as the seed prompts.
    
    \item \textbf{Fine-tuning LLMs specifically for CE}; Fine-tuning is necessary to improve the quality of CE cycles and expand supported LLM types. \textsc{ChaosEater}'s outputs may be used as the instruction-tuning data.
    
    \item \textbf{Evaluation frameworks}; As there are currently no datasets and benchmarks for the systematic CE cycle, we will construct them to enable more solid validation of \textsc{ChaosEater}. Besides, we plan to propose new metrics for quantitatively evaluating CE cycles conducted by \textsc{ChaosEater}. This is not easy because, even in cases where no improvements are made, CE cycles can still provide valuable insights. Therefore, the quality of CE cycles should not be judged solely based on whether improvements were made; metrics that consider its philosophical aspects are also necessary.
    
    \item \textbf{Toward larger and more complex systems}; We need to incorporate the recent advances in the combination of LLMs and graphs to extract necessary sub-graphs from large system graphs. This sub-graph extraction is important to organize the agent's inputs in each phase.

    \item \textbf{Full automation of long-term multiple CE cycles}; By using the \textsc{ChaosEater}'s output as input for the next CE cycle, we can automate multiple CE cycles even with the current \textsc{ChaosEater}. However, we additionally need to develop techniques to manage the long-term history of completed CE cycles and continuous learning (if LLMs are fine-tuned).
    
\end{enumerate}

\section{Evaluation Details}
\label{adx:evaluation}

\subsection{Parameters of LLM agents}
The parameters of LLM agents are as follows: the temperature of LLMs to 0 with a random seed of 42;
the maximum number of proposed steady states is set to 2;
and the maximum number of iterations for both the verification improvement loops is set to 3.
All other parameters of LLMs are set to their default values.

\subsection{Quantitative Evaluation}
\paragraph{Tokens and API costs}
We first count the input and output tokens using the tokenizer of GPT-4o (\texttt{o200k\_base}). 
We then calculate the monetary costs based on the official OpenAI API pricing table in September 2024: \$2.50 per 1 million input tokens and \$10.00 per 1 million output tokens.

\paragraph{Running time}
We measure the running time using the \texttt{time} module in Python.
We implement each phase as a separate Python class and measure the running time as the duration from the start to the termination of each class call.
The specifications of the device used for the evaluation are as follows: AMD Epyc Milian x 2 (CPU) and MEM-DR432LC-ER32 DDR4-3200 x 32 (Memory). 

\subsection{Qualitative Evaluation}
\begin{table*}[!h]
    \caption{Evaluation scale for each phase. The criteria are designed so that a score of 3 or higher is positive.}
    \begin{center}
    \renewcommand{\multirowsetup}{\centering}
    \renewcommand{\arraystretch}{1.5}
    \small
    \setlength{\tabcolsep}{4.2px}
    \begin{tabular}{c|c|p{11 cm}}
        \toprule
        Phase &Score &Criteria \\
        \midrule
        \multirow{7}{*}{Overall} &\multirow{2}{*}{5} &The cycle fixes critical issues in the system and offers meaningful insights for the next cycle according to the experiments conducted.\\
        \cline{2-3}
        &4 &The cycle fixes critical issues in the system. \\
        \cline{2-3}
        &\multirow{2}{*}{3} &The cycle fixes minor issues in the system or offers meaningful insights for the next cycle according to the experiments conducted. \\
        \cline{2-3}
        &\multirow{2}{*}{2} &The cycle neither changes the system nor offers meaningful insights for the next cycle according to the experiments conducted. \\
        \cline{2-3}
        &1 &The cycle worsens the system's resiliency or adds meaningless resiliency.\\
        \midrule

        \multirow{8}{*}{Hypothesis} &\multirow{2}{*}{5} &The hypothesis is relevant to the system and meaningful. Additioanlly, the hypothesis leads to system improvement and offers meaningful insights for the next cycle.\\
        \cline{2-3}
        &\multirow{2}{*}{4} &The hypothesis is relevant to the system and meaningful. Additionally, the hypothesis leads to system improvement or offers meaningful insights for the next cycle. \\
        \cline{2-3}
        &\multirow{2}{*}{3} &The hypothesis is relevant to the system and meaningful. However, the hypothesis neither leads to system improvement nor offers meaningful insights for the next cycle. \\
        \cline{2-3}
        &\multirow{2}{*}{2} &The hypothesis is relevant to the system, but is trivial and meaningless (e.g., hypothesis that is to be obviously satisfied). \\
        \cline{2-3}
        &1 &The hypothesis is irrelevant to the system.\\
        \midrule
        
        \multirow{7}{*}{Experiment} &\multirow{2}{*}{5} &The experiment plan correctly serves to validate the hypothesis. Additionally, it is set up considering an complex, actual failure scenario.\\
        \cline{2-3}
        &\multirow{2}{*}{4} &The experiment plan correctly serves to validate the hypothesis. Additionally, it is set up considering an actual failure scenario. \\
        \cline{2-3}
        &3 &The experiment plan correctly serves to validate the hypothesis. \\
        \cline{2-3}
        &\multirow{2}{*}{2} &The experiment plan mostly serves to validate the hypothesis. However, there are some missed components. \\
        \cline{2-3}
        &1 &The experiment plan does serve to validate the hypothesis at all. \\
        \midrule
        
        \multirow{7}{*}{Analysis} &\multirow{2}{*}{5} &The analysis reports correct and meaningful information. Additioanlly, it provides some meaningful insights for the improvement. \\
        \cline{2-3}
        &\multirow{2}{*}{4} &The analysis reports correct and meaningful information. Additioanlly, it provides some insights for the improvement. \\
        \cline{2-3}
        &3 &The analysis reports correct and meaningful information. \\
        \cline{2-3}
        &2 &The analysis reports meaningless information. \\
        \cline{2-3}
        &1 &The analysis reports information that is not factual. \\
        \midrule
        
        \multirow{7}{*}{Improvement} &\multirow{2}{*}{5} &The improvement succesully changes the system to satisfy the hypothesis in the first attempt. \\
        \cline{2-3}
        &\multirow{2}{*}{4} &The improvement succesully changes the system to satisfy the hypothesis over two or more iterations. \\
        \cline{2-3}
        &3 &The improvement does not change the system. \\
        \cline{2-3}
        &2 &The improvement exceeded the number of attempts and stopped midway. \\
        \cline{2-3}
        &1 &The improvement worsens the system's resiliency or adds meaningless resiliency.\\
        \bottomrule
    \end{tabular}    
    \label{tab:eval_scale}
    \end{center}
\end{table*}
\paragraph{Evaluation scale}
Table \ref{tab:eval_scale} shows the evaluation scale. 
The criteria are designed so that a score of 3 or above is considered a positive rating.
For each phase, a score of 3 corresponds to meeting the minimum required criteria, while scores of 4 and above are incrementally awarded for additional elements, such as contributions to the next CE cycle, quality or efficiency.
On the other hand, scores of 1 or 2 correspond to operations that are either meaningless or that lead to undesirable modifications to the system.
To determine these concrete criteria, we referred to the essential elements of CE mentioned in \cite{ce}.

\paragraph{Evaluators}
We employ two human engineers and three state-of-the-art LLMs.
The human engineers are recruited from external IT companies across different countries. 
The LLM evaluators include gpt-4o-2024-0908-06 (GPT-4o), claude-3-5-sonnet-20240620 (Claude Sonnet 3.5), and gemini-1.5-pro (Gemini 1.5 pro).
Following \cite{surveyllmasajudge}, the temperature is set to 0 for all LLMs to improve the consistency and reproducibility of the evaluations. 
As mentioned earlier, all other parameters remain at their default values.

\paragraph{Evaluation instruction}
We give a system prompt for this qualitative evaluation to LLMs, where they are instructed to output the score and its corresponding reason for each phase in JSON format.
In this format, the reason comes before the score.
Prompt \ref{lst:review_prompt} and Example \ref{lst:ce_cycle_overview} show the system prompt and an example of text embedded in the placeholder \texttt{\textcolor{blue}{\{ce\_cycle\_overview\}}}, respectively.
This embedded overview is generated by embedding the outputs of \textsc{ChaosEater} into a static template.
We give almost the same instruction to the human engineers.
Additionally, before the evaluation, we obtained their consent to use the evaluation results for the purpose of evaluating \textsc{ChaosEater} and for inclusion in this paper.
They were also cautioned to avoid any evaluation bias that might arise from disclosing the purpose in advance.
\newtcblisting[auto counter]{reviewprompt}[2][]{
    colback=gray!5,
    listing only,
    title={Prompt \thetcbcounter: #1},
    listing options={
        language=sysprompt,
        basicstyle=\scriptsize\ttfamily,
        breaklines=true,
        breakindent=0pt,
        showstringspaces=false,
        literate={µ}{{$\mu$}}1,
        escapechar=|,
    },
    boxrule=0.3mm,
    coltitle=black,
    colbacktitle=gray!20,
    mybreakable,
    label=#2
}

\begin{reviewprompt}[System prompt for reviewing each phase of completed CE cycles]{lst:review_prompt}
//System://
You are a professional reviewer for Chaos Engineering.
Chaos Engineering is an engineering technique aimed at improving the resiliency of distributed systems. It involves artificially injecting specific failures into a distributed system and observing its behavior in response. Based on the observation, the system can be proactively improved to handle those failures.
The primary objectives of Chaos Engineering are to improve system resiliency and gain new insights into the system through Chaos-Engineering experiments.
Systematically, Chaos Engineering cycles through four phases: hypothesis, experiment, analysis, and improvement phases.
  1) Hypothesis: Define steady states (i.e., normal behavior) of the system and injected failures (i.e., faults). Then, make a hypothesis that "the steady states are maintained in the system even when the failures are injected".
  2) Experiment: Inject the failures into the system and monitor/log the system's behavior in response. 
  3) Analysis: Analyze the logged data and check if the hypothesis is satisfied. If so, one CE cycle is finished here. If not, move to (4)
  4) Improvement: Reconfigure the system to satisfy the hypothesis. The reconfigured system is tested again in (2) and (3), i.e., repeat (2) to (4) until the hypothesis is satisfied.
Given a Chaos Engineering cycle, you will carefully review it according to the following rules:
- The review must be specific, constructive, and insightful.
- The output should be formatted as a JSON instance that conforms to the JSON schema below.

As an example, for the schema {"properties": {"foo": {"title": "Foo", "description": "a list of strings", "type": "array", "items": {"type": "string"}}}, "required": ["foo"]}
the object {"foo": ["bar", "baz"]} is a well-formatted instance of the schema. The object {"properties": {"foo": ["bar", "baz"]}} is not well-formatted.

Here is the output schema:
```
{
  "properties": {
    "hypothesis": {
      "title": "Hypothesis",
      "description": "The review of the hypothesis phase.",
      "allOf": [{"$ref": "#/definitions/HypothesisReview"}]
    }, 
    "experiment": {
      "title": "Experiment",
      "description": "The review of the experiment (planning) phase.",
      "allOf": [{"$ref": "#/definitions/ExperimentReview"}]
    },
    "analysis": {
      "title": "Analysis",
      "description": "The review of the analysis phase.",
      "allOf": [{"$ref": "#/definitions/AnalysisReview"}]
    },
    "improvement": {
      "title": "Improvement",
      "description": "The review of the improvement phase.",
      "allOf": [{"$ref": "#/definitions/ImprovementReview"}]
    }, 
    "overall": {
      "title": "Overall",
      "description": "The review of the entire Chaos Engineering cycle.",
      "allOf": [{"$ref": "#/definitions/OverallReview"}]
    }
  },
  "required": ["hypothesis", "experiment", "analysis", "improvement", "overall"],
  "definitions": {
    "HypothesisReview": {
      "title": "HypothesisReview",
      "type": "object",
      "properties": {
        "summary": {
          "title": "Summary",
          "description": "Summarize the overall of the hypothesis.",
          "type": "string"
        }, 
        "strengths": {
          "title": "Strengths",
          "description": "List the strengths of the hypothesis. Write them in bullet points (3 - 5 items).",
          "type": "string"
        },
        "weaknesses": {
          "title": "Weaknesses",
          "description": "List the weaknesses of the hypothesis. Write them in bullet points (3 - 5 items).",
          "type": "string"
        }, 
        "score_reason": {
          "title": "Score Reason",
          "description": "Before write the score, please describe why you choose the score according to the rating scale table.",
          "type": "string"
        },
        "score": {
          "title": "Score",
          "description": "Here is the rating scale (descending order). You must choose the score from [1, 5].\nThe higher the score, the better the hypothesis. The scores 1, 2 are negative, The score 3 is accetable, the score 4 is positive, the score 5 is very positive.\nScore: criteria\n5: The hypothesis is relevant to the system and meaningful. Additioanlly, the hypothesis leads to system improvement and offers meaningful insights for the next cycle. \n4: The hypothesis is relevant to the system and meaningful. Additionally, the hypothesis leads to system improvement or offers meaningful insights for the next cycle.\n3: The hypothesis is relevant to the system and meaningful. However, the hypothesis neither leads to system improvement nor offers meaningful insights for the next cycle.\n2: The hypothesis is relevant to the system, but is trivial and meaningless (e.g., hypothesis that is to be obviously satisfied).\n1: The hypothesis is irrelevant to the system.\n",
          "enum": [1, 2, 3, 4, 5],
          "type": "integer"
        }
      },
      "required": ["summary", "strengths", "weaknesses", "score_reason", "score"]
    },
    "ExperimentReview": {
      "title": "ExperimentReview",
      "type": "object",
      "properties": {
        "summary": {
          "title": "Summary",
          "description": "Summarize the overall of the experiment plan.",
          "type": "string"
        },
        "strengths": {
          "title": "Strengths",
          "description": "List the strengths of the experiment plan. Write them in bullet points (3 - 5 items).",
          "type": "string"
        },
        "weaknesses": {
          "title": "Weaknesses",
          "description": "List the weaknesses of the experiment plan. Write them in bullet points (3 - 5 items).",
          "type": "string"
        },
        "score_reason": {
          "title": "Score Reason",
          "description": "Before write the score, please describe why you choose the score according to the rating scale table.",
          "type": "string"
        },
        "score": {
          "title": "Score",
          "description": "Here is the rating scale (descending order). You must choose the score from [1, 5].\nThe higher the score, the better the experiment plan.\n5: The experiment plan correctly serves to validate the hypothesis. Additionally, it is set up considering an complex, actual failure scenario.\n4: The experiment plan correctly serves to validate the hypothesis. Additionally, it is set up considering an actual failure scenario.\n3: The experiment plan correctly serves to validate the hypothesis.\n2: The experiment plan mostly serves to validate the hypothesis. However, there are some missed components.\n1: The experiment plan does serve to validate the hypothesis at all.\n",
          "enum": [1, 2, 3, 4, 5],
          "type": "integer"
        }
      },
      "required": ["summary", "strengths", "weaknesses", "score_reason", "score"]
    },
    "AnalysisReview": {
      "title": "AnalysisReview",
      "type": "object",
      "properties": {
        "summary": {
          "title": "Summary",
          "description": "Summarize the overall of the analysis.",
          "type": "string"
        },
        "strengths": {
          "title": "Strengths",
          "description": "List the strengths of the analysis. Write them in bullet points (3 - 5 items).",
          "type": "string"
        },
        "weaknesses": {
          "title": "Weaknesses",
          "description": "List the weaknesses of the analysis. Write them in bullet points (3 - 5 items).",
          "type": "string"
        },
        "score_reason": {
          "title": "Score Reason",
          "description": "Before write the score, please describe why you choose the score according to the rating scale table.",
          "type": "string"
        },
        "score": {
          "title": "Score",
          "description": "Here is the rating scale (descending order). You must choose the score from [1, 5].\nThe higher the score, the better the analysis.\n5: The analysis reports correct and meaningful information. Additioanlly, it provides some meaningful insights for the improvement.\n4: The analysis reports correct and meaningful information. Additioanlly, it provides some insights for the improvement.\n3: The analysis reports correct and meaningful information.\n2: The analysis reports meaningless information.\n1: The analysis reports information that is not factual.\n",
          "enum": [1, 2, 3, 4, 5],
          "type": "integer"
        }
      },
      "required": ["summary", "strengths", "weaknesses", "score_reason", "score"]
    }, 
    "ImprovementReview": {
      "title": "ImprovementReview",
      "type": "object",
      "properties": {
        "summary": {
          "title": "Summary",
          "description": "Summarize the overall of the improvement.",
          "type": "string"
        },
        "strengths": {
          "title": "Strengths",
          "description": "List the strengths of the improvement. Write them in bullet points (3 - 5 items).",
          "type": "string"
        },
        "weaknesses": {
          "title": "Weaknesses",
          "description": "List the weaknesses of the improvement. Write them in bullet points (3 - 5 items).",
          "type": "string"
        },
        "score_reason": {
          "title": "Score Reason",
          "description": "Before write the score, please describe why you choose the score according to the rating scale table.",
          "type": "string"
        },
        "score": {
          "title": "Score",
          "description": "Here is the rating scale (descending order). You must choose the score from [1, 5].\nThe higher the score, the better the improvement. \n5: The improvement succesully changes the system to satisfy the hypothesis in the first attempt.\n4: The improvement succesully changes the system to satisfy the hypothesis over two or more iterations.\n3: The improvement does not change the system.\n2: The improvement exceeded the number of attempts and stopped midway.\n1: The improvement worsens the system's resiliency or adds meaningless resiliency.\n",
          "enum": [1, 2, 3, 4, 5],
          "type": "integer"
        }
      }, 
      "required": ["summary", "strengths", "weaknesses", "score_reason", "score"]
    }, 
    "OverallReview": {
      "title": "OverallReview",
      "type": "object",
      "properties": {
        "summary": {
          "title": "Summary",
          "description": "Summarize the overall of the Chaos Engineering cycle.",
          "type": "string"
        },
        "strengths": {
          "title": "Strengths",
          "description": "List the strengths of the Chaos Engineering cycle. Write them in bullet points (3 - 5 items).",
          "type": "string"
        },
        "weaknesses": {
          "title": "Weaknesses",
          "description": "List the weaknesses of the Chaos Engineering cycle. Write them in bullet points (3 - 5 items).",
          "type": "string"
        },
        "score_reason": {
          "title": "Score Reason",
          "description": "Before write the score, please describe why you choose the score according to the rating scale table.",
          "type": "string"
        },
        "score": {
          "title": "Score",
          "description": "Here is the rating scale (descending order). You must choose the score from [1, 5]. \nThe higher the score, the better the CE cycle. The scores 1, 2 are negative, The score 3 is accetable, the score 4 is positive, the score 5 is very positive.\nScore: criteria\n5: The cycle fixes critical issues in the system and offers meaningful insights for the next cycle according to the experiments conducted\n4: The cycle fixes critical issues in the system\n3: The cycle fixes minor issues in the system or offers meaningful insights for the next cycle according to the experiments conducted\n2: The cycle neither changes the system nor offers meaningful insights for the next cycle according to the experiments conducted\n1: The cycle worsens the system's resiliency or adds meaningless resiliency.\n",
          "enum": [1, 2, 3, 4, 5],
          "type": "integer"
        }
      },
      "required": ["summary", "strengths", "weaknesses", "score_reason", "score"]
    }
  }
}
```

//User://
Here is the overview of a Chaos Engineering cycle to be reviewed:
//@{ce_cycle_overview}@//

Please review the above Chaos Engineering cycle.
\end{reviewprompt}

\begin{exampledata}[\texttt{ce\_cycle\_overview}]
# Here is a Chaos Engineering cycle |\label{lst:ce_cycle_overview}|
## Step 0. User-input understanding
### Here is the overview of user inputs:
The system consists of the following K8s manifest(s):K8s manifest: nginx/pod.yaml
```yaml
apiVersion: v1
kind: Pod
metadata:
  name: example-pod
  labels:
    app: example
spec:
  restartPolicy: Never
  containers:
  - name: example-container
    image: nginx:1.17.1
    ports:
    - containerPort: 80
```
Summary of nginx/pod.yaml:
- This manifest defines a Kubernetes Pod.
- The Pod is named 'example-pod'.
- It includes metadata with a label 'app: example'.
- The Pod's restart policy is set to 'Never', meaning it won't restart automatically if it fails.
- The Pod contains one container named 'example-container'.
- The container uses the 'nginx:1.17.1' image, which is a specific version of the Nginx web server.
- The container exposes port 80, which is commonly used for HTTP traffic.

K8s manifest: nginx/service.yaml
```yaml
apiVersion: v1
kind: Service
metadata:
  name: example-service
spec:
  selector:
    app: example
  ports:
    - protocol: TCP
      port: 80
      targetPort: 80
```
Summary of nginx/service.yaml:
- This manifest defines a Kubernetes Service.
- The Service is named 'example-service'.
- It uses the 'v1' API version.
- The Service selects pods with the label 'app: example'.
- It exposes the Service on port 80 using the TCP protocol.
- The Service forwards traffic to the target port 80 on the selected pods.

The resiliency issues/weaknesses in the system are as follows:
Issue #0: Pod Restart Policy
  - details: The Pod will not restart automatically if it fails, which can lead to downtime.
  - manifests having the issues: ['nginx/pod.yaml']
  - problematic config: restartPolicy: Never

Issue #1: Single Pod Deployment
  - details: Having a single Pod means there is no redundancy. If the Pod fails, the service will be unavailable.
  - manifests having the issues: ['nginx/pod.yaml']
  - problematic config: Only one Pod is defined without a Deployment or ReplicaSet.

The expected type of application on the system (i.e., K8s manfests):
A simple web server application using Nginx to serve HTTP content.; The manifests provided define a Kubernetes Pod and a Service. The Pod runs an Nginx container, which is a popular web server used to serve static content, reverse proxy, or load balance HTTP traffic. The Service is configured to expose this Pod on port 80, which is the default port for HTTP traffic. Given the use of Nginx and the configuration of the Service, it is logical to assume that these manifests are intended to deploy a simple web server application. The file names and the use of Nginx further support this assumption.

Chaos-Engineering instructions for the system are as follows: - The Chaos-Engineering experiment must be completed within 1 minute.

## Step 1. Hypothesis definition
### Here is the overview of the hypothesis for the system:
The hypothesis is "The steady states of the sytem are maintained even when the fault scenario occurs (i.e., when the faults are injected)".
The steady states here are as follows:
2 steady states are defined.
1st steady state:
- Name: example-pod-running
- Description: The first issue to address is the Pod's restart policy set to 'Never'. This is a critical issue because if the Pod fails, it will not restart, leading to potential downtime. Therefore, the steady state should ensure that the Pod is running and available. A measurable output for this steady state is the number of running Pods, which should be 1, as there is only one Pod defined in the manifest. This steady state will help verify that the Pod is up and running, which is crucial given the restart policy configuration.
- Threshold for the steady state: The pod should be running at least 90% of the time during the check period.; The steady state we are considering is whether the 'example-pod' is running. The current state shows that the pod was checked 5 times over a duration of 5 seconds, and it was running each time, resulting in a running count of 5. This indicates that the pod is consistently running during the check period. Given the constraints of the chaos engineering experiment, which must be completed within 1 minute, we can set a threshold that allows for some tolerance in case of brief fluctuations. A reasonable threshold would be that the pod should be running at least 90% of the time during the check period. This allows for a small margin of error while still ensuring that the pod is generally available and running.
- Whether the steady state meets the threshold is determined by the following Python script with K8s API:
```
import os
import time
import argparse
from kubernetes import client, config
from unittest_base import K8sAPIBase

class TestPodRunningState(K8sAPIBase):
    def __init__(self):
        super().__init__()

    def check_pod_status(self, namespace, pod_name):
        try:
            pod = self.v1.read_namespaced_pod(name=pod_name, namespace=namespace)
            return pod.status.phase == 'Running'
        except client.exceptions.ApiException as e:
            print(f"Exception when calling CoreV1Api->read_namespaced_pod: {e}")
            return False

    def test_pod_running_state(self, duration):
        namespace = 'default'
        pod_name = 'example-pod'
        running_count = 0

        # Check the pod status every second for the specified duration
        for _ in range(duration):
            if self.check_pod_status(namespace, pod_name):
                running_count += 1
            time.sleep(1)

        # Calculate the running percentage
        running_percentage = (running_count / duration) * 100

        # Assert that the running percentage is at least 90%
        assert running_percentage >= 90, f"Pod '{pod_name}' running percentage is below threshold: {running_percentage}%"

        print(f"Pod '{pod_name}' running status checked {duration} times. Running percentage: {running_percentage}%.")

def main():
    parser = argparse.ArgumentParser(description='Test if a pod is running at least 90% of the time.')
    parser.add_argument('--duration', type=int, default=5, help='Duration to check the pod status in seconds.')
    args = parser.parse_args()

    test = TestPodRunningState()
    test.test_pod_running_state(args.duration)

if __name__ == '__main__':
    main()
```

2nd steady state:
- Name: example-service-availability
- Description: The next issue to address is the 'Single Pod Deployment', which is related to the lack of redundancy. This is a significant issue because if the single Pod fails, the service will be unavailable. To verify this, we can define a steady state that checks the availability of the service itself. A measurable output for this steady state is the service's response time or availability. Since the service is exposed on port 80, we can check if the service is responding to HTTP requests. This steady state will help verify that the service is available and responsive, which is crucial given the single Pod deployment configuration.
- Threshold for the steady state: Service availability should be at least 99.9% with a response status of 200.; The steady state we are considering is the availability of the 'example-service'. The k6 test results show that the service is currently responding with a 200 status code for all requests, indicating 100% availability. Given that the system consists of a single Pod, any failure in the Pod would result in the service being unavailable. Therefore, the threshold should ensure that the service remains available and responsive. To account for minor fluctuations and network latency, a reasonable threshold would be to maintain a high availability percentage, slightly below 100% to allow for brief, non-critical failures. A threshold of 99.9% availability is a common standard for web services, allowing for some tolerance while still ensuring high reliability.
- Whether the steady state meets the threshold is determined by the following K6 Javascript:
```
import http from 'k6/http';
import { check } from 'k6';

export const options = {
  vus: 1,
  duration: '5s',
  thresholds: {
    // Ensure that the service availability is at least 99.9%
    'http_req_failed': ['rate<=0.001'], // 0.1% failure rate corresponds to 99.9% availability
  },
};

export default function () {
  const res = http.get('http://example-service.default.svc.cluster.local:80');
  check(res, {
    'status is 200': (r) => r.status === 200,
  });
}
```

The fault scenario here is as follows:

An assumed fault scenario is as follows:
- Event: Cyber Attack
- Used Chaos Engineering tool: Chaos Mesh
- Faults to simulate the event: [[Fault(name='PodChaos', name_id=0, params={'action': 'pod-kill', 'mode': 'one', 'selector': {'namespaces': ['default'], 'labelSelectors': {'app': 'example'}}})], [Fault(name='NetworkChaos', name_id=1, params={'action': 'delay', 'direction': 'to', 'target': {'mode': 'all', 'selector': {'namespaces': ['default'], 'labelSelectors': {'app': 'example'}}}, 'mode': 'all', 'selector': {'namespaces': ['default'], 'labelSelectors': {'app': 'example'}}, 'device': 'eth0', 'delay': {'latency': '100ms', 'jitter': '10ms', 'correlation': '50'}})]]
- Description: Given the system's weaknesses, a cyber attack targeting the single Pod and its network could be highly impactful. The Pod's restart policy set to 'Never' means that if the Pod fails, it will not restart, leading to downtime. Additionally, the single Pod deployment means there is no redundancy, so any failure will make the service unavailable. To simulate a cyber attack, we can start by injecting a PodChaos fault to kill the Pod, testing the system's ability to handle Pod failures. This will directly exploit the lack of redundancy and the restart policy issue. Next, we can simulate a network attack using NetworkChaos to introduce network latency, testing the service's ability to maintain availability under network stress. This sequence simulates a cyber attack by first taking down the Pod and then stressing the network, revealing the system's vulnerabilities in handling such scenarios.

## Step 2.1. Chaos-Engineering experiment
### Here is the overview of my Chaos-Engineering experiment to verify the hypothesis:
The entire time schedule of the Chaos-Engineering experiment is as follows (The experiment is divided into three phases: pre-validation, fault-injection, and post-validation phases):
Given the constraints of the chaos engineering experiment, which must be completed within 1 minute, we need to allocate time efficiently across the three phases: pre-validation, fault-injection, and post-validation. The pre-validation phase is crucial to ensure that the system is in a steady state before we introduce any faults. Since we have two steady states to validate, we should allocate a reasonable amount of time to check both the pod's running status and the service's availability. A duration of 15 seconds should be sufficient for pre-validation, allowing us to run the necessary checks multiple times. The fault-injection phase is where we introduce the chaos to observe the system's behavior under stress. Given the complexity of the faults (PodChaos and NetworkChaos), we should allocate the majority of the time to this phase to ensure that the faults have enough time to manifest and impact the system. A duration of 30 seconds is appropriate for fault injection, allowing us to observe the system's response to both pod failure and network latency. Finally, the post-validation phase is essential to verify that the system returns to its steady states after the faults are removed. We should allocate 15 seconds for post-validation, similar to the pre-validation phase, to ensure that the system stabilizes and meets the defined thresholds for steady states. This allocation results in a total experiment time of 60 seconds, which fits within the 1-minute constraint.
- Total experiment phase: 60s
- Pre-validation phase: 15s
- Fault-injection phase: 30s
- Post-validation phase: 15s

The details of the three phases are as follows:
Pre-validation Phase (15s):
In the pre-validation phase, we need to ensure that the system is in its expected steady state before we proceed with fault injection. Given the constraints of a 15-second total time for this phase, we will conduct two unit tests to verify the steady states: one for the pod's running status and another for the service's availability. These tests will be executed sequentially due to the short duration available, ensuring that each steady state is verified independently and thoroughly. The first test will check if the 'example-pod' is running at least 90% of the time over a 5-second period. This is crucial because the pod's restart policy is set to 'Never', and we need to confirm its availability before introducing any faults. The second test will verify the 'example-service' availability, ensuring it responds with a 200 status code at least 99.9% of the time over another 5-second period. This test is essential to confirm that the service is operational and responsive, given the single pod deployment. By staggering these tests, we can focus on each steady state individually, allowing us to identify any issues before proceeding to the fault injection phase.

Fault-injection Phase (30s):
In this fault-injection phase, we aim to simulate a cyber attack by injecting two types of faults: PodChaos and NetworkChaos. The total duration for this phase is 30 seconds, so we need to carefully schedule the faults and unit tests to fit within this timeframe. 

First, we will inject the PodChaos fault to simulate a pod failure. This fault will be injected at the start of the phase (grace period of 0s) and will last for 10 seconds. This duration is chosen to allow enough time for the system to experience the impact of the pod being killed, given the pod's restart policy is set to 'Never'.

Simultaneously, we will run the unit test for the 'example-pod-running' steady state to verify if the pod is running at least 90% of the time during the fault injection. This test will also start at 0s and run for 10 seconds, aligning with the PodChaos duration.

Next, we will inject the NetworkChaos fault to simulate network latency. This fault will start at 10 seconds (after the PodChaos fault ends) and will last for 20 seconds. This staggered approach allows us to observe the system's behavior under network stress after the pod failure has been simulated.

During the NetworkChaos fault, we will run the unit test for the 'example-service-availability' steady state. This test will start at 10 seconds and run for 20 seconds, matching the NetworkChaos duration. This ensures we are checking the service's availability and response time while the network is under stress.

By staggering the faults and aligning the unit tests with the fault durations, we can effectively observe the system's behavior under each fault condition and verify if the steady states are maintained.

Post-validation Phase (15s):
In the post-validation phase, we need to ensure that the system has returned to its steady states after the fault injection. Given the 15-second time constraint, we will perform quick checks to verify the steady states. The two steady states to verify are: 1) the 'example-pod' is running, and 2) the 'example-service' is available. We will execute these checks sequentially due to the short duration, ensuring each test has enough time to gather meaningful data. The first test will check the pod's running status, followed by the service availability test. This order is logical because the pod must be running for the service to be available. Each test will have a brief grace period to allow the system to stabilize after the fault injection, followed by a short duration to perform the checks.

The summary of the above experiment plan:
The chaos engineering experiment is structured into three phases: pre-validation, fault-injection, and post-validation, all to be completed within a total of 60 seconds. 

In the pre-validation phase, which lasts for 15 seconds, two unit tests are conducted sequentially to ensure the system is in a steady state before fault injection. The first test, named 'pre-unittest-example-pod-running', checks the 'example-pod' running status. It starts immediately at the beginning of the phase and runs for 5 seconds. Following this, the second test, 'pre-unittest-example-service-availability', begins at the 5-second mark and also runs for 5 seconds, verifying the service's availability.

The fault-injection phase spans 30 seconds and involves two types of faults: PodChaos and NetworkChaos. Initially, the PodChaos fault, named 'fault-podchaos', is injected at the start of the phase and lasts for 10 seconds. Concurrently, the 'fault-unittest-example-pod-running' unit test runs for the same duration to verify the pod's status during the fault. After the PodChaos fault concludes, the NetworkChaos fault, named 'fault-networkchaos', begins at the 10-second mark and continues for 20 seconds. Simultaneously, the 'fault-unittest-example-service-availability' test runs for 20 seconds, starting at the same time as the NetworkChaos fault, to check the service's availability under network stress.

Finally, the post-validation phase, also 15 seconds long, ensures the system returns to its steady states. The 'post-unittest-example-pod-running' test starts after a 2-second grace period and runs for 6 seconds to verify the pod's status. Subsequently, the 'post-unittest-example-service-availability' test begins at the 8-second mark and runs for 5 seconds, checking the service's availability. This sequential execution allows for a brief stabilization period before each test.

To automatically conduct the above experiment plan with Chaos Mesh, the following Chaos-Mesh-Worfklow file was created (by applying it to the cluster, the experiment plan will be automatically executed according to the Chaos-Mesh-Worfklow file):
```yaml
|\verbatiminput{system_prompts/workflow.yaml}|
```

## Step 2.2, 3, 4. Experiment execution, analysis and improvement (reconfiguring the system to satisfy the hypothesis)
### Here is the improvement history:
### Experiment result (1st try)
Passed unittests:
- pre-unittest-example-pod-running
- pre-unittest-example-service-availability
Failed unittests:
- fault-unittest-example-pod-running
```log
Exception when calling CoreV1Api->read_namespaced_pod: (404)
Reason: Not Found
HTTP response headers: HTTPHeaderDict({'Audit-Id': '71a5fb6f-8e8c-4e31-9bc2-db80df08c498', 'Cache-Control': 'no-cache, private', 'Content-Type': 'application/json', 'X-Kubernetes-Pf-Flowschema-Uid': 'c4624bd9-7fc7-42c6-bcb8-4235110a860d', ... 'Content-Length': '190'})
HTTP response body: {"kind":"Status"...-Control': 'no-cache, private', 'Content-Type': 'application/json', 'X-Kubernetes-Pf-Flowschema-Uid': 'c4624bd9-7fc7-42c6-bcb8-4235110a860d', 'X-Kubernetes-Pf-Prioritylevel-Uid': '4706085f-6263-43ae-93f5-b4a61de8b6be', 'Date': 'Sun, 24 Nov 2024 13:25:27 GMT', 'Content-Length': '190'})
HTTP response body: {"kind":"Status","apiVersion":"v1","metadata":{},"status":"Failure","message":"pods \"example-pod\" not found","reason":"NotFound","details":{"name":"example-pod","kind":"pods"},"code":404}
```

- fault-unittest-example-service-availability
```log
time="2024-11-24T13:25:29Z" level=warning msg="Request Failed" error="Get \"http://example-service.default.svc.cluster.local... 
  http_reqs............: 26 1.269887/s
  iteration_duration...: avg=787.43ms min=287.51µs med=1.02s max=1.02s p(90)=1.02s p(95)=1.02s
  iterations...........: 26 1.269887/s
  vus..................: 1  min=1 max=1
  vus_max..............: 1  min=1 max=1
time="2024-11-24T13:25:49Z" level=error msg="thresholds on metrics 'http_req_failed' have been crossed"
```

- post-unittest-example-pod-running
```log
Traceback (most recent call last):
  File "/chaos-eater/sandbox/cycle_20241124_132128/hypothesis/unittest_example-pod-running_mod0.py", line 49, in <module>
Exception when calling CoreV1Api->read_namespaced_pod: (404)
Reason: Not Found
HTTP response headers: HTTPHeaderDict({'Audit-Id': '8a587dad-5e7d-44d6-b9a5-dd6a14dc6125', 'Cache-Control': 'no-cache, private', 'Content-Type': 'application/json', 'X-Kubernetes-Pf-Flowschema-Uid': 'c4624bd9-7fc7-42c6-bcb8-4235110a860d', 'X-Kubernetes-Pf-Priori...-Control': 'no-cache, private', 'Content-Type': 'application/json', 'X-Kubernetes-Pf-Flowschema-Uid': 'c4624bd9-7fc7-42c6-bcb8-4235110a860d', ..., 'Content-Length': '190'})
HTTP response body: {"kind":"Status","apiVersion":"v1","metadata":{},"status":"Failure","message":"pods \"example-pod\" not found","reason":"NotFound","details":{"name":"example-pod","kind":"pods"},"code":404}
```

- post-unittest-example-service-availability
```log
time="2024-11-24T13:26:02Z" level=warning msg="Request Failed" error="Get \"http://example-service.default.svc.cluster.local:80\": dial tcp 10.96.152.112:80: connect: connection refused" 
  http_reqs............: 11 2.14294/s
  iteration_duration...: avg=466.61ms min=211.73µs med=3.98ms max=1.03s p(90)=1.02s p(95)=1.02s
  iterations...........: 11 2.14294/s
  vus..................: 1  min=1 max=1
  vus_max..............: 1  min=1 max=1
time="2024-11-24T13:26:08Z" level=error msg="thresholds on metrics 'http_req_failed' have been crossed"
```

### Analysis report (1st try)
The chaos engineering experiment results indicate several critical issues in the system's configuration and its ability to handle faults, particularly in the context of the defined fault scenario. Here is a detailed analysis of the failures observed during the experiment:

1. **Pod Restart Policy and Single Pod Deployment**:
   - The 'fault-unittest-example-pod-running' test failed because the Pod was not found after the PodChaos fault was injected. This is directly related to the Pod's restart policy set to 'Never' in the `nginx/pod.yaml` manifest. When the Pod was killed, it did not restart, leading to a 404 error when attempting to read the Pod's status. This confirms the identified issue #0 (Pod Restart Policy) and issue #1 (Single Pod Deployment), where the lack of redundancy and automatic recovery mechanisms resulted in the Pod being unavailable.

2. **Service Availability**:
   - The 'fault-unittest-example-service-availability' test failed due to connection refusals, indicating that the service was unavailable. This failure is a consequence of the Pod being killed and not restarting, as the Service relies on the Pod to handle requests. The Service's selector in `nginx/service.yaml` targets Pods with the label 'app: example', and with the Pod being down, there were no endpoints available to serve the requests, leading to the connection refused errors.

3. **Post-Validation Failures**:
   - Both 'post-unittest-example-pod-running' and 'post-unittest-example-service-availability' tests failed, showing that the system did not recover to its steady state after the fault injection. The Pod remained unavailable, and consequently, the Service could not respond to HTTP requests. This further highlights the critical impact of the Pod's restart policy and the lack of redundancy.

**Insights and Recommendations**:
- **Pod Restart Policy**: Change the Pod's restart policy from 'Never' to 'Always' or 'OnFailure' to ensure that the Pod automatically restarts if it fails. This will help maintain the Pod's availability and reduce downtime.
- **Redundancy and Scalability**: Implement a Deployment or ReplicaSet to manage the Pod. This will provide redundancy by running multiple replicas of the Pod, ensuring that the Service remains available even if one Pod fails.
- **Service Resilience**: Consider using a LoadBalancer or Ingress to manage traffic to multiple Pods, enhancing the system's ability to handle failures and maintain service availability.

By addressing these issues, the system can improve its resilience against similar fault scenarios in the future, ensuring that the steady states are maintained even when faults are injected.

### Improvement result (1st try)
1 K8s manifests are modified:
- The K8s manifest 'nginx/pod.yaml' was replaced.
Replace the Pod with a Deployment to manage the Pod's lifecycle and provide redundancy. This will ensure that multiple replicas of the Pod are running and can automatically restart if they fail.
```yaml
apiVersion: apps/v1
kind: Deployment
metadata:
  name: example-deployment
  labels:
    app: example
spec:
  replicas: 3
  selector:
    matchLabels:
      app: example
  template:
    metadata:
      labels:
        app: example
    spec:
      containers:
      - name: example-container
        image: nginx:1.17.1
        ports:
        - containerPort: 80
```

### Experiment result (2nd try)
Passed unittests:
- pre-unittest-example-pod-running
- pre-unittest-example-service-availability
- fault-unittest-example-pod-running
- fault-unittest-example-service-availability
- post-unittest-example-pod-running
- post-unittest-example-service-availability
Failed unittests:
\end{exampledata}

\begin{figure*}[!htbp] \centering
    \includegraphics[width=\textwidth]{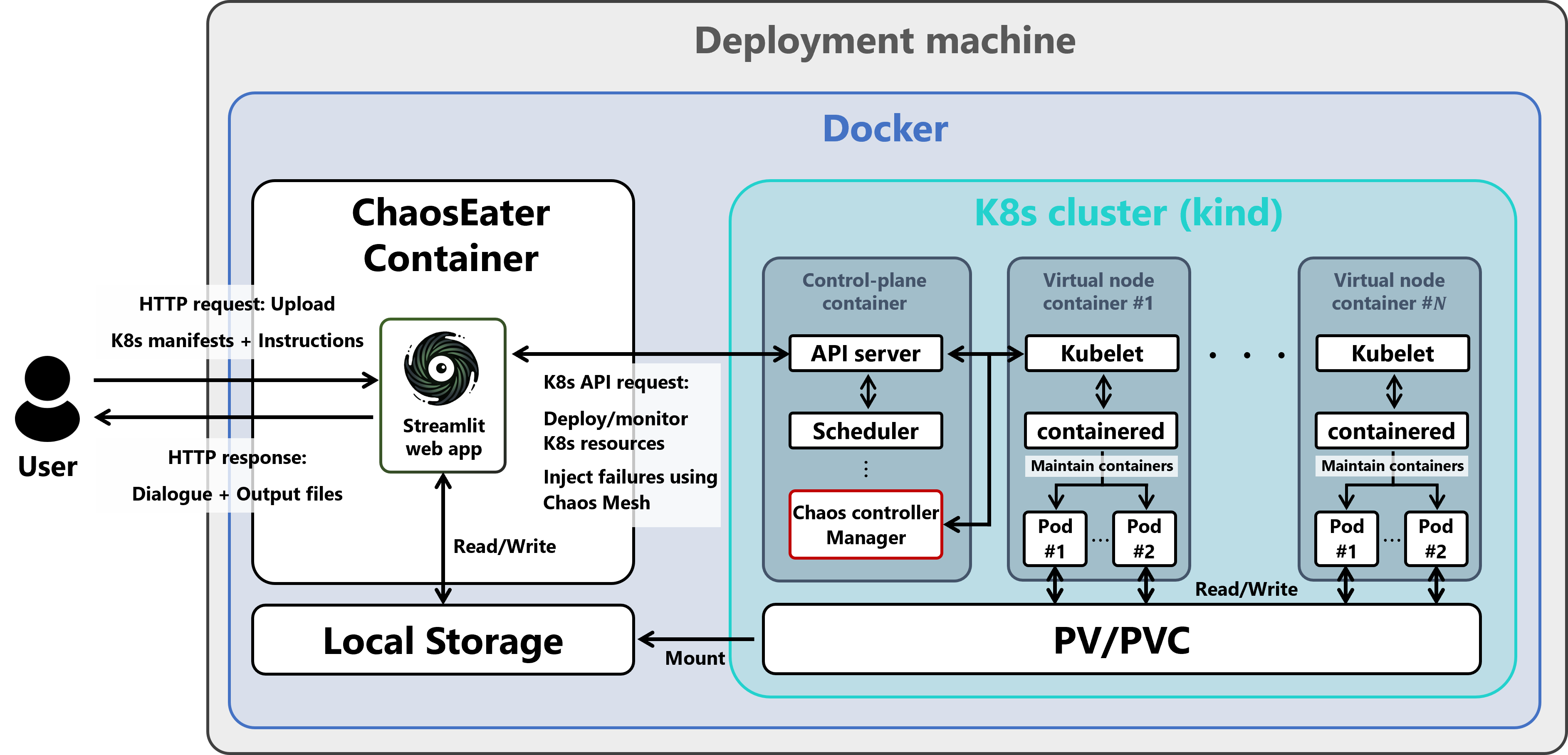}
    \caption{The deployment environment of \textsc{ChaosEater}.}
    \label{fig:deployment}
\end{figure*}

\section{Implementation Details}
\label{adx:implementation}
\subsection{System Deployment}
Figure \ref{fig:deployment} illustrates the deployment environment of \textsc{ChaosEater}.
\textsc{ChaosEater} currently supports only development environments, where K8s clusters are constructed on a single machine using kind \citep{kind}.
The web application of \textsc{ChaosEater} is implemented with Streamlit \citep{streamlit} and is deployed in a Docker container.
Users can interact with it via the Graphical User Interface (GUI).
After receiving the user inputs, \textsc{ChaosEater} autonomously completes a CE cycle for them while calling LLM API. 
\textsc{ChaosEater} interacts with the K8s (kind) clusters via K8s API, requesting resource deployment, monitoring, and failure injection.
See \url{https://github.com/ntt-dkiku/chaos-eater/blob/main/README.md#-quick-start} for specific instructions on setting up this environment.

\begin{figure*}[tb] \centering
    \includegraphics[width=0.96\textwidth]{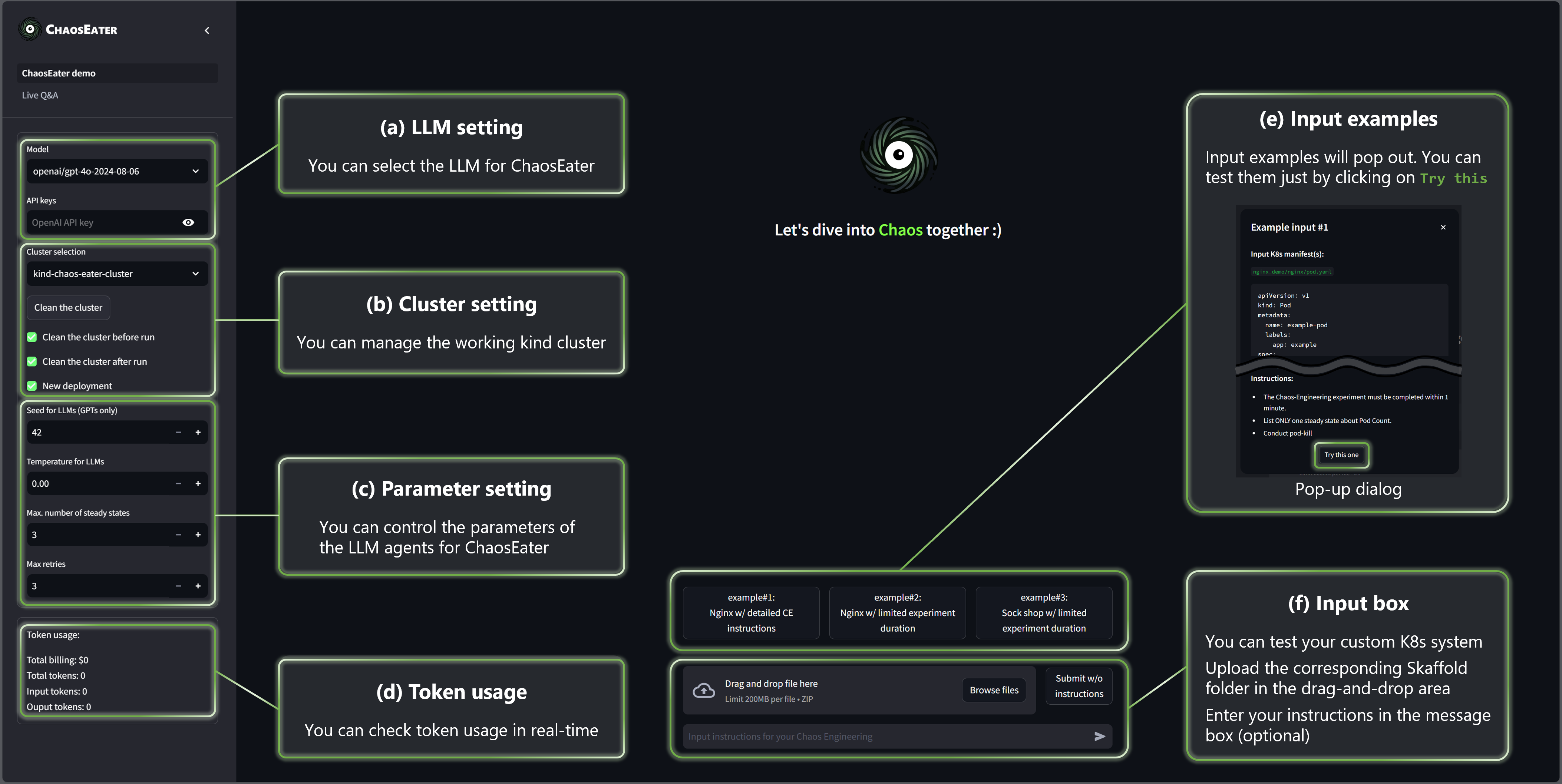}
    \caption{The GUI of \textsc{ChaosEater}}
    \label{fig:gui}
\end{figure*}
\subsection{Graphical User Interface}
Figure \ref{fig:gui} shows the GUI of \textsc{ChaosEater}.
The GUI resembles a typical chatbot interface, with a sidebar that allows for detailed parameter settings.
At a minimum, all you need to do is upload the K8s system files via the file uploader. Optionally, you can enter Chaos Engineering instructions in the chat box and control some parameters.
The details of the GUI controls are as follows.

\paragraph{(a) LLM setting}
You may change the LLMs used by \textsc{ChaosEater} from the \texttt{model} dropdown button. The currently supported LLMs are GPT-4o (gpt-4o-2024-08-06 and gpt-4o-2024-05-13), Claude (claude-3-5-sonnet-20240620), Gemini (google/gemini-1.5-pro).

\paragraph{(b) Cluster setting}
Currently available clusters are listed in the \texttt{Cluster selection} dropdown button. When there are multiple kind clusters, you may change the working kind cluster from here. While the GUI browser is open, the selected cluster will be occupied, and other users will not see the same cluster in the dropdown button.
If you check \texttt{Clean the cluster before/after run}, all resources in the selected cluster, excect for \textsc{ChaosEater}'s, will be removed before/after running every single CE cycle.
If you check \texttt{New deployment}, the input K8s system will be deployed in the preprocessing phase. If it is already deployed, you may uncheck it to skip the deployment.

\paragraph{(c) Parameter setting}
You can control the parameters of the LLM agents for \textsc{ChaosEater}.
\texttt{Seed for LLMs} sets the random seed for the LLMs (this is only effective when using OpenAI models that support seed setting, such as GPT-4o).
\texttt{Temperature for LLMs} sets the temperature of the LLMs.
\texttt{Max. number of steady states} sets the maximum number of steady states proposed during the hypothesis phase.
\texttt{Max retries} sets the maximum number of iterations for the verification loop and improvement loop. If the loop exceeds this limit, an assertion error will occur, immediately terminating the app at that point.

\paragraph{(d) Token usage}
You can monitor token usage in real-time. The total cost is calculated based on the official pricing tables as of September 2024.

\paragraph{(e) Input examples}
We prepare three types of input examples. When you press each button, the content of the K8s manifests to be input and the instructions will be displayed in a dialog. Click the \texttt{Try this one} button for the example you wanna try, and a CE cycle will start for that input example.

\paragraph{(f) Input box}
You can try your custom system by inputting its data to the input box. First, input a zipped folder to the file uploader box following the input format instruction below (this step is mandatory). If you don't have any instructions for the CE cycle, click the \texttt{Submit w/o instructions} button, and a CE cycle will start for that input system. If you do, write your instructions in the chat box and click the send icon $\blacktriangleright$ or \texttt{Enter}. Then, a CE cycle that follows the instructions will start for that input system.

\paragraph{Input format}
As input, \textsc{ChaosEater} currently supports only a zipped Skaffold project folder, which involves of a Skaffold configuration file and K8s manifests. The Skaffold configuration file must be placed in the root directory of the folder. The K8s manifests can be placed anywhere, but ensure that their relative paths are correctly specified in the manifests section of the Skaffold configuration file.

\begin{table*}[tb]
    \caption{Chaos Mesh failure types supported by \textsc{ChaosEater}. The descriptions are quoted from the official document.\protect\footnotemark}
    \begin{center}
    \renewcommand{\arraystretch}{1.5}
    \small
    \setlength{\tabcolsep}{4.2px}
    \begin{tabular}{lp{11 cm}}
        \toprule
        Failure type &Description \\
        \midrule
        \texttt{PodChaos} &It simulates Pod failures, such as Pod node restart, Pod's persistent unavailability, and certain container failures in a specific Pod.\\
        \hline
        \texttt{NetworkChaos} &It simulates network failures, such as network latency, packet loss, packet disorder, and network partitions. \\
        \hline
        \texttt{DNSChaos} &It simulates DNS failures, such as the parsing failure of DNS domain name and the wrong IP address returned. \\
        \hline
        \texttt{HTTPChaos} &It simulates HTTP communication failures, such as HTTP communication latency. \\
        \hline
        \texttt{StressChaos} &It simulates CPU race or memory race.\\
        \hline
        \texttt{IOChaos} &It simulates the I/O failure of an application file, such as I/O delays, read and write failures.\\
        \hline
        \texttt{TimeChaos} &It simulates the time jump exception.\\
        \bottomrule
    \end{tabular}    
    \label{tab:failure-types}
    \end{center}
\end{table*}
\footnotetext{\url{https://chaos-mesh.org/docs/basic-features/}}

\subsection{Supported Failures}
\label{adx:supported_failures}
\textsc{ChaosEater} supports most of the Chaos Mesh \citep{chaosmesh} failure types except for \texttt{KernelChaos}.
Table \ref{tab:failure-types} shows the supported failure types and their short descriptions. In the following, we provide some remarks on the supported failure types.

\texttt{KernelChaos} is one of the strongest failures and can affect other \texttt{Pod}s that share the same kernel as the target \texttt{Pod}.
We believe that various failure scenarios can be sufficiently simulated with only other types of failures.
Therefore, we have currently put the somewhat too strong \texttt{KernelChaos} on hold. 
We may consider adding \texttt{KernelChaos} in the future.

\texttt{PodChaos}, \texttt{HTTPChaos}, \texttt{StressChaos}, and \texttt{IOChaos} include the \texttt{duration} parameter, which specifies the duration of the failure injection.
However, Chaos Mesh workflow manifests do not currently support the \texttt{duration} parameter.
Therefore, we remove the \texttt{duration} parameter from the JSON output instructions when detailing the failure parameters in the failure definition of the \textit{hypothesis} phase.
Alternatively, \textsc{ChaosEater} specifies the duration of the failure injection using the \texttt{deadline} parameter in the experiment planning of the \textit{experiment} phase. This replacement is officially recommended.\footnote{\url{https://chaos-mesh.org/docs/create-chaos-mesh-workflow/\#template-field-description}}

The original \texttt{PodChaos} supports three sub-types: pod-kill, pod-failure, and container-kill.
On the other hand, \textsc{ChaosEater} supports only pod-kill and container-kill, but not pod-failure.
Regarding the pod-failure, the official document says ``\textit{Pod Failure Chaos Experiment would change the image of each container in the target Pod to the 'pause image', which is a special image that does not perform any operations. if the container is configured without command, livenessProbe and readinessProbe, the container would be inspected as Running and Ready, although it had been changed to the 'pause image', and actually does not provide functionalities as normal or not-available.}"\footnote{\url{https://chaos-mesh.org/docs/simulate-pod-chaos-on-kubernetes/}}
Therefore, the state changes of a \texttt{Pod} caused by pod-failure depend on the configuration of livenessProbe and readinessProbe. Depending on their interval settings, it is possible for the \texttt{Pod} to be recognized as being in the Running state even while pod-failure is being injected. Considering this issue and some reported bugs\footnote{\url{https://github.com/chaos-mesh/chaos-mesh/issues/446}}\footnote{\url{https://github.com/chaos-mesh/chaos-mesh/issues/2523}}, we have currently put pod-failure on hold. 

\begin{figure*}[!h] \centering
    \includegraphics[width=\textwidth]{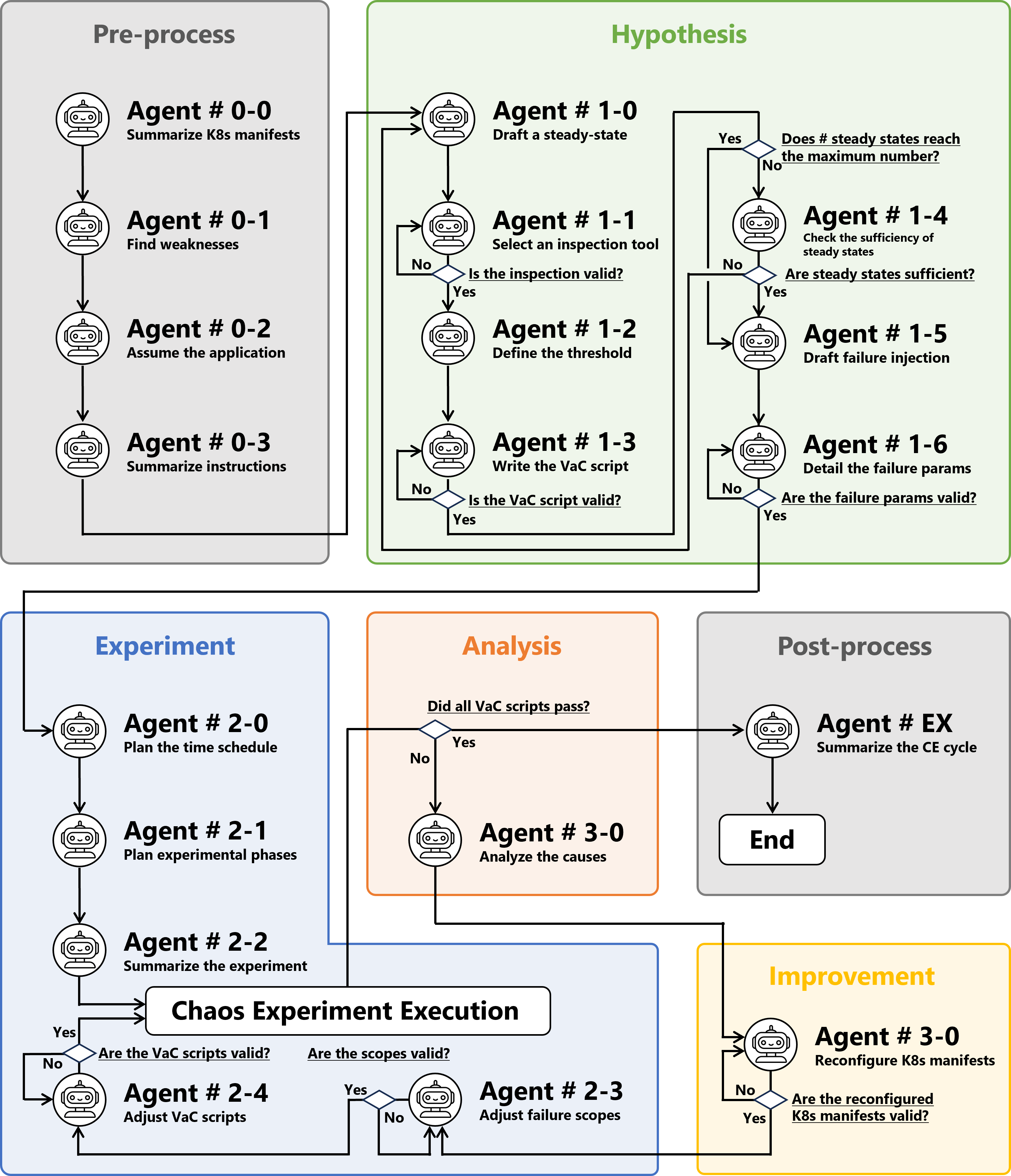}
    \caption{The agentic workflow of \textsc{ChaosEater}}
    \label{fig:agent-workflow}
\end{figure*}

\subsection{Agentic Workflow}
\subsubsection{Overall Design}
To ensure that the LLM agents perform as intended, \textsc{ChaosEater} fixes the general workflow according to the systematic CE cycle.
It then guides LLM agents by assigning them subdivided CE operations within this workflow. 
\textsc{ChaosEater} prepares prompt templates for each agent,\footnote{Instead of simply appending previous data and agent outputs to the conversation history to create the next
agent’s prompt, we create a new conversation for each agent every time and embed the organized previous data and agents' outputs within it. However, the verification loop, which will be discussed later, is an exception.} which include placeholders where text is dynamically embedded.
Therefore, once the user inputs the data, prompts for each agent are dynamically generated internally by embedding the input data and its intermediate outputs into the templates, and the agents that receive these prompts autonomously complete the workflow (i.e., CE cycle).
This is how full automation is achieved.
To facilitate data processing within \textsc{ChaosEater}, all agents output JSON data. This is achieved by instructing agents in their input prompts to output text in JSON format, and then parsing the output text as JSON data. 
\textsc{ChaosEater} uses the JSON output instruction and parser of LangChain \cite{langchain}.

Figure \ref{fig:agent-workflow} shows the agentic workflow of \textsc{ChaosEater}.
It has 20 agents, and they are chained in series according to the systematic CE cycle, with rule-based algorithms and verification loops at several points.
See Appendix \ref{adx:prompts} for the system prompt templates for each agent.
The advantages of dividing operations into smaller tasks and organizing the agentic workflow as shown in Figure \ref{fig:agent-workflow} are as follows:
\begin{enumerate}
    \item \textbf{Task performance improvement};
    It is generally known that dividing complex tasks into smaller sub-tasks enhances the performance of LLMs in solving them\footnote{\url{https://docs.anthropic.com/en/docs/build-with-claude/prompt-engineering/chain-prompts}} \citep{khot2023decomposed}.
    In our case, this is also important for managing context length and ensuring the accuracy of JSON output.
    By dividing tasks, the required input context for each agent can be minimized, mitigating issues such as information loss in long contexts.
    Furthermore, the JSON output structure for each agent can be minimized to the necessary complexity, reducing improperly formatted JSON outputs.
    \item \textbf{Flexibility and extensibility through agent modularization};
    By modularizing agents for each divided task, it becomes easier to make partial system modifications, such as replacing specific agents.
    This modularization also makes it easier for team members to modify agents collaboratively during development.
    \item \textbf{Towards an interactivity system};
    The current \textsc{ChaosEater} is a fully automated system with no user interaction during the CE cycle. However, we plan to add interactive functionalities in the future. 
    When refining outputs based on user feedback, the modularization allows querying only the minimum necessary agents, thereby improving the quality of re-generated outputs, similar to the first advantage.    
\end{enumerate}
Considering the above advantages, we empirically and manually optimized the workflow of agents and their system prompt templates.

On the other hand, agent modularization has some disadvantages, such as increased prompt management costs due to the greater number of prompts and the challenge of maintaining control over all agents as a unified system.
Regarding the latter, in the case study of \textsc{SockShop}, different behaviors were observed despite the fixed seed and a temperature setting of 0. 
This implies the complex behavior of multi-LLM agents and the difficulty of their control.
We believe that automatic prompt tuning \citep{llmasopt,automaticpromptoptimization,pryzant2023automatic,stableprompt,automaticprompteng} and automatic workflow optimization \citep{autoflow,gptswarm,symboliclearningenablesselfevolving,automateddesignagenticsystems,aflow} are promising for overcoming these challenges.

In the following, we provide a detailed explanation of the workflow design for each phase, which could not be fully covered on the main page.

\subsubsection{Phase 0: Pre-processing}
Given the user input, \textsc{ChaosEater} first deploys the user's system to the K8s cluster by running the Skaffold configuration file. Then, each agent sequentially processes the user inputs as follows:
\begin{enumerate}
    \item Summarize each of the input K8s manifests separately.
    \item Identify potential issues for resiliency and redundancy in the K8s manifests.
    \item Assume a possible application of the K8s manifests.
    \item Summarize user instructions for the CE cycle if provided. At the same time, filter out suspicious prompts, e.g., jailbreak prompts.
\end{enumerate}
This phase is for deploying the user's system and explicitly filling in the implicit context of the user's input. In the subsequent phases, this added context will also be provided as input.

\subsubsection{Phase 1: Hypothesis}
\label{adx:hypo}
\paragraph{Steady-state definition}
Given the pre-processed user inputs, each agent defines steady states as follows:
\begin{enumerate}
    \item Select a measurable states critical to maintaining the system's application. If any weak configurations are identified from the K8s manifests, their related states are preferentially selected.
    \item Select a tool to inspect the state. K8s API and k6 \citep{k6} are supported as the tool. Then, write the corresponding inspection script and inspect the current (normal) value of the state in the system by running the script.
    \item Define the threshold for the state based on the inspected value. Note that, according to the definition of a steady state, the threshold must be satisfied under the current conditions. 
    \item Write a unit-test script that validates whether the steady state (i.e., the pair of the state and its threshold) is satisfied by adding threshold-based assertions to the corresponding inspection script.
    \item Check whether the currently defined steady states are sufficient. If they are, the steady-state definition is complete here. Otherwise, return to the first step and define additional steady states. 
\end{enumerate}

\begin{figure*}[!htb]
  \begin{minipage}{0.545\textwidth}
    \begin{tcblisting}{colback=white, listing only, title={\tiny (a) VaC script for K8s API (Python)},
                       listing options={language=Python, basicstyle=\tiny\ttfamily, breaklines=true,
                                        numbers=left, numberstyle=\tiny, stepnumber=1, numbersep=3pt, showstringspaces=false},
                       boxrule=0.3mm,
                       left=3mm, right=-3mm, top=-2mm, bottom=-2mm}
def check_podcount(label, expected_count, duration):
  consistent_count = True
  for i in range(duration):
    pods = self.v1.list_namespaced_pod(
      namespace='default',
      label_selector=label)
    pod_count = len(pods.items)
    print(f"current pod count: {pod_count}")
    consistent_count = pod_count == expected_count
    if not consistent_count:
      break
    time.sleep(1)
  assert consistent_count, "Pod count was inconsistent."
...
    \end{tcblisting}
  \end{minipage}\hfill
  \begin{minipage}{0.45\textwidth}
    \begin{tcblisting}{
        colback=white,
        listing only,
        title={\tiny (b) VaC script for k6 (Javascript)},
        listing options={
            language=java,
            basicstyle=\tiny\ttfamily,
            breaklines=true,
            numbers=left,
            numberstyle=\tiny,
            stepnumber=1,
            numbersep=3pt,
            showstringspaces=false,
            escapeinside={(*@}{@*)},
        },
        boxrule=0.3mm,
        left=3mm,
        right=-1mm,
        top=-2mm,
        bottom=-2mm
    }
export const options = {
  vus: 10,
  duration: '10s',
  thresholds: {
    http_req_duration: ['p(95)<500'],
  },
};

export default function () {
  const res = http.get('(*@http://example.com@*)');
  check(res, {'status was 200': (r) => 
    r.status == 200 });
  sleep(1);
}
    \end{tcblisting}
  \end{minipage}
  \caption{Examples of unit-test scripts to validate steady states.}
  \label{fig:vac_scripts}
\end{figure*}

The unit-test scripts are used in the \textit{experiment} phase to mechanically validate the steady states during chaos experiments.
As mentioned in the main page, we here call this unit-test-based validation approach \textit{Validation as Code} (VaC).
Na\"{i}ve approaches, such as validating steady states using LLMs that take log data directly, do not guarantee consistency in the validation process and may even lead to incorrect judgments.
On the other hand, with VaC, the validation process becomes fixed once a unit test is written, guaranteeing its consistency. Furthermore, the explicit definition of the process in code enhances its transparency.
Figure \ref{fig:vac_scripts} shows examples of VaC scripts for K8s API (Python) and k6 (Javascript).
k6 can collect communication metrics (e.g., response times, error rates, etc) while conducting load tests.
In VaC, k6 is used to inspect the communication metrics, while K8s API is used to inspect the other states of K8s resources.
Both scripts allow for adjusting test durations through command-line arguments. For k6, the script also sets an appropriate number of virtual users for the load tests.

In steps 2 and 4, scripts are repeatedly debugged until they terminate successfully. 
In this verification loop, as an exception, \textsc{ChaosEater} simply appends the previous agent's output and the resulting error messages to the initial conversation as conversation history, and uses it as the agent's prompt in the next loop.
The verification loops discussed later also follow the same process.

\paragraph{Failure definition}
Given the pre-processed user inputs and the steady states, each agent defines failures that may occur in the system as follows: 
\begin{enumerate}
    \item Assume a failure scenario (e.g., a surge in access due to a promotional campaign, cyber attack, etc.) that may occur in the system. Then, define the sequence of failures that simulates the scenario and may affect the defined steady states. The failures are selected from the failure types supported in Chaos Mesh.
    \item Define detailed parameters for each failure, such as the scope of the failure injection, the failure sub-type, the failure strength, etc.
\end{enumerate}

In step 1, the agent outputs a 2D list of Chaos Mesh failure type names, arranged in the order of insertion.
The inner lists involve concurrent failures, and the outer list represents the injection order of each concurrent failure set.
For example, [[\texttt{StressChaos}, \texttt{NetworkChaos}], [\texttt{PodChaos}]] represents that \texttt{PodChaos} is injected after simultaneously injecting \texttt{StressChaos} and \texttt{NetworkChaos}.

\begin{figure}[tb]
    \centering
    \begin{minipage}{0.3\textwidth}
    \begin{tcblisting}{colback=white, listing only, title={\tiny PodChaos parameters},
                       listing options={language=Python, basicstyle=\tiny\ttfamily, breaklines=true,
                                        numbers=left, numberstyle=\tiny, stepnumber=1, numbersep=3pt},
                       boxrule=0.3mm,
                       left=2mm, right=0mm, top=0mm, bottom=0mm}
action: pod-kill
  mode: one
  selector:
    labelSelectors:
      app: example
    namespaces:
      - default
    \end{tcblisting}
    \end{minipage}
    \caption{An example of detailed parameters.}
    \label{fig:podchaos_params}
\end{figure}

In step 2, the agent separately defines the detailed parameters of each failure.
Each failure type requires a different parameter set. 
Therefore, given a failure type name, \textsc{ChaosEater} dynamically selects the corresponding JSON output instruction.
Instructions for each of the seven supported failures are prepared in advance based on the official Chaos Mesh documentation.
The agent then outputs the corresponding parameter set. 
Figure \ref{fig:podchaos_params} shows an example of the parameter set of \texttt{PodChaos}.
See also Dynamic instruction \ref{lst:first_instruct}--\ref{lst:last_instruct} for detailed parameter instruction for each failure type. 
The parameter sets output by the agent are verified through a verification loop, which repeatedly debugs them until their Chaos Mesh manifests pass the \texttt{kubectl apply --dry-run=server} command.
The failure injection duration and more detailed injection timing are defined in the next chaos experiment planning (see next section), along with the duration and timing for running the VaC scripts.

\newtcblisting{nodetemplates}[1][]{
    colback=white,
    listing only,
    title={\tiny #1},
    listing options={
        language=sysprompt,
        basicstyle=\tiny\ttfamily,
        breaklines=true,
        numbers=left,
        numberstyle=\tiny,
        stepnumber=1,
        numbersep=3pt,
        showstringspaces=false
    },
    boxrule=0.3mm,
    left=3mm,
    right=-3mm,
    top=-2mm,
    bottom=-2mm
}

\begin{figure*}[!htb]
  \begin{minipage}{0.545\textwidth}
    \begin{nodetemplates}[Task node (K8s API)]
- name: @//{{ node_name }}//@
  templateType: Task
  deadline: @//{{ duration }}//@
  task:
    container:
      name: @//{{ node_name }}//@-container
      image: chaos-eater/k8sapi:1.0
      imagePullPolicy: IfNotPresent
      command: ["/bin/bash", "-c"]
      args: [
        "python /chaos-eater/@//{{ path_to_vac_script }}//@ 
        --duration @//{{ duration }}//@"
      ]
...
    \end{nodetemplates}
    \vspace{1.3mm}
    \begin{nodetemplates}[Task node (k6)]
- name: @//{{ node_name }}//@
  templateType: Task
  deadline: @//{{ duration }}//@
  task:
    container:
      name: @//{{ node_name }}//@-container
      image: grafana/k6:latest
      command: [
        "k6", "run", "--duration", "@//{{ duration }}//@", 
        "--quiet", "/chaos-eater/@//{{ path_to_vac_script }}//@"
      ] 
...
    \end{nodetemplates}
  \end{minipage}\hfill
  \begin{minipage}{0.45\textwidth}
    \begin{nodetemplates}[Failure node]
- name: @//{{ node_name }}//@
  templateType: @//{{ name }}//@
  deadline: @//{{ duration }}//@
  @//{{ name }}//@:
    @//{{ failure_params }}//@
    \end{nodetemplates}
    
    \begin{nodetemplates}[Suspend node]
- name: @//{{ node_name }}//@
  templateType: Suspend
  deadline: @//{{ duration }}//@
    \end{nodetemplates}

    \begin{nodetemplates}[Serial group node]
- name: @//{{ node_name }}//@
  templateType: Serial
  deadline: @//{{ duration }}//@
  children:
    @//{{ list_of_grouped_node_names }}//@
@//{{ code_snippets_of_grouped_nodes }}//@
    \end{nodetemplates}

    \begin{nodetemplates}[Parallel group node]
- name: @//{{ node_name }}//@
  templateType: Parallel
  deadline: @//{{ duration }}//@
  children:
    @//{{ list_of_grouped_node_names }}//@
@//{{ code_snippets_of_grouped_nodes }}//@
    \end{nodetemplates}
  \end{minipage}
  \caption{YAML code snippets for each node type. The blue double curly braces \textcolor{blue}{\{\{\}\}} represent placeholders. \texttt{node\_name} is a node identifier automatically generated from \texttt{name}. \texttt{code\_snippets\_of\_grouped\_nodes} is filled with the complete code snippets of nodes that are grouped either serially or in parallel. The other placeholders are filled with values from the items of the schedule list.}
  \label{fig:node-templates}
\end{figure*}

\begin{algorithm*}
    \caption{Hierarchical node grouping algorithm}
    \label{alg:node_grouping}
    \small
    \SetKwInOut{Input}{Input}
    \SetKwInOut{Output}{Output}
    \Input{A schedule list output by the agent \texttt{schedule\_list} (dict of list of dict); \newline
    e.g., $\texttt{schedule\_list} = \{$\newline
    \hspace*{3em}$\texttt{"pre-valid"}: [\{\texttt{"name"}: \texttt{"pod-running"}, \texttt{"duration"}: \texttt{"20s"}, ...\}, ...],$\newline
    \hspace*{3em}$\texttt{"failure-injection"}: [\{\texttt{"name"}: \texttt{"PodChaos"}, \texttt{"duration"}: \texttt{"20s"}, ...\}, ...],$\newline
    \hspace*{3em}$\texttt{"post-valid"}: [\{\texttt{"name"}: \texttt{"pod-running"}, \texttt{"duration"}: \texttt{"20s"}, ...\}, ...]$\newline
    \hspace*{2em}$\}$
    }
    \Output{A Chaos Mesh workflow node consisting of hierarchically grouped nodes $\texttt{hierarchical\_node}$}
    \SetNlSty{textbf}{}{:}
    $\texttt{all\_group} \leftarrow [\,]$\\
    \For{$\texttt{stage} \in [\texttt{"pre-valid"}, \texttt{"failure-injection"}, \texttt{"post-valid"}]$}{
        $\texttt{parallel\_group} \leftarrow [\,]$\\
        \For{$\texttt{item} \in \texttt{schedule\_list.stage}$}{
            \uIf{$\texttt{IsFailureInjection(item)}$}{
                $\texttt{node} \leftarrow \texttt{CreateFailureNode}(\texttt{item.name}, \texttt{item.duration}, \texttt{item.failure\_params})$
            }
            \Else{
                $\texttt{node} \leftarrow \texttt{CreateTaskNode}(\texttt{item.name}, \texttt{item.duration}, \texttt{item.vac\_params})$
            }
            \uIf{$\texttt{item.grace\_period} > 0$}{
                $\texttt{suspend} \leftarrow \texttt{CreateSuspendNode}(\texttt{item.grace\_period})$\;
                $\texttt{serial\_group\_node} \leftarrow \texttt{CreateSerialGroupNode}([\texttt{suspend}, \texttt{node}])$\;
                $\texttt{parallel\_group.append(serial\_group\_node)}$\;
            }
            \Else{
                $\texttt{parallel\_group.append(node)}$\;
            }
        }
        $\texttt{parallel\_group\_node} \leftarrow \texttt{CreateParallelGroupNode()}$\\
        \texttt{all\_group.append(parallel\_group\_node)}\\
    }
    $\texttt{hierarchical\_node} \leftarrow \texttt{CreateSerialGroupNode}(\texttt{all\_group})$\\
    \Return $\texttt{hierarchical\_node}$
\end{algorithm*}

\subsubsection{Phase 2: (Chaos) Experiment}
\label{adx:chaosmesh_workflow}
\paragraph{Experiment planning}
Given the pre-processed user inputs and the hypothesis, each agent plans a chaos experiment that validates the hypothesis by dividing it into the three stages (i.e., pre-validation, failure-injection, and post-validation stages) as follows:
\begin{enumerate}
    \item Determine the duration of each stage.
    \item Determine the VaC scripts and failure injections to be executed in each stage. For each of them, specify the duration and grace period within a range that does not exceed the duration of the stage.
    \item Summarize the timeline of the chaos experiment in detail. This summary is referred to when analyzing the experiment results.
\end{enumerate}
In step 2, the agent outputs a list of dictionaries (i.e., schedule list) separately for each stage, with each dictionary containing three keys: \texttt{name}, \texttt{grace\_period}, and \texttt{duration}. 
The \texttt{name} is either a steady state name or a failure type name, and each corresponds one-to-one with the VaC script or failure injection defined in the hypothesis. 
The \texttt{grace\_period} is the waiting time from the start of each stage until the execution of the VaC script or failure injection, allowing flexible adjustment of the execution timing. 
The \texttt{duration} is the execution period after the grace period.

After appending the corresponding failure and VaC parameters to each item in the schedule lists, \textsc{ChaosEater} converts these lists into a Chaos Mesh workflow manifest.
This manifest enables the scheduling of VaC script execution and failure injection by constructing a workflow composed of the following three types of nodes: \texttt{failure} node for executing failure injection, \texttt{task} node for executing VaC scripts, and \texttt{suspend} node for waiting a specified duration.
\textsc{ChaosEater} hierarchically groups these nodes to construct complex workflows.
For convenience, we define each node as the corresponding code snippet shown in Figure \ref{fig:node-templates}.
Therefore, the functions that convert items of the schedule list into nodes, as well as those that group multiple nodes, correspond to operations that fill the placeholder of these code snippets with the argument data.

\begin{figure*}[tb]
    \centering
    \includegraphics[width=0.975\textwidth]{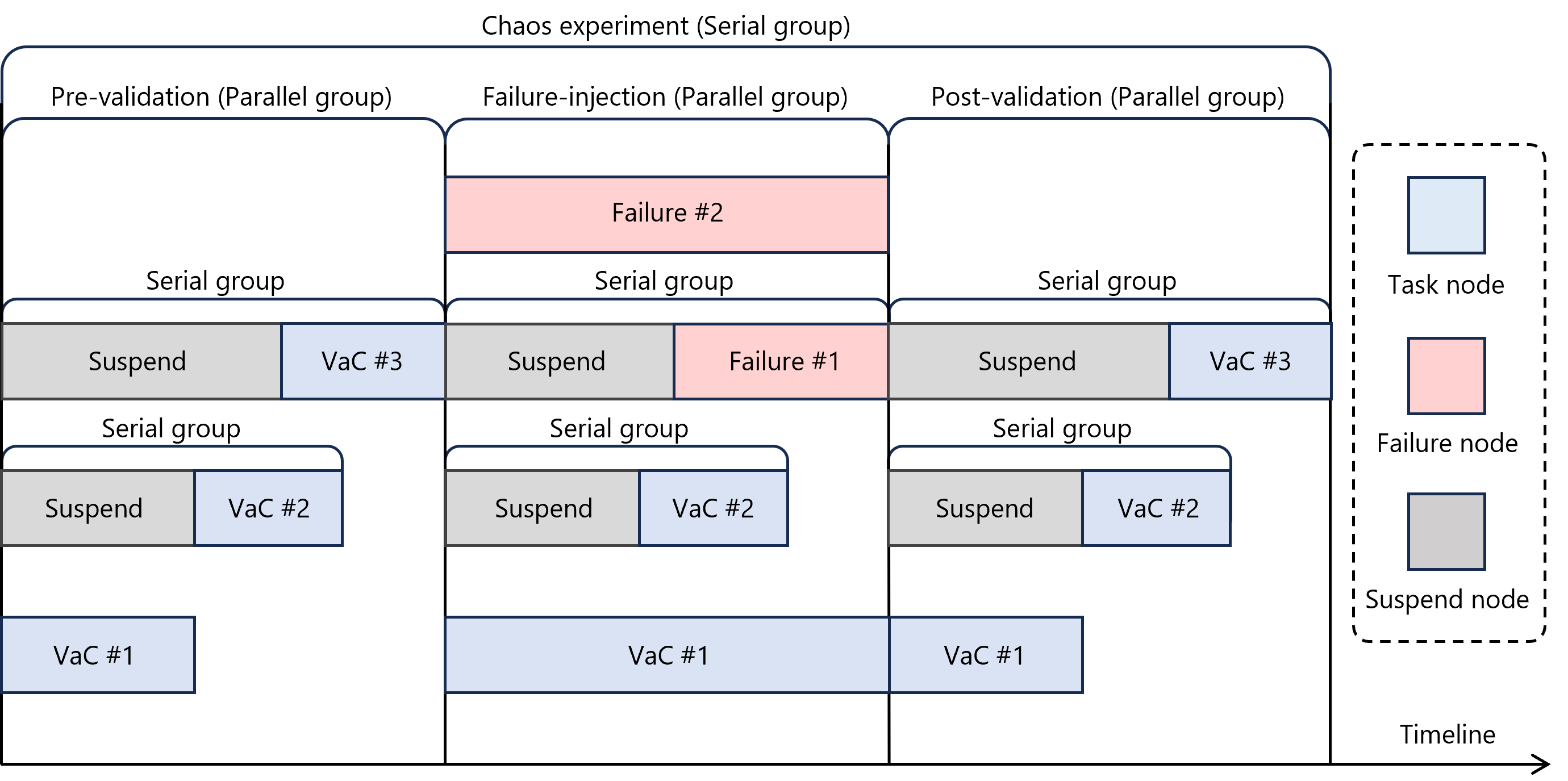}
    \caption{Hierarchical grouping for implementing a complex chaos experiment plan in Chaos Mesh.} 
    \label{fig:planning}
\end{figure*}

Algorithm \ref{alg:node_grouping} shows the hierarchical node grouping algorithm.
The input is a directory that organizes the schedule lists for each stage.
The algorithm first iterates over the three stages. 
For each state, each item in the stage's schedule list is converted to the corresponding type of node (line 5--8).
Next, each node that has a grace period greater than zero is serially grouped with a \texttt{suspend} node.
This \texttt{suspend} node is placed before the grouped node and waits for the duration of the grace period of that node (line 10--12).
Then, all the serially grouped nodes and remaining nodes are grouped in parallel (line 15).
Finally, the nodes grouped in parallel in each stage are grouped serially (line 17).
The output is the hierarchical group node, and by adding a header and other necessary elements to its code snippet, a Chaos Mesh workflow manifest can be generated. 
Figure \ref{fig:planning} shows an example of hierarchically grouped nodes in this algorithm.

\paragraph{Experiment replanning (within the improvement loop)}
Resource types and metadata defined in the K8s manifests may be changed during the \textit{improvement} phase. Therefore, replanning inspection targets in VaC scripts and the scope of failure injections is required between the \textit{improvement} phase and the next experiment execution. 
Given the original and reconfigured K8s manifests, as well as the previous VaC scripts, an agent adjusts or retains the inspection-target specifications in the VaC scripts. The inspection targets refer to resource specifications (line 11 and 12 in (a)), the request DNS (line 13 in (b)), etc., in Figure \ref{fig:vac_scripts}.
Given the original and reconfigured K8s manifests, as well as the previous failure-injection scope, another agent adjusts or retains the scope for the reconfigured manifests.
The scope refers to the \texttt{selector} filed in Figure \ref{fig:podchaos_params}.
These adjustments are also debugged through a verification loop.
After the adjustments, \textsc{ChaosEater} regenerates a new ChaosMesh workflow manifest by replacing only the path of VaC scripts of \texttt{task} nodes and the \texttt{selector} field of \texttt{failure} nodes with adjusted ones.
Note that this replanning only makes minor adjustments to reflect the changes in the K8s manifests, without altering the chaos experiment's original intent.

\subsubsection{Phase 4: Improvement}
Given the K8s manifests, the hypothesis, the experiment plan, and the improvement loop history, an agent reconfigures the K8s manifests so that all the VaC scripts pass in the chaos experiment. 
Here, the agent is instructed to make only minimal changes in a single improvement step, aiming for incremental improvement.
This approach helps prevent undesired improvements, such as resolving issues through excessive resource allocation.
The improvement loop history stores the history of the experiment results, their analysis reports, and their reconfigurations, within the improvement loop.
The history suppresses the repetition of the same reconfiguration.

\begin{figure}[tb]
    \centering
    \includegraphics[width=\linewidth]{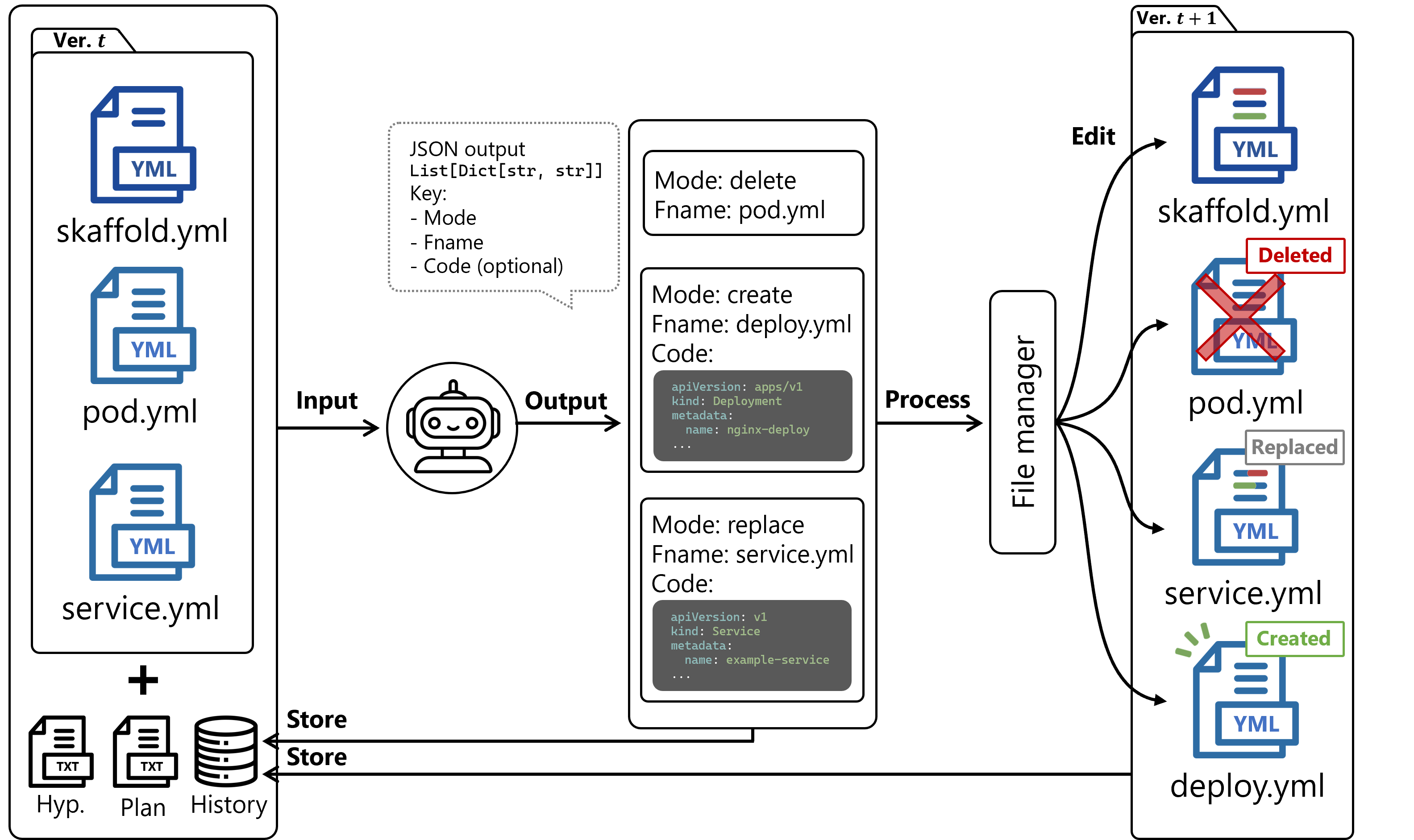}
    \caption{Reconfiguration process by the agent and a file management algorithm.} 
    \label{fig:reconfig}
\end{figure}

There are three reconfiguration modes: \texttt{create}, \texttt{delete}, and \texttt{replace}.
The agent first selects the reconfiguration modes while specifying file names, and then writes the reconfigured K8s manifests only for the \texttt{create} and \texttt{replace} modes.
The file manager of \textsc{ChaosEater} then edits the folder from the previous improvement loop (in the first improvement, it corresponds to the user's input folder) according to the agent's output.
Figure \ref{fig:reconfig} illustrates this reconfiguration process.

The verification loop is also conducted here: the agent's output is debugged repeatedly until all the K8s manifests in the edited folder are correctly applied to the K8s cluster without any errors.

\subsection{System prompt templates}
\label{adx:prompts}
In this section, we share all prompt templates for LLM agents of \textsc{ChaosEater}.
Words enclosed in blue curly braces \textbf{\textcolor{blue}{\{\}}} denote placeholders that are replaced with user input or the previous agent's outputs.
Each placeholder has a unique variable name, and placeholders with the same variable name across different prompt templates will have the same text embedded.
Examples of the text embedded in each placeholder are also provided right after their associated prompt templates.
On the other hand, words enclosed in red curly braces \textbf{\textcolor{red}{\{\}}} denote placeholders that are replaced with pre-defined dynamic instructions. These pre-defined instructions are dynamically selected and embedded in the placeholders according to the conditions. These dynamic instructions are also provided right after their associated prompt templates.
Words enclosed in green curly braces \textbf{\textcolor{green}{\{\}}} denote placeholders that are replaced with the same text as previously introduced, in order to avoid repetition.

Although it is omitted in each prompt template, a prefill will be inserted at the end of every prompt template.
Specifically, the first key in the JSON output will be added as an AI's message, as shown in Figure \ref{fig:prefill}.
It eliminates redundant outputs and improves the stability of the JSON outputs.

The term ``fault" appears in the prompt templates and the output of \textsc{ChaosEater}.
this is aligned with the terminology used in Chaos Mesh and has the same meaning as ``failure".

\begin{figure}[!htbp]  
    \begin{customdialogue}[The prefill for the Agent \#0-0]
<<AI>>:
```json
{\"k8s_summary\":
    \end{customdialogue}
  \caption{An example of prefills.}
  \label{fig:prefill}
\end{figure}

\subsubsection{Pre-processing}
\newtcblisting[use counter from=reviewprompt]{preprocessprompt}[1][]{
    colback=gray!5,
    listing only,
    title={Prompt \thetcbcounter: #1},
    listing options={
        language=sysprompt,
        basicstyle=\scriptsize\ttfamily,
        breaklines=true,
        breakindent=0pt,
        showstringspaces=false,
    },
    boxrule=0.3mm,
    coltitle=black,
    colbacktitle=gray!20,
    mybreakable,
}

\begin{preprocessprompt}[Agent \# 0-0 for summarizing K8s manifests]
//System://
System: You are a professional Kubernetes (k8s) engineer.
Given a K8s manifest, please summarize it according to the following rules:
- The summary must be written in bullet points.
- Summarize the functions of the K8s manifest in a way that is understandable to even beginners.
- The output should be formatted as a JSON instance that conforms to the JSON schema below.

As an example, for the schema {\"properties\": {\"foo\": {\"title\": \"Foo\", \"description\": \"a list of strings\", \"type\": \"array\", \"items\": {\"type\": \"string\"}}}, \"required\": [\"foo\"]}\nthe object {\"foo\": [\"bar\", \"baz\"]} is a well-formatted instance of the schema. The object {\"properties\": {\"foo\": [\"bar\", \"baz\"]}} is not well-formatted.

Here is the output schema:
```
{
  "properties": {
    "k8s_summary": {
      "title": "K8S Summary",
      "description": "Summary of the K8s manifest. Summarize it in bullet points like '- the 1st line\n- the second line...'",
      "type": "string"
    }
  },
  "required": [
    "k8s_summary"
  ]
}
```

//Human://
# K8s manifest
//@{k8s_yaml}@//

Please summarize the above K8s manifest.
\end{preprocessprompt}

\begin{exampledata}[\texttt{\textbf{k8s\_yaml}}]
```nginx/pod.yaml
apiVersion: v1
kind: Pod
metadata:
  name: example-pod
  labels:
    app: example
spec:
  restartPolicy: Never
  containers:
  - name: example-container
    image: nginx:1.17.1
    ports:
    - containerPort: 80
```
\end{exampledata}

\begin{preprocessprompt}[Agent \# 0-1 for finding potential weaknesses]
//System://
You are a professional Kubernetes (K8s) engineer.
Given K8s manifests for a system, you will identify their potential issues for resiliency and redundancy when failures occur in the system.
Always keep the following rules:
- List each issue with its name, associated K8s manifest(s), potential issues due to fault injection, and the configuration causing the issues (no need to suggest improvements).
- If the same issue exists in different manifests, merge them into a single issue, specifying all the associated manifest names.
- The output should be formatted as a JSON instance that conforms to the JSON schema below.

As an example, for the schema {\"properties\": {\"foo\": {\"title\": \"Foo\", \"description\": \"a list of strings\", \"type\": \"array\", \"items\": {\"type\": \"string\"}}}, \"required\": [\"foo\"]}\nthe object {\"foo\": [\"bar\", \"baz\"]} is a well-formatted instance of the schema. The object {\"properties\": {\"foo\": [\"bar\", \"baz\"]}} is not well-formatted.

Here is the output schema:
```
{
  "properties": {
    "issues": {
      "title": "Issues",
      "description": "List issues with its name, potential issues due to fault injection, and manifest configuration causing the issues (no need to suggest improvements).",
      "type": "array",
      "items": {
        "$ref": "#/definitions/K8sIssue"
      }
    }
  },
  "required": [
    "issues"
  ],
  "definitions": {
    "K8sIssue": {
      "title": "K8sIssue",
      "type": "object",
      "properties": {
        "issue_name": {
          "title": "Issue Name",
          "description": "Issue name",
          "type": "string"
        },
        "issue_details": {
          "title": "Issue Details",
          "description": "potential issues due to fault injection",
          "type": "string"
        },
        "manifests": {
          "title": "Manifests",
          "description": "manifest names having the issues",
          "type": "array",
          "items": {
            "type": "string"
          }
        },
        "problematic_config": {
          "title": "Problematic Config",
          "description": "problematic configuration causing the issues (no need to suggest improvements).",
          "type": "string"
        }
      },
      "required": [
        "issue_name",
        "issue_details",
        "manifests",
        "problematic_config"
      ]
    }
  }
}
```

//Human:// 
# Here are the K8s manifests for my system.
//@{k8s_yamls}@//

Please list issues for each K8s manifest.
\end{preprocessprompt}

\begin{exampledata}[\texttt{\textbf{k8s\_yamls}}]
```nginx/pod.yaml
apiVersion: v1
kind: Pod
metadata:
  name: example-pod
  labels:
    app: example
spec:
  restartPolicy: Never
  containers:
  - name: example-container
    image: nginx:1.17.1
    ports:
    - containerPort: 80
```

```nginx/service.yaml
apiVersion: v1
kind: Service
metadata:
  name: example-service
spec:
  selector:
    app: example
  ports:
    - protocol: TCP
      port: 80
      targetPort: 80
```
\end{exampledata}

\begin{preprocessprompt}[Agent \# 0-2 for assuming an application]
//System://
You are a professional Kubernetes (k8s) engineer.
Given k8s manifests and dependencies between them, please assume a real-world application (service) of the manifests according to the following rules:
- If the application is explicitly specified in the instructions, assume it.
- You can leverage any given information, including file name, manifests, and dependencies, to guess the purpose of the manifests.
- The output should be formatted as a JSON instance that conforms to the JSON schema below.

As an example, for the schema {\"properties\": {\"foo\": {\"title\": \"Foo\", \"description\": \"a list of strings\", \"type\": \"array\", \"items\": {\"type\": \"string\"}}}, \"required\": [\"foo\"]}\nthe object {\"foo\": [\"bar\", \"baz\"]} is a well-formatted instance of the schema. The object {\"properties\": {\"foo\": [\"bar\", \"baz\"]}} is not well-formatted.

Here is the output schema:
```
{
  "properties": {
    "thought": {
      "title": "Thought",
      "description": "Before assuming an application, reason logically why you assume it for the given manifests. e.g., from file name, instructions, or other elements?",
      "type": "string"
    },
    "k8s_application": {
      "title": "K8S Application",
      "description": "Specify what the service (application) offers to users.",
      "type": "string"
    }
  },
  "required": [
    "thought",
    "k8s_application"
  ]
}
```

//Human:// 
//@{user_input}@//

Please assume a real-world application of the manifests.
\end{preprocessprompt}

\begin{exampledata}[\texttt{\textbf{user\_input}}]
# K8s manifest:
```nginx/pod.yaml
apiVersion: v1
kind: Pod
metadata:
  name: example-pod
  labels:
    app: example
spec:
  restartPolicy: Never
  containers:
  - name: example-container
    image: nginx:1.17.1
    ports:
    - containerPort: 80
```
# Summary of nginx/pod.yaml:
- This manifest defines a Kubernetes Pod.
- The Pod is named 'example-pod'.
- It includes metadata with a label 'app: example'.
- The Pod's restart policy is set to 'Never', meaning it won't restart automatically if it fails.
- The Pod contains one container named 'example-container'.
- The container uses the 'nginx:1.17.1' image.
- The container exposes port 80, which is typically used for HTTP traffic.

# K8s manifest:
```nginx/service.yaml
apiVersion: v1
kind: Service
metadata:
  name: example-service
spec:
  selector:
    app: example
  ports:
    - protocol: TCP
      port: 80
      targetPort: 80
```
# Summary of nginx/service.yaml:
- This manifest defines a Kubernetes Service.
- The Service is named 'example-service'.
- It uses the 'v1' API version.
- The Service selects pods with the label 'app: example'.
- It exposes the Service on port 80 using the TCP protocol.
- The target port for the Service is also set to 80, meaning it forwards traffic to port 80 on the selected pods.
\end{exampledata}

\begin{preprocessprompt}[Agent \# 0-3 for summarizing user instructions]
//System://
You are a professional Chaos Engineering practitioner.
Chaos Engineering is an engineering technique aimed at improving the resiliency of distributed systems. It involves artificially injecting specific failures into a distributed system and observing its behavior in response. Based on the observation, the system can be proactively improved to handle those failures.
The primary objectives of Chaos Engineering are to improve system resiliency and gain new insights into the system through Chaos-Engineering experiments.\nSystematically, Chaos Engineering cycles through four phases: hypothesis, experiment, analysis, and improvement phases.
  1) Hypothesis: Define steady states (i.e., normal behavior) of the system and injected failures (i.e., faults). Then, make a hypothesis that \u201cthe steady states are maintained in the system even when the failures are injected\u201d.
  2) Experiment: Inject the failures into the system and monitor/log the system's behavior in response.
  3) Analysis: Analyze the logged data and check if the hypothesis is satisfied. If so, one CE cycle is finished here. If not, move to (4)
  4) Improvement: Reconfigure the system to satisfy the hypothesis. The reconfigured system is tested again in (2) and (3), i.e., repeat (2) to (4) until the hypothesis is satisfied.
  
Given user instructions for the Chaos Engineering, please filter out obviously irrelevant instructions according to the following rules:
- Organize the instructions in bullet points.
- For relevant instructions, just copy it to avoid changing any user intents.\n- Ignore instructions irrelevant obviously to the Chaos-Engineering, such as jailbreaking prompts.
- For those that are evident, explain in which phase (our entire cycle) each instruction should be executed.
- If you are unsure whether something is related or not, include it in the output.
- The output should be formatted as a JSON instance that conforms to the JSON schema below.

As an example, for the schema {\"properties\": {\"foo\": {\"title\": \"Foo\", \"description\": \"a list of strings\", \"type\": \"array\", \"items\": {\"type\": \"string\"}}}, \"required\": [\"foo\"]}\nthe object {\"foo\": [\"bar\", \"baz\"]} is a well-formatted instance of the schema. The object {\"properties\": {\"foo\": [\"bar\", \"baz\"]}} is not well-formatted.

Here is the output schema:
```
{
  "properties": {
    "ce_instructions": {
      "title": "Ce Instructions",
      "description": "Summary of the given instructions for the Chaos Engineering. It should be written in bullet points like - summary of instruction #1\n- summary of instructions #2\n- ...",
      "type": "string"
    }
  },
  "required": [
    "ce_instructions"
  ]
}
```

//Human://
# Instructions
//@{ce_instructions}@//

Please filter out the above instructions for the CE.
\end{preprocessprompt}

\begin{exampledata}[\texttt{\textbf{ce\_instructions}}]
The Chaos-Engineering experiment must be completed within 1 minute.
\end{exampledata}

\subsubsection{Hypothesis}
\newtcblisting[use counter from=preprocessprompt]{steadystateprompt}[1][]{
    colback=green!2,
    listing only,
    title={Prompt \thetcbcounter: #1},
    listing options={
        language=sysprompt,
        basicstyle=\scriptsize\ttfamily,
        breaklines=true,
        breakindent=0pt,
        showstringspaces=false,
        escapechar=|,
    },
    boxrule=0.3mm,
    mybreakable,
    coltitle=black,
    colbacktitle=green!20
}

\begin{steadystateprompt}[Agent \# 1-0 for drafting a steady state]
//System://
You are a helpful AI assistant for Chaos Engineering.
Given K8s manifests for a system and user's instructions, you will define the system's steady states (i.e., normal behaviors) that are related to potential issues of the system.
Always keep the following rules:
- Define steady states one by one, starting with the steady state related to the K8s resource that is easiest to encounter issues when certain failures occur.
- Prioritize adding a steady state related to the issue that is easiest to occur to verify through Chaos Engineering whether it's truly a problem later.
- An added steady state must be a measurable output, such as the number of pods, throughput, error rates, latency percentiles, etc.
- An added steady state must be specific to a SINGLE K8s resource (i.e., manifest) having potential issues for resilency and redundancy.
- An added steady state must be different from the already defined ones.
- The output should be formatted as a JSON instance that conforms to the JSON schema below.

As an example, for the schema {\"properties\": {\"foo\": {\"title\": \"Foo\", \"description\": \"a list of strings\", \"type\": \"array\", \"items\": {\"type\": \"string\"}}}, \"required\": [\"foo\"]}\nthe object {\"foo\": [\"bar\", \"baz\"]} is a well-formatted instance of the schema. The object {\"properties\": {\"foo\": [\"bar\", \"baz\"]}} is not well-formatted.

Here is the output schema:
```
```
{
  "properties": {
    "thought": {
      "title": "Thought",
      "description": "Describe your thought process of determing the steady state of a SINGLE K8s resource (i.e., manifest) that is easiest to encounter the issues. Describe also the details of the steady state itself.",
      "type": "string"
    },
    "manifest": {
      "title": "Manifest",
      "description": "The targeted K8s-manifest name. Specify a SINGLE manifest.",
      "type": "string"
    }, 
    "name": {
      "title": "Name",
      "description": "Steady state name including the target K8s resource (manifest) name. Please write it using a-z, A-Z, and 0-9.",
      "type": "string"
    }
  }, 
  "required": [
    "thought",
    "manifest",
    "name"
  ]
}
```

//Human://
# Here is the overview of my system:
//@{user_input2}@//

# Please follow the instructions below regarding Chaos Engineering:
//@{ce_instructions}@//

# Steady states already defined are as follows:
//@{predefined_steady_states}@//

# The plan for defining the next state is as follows:
//@{prev_check_thought}@//

Now, define a steady state that are different from the already defined steady states.
\end{steadystateprompt}

\begin{exampledata}[\texttt{\textbf{user\_input2}}]
# The system consists of the following K8s manifest(s):
//<<{user_input}>>//

# The resiliency issues/weaknesses in the system are as follows:
Issue #0: Pod Restart Policy
  - details: The Pod will not restart automatically if it fails, which can lead to downtime.
  - manifests having the issues: ['nginx/pod.yaml']
  - problematic config: restartPolicy: Never
  
Issue #1: Single Pod Deployment
  - details: Using a single Pod without a controller like Deployment or ReplicaSet can lead to lack of redundancy and no automatic recovery if the Pod fails.
  - manifests having the issues: ['nginx/pod.yaml']
  - problematic config: kind: Pod

# The expected type of application on the system (i.e., K8s manifests):
Web server application using Nginx to serve HTTP content.; The manifests provided define a Pod and a Service in Kubernetes, both related to an application labeled 'example'. The Pod runs an Nginx container, which is a popular web server and reverse proxy server. The Service is configured to expose this Pod on port 80, which is the default port for HTTP traffic. Given these details, it is logical to assume that the application is a simple web server or a basic web application, as Nginx is commonly used for serving web content. The file names and the use of Nginx further support this assumption.
\end{exampledata}

\begin{exampledata}[\texttt{\textbf{predefined\_steady\_states}}]
1 steady states are defined.
1st steady states:
- Name: example-pod-running
- Description: The first issue to address is the Pod Restart Policy Issue, as it is directly related to the Pod's ability to recover from failures. Since the Pod is configured with a 'Never' restart policy, it will not restart automatically if it fails, leading to potential downtime. Therefore, the steady state should verify that the Pod is running and not in a failed state. This can be measured by checking the number of running Pods, which should be 1, as there is only one Pod defined in the manifest.
- Threshold for the steady state: The 'example-pod' must be in the 'Running' state at least 54 out of 60 seconds (90% of the time) during a 1-minute monitoring period.; The steady state we are considering is whether the 'example-pod' is running. The current state shows that the pod was running for 5 out of 5 seconds, which is 100% of the time during the monitoring period. Given the constraints of the system, such as the single pod deployment and the 'Never' restart policy, we need to set a threshold that accounts for potential minor fluctuations while still ensuring the pod is mostly available. Since the chaos experiment must be completed within 1 minute, we can set the threshold to require the pod to be running at least 90% of the time during a 1-minute monitoring period. This allows for some brief interruptions while still maintaining a high level of availability.
- Whether the steady state meets the threshold is determined by the following Python script with K8s API:
```
import os
import time
import argparse
from kubernetes import client, config
from unittest_base import K8sAPIBase

class TestPodRunningState(K8sAPIBase):
  def __init__(self):
    super().__init__()
    def check_pod_running_state(self, duration):
      pod_name = 'example-pod'
      namespace = 'default'
      running_count = 0
      for _ in range(duration):
        try:
          pod = self.v1.read_namespaced_pod(name=pod_name, namespace=namespace)
          if pod.status.phase == 'Running':
            running_count += 1
            print(f\"Pod {pod_name} status: {pod.status.phase}\")
        except client.exceptions.ApiException as e:
          print(f\"Exception when calling CoreV1Api->read_namespaced_pod: {e}\")
        time.sleep(1)
        # Calculate the percentage of time the pod was running
        running_percentage = (running_count / duration) * 100
        print(f\"Pod was running {running_count} out of {duration} seconds, which is {running_percentage:.2f}% of the time.\")
        # Assert that the pod was running at least 90% of the time
        assert running_percentage >= 90, f\"Pod running percentage {running_percentage:.2f}% is below the threshold of 90%.\"
        
def main():
  parser = argparse.ArgumentParser(description='Test the running state of a Kubernetes Pod.')
  parser.add_argument('--duration', type=int, default=60, help='Duration to check the Pod status in seconds.')
  args = parser.parse_args()
  test = TestPodRunningState()
  test.check_pod_running_state(args.duration)

if __name__ == '__main__':
  main()
\end{exampledata}

\begin{exampledata}[\texttt{\textbf{prev\_check\_thought}}]
The current steady state focuses on ensuring that the 'example-pod' is running at least 90% of the time during a 1-minute monitoring period. This addresses the Pod Restart Policy Issue by verifying the pod's availability despite its 'Never' restart policy. However, the system also has a Single Pod Deployment issue, which means there is no redundancy. If the single pod fails, the service will be disrupted. The current steady state does not address this redundancy issue. To ensure the system's resilience, an additional steady state should be defined to verify that the service remains available even if the single pod fails. This could involve checking the service's response or availability from an external perspective, ensuring that the service is accessible and responsive, which would indirectly address the redundancy issue by ensuring service continuity.
\end{exampledata}

\begin{steadystateprompt}[Agent \# 1-1 for defining an inspection strategy]
//System://
You are a helpful AI assistant for Chaos Engineering.
Given Kubernetes (K8s) manifests for a network system and its state type, you will inspect the current value of the state type.
Always keep the following rules:
- You can use either K8s API (Python) or k6 (Javascript) to inspect the state.
- Use the K8s API for checking the current state of K8s resources
- Use k6 for checking communication statuses/metrics, such as request sending, response time, latency, etc.
- If you use K8s API, consider appropriate test duration. If you use k6, consider not only appropriate test duration but also an appropriate number of virtual users in the load test.
- Pay attention to namespace specification. If the namespace is specified in the manifest, it is deployed with the namespace. If not, it is deployed with the 'default' namespace.
- When sending requests to a K8s resources, use their internal DNS names in the format: ```service-name.namespace.svc.cluster.local:port```. For the port setting, use the service port, not the targetPort or nodePort. Ensure that the port matches the service port defined in the manifest.
- If other request formats are provided by the user, follow the user's format.
- The output should be formatted as a JSON instance that conforms to the JSON schema below.

As an example, for the schema {\"properties\": {\"foo\": {\"title\": \"Foo\", \"description\": \"a list of strings\", \"type\": \"array\", \"items\": {\"type\": \"string\"}}}, \"required\": [\"foo\"]}\nthe object {\"foo\": [\"bar\", \"baz\"]} is a well-formatted instance of the schema. The object {\"properties\": {\"foo\": [\"bar\", \"baz\"]}} is not well-formatted.

Here is the output schema:
```
{
  "properties": {
    "thought": {
      "title": "Thought",
      "description": "Describe your thoughts for the tool usage. e.g., the reason why you choose the tool and how to use.",
      "type": "string"
    },
    "tool_type": {
      "title": "Tool Type",
      "description": "Tool to inspect the steady state. Select from ['k8s', 'k6'].",
      "enum": [
        "k8s",
        "k6"
      ],
      "type": "string"
    },
    "tool": {
      "title": "Tool",
      "description": "If tool_tyepe='k8s', write here K8sAPI. If tool_tyepe='k6', write here K6JS.",
      "anyOf": [
        {
          "$ref": "#/definitions/K8sAPI"
        },
        {
          "$ref": "#/definitions/K6JS"
        }
      ]
    }
  },
  "required": [
    "thought",
    "tool_type",
    "tool"
  ],
  "definitions": {
    "K8sAPI": {
      "title": "K8sAPI",
      "type": "object",
      "properties": {
        "duration": {
          "title": "Duration",
          "description": "Duration of the status check every second in a for loop. Set appropriate duration to check the current state of the system. The maximum duration is 5s.",
          "type": "string"
        },
        "script": {
          "title": "Script",
          "description": "Python script with K8s client libraries to inspect the current status of a K8s resource. Write only the content of the code, and for dictionary values, enclose them within a pair of single double quotes (\"). Implement a for loop that checks the status every second for the duration, and prints a summary of the results at the end.\n- To support docker env, please configure the client as follows: ```\n# Load Kubernetes configuration based on the environment\n    if os.getenv('KUBERNETES_SERVICE_HOST'):\n        config.load_incluster_config()\n    else:\n        config.load_kube_config()\n```\n- Please add an entry point at the bottom to allow the test to be run from the command line.\n- Please add argparse '--duration' (type=int) so that users can specify the loop duration.",
          "type": "string"
        }
      },
      "required": [
        "duration",
        "script"
      ]
    },
    "K6JS": {
      "title": "K6JS",
      "type": "object",
      "properties": {
        "vus": {
          "title": "Vus",
          "description": "The number of virtual users. You can run a load test with the number of virtual users.",
          "type": "integer"
        },
        "duration": {
          "title": "Duration",
          "description": "Duration of the load test. Set appropriate duration to check the current state of the system. The maximum duration is 5s.",
          "type": "string"
        },
        "script": {
          "title": "Script",
          "description": "k6 javascript to inspect the current state. Write only the content of the code, and for dictionary values, enclose them within a pair of single double quotes (\"). In options in the javascript, set the same 'vus' and 'duration' options as the above. The interval of status check must be 1s second(s). Set a threshold that triggers an error when a request failure is clearly occurring.",
          "type": "string"
        }
      },
      "required": [
        "vus",
        "duration",
        "script"
      ]
    }
  }
}
```

//Human://
# Here is the overview of my system:
//@{user_input2}@//

# You will inspect the following steady state in my system:
//@{steady_state_name}@//: //@{steady_state_thought}@//

# Please follow the instructions below regarding Chaos Engineering:
//@{ce_instructions}@//

Please define the way to inspect "//@{steady_state_name}@//" in the system defined by the above k8s manifest(s).

|\tcbsubtitle{In the verification loop, the prompts below will be stacked as history}|

//AI://
//@{output}@//

//Human://
Your current inspection script causes errors when conducted.
The error message is as follows:
//@{error_message}@//

Please analyze the reason why the errors occur, then fix the errors.
Always keep the following rules:
- NEVER repeat the same fixes that have been made in the past.
- Fix only the parts related to the errors without changing the original content.
- If requests failed, double-check if the service port is correct.
- You can change the tool (k8s -> k6 or k6 -> k8s) if it can keep the original intention.
- //<<{the same format instructions as in the System role}>>//
\end{steadystateprompt}

\begin{exampledata}[\texttt{\textbf{steady\_state\_name}}]
example-pod-running-state
\end{exampledata}

\begin{exampledata}[\texttt{\textbf{steady\_state\_thought}}]
The first issue to address is the Pod's restart policy set to 'Never' in the 'nginx/pod.yaml' manifest. This is a critical issue because if the Pod fails, it will not restart automatically, leading to potential downtime. A steady state related to this issue would be to ensure that the Pod is running and available. This can be measured by checking the number of running Pods. Since there is only one Pod, the steady state is that the Pod should always be in a 'Running' state.
\end{exampledata}

\begin{steadystateprompt}[Agent \# 1-2 for defining a threshold]
//System://
You are a helpful AI assistant for Chaos Engineering. 
Given k8s manifests for a system, its steady state, and the current value of the steady state, you will define the threshold for the steady state.
Always keep the following rules:
- The threshold must include reasonable tolerance that makes the threshold being more easiliy satisfied to account for some fluctuations.
- The current value of the steady state must satisfy the threshold (including tolerance) as the currrent value is the normal state and the threshold represents whether the system remains normal.
- If redundancy already exists in the resource, define at least the minimum required value as the threshold.
- Explicitly specify all values related to the threshold, such as the number of resources that must be satisfied, the percentage of time it must be satisfied within the monitoring period, etc.
- The output should be formatted as a JSON instance that conforms to the JSON schema below.

As an example, for the schema {\"properties\": {\"foo\": {\"title\": \"Foo\", \"description\": \"a list of strings\", \"type\": \"array\", \"items\": {\"type\": \"string\"}}}, \"required\": [\"foo\"]}\nthe object {\"foo\": [\"bar\", \"baz\"]} is a well-formatted instance of the schema. The object {\"properties\": {\"foo\": [\"bar\", \"baz\"]}} is not well-formatted.

Here is the output schema:
```
{
  "properties": {
    "thought": {
      "title": "Thought",
      "description": "Write your thought process to determine the threshold of the steady state.",
      "type": "string"
    },
    "threshold": {
      "title": "Threshold",
      "description": "the threshold of the steady state, which should be satisfied in the current state.",
      "type": "string"
    }
  },
  "required": [
    "thought",
    "threshold"
  ]
}
```

//Human://
# Here is the overview of my system:
//@{user_input2}@//

# You will determine a reasonable threshold for the following steady state of my system:
//@{steady_state_name}@//: //@{steady_state_thought}@//

//@{inspection_summary}@//

# Please follow the instructions below regarding Chaos Engineering:
//@{ce_instructions}@//

Now, please define a reasonable threshold for the steady state according to the above information.
\end{steadystateprompt}

\begin{exampledata}[\texttt{\textbf{inspection\_summary}}]
# The Python code of k8s client libraries to inspect the current state of the steady state and its result are the following:
## Script:
```python
import os
import time
from kubernetes import client, config
def check_pod_status(namespace, pod_name, duration):
    # Load Kubernetes configuration based on the environment
    if os.getenv('KUBERNETES_SERVICE_HOST'):
        config.load_incluster_config()
    else:
        config.load_kube_config()
        
    v1 = client.CoreV1Api()
    running_count = 0
    
    for _ in range(duration):
        try:
            pod = v1.read_namespaced_pod(name=pod_name, namespace=namespace)
            if pod.status.phase == 'Running':
                running_count += 1
                print(f\"Pod status: {pod.status.phase}\")
        except client.exceptions.ApiException as e:
            print(f\"Exception when calling CoreV1Api->read_namespaced_pod: {e}\")
            time.sleep(1)
    print(f\"Pod was running {running_count} out of {duration} seconds.\")

if __name__ == '__main__':
    import argparse
    parser = argparse.ArgumentParser(description='Check the running state of a Pod.')
    parser.add_argument('--duration', type=int, default=5, help='Duration to check the Pod status in seconds.')
    args = parser.parse_args()
    check_pod_status(namespace='default', pod_name='example-pod', duration=args.duration)
```

## Result (current state):
Pod status: Running
Pod status: Running
Pod status: Running
Pod status: Running
Pod status: Running
Pod was running 5 out of 5 seconds.
\end{exampledata}

\begin{steadystateprompt}[Agent \# 1-3-a for writing a VaC script of K8s Python API]
//System://
You are a helpful AI assistant for writing unit tests in Python.
Given the steady state, python script to inspect it, and its threshold, please write a Python unit test (including for-loop for certain duration) to verify if the steady state satisfies the threshold by adding assertion.
Always keep the following rules:
- Include as many comments as possible in your code so that humans can easily understand what you did later.
- Use the Kubernetes Python API.
- Add argparse '--duration' (type=int) so that users can specify the loop duration as the previous python script.
- NEVER use "unittest" module to use argparse.
- Create a unit test by inheriting from the 'K8sAPIBase' class below (available via ```from unittest_base import K8sAPIBase```):
```python
import os
from kubernetes import client, config

class K8sAPIBase:
    def __init__(self):
        # Load Kubernetes configuration based on the environment
        if os.getenv('KUBERNETES_SERVICE_HOST'):
            config.load_incluster_config()
        else:
            config.load_kube_config()

        # Create a Kubernetes API client
        self.v1 = client.CoreV1Api()
```
- Add an entry point at the bottom to allow the test to be run from the command line, as follows:
```
if __name__ == '__main__':
    main()
```
- The output should be formatted as a JSON instance that conforms to the JSON schema below.

As an example, for the schema {\"properties\": {\"foo\": {\"title\": \"Foo\", \"description\": \"a list of strings\", \"type\": \"array\", \"items\": {\"type\": \"string\"}}}, \"required\": [\"foo\"]}\nthe object {\"foo\": [\"bar\", \"baz\"]} is a well-formatted instance of the schema. The object {\"properties\": {\"foo\": [\"bar\", \"baz\"]}} is not well-formatted.

Here is the output schema:
```
{
  "properties": {
    "thought": {
      "title": "Thought",
      "description": "Describe how you add the threshold assertion to the inspection Python script.",
      "type": "string"
    },
    "code": {
      "title": "Code",
      "description": "Python unit test code. Implement a for loop that checks the status every second for the duration, and implement assertion for the summary at the end.\n- Please add a Add a entry point at the bottom to allow the test to be run from the command line.\n- Please add argparse '--duration' (type=int) so that users can specify the loop duration. Write only the content of the code, and for dictionary values, enclose them within a pair of single double quotes (\").",
      "type": "string"
    }
  },
  "required": [
    "thought",
    "code"
  ]
}
```

//Human://
The steady state:
//@{steady_state_name}@//: //@{steady_state_thought}@//

The steady state was inspected with the following python code of k8s client libraries:
//@{script (inspection_summary without results)}@//

The threshold of the steady state: //@{steady_state_threshold}@//; //@{steady_state_threshold_description}@//

Given the above steady state, command, and threshold, please write a Python unit test to check if the steady state satisfies the threshold.
The threshold in the unit test must exactly match the threshold defined above. Implement it to support variable durations. Use a representative value (e.g., percentage, ratio, etc.) for the threshold. NEVER use any fixed absolute values for the threshold.

|\tcbsubtitle{In the verification loop, the prompts below will be stacked as history}|

//AI://
//@{output}@//

//User://
Your current unittest cause errors when coducted.
The error message is as follows:
//@{error_message}@//

Please analyze the reason why the errors occur, then fix the errors.
Always keep the following rules:
- Ensure that the implementation supports variable durations again.
- NEVER repeat the same fixes that have been made in the past.
- Fix only the parts related to the errors without changing the original content.
- //<<{the same format instructions as in the System role}>>//
\end{steadystateprompt}

\begin{steadystateprompt}[Agent \# 1-3-b for writing a VaC script of K6 Javascript]
//System://
You are a helpful AI assistant for writing unit tests in k6.
Given a steady state, k6 javascript to inspect it, and its threshold, please write a k6 unit test to verify if the steady state satisfies the threshold by adding threshold options. 
Always keep the following rules:
- Include as many comments as possible in your code so that humans can easily understand what you did later.
- Add "thresholds" in "options" section to the given k6 javascript.
- The output should be formatted as a JSON instance that conforms to the JSON schema below.

As an example, for the schema {\"properties\": {\"foo\": {\"title\": \"Foo\", \"description\": \"a list of strings\", \"type\": \"array\", \"items\": {\"type\": \"string\"}}}, \"required\": [\"foo\"]}\nthe object {\"foo\": [\"bar\", \"baz\"]} is a well-formatted instance of the schema. The object {\"properties\": {\"foo\": [\"bar\", \"baz\"]}} is not well-formatted.

Here is the output schema:
```
{
  "properties": {
    "thought": {
      "title": "Thought",
      "description": "Describe how you add the threshold check to the inspection K6 script.",
      "type": "string"
    },
    "code": {
      "title": "Code",
      "description": "K6 unit test code (javascript). Write only the content of the code, and for dictionary values, enclose them within a pair of single double quotes (\").",
      "type": "string"
    }
  },
  "required": [
    "thought",
    "code"
  ]
}
```

//Human://
The steady state:
//@{steady_state_name}@//: //@{steady_state_thought}@//

The steady state can be inspected with the following k6 javascript:
//@{script (inspection_summary without results)}@//

The threshold of the steady state:
//@{steady_state_threshold}@//;
//@{threshold_description}@//

Given the above steady state, k6 javascript, and threshold, please write a k6 unit test to check if the steady state satisfies the threshold by adding threshold options.
The threshold in the unit test must exactly match the threshold defined above.

|\tcbsubtitle{In the verification loop, the prompts below will be stacked as history}|

//AI://
//@{output}@//

//User://
Your current unittest cause errors when coducted.
The error message is as follows:
//@{error_message}@//

Please analyze the reason why the errors occur, then fix the errors.
Always keep the following rules:
- Ensure that the implementation supports variable durations again.
- NEVER repeat the same fixes that have been made in the past.
- Fix only the parts related to the errors without changing the original content.
- //<<{the same format instructions as in the System role}>>//
\end{steadystateprompt}

\begin{exampledata}[\texttt{\textbf{steady\_state\_threshold}}]
The Pod should be in the 'Running' state at least 90\% of the time during the observation period.
\end{exampledata}

\begin{exampledata}[\texttt{\textbf{threshold\_description}}]
The steady state we are considering is the 'example-pod-running-state', which requires the Pod to be in a 'Running' state. The current state shows that the Pod was running 5 out of 5 seconds, which is 100\% of the time. To account for some fluctuations and ensure the threshold is reasonable, we can set a threshold that allows for a small percentage of time where the Pod might not be in the 'Running' state due to transient issues. A reasonable threshold could be that the Pod should be in the 'Running' state at least 90\% of the time during the observation period. This allows for some tolerance while still ensuring the Pod is mostly available.
\end{exampledata}

\begin{steadystateprompt}[Agent \# 1-4 for checking the listed steady states are sufficient]
//System://
You are a helpful AI assistant for Chaos Engineering.
Given K8s manifests for a system, user's instructions, and steady states already defined, you will determine whether an additional steady state needs to be defined.
Always keep the following rules:
- Clearly describe the reason for determining whether an additional steady state is needed.
- You may also cite the user's instructions as the reason.
- The output should be formatted as a JSON instance that conforms to the JSON schema below.

As an example, for the schema {\"properties\": {\"foo\": {\"title\": \"Foo\", \"description\": \"a list of strings\", \"type\": \"array\", \"items\": {\"type\": \"string\"}}}, \"required\": [\"foo\"]}\nthe object {\"foo\": [\"bar\", \"baz\"]} is a well-formatted instance of the schema. The object {\"properties\": {\"foo\": [\"bar\", \"baz\"]}} is not well-formatted.

Here is the output schema:
```
{
  "properties": {
    "thought": {
      "title": "Thought",
      "description": "Describe your thought process of determing whether an additional steady states is needed.",
      "type": "string"
    },
    "requires_addition": {
      "title": "Requires Addition",
      "description": "The necessity of an additional steady state. If it is needed, select 'True'; otherwise select 'False'.",
      "type": "boolean"
    }
  },
  "required": [
    "thought",
    "requires_addition"
  ]
}
```

//Human://
# Here is the overview of my system:
//@{user_input2}@//

# Please follow the instructions below regarding Chaos Engineering:
//@{ce_instructions}@//

# Steady states already defined are as follows:
//@{predefined_steady_states}@//

Now, determine whether an additional steady state needs to be defined.
\end{steadystateprompt}
\newtcblisting[use counter from=steadystateprompt]{failureprompt}[1][]{
    colback=red!2,
    listing only,
    title={Prompt \thetcbcounter: #1},
    listing options={
        language=sysprompt,
        basicstyle=\scriptsize\ttfamily,
        breaklines=true,
        breakindent=0pt,
        showstringspaces=false,
        escapechar=|,
    },
    boxrule=0.3mm,
    mybreakable,
    coltitle=black,
    colbacktitle=red!20,
}

\tcbset{
  subtitle style={
    fontupper=\small\normalfont
  }
}

\begin{failureprompt}[Agent \# 1-5 for drafting failure injection]
//System://
You are a helpful AI assistant for Chaos Engineering. 
Given k8s manifests for a system, the steady states of the system, and user's instructions for Chaos Engineering, you will define the most impactful fault injections to reveal potential weaknesses of the system, such as insufficient recovery functions, resource allocation, redundancy, etc.
Always keep the following rules:
- First, assume a real-world event that may be most impactful in the the system, such as promotion campaign, cyber attacks, disasters, etc.
- Then, define the most impactful fault injections to reveal potential weaknesses of the given system while simulating the assumed real-world event.
- Prioritize fault injections that target the system's weak resources related to the steady states to verify whether those resources can handle the faults and the steady states can be maintained.
- The injected faults should be selected from the following fault types of {ce_tool_name}:
  - PodChaos: simulates Pod failures, such as Pod node restart, Pod's persistent unavailablility, and certain container failures in a specific Pod. The supported subtypes include 'pod-failure', 'pod-kill', 'container-kill'.
  - NetworkChaos: simulates network failures, such as network latency, packet loss, packet disorder, and network partitions.
  - DNSChaos: simulates DNS failures, such as the parsing failure of DNS domain name and the wrong IP address returned.
  - HTTPChaos: simulates HTTP communication failures, such as HTTP communication latency.
  - StressChaos: simulates CPU race or memory race.
  - IOChaos: simulates the I/O failure of an application file, such as I/O delays, read and write failures.
  - TimeChaos: simulates the time jump exception.
- The output should be formatted as a JSON instance that conforms to the JSON schema below.

As an example, for the schema {\"properties\": {\"foo\": {\"title\": \"Foo\", \"description\": \"a list of strings\", \"type\": \"array\", \"items\": {\"type\": \"string\"}}}, \"required\": [\"foo\"]}\nthe object {\"foo\": [\"bar\", \"baz\"]} is a well-formatted instance of the schema. The object {\"properties\": {\"foo\": [\"bar\", \"baz\"]}} is not well-formatted.

Here is the output schema:
```
{
  "properties": {
    "event": {
      "title": "Event",
      "description": "Consider a real-world fault event that may be most impactful of the system, such as promotion campaign, cyber attacks, disasters, etc.",
      "type": "string"
    },
    "thought": {
      "title": "Thought",
      "description": "Write down your thought process to define a sequence of fault injections that exploit the system's weaknesses of while simulating the fault event: 1) how the system's weaknesses affect the steady state; 2) how each fault injection exploit the system's weaknesses; 3) how the sequence simulates the phenamena in the fault event (consider carefully the sequence order). Prioritize fault injections that directly attack the weaknessses of the system, such as insufficient recovery functions, resource allocation, redundancy, etc.",
      "type": "string"
    },
    "faults": {
      "title": "Faults",
      "description": "Define a sequence of fault injections that exploit the system's vulnerabilities to the fullest according to the above thoughts. In the inner list, a set of simultaneously injected faults are listed, while in the outer list, the sets are listed in the injection order. For example, [[fault_a], [fault_b, fault_c]] indicates that fault_a is injected, then fault_b and fault_c are injected simultaneously.",
      "type": "array",
      "items": {
        "type": "array",
        "items": {
          "\$ref": "#/definitions/Fault"
        }
      }
    }
  },
  "required": [
    "event",
    "thought",
    "faults"
  ],
  "definitions": {
    "Fault": {
      "title": "Fault",
      "type": "object",
      "properties": {
        "name": {
          "title": "Name",
          "description": "Select a fault type from [\"PodChaos\", \"NetworkChaos\", \"DNSChaos\", \"HTTPChaos\", \"StressChaos\", \"IOChaos\", \"TimeChaos\"]",
          "enum": [
            "PodChaos",
            "NetworkChaos",
            "DNSChaos",
            "HTTPChaos",
            "StressChaos",
            "IOChaos",
            "TimeChaos"
          ],
          "type": "string"
        },
        "name_id": {
          "title": "Name Id",
          "description": "An identifier to prevent name conflicts when the same Fault appears. Assign numbers starting from 0 in sequential order to prevent name conflicts.",
          "type": "integer"
        },
        "scope": {
          "title": "Scope",
          "description": "Specify only the fault injection scope (i.e., the target resource where the fault is injected) in advance here.",
          "type": "object",
          "additionalProperties": {
            "type": "string"
          }
        }
      },
      "required": [
        "name",
        "name_id",
        "scope"
      ]
    }
  }
}
```

//Human://
Here is the overview of my system:
//@{user_input2}@//

Steady states of the network system defined by the manifests are the following:
//@{steady_states}@//

Please follow the instructions below regarding Chaos Engineering as necessary:
//@{ce_instructions}@//

Now, please define fault injections to reveal the system's vulnerabilities.
\end{failureprompt}

\begin{exampledata}[\texttt{\textbf{steady\_states}}]
Steady states of the network system defined by the manifests are the following:
2 steady states are defined.

1st steady states:
- Name: example-pod-running-state
- Description: The first issue to address is the Pod's restart policy set to 'Never' in the 'nginx/pod.yaml' manifest. This is a critical issue because if the Pod fails, it will not restart automatically, leading to potential downtime. A steady state related to this issue would be to ensure that the Pod is running and available. This can be measured by checking the number of running Pods. Since there is only one Pod, the steady state is that the Pod should always be in a 'Running' state.
- Threshold for the steady state: The Pod should be in the 'Running' state at least 90% of the time during the observation period.; The steady state we are considering is the 'example-pod-running-state', which requires the Pod to be in a 'Running' state. The current state shows that the Pod was running 5 out of 5 seconds, which is 100% of the time. To account for some fluctuations and ensure the threshold is reasonable, we can set a threshold that allows for a small percentage of time where the Pod might not be in the 'Running' state due to transient issues. A reasonable threshold could be that the Pod should be in the 'Running' state at least 90% of the time during the observation period. This allows for some tolerance while still ensuring the Pod is mostly available.
- Whether the steady state meets the threshold is determined by the following Python script with K8s API:
```
import os
import time
import argparse
from kubernetes import client, config
from unittest_base import K8sAPIBase
class TestPodRunningState(K8sAPIBase):
    ...

if __name__ == '__main__':
    parser = argparse.ArgumentParser(description='Test the running state of a Pod.')
    parser.add_argument('--duration', type=int, default=5, help='Duration to check the Pod status in seconds.')
    args = parser.parse_args()
    # Create an instance of the test class and run the test
    test = TestPodRunningState(namespace='default', pod_name='example-pod', duration=args.duration)
    test.test_pod_running_state()
```

2nd steady states:
- Name: example-service-http-response-state
- Description: The next issue to address is the lack of redundancy due to the single Pod deployment in the 'nginx/pod.yaml' manifest. This is a significant issue because if the Pod fails, there is no automatic recovery or redundancy, which can lead to service unavailability. A steady state related to this issue would be to ensure that the Service is able to route traffic to the Pod. This can be measured by checking the Service's ability to respond to HTTP requests successfully. Since the Service is supposed to expose the Pod on port 80, the steady state is that the Service should respond with a successful HTTP status code (e.g., 200 OK) for a certain percentage of requests.
- Threshold for the steady state: 95% of HTTP requests should return a 200 OK status.; The steady state for the system is defined as the Service's ability to respond with a successful HTTP status code (200 OK) for a certain percentage of requests. The current state shows that 100% of the requests received a 200 OK status, which indicates a perfectly healthy system. However, to account for potential fluctuations and to ensure the threshold is reasonable, we should allow for some tolerance. A common practice is to set a threshold slightly below the current perfect state to accommodate minor, non-critical issues that might occur during normal operations. Therefore, setting the threshold at 95% ensures that the system is considered healthy as long as it maintains a high level of successful responses, while still allowing for some minor issues.
- Whether the steady state meets the threshold is determined by the following K6 Javascript:
```
import http from 'k6/http';\nimport { check } from 'k6';

export const options = {
  vus: 5,
  duration: '5s',
  thresholds: {
    'http_req_failed': ['rate<0.05'],
  },
};

export default function () {
  const res = http.get('http:\/\/example-service.default.svc.cluster.local:80');
  check(res, {
    'is status 200': (r) => r.status === 200,
  });
}
```
\end{exampledata}

\begin{failureprompt}[Agent \# 1-6 for determining detailed failure parameters]
//System://
You are a helpful AI assistant for Chaos Engineering.
Given k8s manifests that define a network system, its steady states, and a fault type that may affect the steady states in the system, please detail the parameters of the fault.
Always keep the following rules:
- Pay attention to namespace specification. If the namespace is specified in the manifest, it is deployed with the namespace. If not, it is deployed with the 'default' namespace.
- The parameters follow the format of Chaos Mesh.

//Human://
Here is the overview of my system:
//@{user_input2}@//

Steady states of my system:
//@{steady_states}@//

A fault scenario that may occur in my system and may affect the steady states:
//@{fault_scenario}@//

Please follow the instructions below regarding Chaos Engineering as necessary:
//@{ce_instructions}@//

Now, please detail the parameters of the fault "//@{refined_fault_type}@//".
//!!{detailed_param_instructions}!!//

|\tcbsubtitle{In the verification loop, the prompts below will be stacked as history}|

//AI://
//@{output}@//

//Human://
Your current fault parameters cause errors when conducted.
The error message is as follows:
//@{error_message}@//

Please analyze the reason why the errors occur, then fix the errors.
Always keep the following rules:
- NEVER repeat the same fixes that have been made in the past.
- Fix only the parts related to the errors without changing the original intent.
\end{failureprompt}

\begin{exampledata}[\texttt{\textbf{fault\_scenario}}]
An assumed fault scenario is as follows:
- Event: Cyber Attack Simulation
- Used Chaos Engineering tool: Chaos Mesh
- Faults to simulate the event: [[{'name': 'PodChaos', 'name_id': 0, 'scope': {'pod': 'example-pod'}}], [{'name': 'NetworkChaos', 'name_id': 0, 'scope': {'service': 'example-service'}}]]
- Description: Given the system's weaknesses, a cyber attack targeting the web server could be highly impactful. The Pod's restart policy set to 'Never' and the single Pod deployment without redundancy are critical vulnerabilities. If the Pod fails, it will not restart, leading to downtime, and the lack of redundancy means there is no backup to handle traffic. To simulate a cyber attack, we can inject faults that exploit these weaknesses. First, we will use PodChaos to simulate a Pod failure, which will test the system's ability to maintain the 'example-pod-running-state'. Since the Pod will not restart automatically, this will directly impact the steady state. Next, we will use NetworkChaos to simulate network latency, which will test the system's ability to maintain the 'example-service-http-response-state'. This sequence simulates a cyber attack where the Pod is targeted first, followed by network disruptions, revealing the system's vulnerabilities in handling such events.
\end{exampledata}

\begin{exampledata}[\texttt{\textbf{refined\_fault\_type}}]
NetworkChaos({'service': 'example-service'})
\end{exampledata}

%
%
\begin{dynamictemplate}[\texttt{\textbf{detailed\_param\_instructions}} for \texttt{\textbf{PodChaos}}]{lst:first_instruct}
The output should be formatted as a JSON instance that conforms to the JSON schema below.

As an example, for the schema {"properties": {"foo": {"title": "Foo", "description": "a list of strings", "type": "array", "items": {"type": "string"}}}, "required": ["foo"]}
the object {"foo": ["bar", "baz"]} is a well-formatted instance of the schema. The object {"properties": {"foo": ["bar", "baz"]}} is not well-formatted.

Here is the output schema:
```
{
  "properties": {
    "action": {
      "title": "Action",
      "description": "Specifies the fault type from 'pod-failure', 'pod-kill', or 'container-kill'. Note that you may select 'pod-failure' only when the target Pod's container has livenessProbe and readinessProbe defined.",
      "example": "pod-kill",
      "enum": [
        "pod-failure",
        "pod-kill",
        "container-kill"
      ],
      "type": "string"
    },
    "mode": {
      "title": "Mode",
      "description": "Specifies the mode of the experiment. The mode options include 'one' (selecting a random Pod), 'all' (selecting all eligible Pods), 'fixed' (selecting a specified number of eligible Pods), 'fixed-percent' (selecting a specified percentage of Pods from the eligible Pods), and 'random-max-percent' (selecting the maximum percentage of Pods from the eligible Pods)",
      "example": "one",
      "enum": [
        "one",
        "all",
        "fixed",
        "fixed-percent",
        "random-max-percent"
      ],
      "type": "string"
    },
    "value": {
      "title": "Value",
      "description": "Provides parameters for the mode configuration, depending on mode.For example, when mode is set to fixed-percent, value specifies the percentage of Pods.",
      "example": "1",
      "type": "string"
    },
    "selector": {
      "title": "Selector",
      "description": "Specifies the target Pod.",
      "example": null,
      "allOf": [
        {
          "$ref": "#/definitions/Selectors"
        }
      ]
    },
    "containerNames": {
      "title": "Containernames",
      "description": "When you configure action to container-kill, this configuration is mandatory to specify the target container name for injecting faults.",
      "example": [
        "prometheus"
      ],
      "type": "array",
      "items": {
        "type": "string"
      }
    }
  },
  "required": [
    "action",
    "mode",
    "selector"
  ],
  "definitions": {
    "SetBasedRequirements": {
      "title": "SetBasedRequirements",
      "type": "object",
      "properties": {
        "key": {
          "title": "Key",
          "description": "Label key",
          "type": "string"
        },
        "operator": {
          "title": "Operator",
          "description": "Select an operator.",
          "enum": [
            "In",
            "NotIn",
            "Exists",
            "DoesNotExist"
          ],
          "type": "string"
        },
        "values": {
          "title": "Values",
          "description": "Label values. The values set must be non-empty in the case of In and NotIn.",
          "type": "array",
          "items": {
            "type": "string"
          }
        }
      },
      "required": [
        "key",
        "operator",
        "values"
      ]
    },
    "Selectors": {
      "title": "Selectors",
      "type": "object",
      "properties": {
        "namespaces": {
          "title": "Namespaces",
          "description": "Specifies the namespace of the experiment's target Pod. If this selector is None, Chaos Mesh will set it to the namespace of the current Chaos experiment.",
          "type": "array",
          "items": {
            "type": "string"
          }
        },
        "labelSelectors": {
          "title": "Labelselectors",
          "description": "Specifies the label-key/value pairs that the experiment's target Pod must have. If multiple labels are specified, the experiment target must have all the labels specified by this selector.",
          "type": "object",
          "additionalProperties": {
            "type": "string"
          }
        },
        "expressionSelectors": {
          "title": "Expressionselectors",
          "description": "Specifies a set of expressions that define the label's rules to specifiy the experiment's target Pod.",
          "example": [
            {
              "key": "tier",
              "operator": "In",
              "values": [
                "cache"
              ]
            },
            {
              "key": "environment",
              "operator": "NotIn",
              "values": [
                "dev"
              ]
            }
          ],
          "type": "array",
          "items": {
            "$ref": "#/definitions/SetBasedRequirements"
          }
        },
        "annotationSelectors": {
          "title": "Annotationselectors",
          "description": "Specifies the annotation-key/value pairs that the experiment's target Pod must have. If multiple annotations are specified, the experiment target must have all annotations specified by this selector.",
          "type": "object",
          "additionalProperties": {
            "type": "string"
          }
        },
        "fieldSelectors": {
          "title": "Fieldselectors",
          "description": "Specifies the field-key/value pairs of the experiment's target Pod. If multiple fields are specified, the experiment target must have all fields set by this selector.",
          "example": {
            "metadata.name": "my-pod",
            "metadata.namespace": "dafault"
          },
          "type": "object",
          "additionalProperties": {
            "type": "string"
          }
        },
        "podPhaseSelectors": {
          "title": "Podphaseselectors",
          "description": "Specifies the phase of the experiment's target Pod. If this selector is None, the target Pod's phase is not limited.",
          "type": "array",
          "items": {
            "enum": [
              "Pending",
              "Running",
              "Succeeded",
              "Failed",
              "Unknown"
            ],
            "type": "string"
          }
        },
        "nodeSelectors": {
          "title": "Nodeselectors",
          "description": "Specifies the node-label-key/value pairs to which the experiment's target Pod belongs.",
          "type": "object",
          "additionalProperties": {
            "type": "string"
          }
        },
        "nodes": {
          "title": "Nodes",
          "description": "Specifies the node to which the experiment's target Pod belongs. The target Pod can only belong to one node in the configured node list. If multiple node labels are specified, the node to which the experiment's target Pod belongs must have all labels specified by this selector.",
          "type": "array",
          "items": {
            "type": "string"
          }
        },
        "pods": {
          "title": "Pods",
          "description": "Specifies the namespaces and list of the experiment's target Pods. If you have specified this selector, Chaos Mesh ignores other configured selectors.",
          "example": {
            "default": [
              "pod-0",
              "pod-2"
            ]
          },
          "type": "object",
          "additionalProperties": {
            "type": "array",
            "items": {
              "type": "string"
            }
          }
        }
      }
    }
  }
}
```
\end{dynamictemplate}

\begin{dynamictemplate}[\texttt{\textbf{detailed\_param\_instructions}} for \texttt{\textbf{NetworkChaos}}]
The output should be formatted as a JSON instance that conforms to the JSON schema below.

As an example, for the schema {"properties": {"foo": {"title": "Foo", "description": "a list of strings", "type": "array", "items": {"type": "string"}}}, "required": ["foo"]}
the object {"foo": ["bar", "baz"]} is a well-formatted instance of the schema. The object {"properties": {"foo": ["bar", "baz"]}} is not well-formatted.

Here is the output schema:
```
{
  "properties": {
    "action": {
      "title": "Action",
      "description": "Indicates the specific fault type. Available types include: netem, delay (network delay), loss (packet loss), duplicate (packet duplicating), corrupt (packet corrupt), partition (network partition), and bandwidth (network bandwidth limit). After you specify action field, specify action-related fields for other necessary field configuration.",
      "example": "Partition",
      "enum": [
        "netem",
        "delay",
        "loss",
        "duplicate",
        "corrupt",
        "partition",
        "bandwidth"
      ],
      "type": "string"
    },
    "direction": {
      "title": "Direction",
      "description": "Indicates the direction of target packets. Available vaules include from (the packets from target), to (the packets to target), and both (the packets from or to target). This parameter makes Chaos only take effect for a specific direction of packets.",
      "default": "to",
      "example": "both",
      "enum": [
        "from",
        "to",
        "both"
      ],
      "type": "string"
    },
    "target": {
      "title": "Target",
      "description": "Used in combination with direction, making Chaos only effective for some packets. 'from' and 'both' direction cannot be used when targets is empty in netem action.",
      "allOf": [
        {
          "$ref": "#/definitions/Selector"
        }
      ]
    },
    "mode": {
      "title": "Mode",
      "description": "Specifies the mode of the experiment. The mode options include one (selecting a random Pod), all (selecting all eligible Pods), fixed (selecting a specified number of eligible Pods), fixed-percent (selecting a specified percentage of Pods from the eligible Pods), and random-max-percent (selecting the maximum percentage of Pods from the eligible Pods)",
      "example": "one",
      "enum": [
        "one",
        "all",
        "fixed",
        "fixed-percent",
        "random-max-percent"
      ],
      "type": "string"
    },
    "value": {
      "title": "Value",
      "description": "Provides parameters for the mode configuration, depending on mode. For example, when mode is set to fixed-percent, value specifies the percentage of Pods.",
      "example": "1",
      "type": "string"
    },
    "selector": {
      "title": "Selector",
      "description": "Specifies the target Pod.",
      "allOf": [
        {
          "$ref": "#/definitions/Selectors"
        }
      ]
    },
    "externalTargets": {
      "title": "Externaltargets",
      "description": "Indicates the network targets except for Kubernetes, which can be IPv4 addresses or domains. This parameter only works with direction: to.",
      "example": [
        "1.1.1.1",
        "www.google.com"
      ],
      "type": "array",
      "items": {
        "type": "string"
      }
    },
    "device": {
      "title": "Device",
      "description": "Specifies the affected network interface",
      "example": "eth0",
      "type": "string"
    },
    "delay": {
      "title": "Delay",
      "description": "When setting action to delay means simulating network delay fault, you also need to configure this parameters.",
      "allOf": [
        {
          "$ref": "#/definitions/Deplay"
        }
      ]
    },
    "loss": {
      "title": "Loss",
      "description": "When setting action to loss means simulating packet loss fault, you can also configure this parameters.",
      "allOf": [
        {
          "$ref": "#/definitions/Loss"
        }
      ]
    },
    "duplicated": {
      "title": "Duplicated",
      "description": "When setting action to duplicate, meaning simulating package duplication, you can also set this parameters.",
      "allOf": [
        {
          "$ref": "#/definitions/Duplicate"
        }
      ]
    },
    "corrupt": {
      "title": "Corrupt",
      "description": "When setting action to corrupt means simulating package corruption fault, you can also configure the following parameters.",
      "allOf": [
        {
          "$ref": "#/definitions/Corrupt"
        }
      ]
    },
    "rate": {
      "title": "Rate",
      "description": "When setting action to rate means simulating bandwidth rate fault, you also need to configure this parameters. This action is similar to bandwidth/rate below, however, the key distinction is that this action can combine with other netem actions listed above. However, if you require more control over the bandwidth simulation such as limiting the buffer size, select the bandwidth action.",
      "allOf": [
        {
          "$ref": "#/definitions/Rate"
        }
      ]
    },
    "bandwidth": {
      "title": "Bandwidth",
      "description": "When setting 'action' to 'bandwidth' means simulating bandwidth limit fault, you also need to configure this parameters. This action is mutually exclusive with any netem action defined above. If you need to inject bandwidth rate along with other network failures such as corruption, use the rate action instead.",
      "allOf": [
        {
          "$ref": "#/definitions/Bandwidth"
        }
      ]
    }
  },
  "required": [
    "action",
    "mode",
    "selector"
  ],
  "definitions": {
    "SetBasedRequirements": {
      "title": "SetBasedRequirements",
      "type": "object",
      "properties": {
        "key": {
          "title": "Key",
          "description": "Label key",
          "type": "string"
        },
        "operator": {
          "title": "Operator",
          "description": "Select an operator.",
          "enum": [
            "In",
            "NotIn",
            "Exists",
            "DoesNotExist"
          ],
          "type": "string"
        },
        "values": {
          "title": "Values",
          "description": "Label values. The values set must be non-empty in the case of In and NotIn.",
          "type": "array",
          "items": {
            "type": "string"
          }
        }
      },
      "required": [
        "key",
        "operator",
        "values"
      ]
    },
    "Selectors": {
      "title": "Selectors",
      "type": "object",
      "properties": {
        "namespaces": {
          "title": "Namespaces",
          "description": "Specifies the namespace of the experiment's target Pod. If this selector is None, Chaos Mesh will set it to the namespace of the current Chaos experiment.",
          "type": "array",
          "items": {
            "type": "string"
          }
        },
        "labelSelectors": {
          "title": "Labelselectors",
          "description": "Specifies the label-key/value pairs that the experiment's target Pod must have. If multiple labels are specified, the experiment target must have all the labels specified by this selector.",
          "type": "object",
          "additionalProperties": {
            "type": "string"
          }
        },
        "expressionSelectors": {
          "title": "Expressionselectors",
          "description": "Specifies a set of expressions that define the label's rules to specifiy the experiment's target Pod.",
          "example": [
            {
              "key": "tier",
              "operator": "In",
              "values": [
                "cache"
              ]
            },
            {
              "key": "environment",
              "operator": "NotIn",
              "values": [
                "dev"
              ]
            }
          ],
          "type": "array",
          "items": {
            "$ref": "#/definitions/SetBasedRequirements"
          }
        },
        "annotationSelectors": {
          "title": "Annotationselectors",
          "description": "Specifies the annotation-key/value pairs that the experiment's target Pod must have. If multiple annotations are specified, the experiment target must have all annotations specified by this selector.",
          "type": "object",
          "additionalProperties": {
            "type": "string"
          }
        },
        "fieldSelectors": {
          "title": "Fieldselectors",
          "description": "Specifies the field-key/value pairs of the experiment's target Pod. If multiple fields are specified, the experiment target must have all fields set by this selector.",
          "example": {
            "metadata.name": "my-pod",
            "metadata.namespace": "dafault"
          },
          "type": "object",
          "additionalProperties": {
            "type": "string"
          }
        },
        "podPhaseSelectors": {
          "title": "Podphaseselectors",
          "description": "Specifies the phase of the experiment's target Pod. If this selector is None, the target Pod's phase is not limited.",
          "type": "array",
          "items": {
            "enum": [
              "Pending",
              "Running",
              "Succeeded",
              "Failed",
              "Unknown"
            ],
            "type": "string"
          }
        },
        "nodeSelectors": {
          "title": "Nodeselectors",
          "description": "Specifies the node-label-key/value pairs to which the experiment's target Pod belongs.",
          "type": "object",
          "additionalProperties": {
            "type": "string"
          }
        },
        "nodes": {
          "title": "Nodes",
          "description": "Specifies the node to which the experiment's target Pod belongs. The target Pod can only belong to one node in the configured node list. If multiple node labels are specified, the node to which the experiment's target Pod belongs must have all labels specified by this selector.",
          "type": "array",
          "items": {
            "type": "string"
          }
        },
        "pods": {
          "title": "Pods",
          "description": "Specifies the namespaces and list of the experiment's target Pods. If you have specified this selector, Chaos Mesh ignores other configured selectors.",
          "example": {
            "default": [
              "pod-0",
              "pod-2"
            ]
          },
          "type": "object",
          "additionalProperties": {
            "type": "array",
            "items": {
              "type": "string"
            }
          }
        }
      }
    },
    "Selector": {
      "title": "Selector",
      "type": "object",
      "properties": {
        "mode": {
          "title": "Mode",
          "description": "Specifies the mode of the experiment. The mode options include one (selecting a random Pod), all (selecting all eligible Pods), fixed (selecting a specified number of eligible Pods), fixed-percent (selecting a specified percentage of Pods from the eligible Pods), and random-max-percent (selecting the maximum percentage of Pods from the eligible Pods)",
          "example": "one",
          "enum": [
            "one",
            "all",
            "fixed",
            "fixed-percent",
            "random-max-percent"
          ],
          "type": "string"
        },
        "selector": {
          "title": "Selector",
          "description": "Specifies the target Pod.",
          "example": null,
          "allOf": [
            {
              "$ref": "#/definitions/Selectors"
            }
          ]
        }
      },
      "required": [
        "mode",
        "selector"
      ]
    },
    "Reorder": {
      "title": "Reorder",
      "type": "object",
      "properties": {
        "reorder": {
          "title": "Reorder",
          "description": "Indicates the probability to reorder",
          "default": "0",
          "example": "0.5",
          "type": "string"
        },
        "correlation": {
          "title": "Correlation",
          "description": "Indicates the correlation between this time's length of delay time and the previous time's length of delay time. Range of value: [0, 100]",
          "default": "0",
          "example": "50",
          "type": "string"
        },
        "gap": {
          "title": "Gap",
          "description": "Indicates the gap before and after packet reordering",
          "default": 0,
          "example": 5,
          "type": "integer"
        }
      }
    },
    "Deplay": {
      "title": "Deplay",
      "type": "object",
      "properties": {
        "latency": {
          "title": "Latency",
          "description": "Indicates the network latency",
          "example": "2ms",
          "type": "string"
        },
        "correlation": {
          "title": "Correlation",
          "description": "Indicates the correlation between the current latency and the previous one. Range of value: [0, 100]. Specify only the number. NEVER include any units.",
          "example": "50",
          "type": "string"
        },
        "jitter": {
          "title": "Jitter",
          "description": "Indicates the range of the network latency",
          "example": "1ms",
          "type": "string"
        },
        "reorder": {
          "title": "Reorder",
          "description": "Indicates the status of network packet reordering",
          "allOf": [
            {
              "$ref": "#/definitions/Reorder"
            }
          ]
        }
      }
    },
    "Loss": {
      "title": "Loss",
      "type": "object",
      "properties": {
        "loss": {
          "title": "Loss",
          "description": "Indicates the probability of packet loss. Range of value: [0, 100]. Specify only the number. NEVER include any units.",
          "default": "0",
          "example": "50",
          "type": "string"
        },
        "correlation": {
          "title": "Correlation",
          "description": "Indicates the correlation between the probability of current packet loss and the previous time's packet loss. Range of value: [0, 100]. Specify only the number. NEVER include any units.",
          "default": "0",
          "example": "50",
          "type": "string"
        }
      }
    },
    "Duplicate": {
      "title": "Duplicate",
      "type": "object",
      "properties": {
        "duplicate": {
          "title": "Duplicate",
          "description": "Indicates the probability of packet duplicating. Range of value: [0, 100]. Specify only the number. NEVER include any units.",
          "default": "0",
          "example": "50",
          "type": "string"
        },
        "correlation": {
          "title": "Correlation",
          "description": "Indicates the correlation between the probability of current packet duplicating and the previous time's packet duplicating. Range of value: [0, 100]. Specify only the number. NEVER include any units.",
          "default": "0",
          "example": "50",
          "type": "string"
        }
      }
    },
    "Corrupt": {
      "title": "Corrupt",
      "type": "object",
      "properties": {
        "corrupt": {
          "title": "Corrupt",
          "description": "Indicates the probability of packet corruption. Range of value: [0, 100]. Specify only the number. NEVER include any units.",
          "default": "0",
          "example": "50",
          "type": "string"
        },
        "correlation": {
          "title": "Correlation",
          "description": "Indicates the correlation between the probability of current packet corruption and the previous time's packet corruption. Range of value: [0, 100]. Specify only the number. NEVER include any units.",
          "default": "0",
          "example": "50",
          "type": "string"
        }
      }
    },
    "Rate": {
      "title": "Rate",
      "type": "object",
      "properties": {
        "rate": {
          "title": "Rate",
          "description": "Indicates the rate of bandwidth limit. Allows bit, kbit, mbit, gbit, tbit, bps, kbps, mbps, gbps, tbps unit. bps means bytes per second",
          "example": "1mbps",
          "type": "string"
        }
      }
    },
    "Bandwidth": {
      "title": "Bandwidth",
      "type": "object",
      "properties": {
        "rate": {
          "title": "Rate",
          "description": "Indicates the rate of bandwidth limit. Allows bit, kbit, mbit, gbit, tbit, bps, kbps, mbps, gbps, tbps unit. bps means bytes per second",
          "example": "1mbps",
          "type": "string"
        },
        "limit": {
          "title": "Limit",
          "description": "Indicates the number of bytes waiting in queue",
          "example": 1,
          "type": "integer"
        },
        "buffer": {
          "title": "Buffer",
          "description": "Indicates the maximum number of bytes that can be sent instantaneously",
          "example": 1,
          "type": "integer"
        },
        "peakrate": {
          "title": "Peakrate",
          "description": "Indicates the maximum consumption of bucket (usually not set)",
          "example": 1,
          "type": "integer"
        },
        "minburst": {
          "title": "Minburst",
          "description": "Indicates the size of peakrate bucket (usually not set)",
          "example": 1,
          "type": "integer"
        }
      }
    }
  }
}
```
\end{dynamictemplate}

\begin{dynamictemplate}[\texttt{\textbf{detailed\_param\_instructions}} for \texttt{\textbf{DNSChaos}}]
The output should be formatted as a JSON instance that conforms to the JSON schema below.

As an example, for the schema {"properties": {"foo": {"title": "Foo", "description": "a list of strings", "type": "array", "items": {"type": "string"}}}, "required": ["foo"]}
the object {"foo": ["bar", "baz"]} is a well-formatted instance of the schema. The object {"properties": {"foo": ["bar", "baz"]}} is not well-formatted.

Here is the output schema:
```
{
  "properties": {
    "action": {
      "title": "Action",
      "description": "Defines the behavior of DNS fault from 'random' or 'error'. When the value is random, DNS service returns a random IP address; when the value is error, DNS service returns an error.",
      "example": "random",
      "enum": [
        "random",
        "error"
      ],
      "type": "string"
    },
    "mode": {
      "title": "Mode",
      "description": "Specifies the mode of the experiment. The mode options include 'one' (selecting a random Pod), 'all' (selecting all eligible Pods), 'fixed' (selecting a specified number of eligible Pods), 'fixed-percent' (selecting a specified percentage of Pods from the eligible Pods), and 'random-max-percent' (selecting the maximum percentage of Pods from the eligible Pods)",
      "example": "one",
      "enum": [
        "one",
        "all",
        "fixed",
        "fixed-percent",
        "random-max-percent"
      ],
      "type": "string"
    },
    "value": {
      "title": "Value",
      "description": "Provides parameters for the mode configuration, depending on mode. For example, when mode is set to fixed-percent, value specifies the percentage of Pods.",
      "example": "1",
      "type": "string"
    },
    "patterns": {
      "title": "Patterns",
      "description": "Selects a domain template that matches faults. The fault is applyed to these domains. Placeholder ? and wildcard * are supported, but the wildcard in patterns configuration must be at the end of string. For example, chaos-mes*.org. is an invalid configuration. When patterns is not configured, faults are injected for all domains.",
      "example": "google.com, chaos-mesh.org, github.com",
      "type": "array",
      "items": {
        "type": "string"
      }
    },
    "selector": {
      "title": "Selector",
      "description": "Specifies the target Pod.",
      "example": null,
      "allOf": [
        {
          "$ref": "#/definitions/Selectors"
        }
      ]
    }
  },
  "required": [
    "selector"
  ],
  "definitions": {
    "SetBasedRequirements": {
      "title": "SetBasedRequirements",
      "type": "object",
      "properties": {
        "key": {
          "title": "Key",
          "description": "Label key",
          "type": "string"
        },
        "operator": {
          "title": "Operator",
          "description": "Select an operator.",
          "enum": [
            "In",
            "NotIn",
            "Exists",
            "DoesNotExist"
          ],
          "type": "string"
        },
        "values": {
          "title": "Values",
          "description": "Label values. The values set must be non-empty in the case of In and NotIn.",
          "type": "array",
          "items": {
            "type": "string"
          }
        }
      },
      "required": [
        "key",
        "operator",
        "values"
      ]
    },
    "Selectors": {
      "title": "Selectors",
      "type": "object",
      "properties": {
        "namespaces": {
          "title": "Namespaces",
          "description": "Specifies the namespace of the experiment's target Pod. If this selector is None, Chaos Mesh will set it to the namespace of the current Chaos experiment.",
          "type": "array",
          "items": {
            "type": "string"
          }
        },
        "labelSelectors": {
          "title": "Labelselectors",
          "description": "Specifies the label-key/value pairs that the experiment's target Pod must have. If multiple labels are specified, the experiment target must have all the labels specified by this selector.",
          "type": "object",
          "additionalProperties": {
            "type": "string"
          }
        },
        "expressionSelectors": {
          "title": "Expressionselectors",
          "description": "Specifies a set of expressions that define the label's rules to specifiy the experiment's target Pod.",
          "example": [
            {
              "key": "tier",
              "operator": "In",
              "values": [
                "cache"
              ]
            },
            {
              "key": "environment",
              "operator": "NotIn",
              "values": [
                "dev"
              ]
            }
          ],
          "type": "array",
          "items": {
            "$ref": "#/definitions/SetBasedRequirements"
          }
        },
        "annotationSelectors": {
          "title": "Annotationselectors",
          "description": "Specifies the annotation-key/value pairs that the experiment's target Pod must have. If multiple annotations are specified, the experiment target must have all annotations specified by this selector.",
          "type": "object",
          "additionalProperties": {
            "type": "string"
          }
        },
        "fieldSelectors": {
          "title": "Fieldselectors",
          "description": "Specifies the field-key/value pairs of the experiment's target Pod. If multiple fields are specified, the experiment target must have all fields set by this selector.",
          "example": {
            "metadata.name": "my-pod",
            "metadata.namespace": "dafault"
          },
          "type": "object",
          "additionalProperties": {
            "type": "string"
          }
        },
        "podPhaseSelectors": {
          "title": "Podphaseselectors",
          "description": "Specifies the phase of the experiment's target Pod. If this selector is None, the target Pod's phase is not limited.",
          "type": "array",
          "items": {
            "enum": [
              "Pending",
              "Running",
              "Succeeded",
              "Failed",
              "Unknown"
            ],
            "type": "string"
          }
        },
        "nodeSelectors": {
          "title": "Nodeselectors",
          "description": "Specifies the node-label-key/value pairs to which the experiment's target Pod belongs.",
          "type": "object",
          "additionalProperties": {
            "type": "string"
          }
        },
        "nodes": {
          "title": "Nodes",
          "description": "Specifies the node to which the experiment's target Pod belongs. The target Pod can only belong to one node in the configured node list. If multiple node labels are specified, the node to which the experiment's target Pod belongs must have all labels specified by this selector.",
          "type": "array",
          "items": {
            "type": "string"
          }
        },
        "pods": {
          "title": "Pods",
          "description": "Specifies the namespaces and list of the experiment's target Pods. If you have specified this selector, Chaos Mesh ignores other configured selectors.",
          "example": {
            "default": [
              "pod-0",
              "pod-2"
            ]
          },
          "type": "object",
          "additionalProperties": {
            "type": "array",
            "items": {
              "type": "string"
            }
          }
        }
      }
    }
  }
}
```
\end{dynamictemplate}

\begin{dynamictemplate}[\texttt{\textbf{detailed\_param\_instructions}} for \texttt{\textbf{HTTPChaos}}]
The output should be formatted as a JSON instance that conforms to the JSON schema below.

As an example, for the schema {"properties": {"foo": {"title": "Foo", "description": "a list of strings", "type": "array", "items": {"type": "string"}}}, "required": ["foo"]}
the object {"foo": ["bar", "baz"]} is a well-formatted instance of the schema. The object {"properties": {"foo": ["bar", "baz"]}} is not well-formatted.

Here is the output schema:
```
{
  "properties": {
    "mode": {
      "title": "Mode",
      "description": "Specifies the mode of the experiment. The mode options include one (selecting a random Pod), all (selecting all eligible Pods), fixed (selecting a specified number of eligible Pods), fixed-percent (selecting a specified percentage of Pods from the eligible Pods), and random-max-percent (selecting the maximum percentage of Pods from the eligible Pods)",
      "example": "one",
      "enum": [
        "one",
        "all",
        "fixed",
        "fixed-percent",
        "random-max-percent"
      ],
      "type": "string"
    },
    "value": {
      "title": "Value",
      "description": "Provides parameters for the mode configuration, depending on mode. For example, when mode is set to fixed-percent, value specifies the percentage of Pods.",
      "example": "1",
      "type": "string"
    },
    "target": {
      "title": "Target",
      "description": "Specifies whether the target of fault injection is Request or Response. The target-related fields (replace.path, replace.method, replace.queries, patch.queries) should be configured at the same time.",
      "example": "Request",
      "enum": [
        "Request",
        "Response"
      ],
      "type": "string"
    },
    "port": {
      "title": "Port",
      "description": "The TCP port that the target service listens on.",
      "example": 80,
      "type": "integer"
    },
    "code": {
      "title": "Code",
      "description": "Specifies the status code responded by target. If not specified, the fault takes effect for all status codes by default. This configuration is effective only when the 'target' is set to 'Response'",
      "example": 200,
      "type": "integer"
    },
    "path": {
      "title": "Path",
      "description": "Specify the URI path of the target request. Supports Matching wildcards. If not specified, the fault takes effect on all paths by default.",
      "example": "/api/*",
      "type": "string"
    },
    "method": {
      "title": "Method",
      "description": "Specify the HTTP method of the target request method. If not specified, the fault takes effect for all methods by default.",
      "example": "GET",
      "type": "string"
    },
    "request_headers": {
      "title": "Request Headers",
      "description": "Matches request headers to target.",
      "example": {
        "Content-Type": "application/json"
      },
      "type": "object",
      "additionalProperties": {
        "type": "string"
      }
    },
    "abort": {
      "title": "Abort",
      "description": "Abort fault. Indicates whether to inject the fault that interrupts the connection.",
      "default": false,
      "example": true,
      "type": "boolean"
    },
    "delay": {
      "title": "Delay",
      "description": "Deplay fault. Specifies the time for a latency fault.",
      "default": "0",
      "example": "10s",
      "type": "string"
    },
    "replace": {
      "title": "Replace",
      "description": "Replace fault. Specifies replaced contents.",
      "allOf": [
        {
          "$ref": "#/definitions/Replace"
        }
      ]
    },
    "patch": {
      "title": "Patch",
      "description": "Patch fault. Specifies patch contents.",
      "allOf": [
        {
          "$ref": "#/definitions/Patch"
        }
      ]
    }
  },
  "required": [
    "mode",
    "target",
    "port"
  ],
  "definitions": {
    "Replace": {
      "title": "Replace",
      "type": "object",
      "properties": {
        "headers": {
          "title": "Headers",
          "description": "Specifies the key pair used to replace the request headers or response headers.",
          "example": {
            "Content-Type": "application/xml"
          },
          "type": "object",
          "additionalProperties": {
            "type": "string"
          }
        },
        "body": {
          "title": "Body",
          "description": "Specifies request body or response body to replace the fault (Base64 encoded).",
          "example": "eyJmb28iOiAiYmFyIn0K",
          "type": "string"
        },
        "path": {
          "title": "Path",
          "description": "Specifies the URI path used to replace content.",
          "example": "/api/v2",
          "type": "string"
        },
        "method": {
          "title": "Method",
          "description": "Specifies the replaced content of the HTTP request method.",
          "example": "DELETE",
          "type": "string"
        },
        "queries": {
          "title": "Queries",
          "description": "Specifies the replaced key pair of the URI query.",
          "type": "array",
          "items": {
            "type": "array",
            "items": {
              "type": "string"
            }
          }
        },
        "code": {
          "title": "Code",
          "description": "Specifies the replaced content of the response status code. This configuration is effective only when the 'target' is set to 'Response'.",
          "example": 404,
          "type": "integer"
        }
      }
    },
    "PatchBody": {
      "title": "PatchBody",
      "type": "object",
      "properties": {
        "type": {
          "title": "Type",
          "description": "Specifies the type of patch faults of the request body or response body. Currently, it only supports JSON.",
          "example": "JSON",
          "type": "string"
        },
        "value": {
          "title": "Value",
          "description": "Specifies the fault of the request body or response body with patch faults.",
          "example": "{'foo': 'bar'}",
          "type": "string"
        }
      }
    },
    "Patch": {
      "title": "Patch",
      "type": "object",
      "properties": {
        "headers": {
          "title": "Headers",
          "description": "Specifies the attached key pair of the request headers or response headers with patch faults.",
          "example": [
            [
              "Set-Cookie",
              "one cookie"
            ]
          ],
          "type": "array",
          "items": {
            "type": "array",
            "items": {
              "type": "string"
            }
          }
        },
        "body": {
          "title": "Body",
          "description": "Patch body.",
          "allOf": [
            {
              "$ref": "#/definitions/PatchBody"
            }
          ]
        },
        "queries": {
          "title": "Queries",
          "description": "Specifies the attached key pair of the URI query with patch faults.",
          "example": [
            [
              "foo",
              "bar"
            ]
          ],
          "type": "array",
          "items": {
            "type": "array",
            "items": {
              "type": "string"
            }
          }
        }
      }
    }
  }
}
```
\end{dynamictemplate}

\begin{dynamictemplate}[\texttt{\textbf{detailed\_param\_instructions}} for \texttt{\textbf{StressChaos}}]
The output should be formatted as a JSON instance that conforms to the JSON schema below.

As an example, for the schema {"properties": {"foo": {"title": "Foo", "description": "a list of strings", "type": "array", "items": {"type": "string"}}}, "required": ["foo"]}
the object {"foo": ["bar", "baz"]} is a well-formatted instance of the schema. The object {"properties": {"foo": ["bar", "baz"]}} is not well-formatted.

Here is the output schema:
```
{
  "properties": {
    "mode": {
      "title": "Mode",
      "description": "Specifies the mode of the experiment. The mode options include 'one' (selecting a random Pod), 'all' (selecting all eligible Pods), 'fixed' (selecting a specified number of eligible Pods), 'fixed-percent' (selecting a specified percentage of Pods from the eligible Pods), and 'random-max-percent' (selecting the maximum percentage of Pods from the eligible Pods)",
      "example": "one",
      "enum": [
        "one",
        "all",
        "fixed",
        "fixed-percent",
        "random-max-percent"
      ],
      "type": "string"
    },
    "value": {
      "title": "Value",
      "description": "Provides parameters for the mode configuration, depending on mode.For example, when mode is set to fixed-percent, value specifies the percentage of Pods.",
      "example": "1",
      "type": "string"
    },
    "stressors": {
      "title": "Stressors",
      "description": "Specifies the stress of CPU or memory",
      "dafault": null,
      "allOf": [
        {
          "$ref": "#/definitions/Stressors"
        }
      ]
    },
    "stressngStressors": {
      "title": "Stressngstressors",
      "description": "Specifies the stres-ng parameter to reach richer stress injection",
      "example": "--clone 2",
      "type": "string"
    },
    "containerNames": {
      "title": "Containernames",
      "description": "Specifies the name of the container into which the fault is injected.",
      "example": [
        "nginx"
      ],
      "type": "array",
      "items": {
        "type": "string"
      }
    },
    "selector": {
      "title": "Selector",
      "description": "Specifies the target Pod.",
      "allOf": [
        {
          "$ref": "#/definitions/Selectors"
        }
      ]
    }
  },
  "required": [
    "mode",
    "selector"
  ],
  "definitions": {
    "MemoryStressor": {
      "title": "MemoryStressor",
      "type": "object",
      "properties": {
        "workers": {
          "title": "Workers",
          "description": "Specifies the number of threads that apply memory stress",
          "example": 1,
          "type": "integer"
        },
        "size": {
          "title": "Size",
          "description": "Specifies the memory size to be occupied or a percentage of the total memory size. The final sum of the occupied memory size is size.",
          "example": "256MB",
          "type": "string"
        },
        "oomScoreAdj": {
          "title": "Oomscoreadj",
          "description": "Specifies the oom_score_adj of the stress process.",
          "example": -1000,
          "type": "integer"
        }
      }
    },
    "CPUStressor": {
      "title": "CPUStressor",
      "type": "object",
      "properties": {
        "workers": {
          "title": "Workers",
          "description": "Specifies the number of threads that apply CPU stress",
          "example": 1,
          "type": "integer"
        },
        "load": {
          "title": "Load",
          "description": "Specifies the percentage of CPU occupied. 0 means that no additional CPU is added, and 100 refers to full load. The final sum of CPU load is workers * load.",
          "example": 50,
          "type": "integer"
        }
      }
    },
    "Stressors": {
      "title": "Stressors",
      "type": "object",
      "properties": {
        "memory": {
          "title": "Memory",
          "description": "Specifies the memory stress",
          "allOf": [
            {
              "$ref": "#/definitions/MemoryStressor"
            }
          ]
        },
        "cpu": {
          "title": "Cpu",
          "description": "Specifies the CPU stress",
          "allOf": [
            {
              "$ref": "#/definitions/CPUStressor"
            }
          ]
        }
      }
    },
    "SetBasedRequirements": {
      "title": "SetBasedRequirements",
      "type": "object",
      "properties": {
        "key": {
          "title": "Key",
          "description": "Label key",
          "type": "string"
        },
        "operator": {
          "title": "Operator",
          "description": "Select an operator.",
          "enum": [
            "In",
            "NotIn",
            "Exists",
            "DoesNotExist"
          ],
          "type": "string"
        },
        "values": {
          "title": "Values",
          "description": "Label values. The values set must be non-empty in the case of In and NotIn.",
          "type": "array",
          "items": {
            "type": "string"
          }
        }
      },
      "required": [
        "key",
        "operator",
        "values"
      ]
    },
    "Selectors": {
      "title": "Selectors",
      "type": "object",
      "properties": {
        "namespaces": {
          "title": "Namespaces",
          "description": "Specifies the namespace of the experiment's target Pod. If this selector is None, Chaos Mesh will set it to the namespace of the current Chaos experiment.",
          "type": "array",
          "items": {
            "type": "string"
          }
        },
        "labelSelectors": {
          "title": "Labelselectors",
          "description": "Specifies the label-key/value pairs that the experiment's target Pod must have. If multiple labels are specified, the experiment target must have all the labels specified by this selector.",
          "type": "object",
          "additionalProperties": {
            "type": "string"
          }
        },
        "expressionSelectors": {
          "title": "Expressionselectors",
          "description": "Specifies a set of expressions that define the label's rules to specifiy the experiment's target Pod.",
          "example": [
            {
              "key": "tier",
              "operator": "In",
              "values": [
                "cache"
              ]
            },
            {
              "key": "environment",
              "operator": "NotIn",
              "values": [
                "dev"
              ]
            }
          ],
          "type": "array",
          "items": {
            "$ref": "#/definitions/SetBasedRequirements"
          }
        },
        "annotationSelectors": {
          "title": "Annotationselectors",
          "description": "Specifies the annotation-key/value pairs that the experiment's target Pod must have. If multiple annotations are specified, the experiment target must have all annotations specified by this selector.",
          "type": "object",
          "additionalProperties": {
            "type": "string"
          }
        },
        "fieldSelectors": {
          "title": "Fieldselectors",
          "description": "Specifies the field-key/value pairs of the experiment's target Pod. If multiple fields are specified, the experiment target must have all fields set by this selector.",
          "example": {
            "metadata.name": "my-pod",
            "metadata.namespace": "dafault"
          },
          "type": "object",
          "additionalProperties": {
            "type": "string"
          }
        },
        "podPhaseSelectors": {
          "title": "Podphaseselectors",
          "description": "Specifies the phase of the experiment's target Pod. If this selector is None, the target Pod's phase is not limited.",
          "type": "array",
          "items": {
            "enum": [
              "Pending",
              "Running",
              "Succeeded",
              "Failed",
              "Unknown"
            ],
            "type": "string"
          }
        },
        "nodeSelectors": {
          "title": "Nodeselectors",
          "description": "Specifies the node-label-key/value pairs to which the experiment's target Pod belongs.",
          "type": "object",
          "additionalProperties": {
            "type": "string"
          }
        },
        "nodes": {
          "title": "Nodes",
          "description": "Specifies the node to which the experiment's target Pod belongs. The target Pod can only belong to one node in the configured node list. If multiple node labels are specified, the node to which the experiment's target Pod belongs must have all labels specified by this selector.",
          "type": "array",
          "items": {
            "type": "string"
          }
        },
        "pods": {
          "title": "Pods",
          "description": "Specifies the namespaces and list of the experiment's target Pods. If you have specified this selector, Chaos Mesh ignores other configured selectors.",
          "example": {
            "default": [
              "pod-0",
              "pod-2"
            ]
          },
          "type": "object",
          "additionalProperties": {
            "type": "array",
            "items": {
              "type": "string"
            }
          }
        }
      }
    }
  }
}
```
\end{dynamictemplate}

\begin{dynamictemplate}[\texttt{\textbf{detailed\_param\_instructions}} for \texttt{\textbf{IOChaos}}]
The output should be formatted as a JSON instance that conforms to the JSON schema below.

As an example, for the schema {"properties": {"foo": {"title": "Foo", "description": "a list of strings", "type": "array", "items": {"type": "string"}}}, "required": ["foo"]}
the object {"foo": ["bar", "baz"]} is a well-formatted instance of the schema. The object {"properties": {"foo": ["bar", "baz"]}} is not well-formatted.

Here is the output schema:
```
{
  "properties": {
    "action": {
      "title": "Action",
      "description": "Indicates the specific type of faults. Only latency, fault, attrOverride, and mistake are supported.",
      "example": "latency",
      "enum": [
        "latency",
        "fault",
        "attrOverride",
        "mistake"
      ],
      "type": "string"
    },
    "mode": {
      "title": "Mode",
      "description": "Specifies the mode of the experiment. The mode options include one (selecting a random Pod), all (selecting all eligible Pods), fixed (selecting a specified number of eligible Pods), fixed-percent (selecting a specified percentage of Pods from the eligible Pods), and random-max-percent (selecting the maximum percentage of Pods from the eligible Pods)",
      "example": "one",
      "enum": [
        "one",
        "all",
        "fixed",
        "fixed-percent",
        "random-max-percent"
      ],
      "type": "string"
    },
    "selector": {
      "title": "Selector",
      "description": "Specifies the target Pod.",
      "allOf": [
        {
          "$ref": "#/definitions/Selectors"
        }
      ]
    },
    "value": {
      "title": "Value",
      "description": "Provides parameters for the mode configuration, depending on mode. For example, when mode is set to fixed-percent, value specifies the percentage of Pods.",
      "example": "1",
      "type": "string"
    },
    "volumePath": {
      "title": "Volumepath",
      "description": "The mount point of volume in the target container. Must be the root directory of the mount.",
      "example": "/var/run/etcd",
      "type": "string"
    },
    "path": {
      "title": "Path",
      "description": "The valid range of fault injections, either a wildcard or a single file. If not specified, the fault is valid for all files by default",
      "example": "/var/run/etcd/*/",
      "type": "string"
    },
    "methods": {
      "title": "Methods",
      "description": "Type of the file system call that requires injecting fault. Supported method types: ['lookup', 'forget', 'getattr', 'setattr', 'readlink', 'mknod', 'mkdir', 'unlink', 'rmdir', 'symlink', 'rename', 'link', 'open', 'read', 'write', 'flush', 'release', 'fsync', 'opendir', 'readdir', 'releasedir', 'fsyncdir', 'statfs', 'setxattr', 'getxattr', 'listxattr', 'removexattr', 'access', 'create', 'getlk', 'setlk', 'bmap']. All Types by default.",
      "example": [
        "READ"
      ],
      "type": "array",
      "items": {
        "type": "string"
      }
    },
    "percent": {
      "title": "Percent",
      "description": "Probability of failure per operation, in %.",
      "default": 100,
      "example": 100,
      "type": "integer"
    },
    "containerNames": {
      "title": "Containernames",
      "description": "Specifies the name of the container into which the fault is injected.",
      "type": "array",
      "items": {
        "type": "string"
      }
    },
    "deplay": {
      "title": "Deplay",
      "description": "Specify when the 'action' is set to 'latency'. Specific delay time.",
      "type": "string"
    },
    "errno": {
      "title": "Errno",
      "description": "Specify when the 'action' is set to 'fault'. Returned error number: 1: Operation not permitted, 2: No such file or directory, 5: I/O error, 6: No such device or address, 12: Out of memory, 16: Device or resource busy, 17: File exists, 20: Not a directory, 22: Invalid argument, 24: Too many open files, 28: No space left on device",
      "type": "integer"
    },
    "attr": {
      "title": "Attr",
      "description": "Specify when the 'action' is set to 'attrOverride'. Specific property override rules.",
      "allOf": [
        {
          "$ref": "#/definitions/AttrOverrideSpec"
        }
      ]
    },
    "mistake": {
      "title": "Mistake",
      "description": "Specify when the 'action' is set to 'mistake'. Specific error rules.",
      "allOf": [
        {
          "$ref": "#/definitions/MistakeSpec"
        }
      ]
    }
  },
  "required": [
    "action",
    "mode",
    "volumePath",
    "attr",
    "mistake"
  ],
  "definitions": {
    "SetBasedRequirements": {
      "title": "SetBasedRequirements",
      "type": "object",
      "properties": {
        "key": {
          "title": "Key",
          "description": "Label key",
          "type": "string"
        },
        "operator": {
          "title": "Operator",
          "description": "Select an operator.",
          "enum": [
            "In",
            "NotIn",
            "Exists",
            "DoesNotExist"
          ],
          "type": "string"
        },
        "values": {
          "title": "Values",
          "description": "Label values. The values set must be non-empty in the case of In and NotIn.",
          "type": "array",
          "items": {
            "type": "string"
          }
        }
      },
      "required": [
        "key",
        "operator",
        "values"
      ]
    },
    "Selectors": {
      "title": "Selectors",
      "type": "object",
      "properties": {
        "namespaces": {
          "title": "Namespaces",
          "description": "Specifies the namespace of the experiment's target Pod. If this selector is None, Chaos Mesh will set it to the namespace of the current Chaos experiment.",
          "type": "array",
          "items": {
            "type": "string"
          }
        },
        "labelSelectors": {
          "title": "Labelselectors",
          "description": "Specifies the label-key/value pairs that the experiment's target Pod must have. If multiple labels are specified, the experiment target must have all the labels specified by this selector.",
          "type": "object",
          "additionalProperties": {
            "type": "string"
          }
        },
        "expressionSelectors": {
          "title": "Expressionselectors",
          "description": "Specifies a set of expressions that define the label's rules to specifiy the experiment's target Pod.",
          "example": [
            {
              "key": "tier",
              "operator": "In",
              "values": [
                "cache"
              ]
            },
            {
              "key": "environment",
              "operator": "NotIn",
              "values": [
                "dev"
              ]
            }
          ],
          "type": "array",
          "items": {
            "$ref": "#/definitions/SetBasedRequirements"
          }
        },
        "annotationSelectors": {
          "title": "Annotationselectors",
          "description": "Specifies the annotation-key/value pairs that the experiment's target Pod must have. If multiple annotations are specified, the experiment target must have all annotations specified by this selector.",
          "type": "object",
          "additionalProperties": {
            "type": "string"
          }
        },
        "fieldSelectors": {
          "title": "Fieldselectors",
          "description": "Specifies the field-key/value pairs of the experiment's target Pod. If multiple fields are specified, the experiment target must have all fields set by this selector.",
          "example": {
            "metadata.name": "my-pod",
            "metadata.namespace": "dafault"
          },
          "type": "object",
          "additionalProperties": {
            "type": "string"
          }
        },
        "podPhaseSelectors": {
          "title": "Podphaseselectors",
          "description": "Specifies the phase of the experiment's target Pod. If this selector is None, the target Pod's phase is not limited.",
          "type": "array",
          "items": {
            "enum": [
              "Pending",
              "Running",
              "Succeeded",
              "Failed",
              "Unknown"
            ],
            "type": "string"
          }
        },
        "nodeSelectors": {
          "title": "Nodeselectors",
          "description": "Specifies the node-label-key/value pairs to which the experiment's target Pod belongs.",
          "type": "object",
          "additionalProperties": {
            "type": "string"
          }
        },
        "nodes": {
          "title": "Nodes",
          "description": "Specifies the node to which the experiment's target Pod belongs. The target Pod can only belong to one node in the configured node list. If multiple node labels are specified, the node to which the experiment's target Pod belongs must have all labels specified by this selector.",
          "type": "array",
          "items": {
            "type": "string"
          }
        },
        "pods": {
          "title": "Pods",
          "description": "Specifies the namespaces and list of the experiment's target Pods. If you have specified this selector, Chaos Mesh ignores other configured selectors.",
          "example": {
            "default": [
              "pod-0",
              "pod-2"
            ]
          },
          "type": "object",
          "additionalProperties": {
            "type": "array",
            "items": {
              "type": "string"
            }
          }
        }
      }
    },
    "TimeSpec": {
      "title": "TimeSpec",
      "type": "object",
      "properties": {
        "sec": {
          "title": "Sec",
          "description": "Timestamp in seconds. Specify either sec or nsec.",
          "type": "integer"
        },
        "nsec": {
          "title": "Nsec",
          "description": "Timestamp in nanoseconds. Specify either sec or nsec.",
          "type": "integer"
        }
      }
    },
    "AttrOverrideSpec": {
      "title": "AttrOverrideSpec",
      "type": "object",
      "properties": {
        "ino": {
          "title": "Ino",
          "description": "ino number",
          "type": "integer"
        },
        "size": {
          "title": "Size",
          "description": "File size",
          "type": "integer"
        },
        "blocks": {
          "title": "Blocks",
          "description": "Number of blocks that the file uses",
          "type": "integer"
        },
        "atime": {
          "title": "Atime",
          "description": "Last access time",
          "allOf": [
            {
              "$ref": "#/definitions/TimeSpec"
            }
          ]
        },
        "mtime": {
          "title": "Mtime",
          "description": "Last modified time",
          "allOf": [
            {
              "$ref": "#/definitions/TimeSpec"
            }
          ]
        },
        "ctime": {
          "title": "Ctime",
          "description": "Last status change time",
          "allOf": [
            {
              "$ref": "#/definitions/TimeSpec"
            }
          ]
        },
        "kind": {
          "title": "Kind",
          "description": "File type, see fuser::FileType",
          "type": "string"
        },
        "perm": {
          "title": "Perm",
          "description": "File permissions in decimal",
          "type": "integer"
        },
        "nlink": {
          "title": "Nlink",
          "description": "Number of hard links",
          "type": "integer"
        },
        "uid": {
          "title": "Uid",
          "description": "User ID of the owner",
          "type": "integer"
        },
        "gid": {
          "title": "Gid",
          "description": "Group ID of the owner",
          "type": "integer"
        },
        "rdev": {
          "title": "Rdev",
          "description": "Device ID",
          "type": "integer"
        }
      }
    },
    "MistakeSpec": {
      "title": "MistakeSpec",
      "type": "object",
      "properties": {
        "filling": {
          "title": "Filling",
          "description": "The wrong data to be filled. Only zero (fill 0) or random (fill random bytes) are supported.",
          "type": "string"
        },
        "maxOccurrences": {
          "title": "Maxoccurrences",
          "description": "Maximum number of errors in each operation.",
          "example": 1,
          "type": "integer"
        },
        "maxLength": {
          "title": "Maxlength",
          "description": "Maximum length of each error (in bytes).",
          "example": 1,
          "type": "integer"
        }
      },
      "required": [
        "filling",
        "maxOccurrences",
        "maxLength"
      ]
    }
  }
}
```
\end{dynamictemplate}

\begin{dynamictemplate}[\texttt{\textbf{detailed\_param\_instructions}} for \texttt{\textbf{TimeChaos}}]{lst:last_instruct}
The output should be formatted as a JSON instance that conforms to the JSON schema below.

As an example, for the schema {"properties": {"foo": {"title": "Foo", "description": "a list of strings", "type": "array", "items": {"type": "string"}}}, "required": ["foo"]}
the object {"foo": ["bar", "baz"]} is a well-formatted instance of the schema. The object {"properties": {"foo": ["bar", "baz"]}} is not well-formatted.

Here is the output schema:
```
{
  "properties": {
    "timeOffset": {
      "title": "Timeoffset",
      "description": "Specifies the length of time offset.",
      "example": "-5m",
      "type": "string"
    },
    "clockIds": {
      "title": "Clockids",
      "description": "Specifies the ID of clock that will be offset. See the clock_gettime documentation for details.",
      "default": [
        "CLOCK_REALTIME"
      ],
      "example": [
        "CLOCK_REALTIME",
        "CLOCK_MONOTONIC"
      ],
      "type": "array",
      "items": {
        "type": "string"
      }
    },
    "mode": {
      "title": "Mode",
      "description": "Specifies the mode of the experiment. The mode options include 'one' (selecting a random Pod), 'all' (selecting all eligible Pods), 'fixed' (selecting a specified number of eligible Pods), 'fixed-percent' (selecting a specified percentage of Pods from the eligible Pods), and 'random-max-percent' (selecting the maximum percentage of Pods from the eligible Pods)",
      "example": "one",
      "enum": [
        "one",
        "all",
        "fixed",
        "fixed-percent",
        "random-max-percent"
      ],
      "type": "string"
    },
    "value": {
      "title": "Value",
      "description": "Provides parameters for the mode configuration, depending on mode. For example, when mode is set to fixed-percent, value specifies the percentage of Pods.",
      "example": "1",
      "type": "string"
    },
    "containerNames": {
      "title": "Containernames",
      "description": "Specifies the name of the container into which the fault is injected.",
      "example": [
        "nginx"
      ],
      "type": "array",
      "items": {
        "type": "string"
      }
    },
    "selector": {
      "title": "Selector",
      "description": "Specifies the target Pod.",
      "example": null,
      "allOf": [
        {
          "$ref": "#/definitions/Selectors"
        }
      ]
    }
  },
  "required": [
    "timeOffset",
    "mode",
    "selector"
  ],
  "definitions": {
    "SetBasedRequirements": {
      "title": "SetBasedRequirements",
      "type": "object",
      "properties": {
        "key": {
          "title": "Key",
          "description": "Label key",
          "type": "string"
        },
        "operator": {
          "title": "Operator",
          "description": "Select an operator.",
          "enum": [
            "In",
            "NotIn",
            "Exists",
            "DoesNotExist"
          ],
          "type": "string"
        },
        "values": {
          "title": "Values",
          "description": "Label values. The values set must be non-empty in the case of In and NotIn.",
          "type": "array",
          "items": {
            "type": "string"
          }
        }
      },
      "required": [
        "key",
        "operator",
        "values"
      ]
    },
    "Selectors": {
      "title": "Selectors",
      "type": "object",
      "properties": {
        "namespaces": {
          "title": "Namespaces",
          "description": "Specifies the namespace of the experiment's target Pod. If this selector is None, Chaos Mesh will set it to the namespace of the current Chaos experiment.",
          "type": "array",
          "items": {
            "type": "string"
          }
        },
        "labelSelectors": {
          "title": "Labelselectors",
          "description": "Specifies the label-key/value pairs that the experiment's target Pod must have. If multiple labels are specified, the experiment target must have all the labels specified by this selector.",
          "type": "object",
          "additionalProperties": {
            "type": "string"
          }
        },
        "expressionSelectors": {
          "title": "Expressionselectors",
          "description": "Specifies a set of expressions that define the label's rules to specifiy the experiment's target Pod.",
          "example": [
            {
              "key": "tier",
              "operator": "In",
              "values": [
                "cache"
              ]
            },
            {
              "key": "environment",
              "operator": "NotIn",
              "values": [
                "dev"
              ]
            }
          ],
          "type": "array",
          "items": {
            "$ref": "#/definitions/SetBasedRequirements"
          }
        },
        "annotationSelectors": {
          "title": "Annotationselectors",
          "description": "Specifies the annotation-key/value pairs that the experiment's target Pod must have. If multiple annotations are specified, the experiment target must have all annotations specified by this selector.",
          "type": "object",
          "additionalProperties": {
            "type": "string"
          }
        },
        "fieldSelectors": {
          "title": "Fieldselectors",
          "description": "Specifies the field-key/value pairs of the experiment's target Pod. If multiple fields are specified, the experiment target must have all fields set by this selector.",
          "example": {
            "metadata.name": "my-pod",
            "metadata.namespace": "dafault"
          },
          "type": "object",
          "additionalProperties": {
            "type": "string"
          }
        },
        "podPhaseSelectors": {
          "title": "Podphaseselectors",
          "description": "Specifies the phase of the experiment's target Pod. If this selector is None, the target Pod's phase is not limited.",
          "type": "array",
          "items": {
            "enum": [
              "Pending",
              "Running",
              "Succeeded",
              "Failed",
              "Unknown"
            ],
            "type": "string"
          }
        },
        "nodeSelectors": {
          "title": "Nodeselectors",
          "description": "Specifies the node-label-key/value pairs to which the experiment's target Pod belongs.",
          "type": "object",
          "additionalProperties": {
            "type": "string"
          }
        },
        "nodes": {
          "title": "Nodes",
          "description": "Specifies the node to which the experiment's target Pod belongs. The target Pod can only belong to one node in the configured node list. If multiple node labels are specified, the node to which the experiment's target Pod belongs must have all labels specified by this selector.",
          "type": "array",
          "items": {
            "type": "string"
          }
        },
        "pods": {
          "title": "Pods",
          "description": "Specifies the namespaces and list of the experiment's target Pods. If you have specified this selector, Chaos Mesh ignores other configured selectors.",
          "example": {
            "default": [
              "pod-0",
              "pod-2"
            ]
          },
          "type": "object",
          "additionalProperties": {
            "type": "array",
            "items": {
              "type": "string"
            }
          }
        }
      }
    }
  }
}
```
\end{dynamictemplate}

\subsubsection{Experiment}
\newtcblisting[use counter from=failureprompt]{experimentprompt}[1][]{
    colback=blue!2,
    listing only,
    title={Prompt \thetcbcounter: #1},
    listing options={
        language=sysprompt,
        basicstyle=\scriptsize\ttfamily,
        breaklines=true,
        breakindent=0pt,
        showstringspaces=false,
    },
    boxrule=0.3mm,
    mybreakable,
    coltitle=black,
    colbacktitle=blue!20
}

\newtcblisting[use counter from=experimentprompt]{reexperimentprompt}[1][]{
    colback=violet!3,
    listing only,
    title={Prompt \thetcbcounter: #1},
    listing options={
        language=sysprompt,
        basicstyle=\scriptsize\ttfamily,
        breaklines=true,
        breakindent=0pt,
        breakatwhitespace,
        showstringspaces=false,
        escapechar=|,
    },
    boxrule=0.3mm,
    mybreakable,
    coltitle=black,
    colbacktitle=violet!20
}

\begin{experimentprompt}[Agent \# 2-0 for determining time schedule]
//System://
You are a helpful AI assistant for Chaos Engineering.
Given k8s manifests that define a network system, its steady states, and faults that may affect the steady states in the system, you will design a Chaos Engineering experiment for them.
First, you will determine the time schedule for the Chaos Engineering experiment.
Always keep the following rules:
- The experiment is divided into three phases: pre-validation, fault-injection, and post-validation phases: pre-validation to ensure that the system satisfies the steady states fault injection; fault-injection to observe the system's behavior during fault injection; post-validation to ensure that the system has returned to its steady states after fault injection.
- The output should be formatted as a JSON instance that conforms to the JSON schema below.

As an example, for the schema {\"properties\": {\"foo\": {\"title\": \"Foo\", \"description\": \"a list of strings\", \"type\": \"array\", \"items\": {\"type\": \"string\"}}}, \"required\": [\"foo\"]}\nthe object {\"foo\": [\"bar\", \"baz\"]} is a well-formatted instance of the schema. The object {\"properties\": {\"foo\": [\"bar\", \"baz\"]}} is not well-formatted.

Here is the output schema:
```
{
  "properties": {
    "thought": {
      "title": "Thought",
      "describe": "Think about the total time and the reasonable time allocation for each phase that you are about to design, and explain your thought process in detail.",
      "type": "string"
    },
    "total_time": {
      "title": "Total Time",
      "description": "Total time of the entire chaos experiment. total_time should equal to the sum of pre_validation_time, fault_injection_time, and post_validation_time.",
      "example": "10m",
      "type": "string"
    },
    "pre_validation_time": {
      "title": "Pre Validation Time",
      "description": "Total time of validation before fault injection.",
      "example": "2m",
      "type": "string"
    },
    "fault_injection_time": {
      "title": "Fault Injection Time",
      "description": "Total time of fault injection.",
      "example": "6m",
      "type": "string"
    },
    "post_validation_time": {
      "title": "Post Validation Time",
      "description": "Total time of validation after fault injection.",
      "example": "2m",
      "type": "string"
    }
  },
  "required": [
    "thought",
    "total_time",
    "pre_validation_time",
    "fault_injection_time",
    "post_validation_time"
  ]
}
```

//Human://
# Here is the overview of my system:
//@{user_input2}@//

# Steady states of my system:
//@{steady_states}@//

# A fault scenario that may occur in my system and may affect the steady states:
//@{detailed_fault_scenario}@//

# Please follow the instructions below regarding Chaos Engineering as necessary:
//@{ce_instructions}@//

Now, please plan a Chaos Engineering experiment to check the network system's resiliency that the steady states are remained during fault injection.
\end{experimentprompt}

\begin{exampledata}[\texttt{\textbf{detailed\_fault\_scenario}}]
An assumed fault scenario is as follows:
- Event: Cyber Attack Simulation\n- Used Chaos Engineering tool: Chaos Mesh
- Faults to simulate the event: [[Fault(name='PodChaos', name_id=0, params={'action': 'pod-kill', 'mode': 'one', 'selector': {'namespaces': ['default'], 'labelSelectors': {'app': 'example'}}})], [Fault(name='NetworkChaos', name_id=0, params={'action': 'delay', 'mode': 'all', 'selector': {'namespaces': ['default'], 'labelSelectors': {'app': 'example'}}, 'direction': 'to', 'delay': {'latency': '100ms', 'jitter': '10ms'}})]]
- Description: Given the system's weaknesses, a cyber attack targeting the web server could be highly impactful. The Pod's restart policy set to 'Never' and the single Pod deployment without redundancy are critical vulnerabilities. If the Pod fails, it will not restart, leading to downtime, and the lack of redundancy means there is no backup to handle traffic. To simulate a cyber attack, we can inject faults that exploit these weaknesses. First, we will use PodChaos to simulate a Pod failure, which will test the system's ability to maintain the 'example-pod-running-state'. Since the Pod will not restart automatically, this will directly impact the steady state. Next, we will use NetworkChaos to simulate network latency, which will test the system's ability to maintain the 'example-service-http-response-state'. This sequence simulates a cyber attack where the Pod is targeted first, followed by network disruptions, revealing the system's vulnerabilities in handling such events.
\end{exampledata}

\begin{experimentprompt}[Agent \# 2-1 for scheduling each experiment phase (pre-validation, failure-injection, and post-validation phases)]
//System://
You are a helpful AI assistant for Chaos Engineering.
Given k8s manifests that define a network system, its steady states, and faults that may affect the steady states in the system, you will design a Chaos Engineering experiment for them.
The experiment is divided into three phases: pre-validation, fault-injection, and post-validation phases: pre-validation to ensure that the system satisfies the steady states fault injection; fault-injection to observe the system's behavior during fault injection; post-validation to ensure that the system has returned to its steady states after fault injection.
Here, you will detail the //@{phase_name}@//.
Always keep the following rules:
- //!!{phase_planning_instructions}!!//

//Human://
# Here is the overview of my system:
//@{user_input}@//

# Steady states of my system:
//@{steady_states}@//

# A fault scenario that may occur in my system and may affect the steady states:
//@{detailed_fault_scenario}@//

# Please follow the instructions below regarding Chaos Engineering as necessary:
//@{ce_instructions}@//

Now, please detail the //@{phase_name}@//. Note that the phase's total time is //@{phase_total_time}@//.
\end{experimentprompt}

\begin{exampledata}[\texttt{\textbf{phase\_name}}]
pre-validation phase
\end{exampledata}

\begin{exampledata}[\texttt{\textbf{phase\_total\_time}}]
10s
\end{exampledata}

\begin{dynamictemplate}[\texttt{\textbf{phase\_planning\_instructions}} for the pre-validation and post-validation phases]
The output should be formatted as a JSON instance that conforms to the JSON schema below.

As an example, for the schema {\"properties\": {\"foo\": {\"title\": \"Foo\", \"description\": \"a list of strings\", \"type\": \"array\", \"items\": {\"type\": \"string\"}}}, \"required\": [\"foo\"]}\nthe object {\"foo\": [\"bar\", \"baz\"]} is a well-formatted instance of the schema. The object {\"properties\": {\"foo\": [\"bar\", \"baz\"]}} is not well-formatted.

Here is the output schema:
```
{
  "properties": {
    "thought": {
      "title": "Thought",
      "description": "Describe in detail the timeline for when each fault injection and each unit test (for verifying steady-state) will be executed. For example, explain which fault injections/unit tests will be executed simultaneously, and whether certain fault injections/unit tests will be executed at staggered timings. Additionally, explain the thought process that led you to this approach.",
      "type": "string"
    },
    "unit_tests": {
      "title": "Unit Tests",
      "description": "The list of unit test schedule.",
      "type": "array",
      "items": {
        "$ref": "#/definitions/UnitTest"
      }
    }
  },
  "required": [
    "thought",
    "unit_tests"
  ],
  "definitions": {
    "UnitTest": {
      "title": "UnitTest",
      "type": "object",
      "properties": {
        "name": {
          "title": "Name",
          "description": "Steady state name to be verified by a unit test.",
          "type": "string"
        },
        "grace_period": {
          "title": "Grace Period",
          "description": "Time elapsed from the start of the current phase to the beginning of the unit test.",
          "example": "0s",
          "type": "string"
        },
        "duration": {
          "title": "Duration",
          "description": "Duration of the unit test. (grace_period + duration) should not exceed the current phase's total time.",
          "example": "2m",
          "type": "string"
        }
      },
      "required": [
        "name",
        "grace_period",
        "duration"
      ]
    }
  }
}
```
\end{dynamictemplate}

\begin{dynamictemplate}[\texttt{\textbf{phase\_planning\_instructions}} for the fault-injection phases]
The output should be formatted as a JSON instance that conforms to the JSON schema below.

As an example, for the schema {\"properties\": {\"foo\": {\"title\": \"Foo\", \"description\": \"a list of strings\", \"type\": \"array\", \"items\": {\"type\": \"string\"}}}, \"required\": [\"foo\"]}\nthe object {\"foo\": [\"bar\", \"baz\"]} is a well-formatted instance of the schema. The object {\"properties\": {\"foo\": [\"bar\", \"baz\"]}} is not well-formatted.

Here is the output schema:
```
{
  "properties": {
    "thought": {
      "title": "Thought",
      "description": "Describe in detail the timeline for when each fault injection and each unit test (for verifying steady-state) will be executed. For example, explain which fault injections/unit tests will be executed simultaneously, and whether certain fault injections/unit tests will be executed at staggered timings. Additionally, explain the thought process that led you to this approach.",
      "type": "string"
    },
    "fault_injection": {
      "title": "Fault Injection",
      "description": "The list of fault injection schedules.",
      "type": "array",
      "items": {
        "$ref": "#/definitions/FaultInjection"
      }
    },
    "unit_tests": {
      "title": "Unit Tests",
      "description": "The list of unit test schedule.",
      "type": "array",
      "items": {
        "$ref": "#/definitions/UnitTest"
      }
    }
  },
  "required": [
    "thought",
    "fault_injection",
    "unit_tests"
  ],
  "definitions": {
    "FaultInjection": {
      "title": "FaultInjection",
      "type": "object",
      "properties": {
        "name": {
          "title": "Name",
          "description": "Select a fault type from [\"PodChaos\", \"NetworkChaos\", \"DNSChaos\", \"HTTPChaos\", \"StressChaos\", \"IOChaos\", \"TimeChaos\"]",
          "enum": [
            "PodChaos",
            "NetworkChaos",
            "DNSChaos",
            "HTTPChaos",
            "StressChaos",
            "IOChaos",
            "TimeChaos"
          ],
          "type": "string"
        },
        "name_id": {
          "title": "Name Id",
          "description": "An identifier to prevent name conflicts when the same Fault appears. Assign numbers starting from 0 in sequential order to prevent name conflicts.",
          "type": "integer"
        },
        "grace_period": {
          "title": "Grace Period",
          "description": "Time elapsed from the start of the current phase to the beginning of the fault injection.",
          "example": "0s",
          "type": "string"
        },
        "duration": {
          "title": "Duration",
          "description": "Duration of the unit test. (grace_period + duration) should not exceed the current phase's total time.",
          "example": "2m",
          "type": "string"
        }
      },
      "required": [
        "name",
        "name_id",
        "grace_period",
        "duration"
      ]
    },
    "UnitTest": {
      "title": "UnitTest",
      "type": "object",
      "properties": {
        "name": {
          "title": "Name",
          "description": "Steady state name to be verified by a unit test.",
          "type": "string"
        },
        "grace_period": {
          "title": "Grace Period",
          "description": "Time elapsed from the start of the current phase to the beginning of the unit test.",
          "example": "0s",
          "type": "string"
        },
        "duration": {
          "title": "Duration",
          "description": "Duration of the unit test. (grace_period + duration) should not exceed the current phase's total time.",
          "example": "2m",
          "type": "string"
        }
      },
      "required": [
        "name",
        "grace_period",
        "duration"
      ]
    }
  }
}
```
\end{dynamictemplate}

\begin{experimentprompt}[Agent \# 2-2 for summarizing the planned experiment]
//System://
You are a helpful AI assistant for Chaos Engineering.
Given a Chaos-Engineering-experiment plan, you will summarize it in detail according to the following rules:
- In each phase, describe in detail the timeline for when each fault injection/unit test (for verifying steady-state) will be executed. For example, summarize which fault injections/unit tests will be executed simultaneously, and whether certain fault injections/unit tests will be executed at staggered timings. 
- Be sure to specify both each fault injection/unit test and their corresponding workflow names.
- When explaining the timeline, provide a detailed description using specific values for duration, grace period, etc. Rephrase the specific values in a way that everyone can easily understand.
- The meanings of each value are as follows:
  - Grace Period: Time elapsed from the start of the current phase to the beginning of the fault injection/unit test.
  - Duration: Duration of the fault injection/unit test. (grace_period + duration) should not exceed the corresponding phase's total time.
- Never output bullet points.
- The output should be formatted as a JSON instance that conforms to the JSON schema below.
As an example, for the schema {\"properties\": {\"foo\": {\"title\": \"Foo\", \"description\": \"a list of strings\", \"type\": \"array\", \"items\": {\"type\": \"string\"}}}, \"required\": [\"foo\"]}\nthe object {\"foo\": [\"bar\", \"baz\"]} is a well-formatted instance of the schema. The object {\"properties\": {\"foo\": [\"bar\", \"baz\"]}} is not well-formatted.

Here is the output schema:
```
{
  "properties": {
    "summary": {
      "title": "Summary",
      "description": "The summary of the given Chaos-Engineering-experiment plan.",
      "type": "string"
    }
  },
  "required": [
    "summary"
  ]
}
```

//Human://
# Here is my Chaos-Engineering-experiment plan:
## Time Schedule
//@{time_schedule_overview}@//

## Pre-validation Phase
//@{pre_validation_overview}@//

## Fault-injection Phase 
//@{fault_injection_overview}@//

## Post-validation phase
//@{post_validation_overview}@//

Please summarize the above plan.
\end{experimentprompt}

\begin{exampledata}[\texttt{\textbf{time\_schedule\_overview}}]
Given the constraints of the experiment needing to be completed within 1 minute, we need to carefully allocate time to each phase to ensure that we can effectively validate the system's steady states before and after the fault injection, as well as observe the system's behavior during the fault injection. The pre-validation phase is crucial to establish a baseline that the system is in its expected steady state before any faults are introduced. The fault injection phase is where we introduce the chaos to observe how the system behaves under stress. Finally, the post-validation phase is necessary to ensure that the system returns to its steady state after the faults are removed. Given the short total time of 1 minute, a reasonable allocation could be 10 seconds for pre-validation, 40 seconds for fault injection, and 10 seconds for post-validation. This allocation allows us to have a brief but sufficient observation period for each phase, ensuring that we can gather meaningful insights from the experiment.
\end{exampledata}

\begin{exampledata}[\texttt{\textbf{pre\_validation\_overview}}]
In the pre-validation phase, we need to ensure that the system is in its expected steady states before we proceed with fault injection. Given the constraints, we have 10 seconds to perform these checks. We have two steady states to verify: the 'example-pod-running-state' and the 'example-service-http-response-state'.

The 'example-pod-running-state' requires us to check that the Pod is in the 'Running' state at least 90% of the time. We will use the provided Python script to verify this. Since the script checks the Pod status every second, we can run it for 5 seconds to gather sufficient data for validation.

The 'example-service-http-response-state' requires us to ensure that 95% of HTTP requests return a 200 OK status. We will use the K6 script to simulate HTTP requests to the service. The script is configured to run for 5 seconds with 5 virtual users, which should provide enough data to validate this steady state.

Both unit tests will be executed simultaneously to maximize the use of the 10-second window. This approach ensures that we efficiently validate both steady states within the given time constraint, allowing us to proceed confidently to the fault injection phase.
- Verified Steady State #0: ```example-pod-running-state```
  - Workflow Name: ```pre-unittest-example-pod-running-state```
  - Grace Period: ```0s```  
  - Duration: ```5s```  
- Verified Steady State #1: ```example-service-http-response-state```
  - Workflow Name: ```pre-unittest-example-service-http-response-state```  
  - Grace Period: ```0s```  
  - Duration: ```5s```
\end{exampledata}

\begin{exampledata}[\texttt{\textbf{fault\_injection\_overview}}]
In this fault-injection phase, we aim to simulate a cyber attack by injecting two types of faults: PodChaos and NetworkChaos. The goal is to observe how these faults impact the system's steady states. Given the 40-second time constraint, we will stagger the fault injections to maximize the observation of their effects.

First, we will inject the PodChaos fault to simulate a Pod failure. This will directly test the 'example-pod-running-state' steady state, as the Pod's restart policy is set to 'Never', meaning it will not automatically recover. We will start this fault injection immediately at the beginning of the phase and let it run for 20 seconds.

Simultaneously, we will run the unit test for the 'example-pod-running-state' to verify the impact of the PodChaos fault on the Pod's availability.

After 20 seconds, we will inject the NetworkChaos fault to simulate network latency. This will test the 'example-service-http-response-state' steady state by introducing delays in HTTP responses. We will let this fault run for the remaining 20 seconds of the phase.

During the NetworkChaos fault injection, we will also run the unit test for the 'example-service-http-response-state' to verify the impact of network latency on the Service's ability to respond to HTTP requests.

This staggered approach allows us to isolate the effects of each fault on the respective steady states, providing clear insights into the system's vulnerabilities and resilience.
- Verified Steady State #0: ```example-pod-running-state```  
  - Workflow Name: ```fault-unittest-example-pod-running-state```
  - Grace Period: ```0s```
  - Duration: ```20s```
- Verified Steady State #1: ```example-service-http-response-state```
  - Workflow Name: ```fault-unittest-example-service-http-response-state```
  - Grace Period: ```20s```
  - Duration: ```20s```
  
- Injected Faults #0: ```PodChaos```
  - Workflow Name: ```fault-podchaos```
  - Grace Period: ```0s```
  - Duration: ```20s```

- Injected Faults #1: ```NetworkChaos``` 
  - Workflow Name: ```fault-networkchaos```
  - Grace Period: ```20s```
  - Duration: ```20s```  
\end{exampledata}

\begin{exampledata}[\texttt{\textbf{post\_validation\_overview}}]
In the post-validation phase, we need to ensure that the system has returned to its steady states after the fault injection. Given the 10-second total time for this phase, we will conduct unit tests to verify each steady state. The tests will be executed sequentially due to the short duration, ensuring that each steady state is verified independently. The first test will check the 'example-pod-running-state', ensuring the Pod is running. The second test will verify the 'example-service-http-response-state', ensuring the Service responds with a 200 OK status. Each test will have a brief grace period to allow the system to stabilize after the fault injection, followed by a short duration to perform the checks. This approach ensures that both steady states are validated within the available time, confirming the system's recovery.
- Verified Steady State #0: ```example-pod-running-state```
  - Workflow Name: ```post-unittest-example-pod-running-state```
  - Grace Period: ```1s``` 
  - Duration: ```4s```

- Verified Steady State #1: ```example-service-http-response-state``` 
  - Workflow Name: ```post-unittest-example-service-http-response-state``` 
  - Grace Period: ```5s```  
  - Duration: ```4s``` 
\end{exampledata}

\begin{reexperimentprompt}[Agent \# 2-3 for adjusting a failure scope]
//System://
You are a helpful AI assistant for Chaos Engineering.
Given a previous K8s manifests, a Chaos-Engineering-experiment plan for it, and the current K8s manifests, you will determine whether we need to adujst the scope of fault injections for the current K8s manifests.
Always keep the following rules:
- Cosider how you must change or keep the scope (i.e., target) of the fault injecttion comparing the previous K8s manifests and the current K8s manifests.
- You only make minor adjustments related to resource changes, metadata change, etc, so NEVER make any scope changes that alter the original goal of the chaos experiment.
- The output should be formatted as a JSON instance that conforms to the JSON schema below.

As an example, for the schema {\"properties\": {\"foo\": {\"title\": \"Foo\", \"description\": \"a list of strings\", \"type\": \"array\", \"items\": {\"type\": \"string\"}}}, \"required\": [\"foo\"]}\nthe object {\"foo\": [\"bar\", \"baz\"]} is a well-formatted instance of the schema. The object {\"properties\": {\"foo\": [\"bar\", \"baz\"]}} is not well-formatted.

Here is the output schema:
```
{
  "properties": {
    "thought": {
      "title": "Thought",
      "description": "Describe why you need to change/keep the scope of the fault injection for the current K8s manifests.",
      "type": "string"
    },
    "selector": {
      "title": "Selector",
      "description": "Adjust the scope (target) of the fault injection comparing the differeneces between the current and previous manifests. ",
      "allOf": [
        {
          "$ref": "#/definitions/Selectors"
        }
      ]
    }
  },
  "required": [
    "thought",
    "selector"
  ],
  "definitions": {
    "SetBasedRequirements": {
      "title": "SetBasedRequirements",
      "type": "object",
      "properties": {
        "key": {
          "title": "Key",
          "description": "Label key",
          "type": "string"
        },
        "operator": {
          "title": "Operator",
          "description": "Select an operator.",
          "enum": [
            "In",
            "NotIn",
            "Exists",
            "DoesNotExist"
          ],
          "type": "string"
        },
        "values": {
          "title": "Values",
          "description": "Label values. The values set must be non-empty in the case of In and NotIn.",
          "type": "array",
          "items": {
            "type": "string"
          }
        }
      },
      "required": [
        "key",
        "operator",
        "values"
      ]
    },
    "Selectors": {
      "title": "Selectors",
      "type": "object",
      "properties": {
        "namespaces": {
          "title": "Namespaces",
          "description": "Specifies the namespace of the experiment's target Pod. If this selector is None, Chaos Mesh will set it to the namespace of the current Chaos experiment.",
          "type": "array",
          "items": {
            "type": "string"
          }
        },
        "labelSelectors": {
          "title": "Labelselectors",
          "description": "Specifies the label-key/value pairs that the experiment's target Pod must have. If multiple labels are specified, the experiment target must have all the labels specified by this selector.",
          "type": "object",
          "additionalProperties": {
            "type": "string"
          }
        },
        "expressionSelectors": {
          "title": "Expressionselectors",
          "description": "Specifies a set of expressions that define the label's rules to specifiy the experiment's target Pod.",
          "example": [
            {
              "key": "tier",
              "operator": "In",
              "values": [
                "cache"
              ]
            },
            {
              "key": "environment",
              "operator": "NotIn",
              "values": [
                "dev"
              ]
            }
          ],
          "type": "array",
          "items": {
            "$ref": "#/definitions/SetBasedRequirements"
          }
        },
        "annotationSelectors": {
          "title": "Annotationselectors",
          "description": "Specifies the annotation-key/value pairs that the experiment's target Pod must have. If multiple annotations are specified, the experiment target must have all annotations specified by this selector.",
          "type": "object",
          "additionalProperties": {
            "type": "string"
          }
        },
        "fieldSelectors": {
          "title": "Fieldselectors",
          "description": "Specifies the field-key/value pairs of the experiment's target Pod. If multiple fields are specified, the experiment target must have all fields set by this selector.",
          "example": {
            "metadata.name": "my-pod",
            "metadata.namespace": "dafault"
          },
          "type": "object",
          "additionalProperties": {
            "type": "string"
          }
        },
        "podPhaseSelectors": {
          "title": "Podphaseselectors",
          "description": "Specifies the phase of the experiment's target Pod. If this selector is None, the target Pod's phase is not limited.",
          "type": "array",
          "items": {
            "enum": [
              "Pending",
              "Running",
              "Succeeded",
              "Failed",
              "Unknown"
            ],
            "type": "string"
          }
        },
        "nodeSelectors": {
          "title": "Nodeselectors",
          "description": "Specifies the node-label-key/value pairs to which the experiment's target Pod belongs.",
          "type": "object",
          "additionalProperties": {
            "type": "string"
          }
        },
        "nodes": {
          "title": "Nodes",
          "description": "Specifies the node to which the experiment's target Pod belongs. The target Pod can only belong to one node in the configured node list. If multiple node labels are specified, the node to which the experiment's target Pod belongs must have all labels specified by this selector.",
          "type": "array",
          "items": {
            "type": "string"
          }
        },
        "pods": {
          "title": "Pods",
          "description": "Specifies the namespaces and list of the experiment's target Pods. If you have specified this selector, Chaos Mesh ignores other configured selectors.",
          "example": {
            "default": [
              "pod-0",
              "pod-2"
            ]
          },
          "type": "object",
          "additionalProperties": {
            "type": "array",
            "items": {
              "type": "string"
            }
          }
        }
      }
    }
  }
}
```

//Human://
# Here is the previous K8s manifests of my system:
//@{prev_k8s_yamls}@//

# Here is a planned Chaos Engineering:
//@{experiment_plan_summary}@//

# Here is the current K8s menifests of my system:
//@{curr_k8s_yamls}@//

# Here is the scope of a fault injection for the previous manifests.
//@{curr_fault_injection}@//

Now, please adjust the scope of the fault injection for the current manifests. Note that you here focus on the 'selector' parameter (i.e., scope).
\end{reexperimentprompt}

\begin{reexperimentprompt}[Agent \# 2-4 for adjusting a VaC script]
//System://
You are a helpful AI assistant for Chaos Engineering.
Given the previous K8s manifests, a previous unit test to verify whether the steady state satisfies the threshold, and the reconfigured K8s manifests, you will determine whether the unit test requires adjustment to account for the changes in the reconfigured manifests, and adjust it as necessary.
Always keep the following rules:
- First, consider which K8s manifest resource is the target of the unit test. If there are changes to that manifest, update the unit test as necessary. If there are no changes, the unit test should not require modification.
- You may only make minor adjustments to K8s API, HTTP, or DNS request to account for changes in resource types, parameter seetings, metadata, etc.
- The reconfiguration was made so that the system satisfy the threshold value in the previous unit test, so the threshold value or other parameters must remain unchanged in the new unit test. For example, suppose the number of replicas was reconfigured from 1 to 3 in order to maintain a steady state with more than 1 active pod at all times. In such cases, changing the threshold value from 1 to 3 would alter the intent of this steady state, so the threshold value must remain unchanged (i.e., more than 1 active pod)."
- If redundancy has been newly added, the unit test should verify whether the steady state is maintained by the entire redundancy.
- If the unit test's content needs no changes and only function or variable names need to be changed, leave them as they are to save output costs.
- The output should be formatted as a JSON instance that conforms to the JSON schema below.

As an example, for the schema {\"properties\": {\"foo\": {\"title\": \"Foo\", \"description\": \"a list of strings\", \"type\": \"array\", \"items\": {\"type\": \"string\"}}}, \"required\": [\"foo\"]}\nthe object {\"foo\": [\"bar\", \"baz\"]} is a well-formatted instance of the schema. The object {\"properties\": {\"foo\": [\"bar\", \"baz\"]}} is not well-formatted.

Here is the output schema:
```
{
  "properties": {
    "thought": {
      "title": "Thought",
      "description": "Describe your thought process for determining whether the unit test requires adjustment to account for the changes in the reconfigured manifests: First, consider which K8s manifest resource is the target of the unit test. If there are changes to that manifest, update the unit test as necessary. If there are no changes, the unit test should not require modification. If the unit test needs updating, describe also how you modify the inspection method according to the differences between the previous and reconfigured manifests. If the modification is not required, describe the reason.",
      "type": "string"
    },
    "code": {
      "title": "Code",
      "description": "If the unit test needs updating, write a new unit test code with the inspection method modified. Write only the content of the code without enclosing it in a code block. If not, this field is not required.",
      "type": "string"
    }
  },
  "required": [
    "thought"
  ]
}
```

//Human://
# Here is the previous K8s manifests of my system:
//@{prev_k8s_yamls}@//

# Here is the reconfigured K8s manifests of my system:
//@{curr_k8s_yamls}@//

# Here is the unit test for the previous manifests.
//@{prev_unittest}@//

Now, please determine whether the unit test requires adjustment to account for the changes in the reconfigured manifests, and adjust it as necessary.

|\tcbsubtitle{In the verification loop, the prompts below will be stacked as history}|

//AI://
//@{output}@//

//Human://
Your current unit test causes errors when conducted.
The error message is as follows:
//@{error_message}@//

This unit test should be succeeded.
Please analyze the reason why the errors occur, then fix the errors.
Always keep the following rules:
- NEVER repeat the same fixes that have been made in the past.
- Fix only the parts related to the errors without changing the original intent.
- //<<{the same format instructions as in the System role}>>//
\end{reexperimentprompt}

\begin{exampledata}[\texttt{\textbf{experiment\_plan\_summary}}]
The Chaos Engineering experiment is structured into three phases: pre-validation, fault injection, and post-validation, all to be completed within a total of 1 minute.

In the pre-validation phase, which lasts for 10 seconds, two unit tests are executed simultaneously to verify the system's steady states before any faults are introduced. The 'example-pod-running-state' is checked using a Python script to ensure the Pod is in the 'Running' state at least 90% of the time. This test runs for 5 seconds. Concurrently, the 'example-service-http-response-state' is verified using a K6 script to simulate HTTP requests, ensuring 95% of requests return a 200 OK status. This test also runs for 5 seconds. Both tests start immediately at the beginning of the phase.

The fault injection phase spans 40 seconds and involves two staggered fault injections. Initially, the PodChaos fault is injected to simulate a Pod failure, running for the first 20 seconds. Simultaneously, the 'example-pod-running-state' unit test is conducted to observe the impact of this fault. After 20 seconds, the NetworkChaos fault is introduced to simulate network latency, running for the remaining 20 seconds. During this period, the 'example-service-http-response-state' unit test is executed to assess the effect of network delays. This staggered approach allows for isolated observation of each fault's impact on the system.

Finally, the post-validation phase, lasting 10 seconds, ensures the system returns to its steady states after fault removal. The tests are conducted sequentially. The 'example-pod-running-state' is verified first, with a 1-second grace period followed by a 4-second test duration. Subsequently, the 'example-service-http-response-state' is checked, starting after a 5-second grace period and running for 4 seconds. This sequence confirms the system's recovery to its expected steady states.
\end{exampledata}

\subsubsection{Analysis}
\newtcblisting[use counter from=reexperimentprompt]{analysisprompt}[1][]{
    colback=yellow!3,
    listing only,
    title={Prompt \thetcbcounter: #1},
    listing options={
        language=sysprompt,
        basicstyle=\scriptsize\ttfamily,
        breaklines=true,
        breakindent=0pt,
        showstringspaces=false,
        escapechar=|,
    },
    boxrule=0.3mm,
    mybreakable,
    coltitle=black,
    colbacktitle=yellow!20
}

\begin{analysisprompt}[Agent \# 3-0 for analyzing an experiment results]
//System://
You are a helpful AI assistant for Chaos Engineering.
Given K8s manifests for a network system, its hypothesis, the overview of a Chaos-Engineeering experiment, and the experimental results, you will analyze the experimental results.
Always keep the following rules:
- Analyze step by step why the test(s) failed, based on the system configurations (manifests) and the flow of the experiment.
- Specify the cause while mentioning the corresponding system configurations and the corresponding phenomena in the Chaos-Engineering experiment.
- The analysis report here will be used for reconfiguring the system later to avoid the failures and improve resiliency. Therefore, make carefully the report rich in insights so that it will be helpful at that time.
- When providing insights and reconfiguration recommendations, limit them to areas related to the failed test.
- The output should be formatted as a JSON instance that conforms to the JSON schema below.

As an example, for the schema {\"properties\": {\"foo\": {\"title\": \"Foo\", \"description\": \"a list of strings\", \"type\": \"array\", \"items\": {\"type\": \"string\"}}}, \"required\": [\"foo\"]}\nthe object {\"foo\": [\"bar\", \"baz\"]} is a well-formatted instance of the schema. The object {\"properties\": {\"foo\": [\"bar\", \"baz\"]}} is not well-formatted.

Here is the output schema:
```
{
  "properties": {
    "report": {
      "title": "Report",
      "description": "Analysis of the experiment result.",
      "type": "string"
    }
  },
  "required": [
    "report"
  ]
}
```

//Human://
# Here is the overview of my system:
//@{user_input2}@//

# Here is the hypothesis for my system:
The hypothesis is "The steady states of the sytem are maintained even when the fault scenario occurs (i.e., when the faults are injected)".
The steady states here are as follows:
//@{steady_states}@//

The fault scenario here is as follows:
//@{detailed_fault_scenario}@//

# Here is the overview of my Chaos-Engineering experiment to verify the hypothesis:
//@{experiment_plan_summary}@//

|\tcbsubtitle{For the first analysis, the following prompt is added}|

# The experiment's results are as follows:
//@{experiment_result}@//

Now, please analyze the results and provide an analysis report rich in insights.

|\tcbsubtitle{For the second and subsequent analyses, the following prompt is added}|

# The update history for the above K8s manifests is the following:
//@{reconfig_history}@//

# The experiment's results in the latest K8s manifests are as follows:
//@{experiment_result}@//

Now, please analyze the results and provide an analysis report rich in insights.

\end{analysisprompt}

\begin{exampledata}[\texttt{\textbf{experiment\_result}}]
Passed unittests:
- pre-unittest-example-pod-running-state
- pre-unittest-example-service-http-response-state

Failed unittests:
- fault-unittest-example-pod-running-state
```log
Exception when calling CoreV1Api->read_namespaced_pod: (404)
Reason: Not Found\nHTTP response headers: HTTPHeaderDict({'Audit-Id': '8a1e6c00-ebd9-43ee-9522-6399ce015252', 'Cache-Control': 'no-cache, private', 'Content-Type': 'application/json', 'X-Kubernetes-Pf-Flowschema-Uid': 'c4624bd9-7fc7-42c6-bcb8-4235110a860d', 'X-Kubernetes-Pf-Prioritylevel-Uid': '4706085f-6263-43ae-93f5-b4a61de8b6be', 'Date': 'Sun, 24 Nov 2024 12:06:18 GMT', 'Content-Length': '190'})
HTTP response body: {\"kind\":\"Status\"...', 'X-Kubernetes-Pf-Flowschema-Uid': 'c4624bd9-7fc7-42c6-bcb8-4235110a860d', 'X-Kubernetes-Pf-Prioritylevel-Uid': '4706085f-6263-43ae-93f5-b4a61de8b6be', 'Date': 'Sun, 24 Nov 2024 12:06:37 GMT', 'Content-Length': '190'})\nHTTP response body: {\"kind\":\"Status\",\"apiVersion\":\"v1\",\"metadata\":{},\"status\":\"Failure\",\"message\":\"pods \\\"example-pod\\\" not found\",\"reason\":\"NotFound\",\"details\":{\"name\":\"example-pod\",\"kind\":\"pods\"},\"code\":404}

Pod was running 0 out of 20 seconds, which is 0.00% of the time.
```

- fault-unittest-example-service-http-response-state
```log
time=\"2024-11-24T12:06:38Z\" level=warning msg=\"Request Failed\" error=\"Get \\\"http:\/\/example-service.default.svc.cluster.local:80\\\": dial tcp 10.96.255.84:80: connect: connection refused\"\ntime=\"2024-11-24T12:06:38Z\" level=warning msg=\"Request Failed\" error=\"Get \\\"http:\/\/example-service.default.svc.cluster.local:80\\\": dial tcp 10.96.255.84:80: connect: connection refused\"\ntime=\"2024-11-24T12:06:38Z\" level=warning msg=\"Request Failed\" error=\"Get \\\"http:\/\/example-service.default.svc.cluster.local:8... level=error msg=\"thresholds on metrics 'http_req_failed' have been crossed
```
\end{exampledata}

\subsubsection{Improvement}
\newtcblisting[use counter from=analysisprompt]{improveprompt}[1][]{
    colback=orange!3,
    listing only,
    title={Prompt \thetcbcounter: #1},
    listing options={
        language=sysprompt,
        basicstyle=\scriptsize\ttfamily,
        breaklines=true,
        breakindent=0pt,
        showstringspaces=false,
        escapechar=|,
    },
    boxrule=0.3mm,
    mybreakable,
    coltitle=black,
    colbacktitle=orange!20,
}

\begin{improveprompt}[Agent \# 4-0 for reconfiguring K8s manifests]
//System://
You are a helpful AI assistant for Chaos Engineering.
Given K8s manifests that define a network system, its hypothesis, the overview of a Chaos-Engineeering experiment, and the experiment's results, you will reconfigure the system based on analysis of the experiment's results.
Always keep the following rules:
- NEVER change the original intention (its description) of the original version of the system.
- NEVER do the same reconfiguration as in the history.
- Start with simple reconfiguration, and if the hypothesis is still not satisfied, gradually try more complex reconfigurations.
- The output should be formatted as a JSON instance that conforms to the JSON schema below.

As an example, for the schema {\"properties\": {\"foo\": {\"title\": \"Foo\", \"description\": \"a list of strings\", \"type\": \"array\", \"items\": {\"type\": \"string\"}}}, \"required\": [\"foo\"]}\nthe object {\"foo\": [\"bar\", \"baz\"]} is a well-formatted instance of the schema. The object {\"properties\": {\"foo\": [\"bar\", \"baz\"]}} is not well-formatted.

Here is the output schema:
```
{
  "properties": {
    "thought": {
      "title": "Thought",
      "description": "Describe your plan to modify the K8s manifests.",
      "type": "string"
    },
    "modified_k8s_yamls": {
      "title": "Modified K8S Yamls",
      "description": "The list of modified K8s manifests (yamls). If you create a new manifest to modify resources in an existing manifest, make sure to delete the existing manifest before creating the new one.",
      "type": "array",
      "items": {
        "$ref": "#/definitions/ModK8sYAML"
      }
    }
  },
  "required": [
    "thought",
    "modified_k8s_yamls"
  ],
  "definitions": {
    "ModK8sYAML": {
      "title": "ModK8sYAML",
      "type": "object",
      "properties": {
        "mod_type": {
          "title": "Mod Type",
          "description": "Modification type. Select from ['replace', 'create', 'delete']. The 'replace' replaces/overwites the content of an exisiting yaml. The 'create' creates a new yaml. The 'delete' deletes an existing yaml.",
          "enum": [
            "replace",
            "create",
            "delete"
          ],
          "type": "string"
        },
        "fname": {
          "title": "Fname",
          "description": "The file name of the modified yaml. If mod_type is 'replace' or 'delete', the name must match an existing yaml's name. If mod_type='create', name the file appropriately to avoid overlapping with existing yamls' names.",
          "type": "string"
        },
        "explanation": {
          "title": "Explanation",
          "description": "If mod_type is 'delete', explain why you need to delete the yaml. If mod_type is 'replace', explain which part you should modify from the original conde and why. If mod_type is 'create', explain whether it is a completely new resource or a replacement resouce for an existing resource. If it is a replacement, also explain the differences and the reasons for them, just like with 'replace'.",
          "type": "string"
        },
        "code": {
          "title": "Code",
          "description": "If mod_type is 'delete', this field is not required. Otherwise, write the content of a K8s YAML manifest modified to pass all the unit tests. Write only the content of the code, and for dictionary values, enclose them within a pair of single double quotes (\").",
          "type": "string"
        }
      },
      "required": [
        "mod_type",
        "fname",
        "explanation"
      ]
    }
  }
}
```

//Human://
# Here is the overview of my system (original version):
//@{user_input2}@//

# Here is the hypothesis for my system:
The hypothesis is "The steady states of the sytem are maintained even when the fault scenario occurs (i.e., when the faults are injected)".
The steady states here are as follows:
//@{steady_states}@//

The fault scenario here is as follows:
//@{detailed_fault_scenario}@//

# Here is the overview of my Chaos-Engineering experiment to verify the hypothesis:
//@{experiment_plan_summary}@//

# The experiment's results of the original system are as follows:
//@{experiment_result}@//

First, please analyze the results and provide an analysis report rich in insights.

//AI://
# Here is my analysis report:
//@{analysis_report}@//

//Human://
Then, please reconfigure the system to avoid the fails (improve resiliency).

|\tcbsubtitle{In the improvement loop, the prompts below will be stacked as improvement history}|

//AI://
```json
//@{output}@//
```"""

//Human://
# Here is the K8s menifests of the modified system (version=//@{mod_version}@//):
//@{k8s_yamls_mod}@//

# The experiment's results of the modified system were as follows:
//@{experiment_result_mod}@//

Please analyze the results and provide an analysis report rich in insights again.

//AI://
# Here is my analysis report:
//@{analysis_report_mod}@//

//Human://
Then, please reconfigure the system to avoid the fails (improve resiliency).

|\tcbsubtitle{In the verification loop, the prompts below will be stacked as verification history. Verification history is kept within a single verification loop and is reset in every improvement iteration}|

//AI://
```json
//@{output}@//
```

//User://
Your current unittest causes errors when conducted.
The error message is as follows:
//@{error_message}@//

Please analyze the reason why the errors occur, then fix the errors.
Always keep the following rules:
- Ensure that the implementation supports variable durations again.
- NEVER repeat the same fixes that have been made in the past.
- Fix only the parts related to the errors without changing the original content.
- //<<{the same format instruction as in the System role}>>//
\end{improveprompt}

\begin{exampledata}[\texttt{\textbf{analysis\_report}}]
The Chaos Engineering experiment aimed to test the system's resilience by simulating a cyber attack through PodChaos and NetworkChaos. The experiment results indicate failures in maintaining the defined steady states during and after the fault injection phase. Here is a detailed analysis of the results:

1. **Pre-Validation Phase:**
  - Both pre-unittests, `example-pod-running-state` and `example-service-http-response-state`, passed successfully. This indicates that the system was in a healthy state before the fault injection, with the Pod running 100% of the time and the Service responding with a 200 OK status for all requests.

2. **Fault Injection Phase:**
  - **PodChaos Fault:** The `fault-unittest-example-pod-running-state` failed because the Pod was not found (404 error) during the test. This is expected due to the PodChaos action 'pod-kill', which terminated the Pod. Since the Pod's restart policy is set to 'Never', it did not restart, leading to 0% uptime during the test. This highlights the critical issue of having a single Pod with no redundancy or automatic recovery mechanism.
  - **NetworkChaos Fault:** The `fault-unittest-example-service-http-response-state` also failed. The logs show repeated connection refused errors, indicating that the Service could not route traffic to the Pod, as it was not running. This failure is a direct consequence of the Pod being unavailable, demonstrating the lack of redundancy and the impact of a single point of failure.

3. **Post-Validation Phase:**
  - Both post-unittests, `example-pod-running-state` and `example-service-http-response-state`, failed. The Pod was still not found, and the Service continued to refuse connections. This indicates that the system did not recover to its steady states after the faults were removed, primarily due to the Pod's restart policy and the absence of a controller to manage Pod lifecycle and redundancy.

**Insights and Recommendations:**
  - **Pod Restart Policy:** The Pod's restart policy should be changed from 'Never' to 'Always' or 'OnFailure' to ensure automatic recovery in case of failures.
  - **Redundancy and Scalability:** Implement a Deployment or ReplicaSet to manage the Pod. This will provide redundancy and ensure that a minimum number of Pods are always running, improving the system's resilience to failures.
  - **Service Availability:** Ensure that the Service can handle traffic even if one Pod fails by having multiple replicas. This can be achieved by scaling the Deployment to have more than one replica.
  - **Monitoring and Alerts:** Implement monitoring and alerting mechanisms to detect and respond to Pod failures promptly, minimizing downtime.

By addressing these issues, the system can improve its resilience and maintain its steady states even during fault scenarios.
\end{exampledata}

\subsubsection{Post-processing}
\newtcblisting[use counter from=improveprompt]{postprocessprompt}[1][]{
    colback=gray!5,
    listing only,
    title={Prompt \thetcbcounter: #1},
    listing options={
        language=sysprompt,
        basicstyle=\scriptsize\ttfamily,
        breaklines=true,
        breakindent=0pt,
        showstringspaces=false,
    },
    boxrule=0.3mm,
    mybreakable,
    coltitle=black,
    colbacktitle=gray!20
}

\begin{postprocessprompt}[Agent \# EX for summarizing a completed CE cycle]
//System://
You are a helpful AI assistant for Chaos Engineering.
Given a summary of a Chaos Engineering cycle, please elaborate the summary.
The output should be formatted as a JSON instance that conforms to the JSON schema below.

As an example, for the schema {\"properties\": {\"foo\": {\"title\": \"Foo\", \"description\": \"a list of strings\", \"type\": \"array\", \"items\": {\"type\": \"string\"}}}, \"required\": [\"foo\"]}\nthe object {\"foo\": [\"bar\", \"baz\"]} is a well-formatted instance of the schema. The object {\"properties\": {\"foo\": [\"bar\", \"baz\"]}} is not well-formatted.

Here is the output schema:
```
{
  "properties": {
    "summary": {
      "title": "Summary",
      "type": "string"
    }
  },
  "required": [
    "summary"
  ]
}
```

//Human://
Here is the overview of a Chaos Engineering Cycle:
# Here is a Chaos Engineering cycle
## Step 0. User-input understanding
### Here is the overview of user inputs:
//@{user_input2}@//

## Step 1. Hypothesis definition
### Here is the overview of the hypothesis for the system:
The hypothesis is "The steady states of the system are maintained even when the fault scenario occurs (i.e., when the faults are injected)".
The steady states here are as follows:
//@{steady_states}@//

The fault scenario here is as follows:
//@{detailed_fault_scenario}@//

## Step 2.1. Chaos-Engineering experiment
### Here is the overview of my Chaos-Engineering experiment to verify the hypothesis:
//@{experiment_plan_summary}@//

## Step 2.2, 3, 4. Experiment execution, analysis and improvement (reconfiguring the system to satisfy the hypothesis)
### Here is the improvement history:
//@{improvement_history}@//

Please elaborate the above summary of a Chaos Engineering Cycle.
\end{postprocessprompt}

\begin{exampledata}[\texttt{\textbf{improvement\_history}}]
### Experiment result (1st try)
//<<{experiment_result}>>//

### Analysis report (1st try)
//<<analysis_report>>//

### Improvement result (1st try)
1 K8s manifests are modified:
- The K8s manifest 'nginx/pod.yaml' was replaced.

Replace the Pod manifest with a Deployment manifest to provide redundancy and automatic recovery. The Deployment will manage the Pod lifecycle, ensuring that a specified number of replicas are always running, which addresses both the restart policy and redundancy issues.
```yaml
apiVersion: apps/v1
kind: Deployment
metadata:
  name: example-deployment
  labels:
    app: example
spec:
  replicas: 3
  selector:
    matchLabels:
      app: example
    template:
      metadata:
        labels:
          app: example
      spec:
        containers:
        - name: example-container
          image: nginx:1.17.1
          ports:
          - containerPort: 80
```

### Experiment result (2nd try)
Passed unittests:
- pre-unittest-example-pod-running-state
- pre-unittest-example-service-http-response-state
- fault-unittest-example-pod-running-state
- fault-unittest-example-service-http-response-state
- post-unittest-example-pod-running-state
- post-unittest-example-service-http-response-state

Failed unittests:
\end{exampledata}

\section{Full Results of Case Studies}
\label{adx:case study}
In this section, we provide full inputs and outputs for each run studied in the case study.
The inputs include instructions, Skaffold configuration files, K8s manifests. The outputs include VaC scripts, Chaos Mesh workflow manifests, reconfigured K8s manifests, adjusted VaC scripts in the experiment replanning phase.
In the case study, we pick up the fourth run for \textsc{Nginx} and the fifth run for \textsc{SockShop}.
Table \ref{tab:input_stats} shows the statistics of \textsc{Nginx} and \textsc{SockShop}.
Before showing the full results, we describe the CE cycles for each system referring to their highlight outputs.
\begin{table}[!h]
    \caption{Statistics of the systems for the case study. Tokens are counted by the tokenizer \texttt{cl100k\_base}.}
    \begin{center}
    \small
    \setlength{\tabcolsep}{4.2px}
    \begin{tabular}{l cccc}
        \toprule
        System &\# manifests &\# lines & \# words &\# tokens\\
        \midrule
        \textsc{Nginx} &2 &24 &373 &115 \\
        \textsc{SockShop} &29 &869 &17696 &4605 \\
        \bottomrule
    \end{tabular}
    \label{tab:input_stats}
    \end{center}
\end{table}


\paragraph{Description of the results for \textsc{Nginx}}
In the \textit{hypothesis} phase, \textsc{ChaosEater} first defines two steady states: 1) ``The \texttt{Pod} should be running at least 90\% of the time during the check period"; 2) ``Service availability should be at least 99.9\% with a response status of 200". 
The VaC scripts shown in Figure \ref{fig:highlight} correctly implement these steady states. 
It then defines a failure sequence that injects \texttt{NetworkChaos} (delay) into the Nginx \texttt{Pod} following \texttt{PodChaos} (pod-kill) to simulate a cyberattack.

In the experiment planning, it plans the following chaos experiment: 1) the two steady states are sequentially validated in the pre-validation phase; 2) in the failure-injection phase, the two steady states are sequentially validated alongside the injection of each failure that may affect them; 3) the two steady states are validated sequentially once again in the post-validation phase.

The first chaos experiment reveals that the two steady states are currently not satisfied in both the failure-injection and post-validation phases.
\textsc{ChaosEater} successfully identifies the cause as the Nginx \texttt{Pod}'s never restarting policy and replaces it with a \texttt{Deployment} with three replicas.
It then adjusts the success criteria of the VaC script to ensure that at least one Pod out of three replicas is running and conducts the chaos experiment for the reconfigured system again.
Finally, the additional chaos experiment confirms that the hypothesis is satisfied in the reconfigured system.

Intuitively, these operations and outputs follow best practices, demonstrating that \textsc{ChaosEater} can complete our expected CE cycle for the small-scale system, \textsc{Nginx}, without explicit user instructions.

\paragraph{Description of the results for \textsc{SockShop}}
In the \textit{hypothesis} phase, \textsc{ChaosEater} first defines two steady states: 1) ``At least 1 ready replica 100\% of the time and 2 ready replicas at least 80\% of the time during the monitoring period" for the carts \texttt{Deployment}; 2) ``At least 1 ready replica must be present 100\% of the time during the monitoring period" for the front-end \texttt{Deployment}.
The VaC scripts shown in Figure \ref{fig:highlight} correctly implement these steady states. 

It then defines a failure sequence that injects \texttt{StressChaos} (CPU) into all the carts-db replicas following \texttt{PodChaos} (pod-kill) that targets the single front-end replica to simulate possible problems in a black Friday sale.

In the experiment planning, it plans the following chaos experiment: 1) the two steady states are simultaneously validated in the pre-validation phase; 2) in the failure-injection phase, the two steady states are sequentially validated alongside the injection of each failure that may affect them; 3) the two steady states are validated simultaneously once again in the post-validation phase.

The first chaos experiment reveals that the second steady state is currently not satisfied in both the failure-injection and post-validation phases.
\textsc{ChaosEater} successfully identifies the cause as the single replica setting of the front-end \texttt{Deployment} and increases the number of replicas to two.
It then conducts the chaos experiment for the reconfigured system again without experiment adjustments (no adjustment is needed as the resource type remains unchanged).
Finally, the additional chaos experiment confirms that the hypothesis is satisfied in the reconfigured system.

This case also broadly follows best practices, demonstrating that \textsc{ChaosEater} can complete our expected CE cycle even for the comparatively large system, \textsc{SockShop}, without explicit user instructions.

\subsection{Nginx}
\label{adx:nginx}
\subsubsection{Inputs}
\begin{custominput}[Instructions]
The Chaos-Engineering experiment must be completed within 1 minute.
\end{custominput}

\begin{custominput}[skaffold.yaml]
apiVersion: skaffold/v3
kind: Config
metadata:
  name: nginx-app
manifests:
  rawYaml:
    - pod.yaml
    - service.yaml
\end{custominput}

\begin{custominput}[pod.yaml]
apiVersion: v1
kind: Pod
metadata:
  name: example-pod
  labels:
    app: example
spec:
  restartPolicy: Never
  containers:
  - name: example-container
    image: nginx:1.17.1
    ports:
    - containerPort: 80
\end{custominput}

\begin{custominput}[service.yaml]
apiVersion: v1
kind: Service
metadata:
  name: example-service
spec:
  selector:
    app: example
  ports:
    - protocol: TCP
      port: 80
      targetPort: 80
\end{custominput}

\subsubsection{CE cycle completed by \textsc{ChaosEater}}
\begin{customdialogue}[Preprocessing]
<<Summaries of each manifest>>
nginx/pod.yaml
- This manifest defines a Kubernetes Pod.
- The Pod is named 'example-pod'.
- It includes metadata with a label 'app: example'.
- The Pod's restart policy is set to 'Never', meaning it won't restart automatically 
  if it fails.
- The Pod contains one container named 'example-container'.
- The container uses the 'nginx:1.17.1' image, which is a specific version of the
  Nginx webserver.
- The container exposes port 80, which is commonly used for HTTP traffic.

nginx/service.yaml
- This manifest defines a Kubernetes Service.
- The Service is named 'example-service'.
- It uses the 'v1' API version.
- The Service selects pods with the label 'app: example'.
- It exposes the Service on port 80 using the TCP protocol.
- The Service forwards traffic to the target port 80 on the selected pods.

<<Resiliency issues/weaknesses in the manifests>>
Issue #0: Pod Restart Policy
details: The Pod will not restart automatically if it fails, which can lead to downtime.
manifests having the issues: ['nginx/pod.yaml']
problematic config: restartPolicy: Never

Issue #1: Single Pod Deployment
details: Having a single Pod means there is no redundancy. If the Pod fails, the service will be unavailable.
manifests having the issues: ['nginx/pod.yaml']
problematic config: Only one Pod is defined without a Deployment or ReplicaSet.

<<Application of the manifests>>
Thoughts:
The manifests provided define a Kubernetes Pod and a Service. The Pod runs an Nginx container, which is a popular web server used to serve static content, reverse proxy, or load balance HTTP traffic. The Service is configured to expose this Pod on port 80, which is the default port for HTTP traffic. Given the use of Nginx and the configuration of the Service, it is logical to assume that these manifests are intended to deploy a simple web server application. The file names and the use of Nginx further support this assumption.

Assumed application:
A simple web server application using Nginx to serve HTTP content.

<<Summary of user instructions>>
- The Chaos-Engineering experiment must be completed within 1 minute.
\end{customdialogue}

\begin{customdialogue}[Steady state \#1: example-pod-running]
//Description//
The first issue to address is the Pod's restart policy set to 'Never'. This is a critical issue because if the Pod fails, it will not restart, leading to potential downtime. Therefore, the steady state should ensure that the Pod is running and available. A measurable output for this steady state is the number of running Pods, which should be 1, as there is only one Pod defined in the manifest. This steady state will help verify that the Pod is up and running, which is crucial given the restart policy configuration.

//Current state inspection//
To inspect whether the 'example-pod' is running, I will use the K8s API. This is because the state to be checked is the status of a Kubernetes resource, specifically the Pod, which is best done using the Kubernetes client libraries. The script will check the status of the Pod every second for a duration of 5 seconds, which is within the allowed maximum duration. This will ensure that the Pod is running and available, addressing the critical issue of the Pod's restart policy being set to 'Never'.
tool: k8s duration: 5s
Inspection script (Python) k8s_example-pod-running.py :
```
import os
import time
from kubernetes import client, config

# Load Kubernetes configuration based on the environment
if os.getenv('KUBERNETES_SERVICE_HOST'):
    config.load_incluster_config()
else:
    config.load_kube_config()

v1 = client.CoreV1Api()

def check_pod_status(namespace, pod_name):
    try:
        pod = v1.read_namespaced_pod(
            name=pod_name, namespace=names)
        return pod.status.phase == 'Running'
    except client.exceptions.ApiException as e:
        print(f"Exception when calling CoreV1Api->read_namespaced_po
        return False
    
def main(duration):
    namespace = 'default'
    pod_name = 'example-pod'
    running_count = 0
    for _ in range(duration):
        if check_pod_status(namespace, pod_name):
            running_count += 1
        time.sleep(1)
    print(f"Pod '{pod_name}' running status checked {duration} times

if __name__ == '__main__':
    import argparse
    parser = argparse.ArgumentParser(description='Check if a pod is
    parser.add_argument('--duration', type=int, default=5, help='Dur
    args = parser.parse_args()
    main(args.duration)
```

The inspection script's results (current states) are as follows:
```
Pod 'example-pod' running status checked 5 times. Running count: 5.
```

//Threshold//
The steady state we are considering is whether the 'example-pod' is running. The current state shows that the pod was checked 5 times over a duration of 5 seconds, and it was running each time, resulting in a running count of 5. This indicates that the pod is consistently running during the check period. Given the constraints of the chaos engineering experiment, which must be completed within 1 minute, we can set a threshold that allows for some tolerance in case of brief fluctuations. A reasonable threshold would be that the pod should be running at least 90% of the time during the
check period. This allows for a small margin of error while still ensuring that the pod is generally available and running.

The pod should be running at least 90% of the time during the check period.

//Unit test to validate the steady state//
To add the threshold assertion to the inspection Python script, I will calculate the percentage of time the pod is running during the specified duration. The threshold is set to 90%, meaning the pod should be running at least 90% of the time during the check period. I will implement a for-loop to check the pod status every second for the specified duration and then assert that the running percentage is greater than or equal to 90%. This ensures that the pod is generally available and running, allowing for a small margin of error.
```
import os
import time
import argparse
from kubernetes import client, config
from unittest_base import K8sAPIBase

class TestPodRunningState(K8sAPIBase):
    def __init__(self):
        super().__init__()

    def check_pod_status(self, namespace, pod_name):
        try:
            pod = self.v1.read_namespaced_pod(
                name=pod_name, namespace=namespace)
            return pod.status.phase == 'Running'
        except client.exceptions.ApiException as e:
            print(f"Exception when calling CoreV1Api->read_namespaced_pod: {e}")
            return False

    def test_pod_running_state(self, duration):
        namespace = 'default'
        pod_name = 'example-pod'
        running_count = 0

        # Check the pod status every second for the specified duration
        for _ in range(duration):
            if self.check_pod_status(namespace, pod_name):
                running_count += 1
            time.sleep(1)

        # Calculate the running percentage
        running_percentage = (running_count / duration) * 100

        # Assert that the running percentage is at least 90%
        assert running_percentage >= 90, f"Pod '{pod_name}' running percentage is below threshold: {running_percentage}%"

        print(f"Pod '{pod_name}' running status checked {duration} times. Running percentage: {running_percentage}%.")


def main():
    parser = argparse.ArgumentParser(description='Test if a pod is running at least 90% of the time.')
    parser.add_argument('--duration', type=int, default=5, help='Duration to check the pod status in seconds.')
    args = parser.parse_args()

    test = TestPodRunningState()
    test.test_pod_running_state(args.duration)


if __name__ == '__main__':
    main()
```

The unit test results are as follows:
```
Pod 'example-pod' running status checked 5 times. Running percentage 100%.
```
\end{customdialogue}

\begin{customdialogue}[Steady state \#2: example-service-availability]
<<Description>>
The next issue to address is the 'Single Pod Deployment', which is related to the lack of redundancy. This is a significant issue because if the single Pod fails, the service will be unavailable. To verify this, we can define a steady state that checks the availability of the service itself. A measurable output for this steady state is the service's response time or availability. Since the service is exposed on port 80, we can check if the service is responding to HTTP requests. This steady state will help verify that the service is available and responsive, which is crucial given the single Pod deployment configuration.

<<Current state inspection>>
To inspect the availability of the 'example-service', we need to check if the service is responding to HTTP requests. Since the service is exposed on port 80, we can use k6 to simulate HTTP requests to the service and measure its response time and availability. This approach allows us to verify the service's responsiveness and availability, which is crucial given the single Pod deployment configuration. We will use a small number of virtual users and a short duration to fit within the 1-minute constraint of the Chaos Engineering experiment.
tool: k6 vus: 1 duration: 5s
Inspection script (Javascript) k6_example-service-availability.js :
```
import http from 'k6/http';
import { check } from 'k6';

export const options = {
  vus: 1,
  duration: '5s',
};

export default function () {
  const res = http.get(
    'http://example-service.default.svc.cluster
     .local'
  )
  check(res, {
    'status is 200': (r) => r.status === 200,
  });
}
```

The inspection script's results (current states) are as follows:
```
OK status is 200
  
checks..........: 100.00% 13915 out of 13915
data_received...: 12 MB 2.4 MB/s
data_sent.......: 1.5 MB 306 kB/s
http_req_blocked...: avg=8.11µs min=1.4µs
                     med=4.19µs max=2.82ms 
                     p(90)=6.47µs
                     p(95)=7.28µs
http_req_connecting...: avg=2.23µs min=0s
                        med=0s max=345.12µs
                        p(90)=0s p(95)=0s      
http_req_duration...: avg=236.57µs 
                      min=90.31µs
                      med=232.08µs
                      max=3.38ms   
                      p(90)=263.84µs
                      p(95)=280.9µs
{expected_response:true}...: avg=236.57µs
                             min=90.31µs
                             med=232.08µs
                             max=3.38ms   
                             p(90)=263.84µs
                             p(95)=280.9µs
http_req_failed...: 0.00% 0 out of 13915
http_req_receiving...: avg=67.9µs
                       min=17.89µs
                       med=61.39µs max=2.36ms   
                       p(90)=91.92µs p(95)=99.14µs
http_req_sending...: avg=18.51µs min=4.66µs
                     med=14.77µs max=2.51ms   
                     p(90)=21.11µs p(95)=23.45µs
http_req_tls_handshaking...: avg=0s min=0s
                             med=0s max=0s      
                             p(90)=0s p(95)=0s
http_req_waiting...: avg=150.16µs
                     min=54.98µs
                     med=151.23µs max=880.47µs 
                     p(90)=168.67µs p(95)=176.19µs
http_reqs...: 13915 2782.666155/s
iteration_duration...: avg=344.61µs
                       min=131.12µs
                       med=333.12µs max=3.98ms   
                       p(90)=377.15µs p(95)=401.15µs
iterations...: 13915 2782.666155/s
vus..........: 1 min=1 max=1
vus_max......: 1 min=1 max=1
```

<<Threshold>>
The steady state we are considering is the availability of the 'example-service'. The k6 test results show that the service is currently responding with a 200 status code for all requests, indicating 100% availability. Given that the system consists of a single Pod, any failure in the Pod would result in the service being unavailable. Therefore, the threshold should ensure that the service remains available and responsive. To account for minor fluctuations and network latency, a reasonable threshold would be to maintain a high availability percentage, slightly below 100% to allow for brief, noncritical failures. A threshold of 99.9% availability is a common standard for web services, allowing for some tolerance while still ensuring high reliability.

Service availability should be at least 99.9% with a response status of 200.

<<Unit test to validate the steady state>>
To add the threshold check to the k6 script, I will include a 'thresholds' property in the 'options' section. This property will specify that the service's availability should be at least 99.9% with a response status of 200. This is done by setting a threshold on the 'http_req_failed' metric, ensuring that the failure rate is less than or equal to 0.1% (which corresponds to 99.9% availability).
```
import http from 'k6/http';
import { check } from 'k6';

export const options = {
  vus: 1,
  duration: '5s',
  thresholds: {
    // Ensure that the service availability is at least 99.9%
    'http_req_failed': ['rate<=0.001'], // 0.1% failure rate corresponds to 99.9% availability
  },
};

export default function () {
  const res = http.get(
    'http://example-service.default.svc
     .cluster.local:80');
  check(res, {
    'status is 200': (r) => r.status === 200,
  });
}
```

The unit test results are as follows:
```
OK status is 200

checks...: 100.00% 14565 out of 14565
data_received.......: 12 MB 2.5 MB/s
data_sent...........: 1.6 MB 320 kB/s
http_req_blocked....: avg=7.92µs min=1.33µs 
                      med=4.1...
http_req_connecting...: avg=2.19µs min=0s 
                        med=0s...
http_req_duration.....: avg=225.1µs
                        min=82.21µs
                        med=226...
{ expected_response:true }...: avg=225.1µs
                               min=82.21µs
                               med=226...
OK http_req_failed...: 0.00% 0 out of 14565
http_req_receiving...: avg=66.97µs 
                       min=15.24µs
                       med=60...
http_req_sending.....: avg=17.27µs
                       min=4.03µs
                       med=13...
http_req_tls_handshaking...: avg=0s min=0s
                             med=0s...
http_req_waiting.....: avg=140.85µs
                       min=53.55µs
                       med=144...
http_reqs............: 14565 2912.709122/s
iteration_duration...: avg=329.52µs min=117.97µs
                       med=323...
iterations...........: 14565 2912.709122/s
vus..................: 1 min=1 max=1
vus_max..............: 1 min=1 max=1
```
\end{customdialogue}

\begin{customdialogue}[Failure senario: Cyber Attack]
<<Description>>
Given the system's weaknesses, a cyber attack targeting the single Pod and its network could be highly impactful. The Pod's restart policy set to 'Never' means that if the Pod fails, it will not restart, leading to downtime. Additionally, the single Pod deployment means there is no redundancy, so any failure will make the service unavailable. To simulate a cyber attack, we can start by injecting a PodChaos fault to kill the Pod, testing the system's ability to handle Pod failures. This will directly exploit the lack of redundancy and the restart policy issue. Next, we can simulate a network attack using NetworkChaos to introduce network latency, testing the service's ability to maintain availability under network stress. This sequence simulates a cyber attack by first taking down the Pod and then stressing the network, revealing the system's vulnerabilities in handling such scenarios.

<<Failure-injection sequence>>
PodChaos ({'pod': 'example-pod'}) -> NetworkChaos ({'service': 'example-service'}):

<<Detailed failure parameters>>
Detailed parameters of PodChaos ({'pod': 'example-pod'})
```
{
  "action": "pod-kill",
  "mode": "one",
  "selector": {
    "namespaces": [
      "default"
    ],
    "labelSelectors": {
      "app": "example"
    }
  }
}
```

Detailed parameters of NetworkChaos ({'service': 'example-service'}):
```
{
  "action": "delay",
  "direction": "to",
  "target": {
    "mode": "all",
    "selector": {
      "namespaces": ["default"],
      "labelSelectors": {
        "app": "example"
      }
    }
  },
  "mode": "all",
  "selector": {
    "namespaces": ["default"],
    "labelSelectors": {
      "app": "example"
    }
  },
  "device": "eth0",
  "delay": {
    "latency": "100ms",
    "jitter": "10ms",
    "correlation": "50"
 }
}
```
\end{customdialogue}

\begin{customdialogue}[Chaos experiment (Planning)]
<<Time schedule>>
Given the constraints of the chaos engineering experiment, which must be completed within 1 minute, we need to allocate time efficiently across the three phases: prevalidation, fault-injection, and post-validation. The pre-validation phase is crucial to ensure that the system is in a steady state before we introduce any faults. Since we have two steady states to validate, we should allocate a reasonable amount of time to check both the pod's running status and the service's availability. A duration of 15 seconds should be sufficient for pre-validation, allowing us to run the necessary checks multiple times. The fault-injection phase is where we introduce the chaos to observe the system's behavior under stress. Given the complexity of the faults (PodChaos and NetworkChaos), we should allocate the majority of the time to this phase to ensure that the faults have enough time to manifest and impact the system. A duration of 30 seconds is appropriate for fault injection, allowing us to observe the system's response to both pod failure and network latency. Finally, the post-validation phase is essential to verify that the system returns to its steady states after the faults are removed. We should allocate 15 seconds for post-validation, similar to the pre-validation phase, to ensure that the system stabilizes and meets the defined thresholds for steady states. This allocation results in a total experiment time of 60 seconds, which fits within the 1-minute constraint.

Total experiment time: 60s
Pre-validation Phase : 15s
Fault-injection Phase: 30s
Post-validation Phase: 15s

<<Pre-validation phase>>
In the pre-validation phase, we need to ensure that the system is in its expected steady state before we proceed with fault injection. Given the constraints of a 15-second total time for this phase, we will conduct two unit tests to verify the steady states: one for the pod's running status and another for the service's availability. These tests will be executed sequentially due to the short duration available, ensuring that each steady state is verified independently and thoroughly. The first test will check if the 'examplepod' is running at least 90% of the time over a 5-second period. This is crucial because
the pod's restart policy is set to 'Never', and we need to confirm its availability before introducing any faults. The second test will verify the 'example-service' availability, ensuring it responds with a 200 status code at least 99.9% of the time over another 5-second period. This test is essential to confirm that the service is operational and responsive, given the single pod deployment. By staggering these tests, we can focus on each steady state individually, allowing us to identify any issues before proceeding to the fault injection phase.

- Verified Steady State #0: example-pod-running
  - Workflow Name: pre-unittest-example-pod-running
  - Grace Period: 0s
  - Duration: 5s

- Verified Steady State #1: example-service-availability
  - Workflow Name: pre-unittest-example-service-availability
  - Grace Period: 5s
  - Duration: 5s

<<Fault-injection phase>>
In this fault-injection phase, we aim to simulate a cyber attack by injecting two types of faults: PodChaos and NetworkChaos. The total duration for this phase is 30 seconds, so we need to carefully schedule the faults and unit tests to fit within this timeframe.

First, we will inject the PodChaos fault to simulate a pod failure. This fault will be injected at the start of the phase (grace period of 0s) and will last for 10 seconds. This duration is chosen to allow enough time for the system to experience the impact of the pod being killed, given the pod's restart policy is set to 'Never'.

Simultaneously, we will run the unit test for the 'example-pod-running' steady state to verify if the pod is running at least 90% of the time during the fault injection. This test will also start at 0s and run for 10 seconds, aligning with the PodChaos duration.

Next, we will inject the NetworkChaos fault to simulate network latency. This fault will start at 10 seconds (aer the PodChaos fault ends) and will last for 20 seconds. This staggered approach allows us to observe the system's behavior under network stress after the pod failure has been simulated.

During the NetworkChaos fault, we will run the unit test for the 'example-serviceavailability' steady state. This test will start at 10 seconds and run for 20 seconds, matching the NetworkChaos duration. This ensures we are checking the service's availability and response time while the network is under stress.

By staggering the faults and aligning the unit tests with the fault durations, we can effectively observe the system's behavior under each fault condition and verify if the steady states are maintained.

- Verified Steady State #0: example-pod-running
  - Workflow Name: fault-unittest-example-pod-running
  - Grace Period: 0s
  - Duration: 10s

- Verified Steady State #1: example-service-availability
  - Workflow Name: fault-unittest-example-service-availability
  - Grace Period: 10s
  - Duration: 20s

- Injected Faults #0: PodChaos
  - Workflow Name: fault-podchaos
  - Grace Period: 0s
  - Duration: 10s

- Injected Faults #1: NetworkChaos
  - Workflow Name: fault-networkchaos
  - Grace Period: 10s
  - Duration: 20s

<<Post-validation phase>>
In the post-validation phase, we need to ensure that the system has returned to its steady states aer the fault injection. Given the 15-second time constraint, we will perform quick checks to verify the steady states. The two steady states to verify are: 1) the 'example-pod' is running, and 2) the 'example-service' is available. We will execute these checks sequentially due to the short duration, ensuring each test has enough time to gather meaningful data. The first test will check the pod's running status, followed by the service availability test. This order is logical because the pod must be running for the service to be available. Each test will have a brief grace period to allow the system to stabilize aer the fault injection, followed by a short duration to perform the checks.

- Verified Steady State #0: example-pod-running
  - Workflow Name: post-unittest-example-pod-running
  - Grace Period: 2s
  - Duration: 6s

- Verified Steady State #1: example-service-availability
  - Workflow Name: post-unittest-example-service-availability
  - Grace Period: 8s
  - Duration: 5s

<<Summary>>
The chaos engineering experiment is structured into three phases: pre-validation, fault-injection, and post-validation, all to be completed within a total of 60 seconds.

In the pre-validation phase, which lasts for 15 seconds, two unit tests are conducted sequentially to ensure the system is in a steady state before fault injection. The first test, named 'pre-unittest-example-pod-running', checks the 'example-pod' running status. It starts immediately at the beginning of the phase and runs for 5 seconds. Following this, the second test, 'pre-unittest-example-service-availability', begins at the 5-second mark and also runs for 5 seconds, verifying the service's availability.

The fault-injection phase spans 30 seconds and involves two types of faults: PodChaos and NetworkChaos. Initially, the PodChaos fault, named 'fault-podchaos', is injected at the start of the phase and lasts for 10 seconds. Concurrently, the 'fault-unittestexample-pod-running' unit test runs for the same duration to verify the pod's status during the fault. Aer the PodChaos fault concludes, the NetworkChaos fault, named 'fault-networkchaos', begins at the 10-second mark and continues for 20 seconds. Simultaneously, the 'fault-unittest-example-service-availability' test runs for 20 seconds, starting at the same time as the NetworkChaos fault, to check the service's availability under network stress.

Finally, the post-validation phase, also 15 seconds long, ensures the system returns to its steady states. The 'post-unittest-example-pod-running' test starts after a 2-second grace period and runs for 6 seconds to verify the pod's status. Subsequently, the 'postunittest-example-service-availability' test begins at the 8-second mark and runs for 5 seconds, checking the service's availability. This sequential execution allows for a brief stabilization period before each test.

The generated Chaos Mesh workflow manifest:
```
apiVersion: chaos-mesh.org/v1alpha1
kind: Workflow
metadata:
  name: chaos-experiment-20241124-132854
spec:
  entry: the-entry
  templates:
    #-------------------------------
    # entry point of whole workflow
    #-------------------------------
    - name: the-entry
      templateType: Serial
      deadline: 30m51s
      children:
        - pre-validation-phase
        - fault-injection-phase
        - post-validation-phase

    #-------------------------------------
    # Entry point of pre-validation-phase
    #-------------------------------------
    - name: pre-validation-phase
      templateType: Serial
      deadline: 10m10s
      children:
        - pre-validation-overlapped-workflows
    
    - name: pre-validation-suspend-workflow
      templateType: Serial
      deadline: 5m10s
      children:
        - pre-validation-suspend
        - pre-unittest-example-service-
          availability

    - name: pre-validation-suspend
      templateType: Suspend
      deadline: 5s

    - name: pre-validation-overlapped-workflows
      templateType: Parallel
      deadline: 5m10s
      children:
        - pre-unittest-example-pod-running
        - pre-validation-suspend-workflow

    # Definitions of children of pre-validation-phase
    - name: pre-unittest-example-pod-running
      templateType: Task
      deadline: 5m5s
      task:
        container:
          name: pre-unittest-example-pod-running-
                container
          image: chaos-eater/k8sapi:1.0
          imagePullPolicy: IfNotPresent
          command: ["/bin/bash", "-c"]
          args: ["python 
            /chaos-eater/sandbox/cycle_20241124
            _132128/unittest_example-pod-running
            _mod0.py --duration 5"]
          volumeMounts:
            - name: pvc-volume
              mountPath: /chaos-eater
        volumes:
          - name: pvc-volume
            persistentVolumeClaim:
              claimName: pvc

    - name: pre-unittest-example-service-availability
      templateType: Task
      deadline: 5m5s
      task:
        container:
          name: pre-unittest-example-service-
                availability-container
          image: grafana/k6:latest
          command: ["k6", "run", 
            "--duration", "5s", "--quiet",
            "/chaos-eater/sandbox/cycle_20241124
             _132128/unittest_example-service-
             availability_mod0.js"]    
          volumeMounts:
            - name: pvc-volume
              mountPath: /chaos-eater
        volumes:
          - name: pvc-volume
            persistentVolumeClaim:
              claimName: pvc
    
    #--------------------------------------
    # Entry point of fault-injection-phase
    #--------------------------------------
    - name: fault-injection-phase
      templateType: Serial
      deadline: 10m30s
      children:
        - fault-injection-overlapped-workflows

    - name: fault-injection-parallel-workflow
      templateType: Parallel
      deadline: 5m10s
      children:
        - fault-unittest-example-pod-running
        - fault-podchaos

    - name: fault-injection-suspend-workflow
      templateType: Serial
      deadline: 5m30s
      children:
        - fault-injection-suspend
        - fault-injection-parallel-workflows

    - name: fault-injection-suspend
      templateType: Suspend
      deadline: 10s

    - name: fault-injection-parallel-workflows
      templateType: Parallel
      deadline: 5m20s
      children:
        - fault-unittest-example-service-
          availability
        - fault-networkchaos

    - name: fault-injection-overlapped-workflows
      templateType: Parallel
      deadline: 5m30s
      children:
        - fault-injection-parallel-workflow
        - fault-injection-suspend-workflow

    # Definitions of children of pre-validation-phase
    # unit tests
    - name: fault-unittest-example-pod-running
      templateType: Task
      deadline: 5m10s
      task:
        container:
          name: fault-unittest-example-pod-running-
                container
          image: chaos-eater/k8sapi:1.0
          imagePullPolicy: IfNotPresent
          command: ["/bin/bash", "-c"]
          args: ["python 
            /chaos-eater/sandbox/cycle_20241124
            _132128/unittest_example-pod-running
            _mod0.py --duration 10"]
          volumeMounts:
            - name: pvc-volume
              mountPath: /chaos-eater
        volumes:
          - name: pvc-volume
            persistentVolumeClaim:
              claimName: pvc

    - name: fault-unittest-example-service-availability
      templateType: Task
      deadline: 5m20s
      task:
        container:
          name: fault-unittest-example-service-
                availability-container
          image: grafana/k6:latest
          command: ["k6", "run", 
            "--duration", "20s", "--quiet",
            "/chaos-eater/sandbox/cycle_20241124
             _132128/unittest_example-service-
             availability_mod0.js"]    
          volumeMounts:
            - name: pvc-volume
              mountPath: /chaos-eater
        volumes:
          - name: pvc-volume
            persistentVolumeClaim:
              claimName: pvc

    # fault_injections
    - name: fault-podchaos
      templateType: PodChaos
      deadline: 10s
      podChaos:
        action: pod-kill
        mode: one
        selector:
          labelSelectors:
            app: example
          namespaces:
            - default


    - name: fault-networkchaos
      templateType: NetworkChaos
      deadline: 20s
      networkChaos:
        action: delay
        delay:
          correlation: '50'
          jitter: 10ms
          latency: 100ms
        device: eth0
        direction: to
        mode: all
        selector:
          labelSelectors:
            app: example
          namespaces:
            - default
        target:
          mode: all
          selector:
            labelSelectors:
              app: example
            namespaces:
              - default

    
    #--------------------------------------
    # Entry point of post-validation phase
    #--------------------------------------
    - name: post-validation-phase
      templateType: Serial
      deadline: 10m11s
      children:
        - post-validation-overlapped-workflows

    - name: post-validation-suspend-workflow
      templateType: Serial
      deadline: 5m8s
      children:
        - post-validation-suspend
        - post-unittest-example-pod-running

    - name: post-validation-suspend
      templateType: Suspend
      deadline: 2s

    - name: post-validation-suspend-workflow2
      templateType: Serial
      deadline: 5m11s
      children:
        - post-validation-suspend2
        - post-unittest-example-service-
          availability

    - name: post-validation-suspend2
      templateType: Suspend
      deadline: 6s

    - name: post-validation-overlapped-workflows
      templateType: Parallel
      deadline: 5m11s
      children:
        - post-validation-suspend-workflow
        - post-validation-suspend-workflow2

    # Definitions of children of pre-validation-phase
    - name: post-unittest-example-pod-running
      templateType: Task
      deadline: 5m6s
      task:
        container:
          name: post-unittest-example-pod-running-
                container
          image: chaos-eater/k8sapi:1.0
          imagePullPolicy: IfNotPresent
          command: ["/bin/bash", "-c"]
          args: ["python
            /chaos-eater/sandbox/cycle_20241124
            _132128/unittest_example-pod-running
            _mod0.py --duration 6"]
          volumeMounts:
            - name: pvc-volume
              mountPath: /chaos-eater
        volumes:
          - name: pvc-volume
            persistentVolumeClaim:
              claimName: pvc

    - name: post-unittest-example-service-availability
      templateType: Task
      deadline: 5m5s
      task:
        container:
          name: post-unittest-example-service-
                availability-container
          image: grafana/k6:latest
          command: ["k6", "run",
            "--duration", "5s", "--quiet",
            "/chaos-eater/sandbox/cycle_20241124
             _132128/unittest_example-service-
             availability_mod0.js"]    
          volumeMounts:
            - name: pvc-volume
              mountPath: /chaos-eater
        volumes:
          - name: pvc-volume
            persistentVolumeClaim:
              claimName: pvc
```
\end{customdialogue}

\begin{customdialogue}[Analysis]
The chaos engineering experiment results indicate several critical issues in the system's configuration and its ability to handle faults, particularly in the context of the defined fault scenario. Here is a detailed analysis of the failures observed during the experiment:

1. Pod Restart Policy and Single Pod Deployment:
The 'fault-unittest-example-pod-running' test failed because the Pod was not found after the PodChaos fault was injected. This is directly related to the Pod's restart policy set to 'Never' in the nginx/pod.yaml manifest. When the Pod was killed, it did not restart, leading to a 404 error when attempting to read the Pod's status. This confirms the identified issue #0 (Pod Restart Policy) and issue #1 (Single Pod Deployment), where the lack of redundancy and automatic recovery mechanisms resulted in the Pod being unavailable.

2. Service Availability:
The 'fault-unittest-example-service-availability' test failed due to connection refusals, indicating that the service was unavailable. This failure is a consequence of the Pod being killed and not restarting, as the Service relies on the Pod to handle requests. The Service's selector in nginx/service.yaml targets Pods with the label 'app: example', and with the Pod being down, there were no endpoints available to serve the requests, leading to the connection refused errors.

3. Post-Validation Failures:
Both 'post-unittest-example-pod-running' and 'post-unittest-example-service-availability' tests failed, showing that the system did not recover to its steady state after the fault injection. The Pod remained unavailable, and consequently, the Service could not respond to HTTP requests. This further highlights the critical impact of the Pod's restart policy and the lack of redundancy.


<<Insights and Recommendations:>>
- Pod Restart Policy: Change the Pod's restart policy from 'Never' to 'Always' or 'OnFailure' to ensure that the Pod automatically restarts if it fails. This will help maintain the Pod's availability and reduce downtime.

- Redundancy and Scalability: Implement a Deployment or ReplicaSet to manage the Pod. This will provide redundancy by running multiple replicas of the Pod, ensuring that the Service remains available even if one Pod fails.

- Service Resilience: Consider using a LoadBalancer or Ingress to manage traffic to multiple Pods, enhancing the system's ability to handle failures and maintain service availability.

By addressing these issues, the system can improve its resilience against similar fault scenarios in the future, ensuring that the steady states are maintained even when faults are injected.
\end{customdialogue}

\begin{customdialogue}[Improvement]
To improve the resiliency of the system, I will replace the Pod with a Deployment to manage the Pod's lifecycle and provide redundancy. This will address both the restart policy issue and the single Pod deployment issue by ensuring that multiple replicas of the Pod are running and can automatically restart if they fail.

Modification_type: replace
File name: nginx/pod.yaml

Replace the Pod with a Deployment to manage the Pod's lifecycle and provide redundancy. This will ensure that multiple replicas of the Pod are running and can automatically restart if they fail.

```
apiVersion: apps/v1
kind: Deployment
metadata:
  name: example-deployment
  labels:
    app: example
spec:
  replicas: 3
  selector:
    matchLabels:
      app: example
  template:
    metadata:
      labels:
        app: example
    spec:
      containers:
      - name: example-container
        image: nginx:1.17.1
        ports:
        - containerPort: 80
```
\end{customdialogue}

\begin{customdialogue}[Experiment replanning]
<<Failure scope of PodChaos>>
The previous K8s manifests targeted a single pod with the label 'app: example'. The current manifests have changed to a Deployment with 3 replicas, all having the same label 'app: example'. The original chaos experiment aimed to simulate a pod failure, which can still be achieved by targeting one of the pods in the Deployment. Therefore, the scope of the fault injection should remain the same, targeting pods with the label 'app: example' in the 'default' namespace. This ensures that the experiment's goal of simulating a pod failure is maintained, even with the change from a single pod to a Deployment with multiple replicas.

<<Failure scope of NetworkChaos>>
The previous K8s manifests had a single pod, and the NetworkChaos was targeting all pods with the label 'app: example' in the 'default' namespace. The current manifests have been updated to a Deployment with 3 replicas, but the label 'app: example' and the namespace 'default' remain the same. Therefore, the scope of the fault injection should remain unchanged, as it will now target all pods created by the Deployment, which aligns with the original goal of the chaos experiment to test the system's behavior under network stress.

<<VaC script for steady state #1>>
The previous unit test targets a Pod resource named 'example-pod'. In the reconfigured manifests, the Pod has been replaced by a Deployment named 'example-deployment' with 3
replicas. This change means that the unit test should now target the Deployment instead of a single Pod. The test should verify that at least one of the Pods created by the Deployment is
running at least 90% of the time. Therefore, the unit test requires modification to account for the change from a Pod to a Deployment with multiple replicas. The test should be updated to check the status of the Pods created by the Deployment, ensuring that the steady state is maintained by the entire redundancy.
```
import os
import time
import argparse
from kubernetes import client, config
from unittest_base import K8sAPIBase

class TestDeploymentRunningState(K8sAPIBase):
  def __init__(self):
    super().__init__()

  def check_deployment_pods_status(
    self, namespace, deployment_name
  ):
    try:
      pods = self.v1.list_namespaced_pod(
        namespace=namespace,
        label_selector=f'app={deployment_name}'
      ).items
      running_pods = [pod for pod in pods if pod.status.phase == 'Running']
      return len(running_pods) > 0
    except client.exceptions.ApiException as e:
      print(f"Exception when calling CoreV1Api->list_namespaced_pod: {e}")
      return False

  def test_deployment_running_state(self, duration):
    namespace = 'default'
    deployment_name = 'example'
    running_count = 0

    # Check the deployment pods status every second for the specified duration
    for _ in range(duration):
      if self.check_deployment_pods_status(
        namespace, deployment_name):
        running_count += 1
      time.sleep(1)

    # Calculate the running percentage
    running_percentage = (running_count / duration) * 100

    # Assert that the running percentage is at least 90%
    assert running_percentage >= 90, f"Deployment '{deployment_name}' running percentage is below threshold: {running_percentage}%"

    print(f"Deployment '{deployment_name}' running status checked {duration} times. Running percentage: {running_percentage}%.")


def main():
  parser = argparse.ArgumentParser(
    description='Test if a deployment has at least one pod running at least 90% of the time.')
  parser.add_argument(
    '--duration', type=int, default=5, help='Duration to check the deployment pods status in seconds.')
  args = parser.parse_args()

  test = TestDeploymentRunningState()
  test.test_deployment_running_state(
    args.duration)

if __name__ == '__main__':
  main()
```

<<VaC script for steady state #2>>
The target of the unit test is the service availability, which is associated with the Service resource in the Kubernetes manifests. In the previous and reconfigured manifests, the Service resource has not changed, so the endpoint and the service configuration remain the same. The unit test checks the availability of the service by sending HTTP requests to the service endpoint and verifying the response status. Since the Service resource has not been modified, the unit test does not require any changes. The reconfiguration of the Pod to a Deployment with 3 replicas is intended to improve the system's ability to meet the availability threshold, but it does not affect the unit test itself, which remains valid for checking the service's availability.
\end{customdialogue}

\begin{customdialogue}[Summary of the completed CE cycle]
The Chaos Engineering cycle begins with understanding the user inputs, which include Kubernetes manifests for a Pod and a Service. The Pod manifest defines a single Nginx container with a restart policy of 'Never', and the Service manifest exposes this Pod on port 80. Two main resiliency issues are identified: the Pod's restart policy and the lack of redundancy due to a single Pod deployment.

The hypothesis for the experiment is that the system's steady states will be maintained even when faults are injected. Two steady states are defined: the Pod should be running at least 90% of the time, and the Service should have 99.9% availability with a response status of 200.

The fault scenario simulates a cyber attack using Chaos Mesh, injecting PodChaos to kill the Pod and NetworkChaos to introduce network latency. The experiment is divided into three phases: pre-validation, fault-injection, and post-validation, each with specific tasks and durations to verify the system's behavior under stress.

In the first experiment attempt, the system fails to maintain the steady states during and after the fault injection. The Pod does not restart due to its 'Never' restart policy, and the Service becomes unavailable, confirming the identified issues. Recommendations include changing the Pod's restart policy and implementing a Deployment for redundancy.

After modifying the system by replacing the Pod with a Deployment, the second experime attempt is successful. All unit tests pass, indicating that the system maintains its steady states even when faults are injected, demonstrating improved resilience.
\end{customdialogue}

\subsection{Reviews}
\begin{customdialogue}[Review by Engineer A]
<<Overall: 4>>
- This is a critical issue because if the Pod fails, it will not restart, leading to potential downtime. If the single Pod fails, the service will be unavailable.
- After modifying the system by replacing the Pod with a Deployment, the second experiment attempt is successful. All unit tests pass, indicating that the system maintains its steady states even when faults are injected, demonstrating improved resilience.
- However, it did not provide significant insights for future cycles, such as testing different replica counts, resource usage or cost efficiency.

<<Hypothesis: 4>>
The hypothesis for the experiment is that the system's steady states will be maintained even when faults are injected. Two steady states are defined: the Pod should be running at least 90\% of the time, and the Service should have 99.9\% availability with a response status of 200.
- Steady state #1: example-pod-running
 + The first issue to address is the Pod's restart policy set to 'Never'. This is a critical issue because if the Pod fails, it will not restart, leading to potential downtime.  
- Steady state #2: example-service-availability
 + The next issue to address is the 'Single Pod Deployment', which is related to the lack of redundancy. This is a significant issue because if the single Pod fails, the service will be unavailable.

Did not explore more failure scenarios, example: simultaneous Pod failure or network latency. (Only Failure senario: Cyber Attack)

<<Experiment: 4>>
The experiment plan correctly serves to validate the hypothesis
Time schedule
- Total experiment time: 60s
- Pre-validation Phase : 15s
 + We need to ensure that the system is in its expected steady state before we proceed with fault injection.  Conduct two unit tests to verify the steady states: one for the pod's running status and another for the service's availability. 
- Fault-injection Phase: 30s
 + Simulate a Cyber Attack by injecting two types of faults: PodChaos and NetworkChaos.
 + First, inject the PodChaos fault to simulate a pod failure. . This fault will be injected at the start of the phase (grace period of 0s) and will last for 10 seconds.
 + Next, inject the NetworkChaos fault to simulate network latency. This fault will start at 10 seconds (aer the PodChaos fault ends) and will last for 20 seconds.
 + By staggering the faults and aligning the unit tests with the fault durations, we can effectively observe the system's behavior under each fault condition and verify if the steady states are maintained.
- Post-validation Phase: 15s
 + Pod must be running for the service to be available.

Additionally, it is set up considering an actual failure scenario.
- The fault scenario simulates a Cyber Attack using Chaos Mesh, injecting PodChaos to kill the Pod and NetworkChaos to introduce network latency. 

<<Analysis: 4>>
The analysis reports correct and meaningful information:
- Pod Restart Policy and Single Pod Deployment: The 'fault-unittest-example-pod-running' test failed because the Pod  was not found after the PodChaos fault was injected. This is directly related to the Pod's restart policy set to 'Never' in the nginx/pod.yaml manifest. When the Pod was killed, it did not restart, leading to a 404 error when attempting to read the Pod's status. This confirms the identified issue #0 (Pod Restart Policy) and issue #1 (Single Pod Deployment), where the lack of redundancy and automatic recovery  mechanisms resulted in the Pod being unavailable.
- Service Availability: The 'fault-unittest-example-service-availability' test failed due to  connection refusals, indicating that the service was unavailable. This  failure is a consequence of the Pod being killed and not restarting, as the Service relies on the Pod to handle requests. The Service's selector in nginx/service.yaml targets Pods with the label 'app: example', and with the Pod being down, there were no endpoints available to serve the requests, leading to the connection refused errors.
- Post-Validation Failures: Both 'post-unittest-example-pod-running' and 'post-unittest-example-service-availability' tests failed, showing that the system did not recover to its steady state after the fault injection. The Pod remained
unavailable, and consequently, the Service could not respond to HTTP requests. This further highlights the critical impact of the Pod's restart policy and the lack of redundancy.

Lacked detailed performance impact analysis, such as checking if the Pod was running at least 90\% of the time and if the Service maintained 99.9\% availability.

Meaningful insights for the improvement
- Pod Restart Policy: Change the Pod's restart policy from 'Never' to  'Always' or 'OnFailure' to ensure that the Pod automatically restarts if it fails. This will help maintain the Pod's availability and reduce downtime.
- Redundancy and Scalability: Implement a Deployment or ReplicaSet to manage the Pod. This will provide redundancy by running multiple replicas of the Pod, ensuring that the Service remains available even if one Pod fails.
- Service Resilience: Consider using a LoadBalancer or Ingress to manage traffic to multiple Pods, enhancing the system's ability to handle failures and maintain service availability.

<<Improvement: 5>>
Successfully fixed issues on the first attempt and enhanced resiliency and satisfied the hypothesis.
- In the first experiment attempt, the system fails to maintain the steady states during and after the fault injection. The Pod does not restart due to its 'Never' restart policy, and the Service becomes unavailable, confirming the identified issues.
- After modifying the system by replacing the Pod with a Deployment, the second experime attempt is successful. All unit tests pass, indicating that the system maintains its steady states even when faults are injected, demonstrating improved resilience.
\end{customdialogue}

\begin{customdialogue}[Review by Engineer B (Translated)]
<<Overall: 5>>
Reason for positive evaluation:
- The issue where the system failed to recover when a Pod went down was resolved.
Meaningful insights and recommendations were obtained regarding service resilience.

The resolution of the system recovery issue and the insights gained on resilience contribute to a high evaluation.

<<Hypothesis: 4>>
Reason for positive evaluation:
- The summaries of each manifest file are accurate.  
- The resiliency issues in the manifests are appropriately pointed out.  
- The content of the image ('nginx') is explained.  
- The assumed purpose of each manifest file's deployment is inferred correctly.  
- The steady-state conditions, such as Pod uptime and HTTP request responsiveness, are properly identified.  
- Service availability is considered with a commercial-grade mindset (e.g., 99.9% uptime), and HTTP response status code 200 is specified.

Room for improvement:
- The API version information in 'nginx/service.yaml' may not be necessary.  
- Since 'nginx:1.17.1' is not the latest version, it would be better if there were comments expressing concern about potential vulnerabilities.  
- Furthermore, incorporating tests from a security vulnerability perspective would be desirable.

The evaluation is high because the manifest summaries and resiliency issue identifications are appropriate. However, it is not a perfect score due to the inclusion of unnecessary information and the lack of comments on the outdated image.

<<Experiment: 5>>
Reason for positive evaluation:
- Pod status and service availability were checked during pre-validation, fault injection, and post-validation phases.

The evaluation is high because the test was conducted within the one-minute time constraint, while confirming Pod status and service availability.

<<Analysis: 5>>
Reason for positive evaluation:
- The occurrence of 404 errors after the fault and the lack of recovery were accurately analyzed.  
- The three proposed recommendations are meaningful for improving resilience.

The evaluation is high because the post-fault situation was accurately analyzed, and valuable suggestions for enhancing resilience were provided.

<<Improvement: 5>>
Reason for positive evaluation:
- Replacing the Pod with a Deployment enables automatic restarts.  
- Setting the number of replicas to 3 provides redundancy.

The evaluation is high because improvements were made by replacing the Pod with a Deployment and introducing redundancy.
\end{customdialogue}

\begin{customdialogue}[Review by GPT-4o]
<<Overall: 4>>
The Chaos Engineering cycle effectively identified and addressed critical issues in the system, leading to significant improvements in resilience.

Strengths:
- Comprehensive approach to identifying and addressing system weaknesses.
- Successfully improved system resilience through targeted improvements.
- Provided meaningful insights for future cycles.

Weaknesses:
- The cycle could have explored additional fault scenarios for a more comprehensive assessment.
- Limited exploration of potential secondary effects of the faults.
- The impact of improvements on overall system performance was not fully quantified.

Reason for the score:
The cycle fixed critical issues in the system and offered meaningful insights for the next cycle, though it could have been more comprehensive in its exploration of potential vulnerabilities.

<<Hypothesis: 4>>
The hypothesis aimed to ensure that the system maintains its steady states even when faults are injected, focusing on pod availability and service responsiveness.

Strengths:
- Clearly defined steady states for both pod and service availability.
- Utilized measurable outputs to verify steady states.
- Addressed critical issues like pod restart policy and single pod deployment.

Weaknesses:
- The hypothesis could have been more specific about the expected behavior under each fault condition.
- Lacked consideration for potential cascading failures beyond the immediate faults.
- Did not account for external dependencies that might affect the system's steady states.

Reason for the score:
The hypothesis was relevant and meaningful, addressing critical issues in the system. It led to system improvement and provided insights for future cycles.

<<Experiment: 4>>
The experiment was well-structured, with clear phases for pre-validation, fault-injection, and post-validation, all within a 60-second timeframe.

Strengths:
- Efficient time allocation across phases.
- Comprehensive fault-injection plan using PodChaos and NetworkChaos.
- Aligned unit tests with fault durations for effective observation.

Weaknesses:
- The experiment could have included more diverse fault scenarios to test other potential vulnerabilities.
- Limited time for each phase might not capture all transient issues.
- The experiment plan could have been more detailed in terms of expected outcomes for each fault.

Reason for the score:
The experiment plan correctly served to validate the hypothesis and was set up considering actual failure scenarios, though it could have been more comprehensive.

<<Analysis: 4>>
The analysis identified critical issues related to pod restart policy and single pod deployment, providing clear insights and recommendations for improvement.

Strengths:
- Thorough identification of issues and their impact on system behavior.
- Provided actionable recommendations for system improvement.
- Highlighted the importance of redundancy and automatic recovery mechanisms.

Weaknesses:
- The analysis could have explored potential secondary effects of the faults.
- Lacked quantitative data on the impact of network latency on service performance.
- Could have included a comparison with baseline performance metrics.

Reason for the score:
The analysis reported correct and meaningful information, providing insights for improvement, though it could have been more detailed in certain areas.

<<Improvement: 5>>
The improvement involved replacing the Pod with a Deployment to provide redundancy and ensure automatic recovery, successfully addressing the identified issues.

Strengths:
- Successfully changed the system to satisfy the hypothesis in the first attempt.
- Addressed both redundancy and restart policy issues effectively.
- Improved system resilience against similar fault scenarios.

Weaknesses:
- The improvement focused solely on the pod configuration, potentially overlooking other areas for enhancement.
- Could have included additional resilience measures, such as load balancing.
- The impact of the improvement on overall system performance was not quantified.

Reason for the score:
The improvement successfully changed the system to satisfy the hypothesis in the first attempt, addressing critical issues effectively.
\end{customdialogue}

\begin{customdialogue}[Review by Claude Sonnet 3.5]
<<Overall: 5>>
This Chaos Engineering cycle effectively identified and addressed critical resiliency issues in the system. It progressed from a well-defined hypothesis through a comprehensive experiment, insightful analysis, and successful improvement, resulting in a more resilient system.

Strengths:
- Clear progression through all phases of the Chaos Engineering cycle
- Effective use of Chaos Mesh for fault injection and automated testing
- Successful identification and resolution of critical system vulnerabilities
- Demonstrable improvement in system resilience, as evidenced by the second experiment run

Weaknesses:
- Limited scope of improvements, focusing mainly on Pod management
- Potential for more comprehensive testing of network-related issues
- Short duration of experiment phases may limit the depth of insights gained

Reason for the score:
The cycle fixes critical issues in the system, specifically addressing the Pod restart policy and lack of redundancy. It offers meaningful insights for future improvements, particularly in the areas of network resilience and more comprehensive testing strategies.

<<Hypothesis: 4>>
The hypothesis aims to verify that the system maintains its steady states even when subjected to a simulated cyber attack. Two steady states are defined: the 'example-pod' running status and the 'example-service' availability.

Strengths:
- Clear definition of steady states with measurable outputs
- Appropriate thresholds set for each steady state (90\% for pod running, 99.9\% for service availability)
- Well-defined fault scenario simulating a cyber attack using PodChaos and NetworkChaos
- Use of both Kubernetes API and k6 for comprehensive testing

Weaknesses:
- The hypothesis doesn't explicitly address the known system weaknesses (restart policy and single pod deployment)
- The steady states don't directly test the system's ability to recover from failures
- The hypothesis doesn't consider potential cascading effects of the simulated cyber attack

Reason for the score:
The hypothesis is relevant to the system and meaningful, addressing key aspects of system resilience. It leads to system improvement by revealing critical vulnerabilities. However, it could be more comprehensive in addressing known weaknesses and recovery capabilities.

<<Experiment: 4>>
The experiment is divided into three phases: pre-validation (15s), fault-injection (30s), and post-validation (15s). It uses Chaos Mesh to inject PodChaos and NetworkChaos faults, simulating a cyber attack. Unit tests are conducted before, during, and after fault injection to verify steady states.

Strengths:
- Well-structured experiment with clear phases and timing
- Use of both PodChaos and NetworkChaos to simulate a complex attack scenario
- Concurrent execution of faults and unit tests during fault injection
- Use of Chaos Mesh Workflow for automated experiment execution

Weaknesses:
- Short duration for pre and post-validation phases may not capture all system behaviors
- No explicit test for system recovery time after fault injection
- Limited variation in fault parameters (e.g., only one type of network chaos)

Reason for the score:
The experiment plan correctly serves to validate the hypothesis and is set up considering an actual failure scenario (cyber attack). It effectively uses Chaos Mesh to inject faults and conduct tests concurrently, providing a comprehensive view of system behavior under stress.

<<Analysis: 5>>
The analysis identified critical issues in the system's configuration, particularly the Pod's restart policy and lack of redundancy. It correctly linked these issues to the failures observed during the experiment, providing clear insights into the system's vulnerabilities.

Strengths:
- Detailed examination of each failed unit test
- Clear connection between observed failures and known system weaknesses
- Identification of cascading effects (Pod failure leading to service unavailability)
- Specific, actionable recommendations for improvement

Weaknesses:
- Limited analysis of the system's behavior during the NetworkChaos fault
- No quantitative analysis of failure rates or downtime duration
- Lack of discussion on potential false positives or negatives in the test results

Reason for the score:
The analysis reports correct and meaningful information, clearly linking experiment results to system weaknesses. It provides valuable insights for improvement, including specific recommendations to enhance system resilience.

<<Improvement: 5>>
The improvement phase replaced the single Pod definition with a Deployment, addressing both the restart policy issue and the lack of redundancy. This change successfully resolved the vulnerabilities identified in the analysis.

Strengths:
- Direct address of the two main issues (restart policy and single pod deployment)
- Use of a Deployment with multiple replicas for improved redundancy
- Simplicity of the solution, requiring changes to only one manifest
- Successful resolution of all failed unit tests in the second experiment run

Weaknesses:
- No explicit mention of changing the restart policy (though implicitly addressed by using a Deployment)
- No additional improvements to address potential network issues identified by NetworkChaos

Reason for the score:
The improvement successfully changes the system to satisfy the hypothesis in the first attempt. All unit tests passed in the second experiment run, demonstrating the effectiveness of the changes made.
\end{customdialogue}

\begin{customdialogue}[Review by Gemini 1.5 Pro]
<<Overall: 5>>
This Chaos Engineering cycle effectively identifies and addresses critical resiliency issues in the system. The cycle demonstrates a clear understanding of the system's weaknesses and utilizes appropriate tools and techniques to validate the hypothesis and implement improvements. The cycle is well-documented and provides valuable insights for future iterations.

Strengths:
- Clear identification of resiliency issues.
- Well-defined hypothesis and experiment plan.
- Effective use of Chaos Mesh for fault injection.
- Accurate analysis of experiment results.
- Successful implementation of improvements.

Weaknesses:
- The cycle could benefit from more comprehensive monitoring and data collection during the experiment. Gathering more data on resource utilization, error rates, and other relevant metrics would provide a more complete understanding of the system's behavior under stress.
- The cycle could be more robust by exploring a wider range of fault scenarios. While the chosen faults are relevant, considering other potential failures would provide a more comprehensive assessment of the system's resilience.
- The cycle could be more iterative by incorporating more feedback loops between the analysis and improvement phases. This would allow for more fine-tuning of the improvements and ensure that the system meets the desired resilience goals.
- The cycle could be more automated by integrating the Chaos Engineering experiments into the CI/CD pipeline. This would allow for continuous resilience testing and ensure that any changes to the system do not introduce new vulnerabilities.

Reason for the score:
The cycle fixes critical issues in the system and offers meaningful insights for the next cycle according to the experiments conducted

<<Hypothesis: 4>>
The hypothesis "The steady states of the system are maintained even when the fault scenario occurs" is well-defined and relevant to the identified resiliency issues. The steady states, 'example-pod-running' and 'example-service-availability', directly address the problems of the Pod's restart policy and the lack of redundancy.

Strengths:
- Clearly defines two relevant steady states.
- Directly addresses the identified resiliency issues.
- Provides measurable outputs for each steady state.
- Sets specific thresholds for each steady state.

Weaknesses:
- The hypothesis could be more specific about the expected behavior under the fault scenario. While it states that steady states should be maintained, it doesn't quantify the acceptable deviation from the normal operating conditions.
- The hypothesis could benefit from considering additional steady states, such as resource utilization or error rates, to provide a more comprehensive view of the system's health.
- The description of the steady states could be more concise.
- The hypothesis doesn't explicitly mention the expected duration of the steady state deviations during the fault injection.

Reason for the score:
The hypothesis is relevant to the system and meaningful. Additionally, the hypothesis leads to system improvement and offers meaningful insights for the next cycle.

<<Experiment: 4>>
The experiment plan is well-structured and effectively targets the identified weaknesses. The three-phase approach (pre-validation, fault-injection, post-validation) is appropriate, and the time allocation for each phase is reasonable. The use of Chaos Mesh to inject faults is a good choice, and the selected faults (PodChaos and NetworkChaos) are relevant to the scenario.

Strengths:
- Well-defined three-phase structure.
- Appropriate time allocation for each phase.
- Relevant fault selection (PodChaos and NetworkChaos).
- Use of Chaos Mesh for automated fault injection.
- Clear description of the fault injection sequence and unit tests.

Weaknesses:
- The experiment plan could benefit from more detailed monitoring metrics beyond just the steady states. Collecting data on resource utilization, error rates, and other relevant metrics would provide a more comprehensive understanding of the system's behavior under stress.
- The experiment plan could be more robust by including a wider range of fault scenarios. While the chosen faults are relevant, exploring other potential failures would provide a more complete picture of the system's resilience.
- The experiment plan could be improved by adding more details about the expected results of the unit tests during the fault injection phase. This would make it easier to interpret the results and draw conclusions about the system's behavior.
- The experiment plan could be more specific about the criteria for determining whether the system has returned to its steady state in the post-validation phase.

Reason for the score:
The experiment plan correctly serves to validate the hypothesis. Additionally, it is set up considering an actual failure scenario.

<<Analysis: 5>>
The analysis accurately identifies the root causes of the failed unit tests and provides specific recommendations for improvement. The analysis correctly correlates the observed failures with the Pod's restart policy and the lack of redundancy.

Strengths:
- Accurate identification of the root causes of failures.
- Clear correlation between failures and the identified resiliency issues.
- Specific and actionable recommendations for improvement.
- Well-organized and easy to understand.

Weaknesses:
- The analysis could be more data-driven. While it describes the failures, it doesn't provide specific metrics or data points to support the observations. Including data from the unit tests and Chaos Mesh would strengthen the analysis.
- The analysis could benefit from a more in-depth discussion of the system's behavior during the fault injection. Describing how the system responded to the faults and the specific impact on the steady states would provide more valuable insights.
- The analysis could be more comprehensive by considering the potential impact of the failures on downstream services or users. While it focuses on the system under test, considering the broader context would provide a more complete picture of the impact of the failures.
- The analysis could be improved by including a discussion of the limitations of the experiment and potential sources of error.

Reason for the score:
The analysis reports correct and meaningful information. Additioanlly, it provides some meaningful insights for the improvement.

<<Improvement: 5>>
The improvement successfully addresses the identified issues by replacing the Pod with a Deployment. This change introduces redundancy and automatic restart capabilities, directly addressing the weaknesses highlighted in the analysis. The improvement is implemented efficiently and effectively.

Strengths:
- Directly addresses the identified weaknesses.
- Introduces redundancy and automatic restart capabilities.
- Effective and efficient implementation.
- Uses a Deployment for robust Pod management.

Weaknesses:
- The improvement could be more comprehensive by addressing other potential weaknesses in the system, such as resource limits or network policies. While the implemented change is significant, considering other aspects of resilience would further strengthen the system.
- The improvement could be more robust by including automated testing to verify the changes. While the subsequent experiment confirms the improvement, incorporating automated tests would provide more confidence in the changes.
- The improvement could be more flexible by allowing for configuration of the number of replicas. While three replicas are a good starting point, providing a way to adjust the replica count based on the system's needs would be beneficial.
- The improvement could be more transparent by providing more details about the decision-making process behind the chosen solution. Explaining why a Deployment was chosen and the rationale behind the configuration would be helpful.

Reason for the score:
The improvement succesully changes the system to satisfy the hypothesis in the first attempt.
\end{customdialogue}

\subsection{SockShop}
\label{adx:sockshop}
\subsubsection{Inputs}
\begin{custominput}[Instructions]
- The Chaos-Engineering experiment must be completed within 1 minute.
- When using k6 in steady-state definition, always select a request URL from the following options (other requests are invalid):
1. http://front-end.sock-shop.svc.cluster.local/
2. http://front-end.sock-shop.svc.cluster.local/catalogue?size=10
3. http://front-end.sock-shop.svc.cluster.local/detail.html?id=<ID>
   Replace <ID> with an available ID: [03fef6ac-1896-4ce8-bd69-b798f85c6e0b, 3395a43e-2d88-40de-b95f-e00e1502085b, 510a0d7e-8e83-4193-b483-e27e09ddc34d, 808a2de1-1aaa-4c25-a9b9-6612e8f29a38, 819e1fbf-8b7e-4f6d-811f-693534916a8b, 837ab141-399e-4c1f-9abc-bace40296bac, a0a4f044-b040-410d-8ead-4de0446aec7e, d3588630-ad8e-49df-bbd7-3167f7efb246, zzz4f044-b040-410d-8ead-4de0446aec7e]
4. http://front-end.sock-shop.svc.cluster.local/category/
5. http://front-end.sock-shop.svc.cluster.local/category?tags=<TAG>
   Replace <TAG> with an available tag: [magic, action, blue, brown, black, sport, formal, red, green, skin, geek]
6. http://front-end.sock-shop.svc.cluster.local/basket.html
\end{custominput}

\begin{custominput}[skaffold.yaml]
apiVersion: skaffold/v3
kind: Config
metadata:
  name: sock-shop-app
manifests:
  rawYaml:
    - manifests/00-sock-shop-ns.yaml
    - manifests/01-carts-dep.yaml
    - manifests/02-carts-svc.yaml
    - manifests/03-carts-db-dep.yaml
    - manifests/04-carts-db-svc.yaml
    - manifests/05-catalogue-dep.yaml
    - manifests/06-catalogue-svc.yaml
    - manifests/07-catalogue-db-dep.yaml
    - manifests/08-catalogue-db-svc.yaml
    - manifests/09-front-end-dep.yaml
    - manifests/10-front-end-svc.yaml
    - manifests/11-orders-dep.yaml
    - manifests/12-orders-svc.yaml
    - manifests/13-orders-db-dep.yaml
    - manifests/14-orders-db-svc.yaml
    - manifests/15-payment-dep.yaml
    - manifests/16-payment-svc.yaml
    - manifests/17-queue-master-dep.yaml
    - manifests/18-queue-master-svc.yaml
    - manifests/19-rabbitmq-dep.yaml
    - manifests/20-rabbitmq-svc.yaml
    - manifests/21-session-db-dep.yaml
    - manifests/22-session-db-svc.yaml
    - manifests/23-shipping-dep.yaml
    - manifests/24-shipping-svc.yaml
    - manifests/25-user-dep.yaml
    - manifests/26-user-svc.yaml
    - manifests/27-user-db-dep.yaml
    - manifests/28-user-db-svc.yaml
\end{custominput}

\begin{custominput}[manifests/00-sock-shop-ns.yaml]
apiVersion: v1
kind: Namespace
metadata:
  name: sock-shop
\end{custominput}

\begin{custominput}[manifests/01-carts-dep.yaml]
apiVersion: apps/v1
kind: Deployment
metadata:
  name: carts
  labels:
    name: carts
  namespace: sock-shop
spec:
  replicas: 2
  selector:
    matchLabels:
      name: carts
  template:
    metadata:
      labels:
        name: carts
    spec:
      containers:
      - name: carts
        image: weaveworksdemos/carts:0.4.8
        env:
         - name: JAVA_OPTS
           value: -Xms64m -Xmx128m -XX:+UseG1GC -Djava.security.egd=file:/dev/urandom -Dspring.zipkin.enabled=false
        resources:
          limits:
            cpu: 300m
            memory: 500Mi
          requests:
            cpu: 100m
            memory: 200Mi
        ports:
        - containerPort: 80
        securityContext:
          runAsNonRoot: true
          runAsUser: 10001
          capabilities:
            drop:
              - all
            add:
              - NET_BIND_SERVICE
          readOnlyRootFilesystem: true
        volumeMounts:
        - mountPath: /tmp
          name: tmp-volume
      volumes:
        - name: tmp-volume
          emptyDir:
            medium: Memory
      nodeSelector:
        beta.kubernetes.io/os: linux
\end{custominput}

\begin{custominput}[manifests/02-carts-svc.yaml]
apiVersion: v1
kind: Service
metadata:
  name: carts
  annotations:
        prometheus.io/scrape: 'true'
  labels:
    name: carts
  namespace: sock-shop
spec:
  ports:
    # the port that this service should serve on
  - port: 80
    targetPort: 80
  selector:
    name: carts
\end{custominput}

\begin{custominput}[manifests/03-carts-db-dep.yaml]
apiVersion: apps/v1
kind: Deployment
metadata:
  name: carts-db
  labels:
    name: carts-db
  namespace: sock-shop
spec:
  replicas: 2
  selector:
    matchLabels:
      name: carts-db
  template:
    metadata:
      labels:
        name: carts-db
    spec:
      containers:
      - name: carts-db
        image: mongo
        ports:
        - name: mongo
          containerPort: 27017
        securityContext:
          capabilities:
            drop:
              - all
            add:
              - CHOWN
              - SETGID
              - SETUID
          readOnlyRootFilesystem: true
        volumeMounts:
        - mountPath: /tmp
          name: tmp-volume
      volumes:
        - name: tmp-volume
          emptyDir:
            medium: Memory
      nodeSelector:
        beta.kubernetes.io/os: linux
\end{custominput}

\begin{custominput}[manifests/04-carts-db-svc.yaml]
apiVersion: v1
kind: Service
metadata:
  name: carts-db
  labels:
    name: carts-db
  namespace: sock-shop
spec:
  ports:
    # the port that this service should serve on
  - port: 27017
    targetPort: 27017
  selector:
    name: carts-db
\end{custominput}

\begin{custominput}[manifests/05-catalogue-dep.yaml]
apiVersion: apps/v1
kind: Deployment
metadata:
  name: catalogue
  labels:
    name: catalogue
  namespace: sock-shop
spec:
  replicas: 2
  selector:
    matchLabels:
      name: catalogue
  template:
    metadata:
      labels:
        name: catalogue
    spec:
      containers:
      - name: catalogue
        image: weaveworksdemos/catalogue:0.3.5
        command: ["/app"]
        args:
        - -port=80
        resources:
          limits:
            cpu: 200m
            memory: 200Mi
          requests:
            cpu: 100m
            memory: 100Mi
        ports:
        - containerPort: 80
        securityContext:
          runAsNonRoot: true
          runAsUser: 10001
          capabilities:
            drop:
              - all
            add:
              - NET_BIND_SERVICE
          readOnlyRootFilesystem: true
        livenessProbe:
          httpGet:
            path: /health
            port: 80
          initialDelaySeconds: 300
          periodSeconds: 3
        readinessProbe:
          httpGet:
            path: /health
            port: 80
          initialDelaySeconds: 180
          periodSeconds: 3
      nodeSelector:
        beta.kubernetes.io/os: linux
\end{custominput}

\begin{custominput}[manifests/06-catalogue-svc.yaml]
apiVersion: v1
kind: Service
metadata:
  name: catalogue
  annotations:
        prometheus.io/scrape: 'true'
  labels:
    name: catalogue
  namespace: sock-shop
spec:
  ports:
    # the port that this service should serve on
  - port: 80
    targetPort: 80
  selector:
    name: catalogue
\end{custominput}

\begin{custominput}[manifests/06-catalogue-svc.yaml]
apiVersion: v1
kind: Service
metadata:
  name: catalogue
  annotations:
        prometheus.io/scrape: 'true'
  labels:
    name: catalogue
  namespace: sock-shop
spec:
  ports:
    # the port that this service should serve on
  - port: 80
    targetPort: 80
  selector:
    name: catalogue
\end{custominput}

\begin{custominput}[manifests/07-catalogue-db-dep.yaml]
apiVersion: apps/v1
kind: Deployment
metadata:
  name: catalogue-db
  labels:
    name: catalogue-db
  namespace: sock-shop
spec:
  replicas: 2
  selector:
    matchLabels:
      name: catalogue-db
  template:
    metadata:
      labels:
        name: catalogue-db
    spec:
      containers:
      - name: catalogue-db
        image: weaveworksdemos/catalogue-db:0.3.0
        env:
          - name: MYSQL_ROOT_PASSWORD
            value: fake_password
          - name: MYSQL_DATABASE
            value: socksdb
        ports:
        - name: mysql
          containerPort: 3306
      nodeSelector:
        beta.kubernetes.io/os: linux
\end{custominput}

\begin{custominput}[manifests/08-catalogue-db-svc.yaml]
apiVersion: v1
kind: Service
metadata:
  name: catalogue-db
  labels:
    name: catalogue-db
  namespace: sock-shop
spec:
  ports:
    # the port that this service should serve on
  - port: 3306
    targetPort: 3306
  selector:
    name: catalogue-db
\end{custominput}

\begin{custominput}[manifests/09-front-end-dep.yaml]
apiVersion: apps/v1
kind: Deployment
metadata:
  name: front-end
  namespace: sock-shop
spec:
  replicas: 1
  selector:
    matchLabels:
      name: front-end
  template:
    metadata:
      labels:
        name: front-end
    spec:
      containers:
      - name: front-end
        image: weaveworksdemos/front-end:0.3.12
        resources:
          limits:
            cpu: 300m
            memory: 1000Mi
          requests:
            cpu: 100m
            memory: 300Mi
        ports:
        - containerPort: 8079
        env:
        - name: SESSION_REDIS
          value: "true"
        securityContext:
          runAsNonRoot: true
          runAsUser: 10001
          capabilities:
            drop:
              - all
          readOnlyRootFilesystem: true
        livenessProbe:
          httpGet:
            path: /
            port: 8079
          initialDelaySeconds: 300
          periodSeconds: 3
        readinessProbe:
          httpGet:
            path: /
            port: 8079
          initialDelaySeconds: 30
          periodSeconds: 3
      nodeSelector:
        beta.kubernetes.io/os: linux
\end{custominput}

\begin{custominput}[manifests/10-front-end-svc.yaml]
apiVersion: v1
kind: Service
metadata:
  name: front-end
  annotations:
        prometheus.io/scrape: 'true'
  labels:
    name: front-end
  namespace: sock-shop
spec:
  type: NodePort
  ports:
  - port: 80
    targetPort: 8079
    nodePort: 30001
  selector:
    name: front-end
\end{custominput}

\begin{custominput}[manifests/11-orders-dep.yaml]
apiVersion: apps/v1
kind: Deployment
metadata:
  name: orders
  labels:
    name: orders
  namespace: sock-shop
spec:
  replicas: 2
  selector:
    matchLabels:
      name: orders
  template:
    metadata:
      labels:
        name: orders
    spec:
      containers:
      - name: orders
        image: weaveworksdemos/orders:0.4.7
        env:
         - name: JAVA_OPTS
           value: -Xms64m -Xmx128m -XX:+UseG1GC -Djava.security.egd=file:/dev/urandom -Dspring.zipkin.enabled=false
        resources:
          limits:
            cpu: 500m
            memory: 500Mi
          requests:
            cpu: 100m
            memory: 300Mi
        ports:
        - containerPort: 80
        securityContext:
          runAsNonRoot: true
          runAsUser: 10001
          capabilities:
            drop:
              - all
            add:
              - NET_BIND_SERVICE
          readOnlyRootFilesystem: true
        volumeMounts:
        - mountPath: /tmp
          name: tmp-volume
      volumes:
        - name: tmp-volume
          emptyDir:
            medium: Memory
      nodeSelector:
        beta.kubernetes.io/os: linux
\end{custominput}

\begin{custominput}[manifests/12-orders-svc.yaml]
apiVersion: v1
kind: Service
metadata:
  name: orders
  annotations:
        prometheus.io/scrape: 'true'
  labels:
    name: orders
  namespace: sock-shop
spec:
  ports:
    # the port that this service should serve on
  - port: 80
    targetPort: 80
  selector:
    name: orders
\end{custominput}

\begin{custominput}[manifests/13-orders-db-dep.yaml]
apiVersion: apps/v1
kind: Deployment
metadata:
  name: orders-db
  labels:
    name: orders-db
  namespace: sock-shop
spec:
  replicas: 2
  selector:
    matchLabels:
      name: orders-db
  template:
    metadata:
      labels:
        name: orders-db
    spec:
      containers:
      - name: orders-db
        image: mongo
        ports:
        - name: mongo
          containerPort: 27017
        securityContext:
          capabilities:
            drop:
              - all
            add:
              - CHOWN
              - SETGID
              - SETUID
          readOnlyRootFilesystem: true
        volumeMounts:
        - mountPath: /tmp
          name: tmp-volume
      volumes:
        - name: tmp-volume
          emptyDir:
            medium: Memory
      nodeSelector:
        beta.kubernetes.io/os: linux
\end{custominput}

\begin{custominput}[manifests/14-orders-db-svc.yaml]
apiVersion: v1
kind: Service
metadata:
  name: orders-db
  labels:
    name: orders-db
  namespace: sock-shop
spec:
  ports:
    # the port that this service should serve on
  - port: 27017
    targetPort: 27017
  selector:
    name: orders-db
\end{custominput}

\begin{custominput}[manifests/15-payment-dep.yaml]
apiVersion: apps/v1
kind: Deployment
metadata:
  name: payment
  labels:
    name: payment
  namespace: sock-shop
spec:
  replicas: 2
  selector:
    matchLabels:
      name: payment
  template:
    metadata:
      labels:
        name: payment
    spec:
      containers:
      - name: payment
        image: weaveworksdemos/payment:0.4.3
        resources:
          limits:
            cpu: 200m
            memory: 200Mi
          requests:
            cpu: 99m
            memory: 100Mi
        ports:
        - containerPort: 80
        securityContext:
          runAsNonRoot: true
          runAsUser: 10001
          capabilities:
            drop:
              - all
            add:
              - NET_BIND_SERVICE
          readOnlyRootFilesystem: true
        livenessProbe:
          httpGet:
            path: /health
            port: 80
          initialDelaySeconds: 300
          periodSeconds: 3
        readinessProbe:
          httpGet:
            path: /health
            port: 80
          initialDelaySeconds: 180
          periodSeconds: 3
      nodeSelector:
        beta.kubernetes.io/os: linux
\end{custominput}

\begin{custominput}[manifests/16-payment-svc.yaml]
apiVersion: v1
kind: Service
metadata:
  name: payment
  annotations:
        prometheus.io/scrape: 'true'
  labels:
    name: payment
  namespace: sock-shop
spec:
  ports:
    # the port that this service should serve on
  - port: 80
    targetPort: 80
  selector:
    name: payment
\end{custominput}

\begin{custominput}[manifests/17-queue-master-dep.yaml]
apiVersion: apps/v1
kind: Deployment
metadata:
  name: queue-master
  labels:
    name: queue-master
  namespace: sock-shop
spec:
  replicas: 2
  selector:
    matchLabels:
      name: queue-master
  template:
    metadata:
      labels:
        name: queue-master
    spec:
      containers:
      - name: queue-master
        image: weaveworksdemos/queue-master:0.3.1
        env:
         - name: JAVA_OPTS
           value: -Xms64m -Xmx128m -XX:+UseG1GC -Djava.security.egd=file:/dev/urandom -Dspring.zipkin.enabled=false
        resources:
          limits:
            cpu: 300m
            memory: 500Mi
          requests:
            cpu: 100m
            memory: 300Mi
        ports:
        - containerPort: 80
      nodeSelector:
        beta.kubernetes.io/os: linux
\end{custominput}

\begin{custominput}[manifests/18-queue-master-svc.yaml]
apiVersion: v1
kind: Service
metadata:
  name: queue-master
  annotations:
        prometheus.io/scrape: 'true'
  labels:
    name: queue-master
  namespace: sock-shop
spec:
  ports:
    # the port that this service should serve on
  - port: 80
    targetPort: 80
  selector:
    name: queue-master
\end{custominput}

\begin{custominput}[manifests/19-rabbitmq-dep.yaml]
apiVersion: apps/v1
kind: Deployment
metadata:
  name: rabbitmq
  labels:
    name: rabbitmq
  namespace: sock-shop
spec:
  replicas: 2
  selector:
    matchLabels:
      name: rabbitmq
  template:
    metadata:
      labels:
        name: rabbitmq
      annotations:
        prometheus.io/scrape: "false"
    spec:
      containers:
      - name: rabbitmq
        image: rabbitmq:3.6.8-management
        ports:
        - containerPort: 15672
          name: management
        - containerPort: 5672
          name: rabbitmq
        securityContext:
          capabilities:
            drop:
              - all
            add:
              - CHOWN
              - SETGID
              - SETUID
              - DAC_OVERRIDE
          readOnlyRootFilesystem: true
      - name: rabbitmq-exporter
        image: kbudde/rabbitmq-exporter
        ports:
        - containerPort: 9090
          name: exporter
      nodeSelector:
        beta.kubernetes.io/os: linux
\end{custominput}

\begin{custominput}[manifests/20-rabbitmq-svc.yaml]
apiVersion: v1
kind: Service
metadata:
  name: rabbitmq
  annotations:
        prometheus.io/scrape: 'true'
        prometheus.io/port: '9090'
  labels:
    name: rabbitmq
  namespace: sock-shop
spec:
  ports:
    # the port that this service should serve on
  - port: 5672
    name: rabbitmq
    targetPort: 5672
  - port: 9090
    name: exporter
    targetPort: exporter
    protocol: TCP
  selector:
    name: rabbitmq
\end{custominput}

\begin{custominput}[manifests/21-session-db-dep.yaml]
apiVersion: apps/v1
kind: Deployment
metadata:
  name: session-db
  labels:
    name: session-db
  namespace: sock-shop
spec:
  replicas: 2
  selector:
    matchLabels:
      name: session-db
  template:
    metadata:
      labels:
        name: session-db
      annotations:
        prometheus.io.scrape: "false"
    spec:
      containers:
      - name: session-db
        image: redis:alpine
        ports:
        - name: redis
          containerPort: 6379
        securityContext:
          capabilities:
            drop:
              - all
            add:
              - CHOWN
              - SETGID
              - SETUID
          readOnlyRootFilesystem: true
      nodeSelector:
        beta.kubernetes.io/os: linux
\end{custominput}

\begin{custominput}[manifests/22-session-db-svc.yaml]
apiVersion: v1
kind: Service
metadata:
  name: session-db
  labels:
    name: session-db
  namespace: sock-shop
spec:
  ports:
    # the port that this service should serve on
  - port: 6379
    targetPort: 6379
  selector:
    name: session-db
\end{custominput}

\begin{custominput}[manifests/23-shipping-dep.yaml]
apiVersion: apps/v1
kind: Deployment
metadata:
  name: shipping
  labels:
    name: shipping
  namespace: sock-shop
spec:
  replicas: 2
  selector:
    matchLabels:
      name: shipping
  template:
    metadata:
      labels:
        name: shipping
    spec:
      containers:
      - name: shipping
        image: weaveworksdemos/shipping:0.4.8
        env:
         - name: ZIPKIN
           value: zipkin.jaeger.svc.cluster.local
         - name: JAVA_OPTS
           value: -Xms64m -Xmx128m -XX:+UseG1GC -Djava.security.egd=file:/dev/urandom -Dspring.zipkin.enabled=false
        resources:
          limits:
            cpu: 300m
            memory: 500Mi
          requests:
            cpu: 100m
            memory: 300Mi
        ports:
        - containerPort: 80
        securityContext:
          runAsNonRoot: true
          runAsUser: 10001
          capabilities:
            drop:
              - all
            add:
              - NET_BIND_SERVICE
          readOnlyRootFilesystem: true
        volumeMounts:
        - mountPath: /tmp
          name: tmp-volume
      volumes:
        - name: tmp-volume
          emptyDir:
            medium: Memory
      nodeSelector:
        beta.kubernetes.io/os: linux
\end{custominput}

\begin{custominput}[manifests/24-shipping-svc.yaml]
apiVersion: v1
kind: Service
metadata:
  name: shipping
  annotations:
        prometheus.io/scrape: 'true'
  labels:
    name: shipping
  namespace: sock-shop
spec:
  ports:
    # the port that this service should serve on
  - port: 80
    targetPort: 80
  selector:
    name: shipping
\end{custominput}

\begin{custominput}[manifests/25-user-dep.yaml]
apiVersion: apps/v1
kind: Deployment
metadata:
  name: user
  labels:
    name: user
  namespace: sock-shop
spec:
  replicas: 2
  selector:
    matchLabels:
      name: user
  template:
    metadata:
      labels:
        name: user
    spec:
      containers:
      - name: user
        image: weaveworksdemos/user:0.4.7
        resources:
          limits:
            cpu: 300m
            memory: 200Mi
          requests:
            cpu: 100m
            memory: 100Mi
        ports:
        - containerPort: 80
        env:
        - name: mongo
          value: user-db:27017
        securityContext:
          runAsNonRoot: true
          runAsUser: 10001
          capabilities:
            drop:
              - all
            add:
              - NET_BIND_SERVICE
          readOnlyRootFilesystem: true
        livenessProbe:
          httpGet:
            path: /health
            port: 80
          initialDelaySeconds: 300
          periodSeconds: 3
        readinessProbe:
          httpGet:
            path: /health
            port: 80
          initialDelaySeconds: 180
          periodSeconds: 3
      nodeSelector:
        beta.kubernetes.io/os: linux
\end{custominput}

\begin{custominput}[manifests/26-user-svc.yaml]
apiVersion: v1
kind: Service
metadata:
  name: user
  annotations:
        prometheus.io/scrape: 'true'
  labels:
    name: user
  namespace: sock-shop
spec:
  ports:
    # the port that this service should serve on
  - port: 80
    targetPort: 80
  selector:
    name: user
\end{custominput}

\begin{custominput}[manifests/27-user-db-dep.yaml]
apiVersion: apps/v1
kind: Deployment
metadata:
  name: user-db
  labels:
    name: user-db
  namespace: sock-shop
spec:
  replicas: 2
  selector:
    matchLabels:
      name: user-db
  template:
    metadata:
      labels:
        name: user-db
    spec:
      containers:
      - name: user-db
        image: weaveworksdemos/user-db:0.3.0
        ports:
        - name: mongo
          containerPort: 27017
        securityContext:
          capabilities:
            drop:
              - all
            add:
              - CHOWN
              - SETGID
              - SETUID
          readOnlyRootFilesystem: true
        volumeMounts:
        - mountPath: /tmp
          name: tmp-volume
      volumes:
        - name: tmp-volume
          emptyDir:
            medium: Memory
      nodeSelector:
        beta.kubernetes.io/os: linux
\end{custominput}

\begin{custominput}[manifests/28-user-db-svc.yaml]
apiVersion: v1
kind: Service
metadata:
  name: user-db
  labels:
    name: user-db
  namespace: sock-shop
spec:
  ports:
    # the port that this service should serve on
  - port: 27017
    targetPort: 27017
  selector:
    name: user-db
\end{custominput}

\subsubsection{CE cycle completed by \textsc{ChaosEater}}
\begin{customdialogue}[Preprocessing]
<<Summaries of each manifest>>
sock-shop-2/manifests/00-sock-shop-ns.yaml
- This manifest defines a Kubernetes Namespace.
- The Namespace is named 'sock-shop'.
- Namespaces are used to organize and manage resources in a Kubernetes cluster.

sock-shop-2/manifests/01-carts-dep.yaml
- This manifest defines a Deployment in Kubernetes.
- The Deployment is named 'carts' and is located in the 'sock-shop' namespace.
- It specifies that there should be 2 replicas of the 'carts' application running.
- The Deployment uses the Docker image 'weaveworksdemos/carts:0.4.8'.
- Environment variables are set for Java options to optimize memory usage and disable 
  certain features.
- Resource limits and requests are defined, with a maximum of 300m CPU and 500Mi
  memory, and a minimum of 100m CPU and 200Mi memory.
- The application listens on port 80 within the container.
- Security settings ensure the container runs as a non-root user with specific
  capabilities and a read-only root filesystem.
- A temporary volume is mounted at '/tmp' using an in-memory empty directory.
- The Deployment is scheduled to run on nodes with the Linux operating system.

sock-shop-2/manifests/02-carts-svc.yaml
- This manifest defines a Kubernetes Service.
- The Service is named 'carts'.
- It is annotated to enable Prometheus scraping with 'prometheus.io/scrape: true'.
- The Service is labeled with 'name: carts'.
- It is deployed in the 'sock-shop' namespace.
- The Service exposes port 80 and directs traffic to the same port on the selected
  pods.
- It uses a selector to target pods with the label 'name: carts'.

sock-shop-2/manifests/03-carts-db-dep.yaml
- This manifest defines a Deployment in Kubernetes.
- The Deployment is named 'carts-db' and is located in the 'sock-shop' namespace.
- It specifies that there should be 2 replicas of the 'carts-db' pod running.
- The pods are selected based on the label 'name: carts-db'.
- Each pod runs a single container using the 'mongo' image.
- The container exposes port 27017, which is the default port for MongoDB.
- Security settings are applied to drop all capabilities and only add CHOWN, SETGID,
  and SETUID.
- The root filesystem of the container is set to read-only for security purposes.
- A temporary volume is mounted at '/tmp' using an in-memory emptyDir volume.
- The pods are scheduled to run on nodes with the operating system labeled as 'linux'.

sock-shop-2/manifests/04-carts-db-svc.yaml
- This manifest defines a Kubernetes Service.
- The Service is named 'carts-db'.
- It is labeled with 'name: carts-db'.
- The Service is created in the 'sock-shop' namespace.
- It exposes port 27017 and directs traffic to the same port on the target pods.
- The Service selects pods with the label 'name: carts-db' to route traffic to them.

sock-shop-2/manifests/05-catalogue-dep.yaml
- This manifest defines a Deployment in Kubernetes.
- The Deployment is named 'catalogue' and is part of the 'sock-shop' namespace.
- It specifies that there should be 2 replicas of the 'catalogue' application running.
- The Deployment uses the Docker image 'weaveworksdemos/catalogue:0.3.5'.
- The application runs with the command '/app' and listens on port 80.
- Resource limits are set to 200m CPU and 200Mi memory, with requests for 100m CPU and
  100Mi memory.
- The container is configured to run as a non-root user with user ID 10001.
- Security settings include dropping all capabilities except 'NET_BIND_SERVICE' and
  using a read-only root filesystem.
- Liveness and readiness probes are configured to check the '/health' endpoint on port
  80, with initial delays of 300 and 180 seconds respectively.
- The Deployment is scheduled to run on nodes with the Linux operating system.

sock-shop-2/manifests/06-catalogue-svc.yaml
- This manifest defines a Kubernetes Service.
- The Service is named 'catalogue'.
- It is annotated to enable Prometheus scraping for monitoring purposes.
- The Service is labeled with 'name: catalogue'.
- It is deployed in the 'sock-shop' namespace.
- The Service exposes port 80 and directs traffic to the same port on the selected
  pods.
- It uses a selector to target pods with the label 'name: catalogue'.

sock-shop-2/manifests/07-catalogue-db-dep.yaml
- This manifest defines a Deployment in Kubernetes.
- The Deployment is named 'catalogue-db'.
- It is located in the 'sock-shop' namespace.
- The Deployment will create 2 replicas of the pod.
- Each pod will run a container from the image 'weaveworksdemos/catalogue-db:0.3.0'.
- The container is configured with environment variables for MySQL, including a root
  password and database name.
- The container exposes port 3306, which is commonly used for MySQL.
- The pods are scheduled to run on nodes with the Linux operating system.

sock-shop-2/manifests/08-catalogue-db-svc.yaml
- This manifest defines a Kubernetes Service.
- The Service is named 'catalogue-db'.
- It is associated with the 'sock-shop' namespace.
- The Service listens on port 3306 and forwards traffic to the same port on the target
  pods.
- It uses a selector to target pods with the label 'name: catalogue-db'.

sock-shop-2/manifests/09-front-end-dep.yaml
- This manifest defines a Deployment in Kubernetes.
- The Deployment is named 'front-end' and is located in the 'sock-shop' namespace.
- It specifies that there should be 1 replica of the front-end application running.
- The Deployment uses a selector to match pods with the label 'name: front-end'.
- The pod template includes a single container named 'front-end'.
- The container uses the image 'weaveworksdemos/front-end:0.3.12'.
- Resource limits are set for the container: 300m CPU and 1000Mi memory.
- Resource requests are set for the container: 100m CPU and 300Mi memory.
- The container exposes port 8079.
- An environment variable 'SESSION_REDIS' is set to 'true'.
- Security context is configured to run the container as a non-root user with user ID
  10001.
- All Linux capabilities are dropped, and the root filesystem is set to read-only.
- A liveness probe is configured to check the '/' path on port 8079, with an initial
  delay of 300 seconds and a period of 3 seconds.
- A readiness probe is also configured to check the '/' path on port 8079, with an
  initial delay of 30 seconds and a period of 3 seconds.
- The node selector ensures that the pod runs on nodes with the operating system
  labeled as Linux.

sock-shop-2/manifests/10-front-end-svc.yaml
- This manifest defines a Kubernetes Service.
- The Service is named 'front-end'.
- It is located in the 'sock-shop' namespace.
- The Service type is 'NodePort', which exposes the service on each Node's IP at a
  static port.
- It listens on port 80 and forwards traffic to target port 8079 on the pods.
- The nodePort is set to 30001, allowing external access to the service.
- The Service is configured to be scraped by Prometheus for monitoring, as indicated
  by the annotation 'prometheus.io/scrape: true'.
- It selects pods with the label 'name: front-end' to route traffic to.

sock-shop-2/manifests/11-orders-dep.yaml
- This manifest defines a Deployment in Kubernetes.
- The Deployment is named 'orders' and is located in the 'sock-shop' namespace.
- It specifies that there should be 2 replicas of the 'orders' application running.
- The Deployment uses the 'weaveworksdemos/orders:0.4.7' Docker image for the
  container.
- Environment variables are set for Java options to optimize memory usage and disable certain features.
- Resource limits and requests are defined, with a maximum of 500m CPU and 500Mi
  memory, and a minimum of 100m CPU and 300Mi memory.
- The container listens on port 80.
- Security context is configured to run the container as a non-root user with specific
  capabilities and a read-only root filesystem.
- A temporary volume is mounted at '/tmp' using an in-memory empty directory.
- The Deployment is scheduled to run on nodes with the Linux operating system.

sock-shop-2/manifests/12-orders-svc.yaml
- This manifest defines a Kubernetes Service.
- The Service is named 'orders'.
- It is annotated to enable Prometheus scraping with 'prometheus.io/scrape: true'.
- The Service is labeled with 'name: orders'.
- It is deployed in the 'sock-shop' namespace.
- The Service exposes port 80 and directs traffic to the same port on the target pods.
- It uses a selector to match pods with the label 'name: orders'.

sock-shop-2/manifests/13-orders-db-dep.yaml
- This manifest defines a Deployment in Kubernetes.
- The Deployment is named 'orders-db' and is located in the 'sock-shop' namespace.
- It specifies 2 replicas of the 'orders-db' pod to be created.
- The pods are labeled with 'name: orders-db' for identification and selection.
- Each pod runs a single container using the 'mongo' image.
- The container exposes port 27017, which is the default port for MongoDB.
- Security settings are applied to drop all capabilities and add only CHOWN, SETGID,
  and SETUID.
- The root filesystem of the container is set to read-only for security purposes.
- A temporary volume is mounted at '/tmp' using an in-memory emptyDir volume.
- The pods are scheduled to run on nodes with the operating system labeled as 'linux'.

sock-shop-2/manifests/14-orders-db-svc.yaml
- This manifest defines a Kubernetes Service.
- The Service is named 'orders-db'.
- It is located in the 'sock-shop' namespace.
- The Service is configured to expose port 27017.
- It targets the same port (27017) on the pods it selects.
- The Service uses a selector to match pods with the label 'name: orders-db'.

sock-shop-2/manifests/15-payment-dep.yaml
- This manifest defines a Deployment in Kubernetes.
- The Deployment is named 'payment' and is part of the 'sock-shop' namespace.
- It specifies that there should be 2 replicas of the 'payment' application running.
- The Deployment uses the Docker image 'weaveworksdemos/payment:0.4.3'.
- Resource limits are set for the container, with a maximum of 200m CPU and 200Mi
  memory, and requests for 99m CPU and 100Mi memory.
- The container listens on port 80.
- Security settings ensure the container runs as a non-root user with user ID 10001,
  drops all capabilities except 'NET_BIND_SERVICE', and uses a read-only root filesystem.
- Liveness and readiness probes are configured to check the '/health' endpoint on port
  80, with initial delays of 300 and 180 seconds respectively, and a period of 3 seconds.
- The Deployment is scheduled to run on nodes with the Linux operating system.

sock-shop-2/manifests/16-payment-svc.yaml
- This manifest defines a Kubernetes Service.
- The Service is named 'payment'.
- It is annotated for Prometheus scraping, which means it is set up for monitoring.
- The Service is labeled with 'name: payment'.
- It is deployed in the 'sock-shop' namespace.
- The Service exposes port 80 and directs traffic to the same port on the selected
  pods.
- The Service selects pods with the label 'name: payment'.

sock-shop-2/manifests/17-queue-master-dep.yaml
- This manifest defines a Deployment in Kubernetes.
- The Deployment is named 'queue-master' and is located in the 'sock-shop' namespace.
- It specifies that there should be 2 replicas (instances) of the 'queue-master'
  application running.
- The Deployment uses a container image 'weaveworksdemos/queue-master:0.3.1'.
- Environment variables are set for the container, including Java options for memory
  management and garbage collection.
- Resource limits and requests are defined, with a CPU limit of 300m and memory limit
  of 500Mi, and requests for 100m CPU and 300Mi memory.
- The container exposes port 80 for network traffic.
- The Deployment is configured to run on nodes with the Linux operating system.

sock-shop-2/manifests/18-queue-master-svc.yaml
- This manifest defines a Kubernetes Service.
- The Service is named 'queue-master'.
- It is annotated for Prometheus scraping with 'prometheus.io/scrape: true'.
- The Service is labeled with 'name: queue-master'.
- It is deployed in the 'sock-shop' namespace.
- The Service exposes port 80 and directs traffic to the same port on the selected
  pods.
- It selects pods with the label 'name: queue-master' to route traffic to.

sock-shop-2/manifests/19-rabbitmq-dep.yaml
- This manifest defines a Deployment for RabbitMQ in Kubernetes.
- It is set to run in the 'sock-shop' namespace.
- The Deployment is named 'rabbitmq' and is labeled accordingly.
- It specifies 2 replicas, meaning there will be 2 instances of RabbitMQ running.
- The Deployment uses a selector to match pods with the label 'name: rabbitmq'.
- The pod template includes two containers: one for RabbitMQ and another for a
  RabbitMQ exporter.
- The RabbitMQ container uses the image 'rabbitmq:3.6.8-management'.
- It exposes two ports: 15672 for management and 5672 for RabbitMQ operations.
- Security context is set to drop all capabilities and add specific ones like CHOWN,
  SETGID, SETUID, and DAC_OVERRIDE.
- The root filesystem is set to read-only for security purposes.
- The RabbitMQ exporter container uses the image 'kbudde/rabbitmq-exporter' and
  exposes port 9090.
- The Deployment is configured to run on nodes with the label 'beta.kubernetes.io/os: 
  linux'.
- Annotations are set to prevent Prometheus from scraping metrics from this deployment.

sock-shop-2/manifests/20-rabbitmq-svc.yaml
- This manifest defines a Kubernetes Service.
- The Service is named 'rabbitmq'.
- It is annotated for Prometheus scraping on port 9090.
- The Service is labeled with 'name: rabbitmq'.
- It is deployed in the 'sock-shop' namespace.
- The Service exposes two ports: 5672 for RabbitMQ and 9090 for an exporter.
- The protocol used for the ports is TCP.
- The Service selects pods with the label 'name: rabbitmq'.

sock-shop-2/manifests/21-session-db-dep.yaml
- This manifest defines a Deployment in Kubernetes.
- The Deployment is named 'session-db' and is located in the 'sock-shop' namespace.
- It specifies that there should be 2 replicas of the 'session-db' pod running.
- The pods are selected based on the label 'name: session-db'.
- Each pod runs a single container using the 'redis' image.
- The container exposes port 6379, which is commonly used by Redis.
- Security settings are applied to drop all capabilities and only add CHOWN, SETGID, and SETUID, with a read-only root filesystem for enhanced security.
- The pods are scheduled to run on nodes with the operating system labeled as Linux.

sock-shop-2/manifests/22-session-db-svc.yaml
- This manifest defines a Kubernetes Service.
- The Service is named 'session-db'.
- It is located in the 'sock-shop' namespace.
- The Service is configured to expose port 6379.
- It targets the same port (6379) on the selected pods.
- The Service uses a selector to match pods with the label 'name: session-db'.

sock-shop-2/manifests/23-shipping-dep.yaml
- This manifest defines a Deployment in Kubernetes.
- The Deployment is named 'shipping' and is located in the 'sock-shop' namespace.
- It specifies that there should be 2 replicas of the 'shipping' application running.
- The Deployment uses the Docker image 'weaveworksdemos/shipping:0.4.8'.
- Environment variables are set for the application, including 'ZIPKIN' and
  'JAVA_OPTS'.
- Resource limits and requests are defined, with limits set to 300m CPU and 500Mi
  memory, and requests set to 100m CPU and 300Mi memory.
- The application listens on port 80.
- Security context is configured to run the container as a non-root user with user ID
  10001, and it drops all capabilities except 'NET_BIND_SERVICE'.
- The root filesystem is set to be read-only.
- A temporary volume is mounted at '/tmp', using an in-memory emptyDir volume.
- The Deployment is scheduled to run on nodes with the operating system labeled as
  'linux'.

sock-shop-2/manifests/24-shipping-svc.yaml
- This is a Kubernetes Service manifest.
- The service is named 'shipping'.
- It is annotated for Prometheus scraping with 'prometheus.io/scrape: true'.
- The service is labeled with 'name: shipping'.
- It is deployed in the 'sock-shop' namespace.
- The service exposes port 80 and directs traffic to the same port on the selected
  pods.
- It selects pods with the label 'name: shipping' to route traffic to.

sock-shop-2/manifests/25-user-dep.yaml
- This manifest defines a Deployment in Kubernetes.
- The Deployment is named 'user' and is located in the 'sock-shop' namespace.
- It specifies that there should be 2 replicas of the 'user' application running.
- The Deployment uses the Docker image 'weaveworksdemos/user:0.4.7'.
- Resource limits are set for the container: 300m CPU and 200Mi memory, with requests 
  for 100m CPU and 100Mi memory.
- The container listens on port 80.
- An environment variable 'mongo' is set with the value 'user-db:27017'.
- Security context is configured to run the container as a non-root user with user ID
  10001.
- The container has a read-only root filesystem and drops all capabilities except
  'NET_BIND_SERVICE'.
- Liveness and readiness probes are configured to check the '/health' endpoint on port
  80.
- The liveness probe starts after 300 seconds and checks every 3 seconds.
- The readiness probe starts after 180 seconds and checks every 3 seconds.
- The Deployment is scheduled to run on nodes with the Linux operating system.

sock-shop-2/manifests/26-user-svc.yaml
- This manifest defines a Kubernetes Service.
- The Service is named 'user'.
- It is annotated for Prometheus scraping, which means it is set up for monitoring.
- The Service is labeled with 'name: user'.
- It is deployed in the 'sock-shop' namespace.
- The Service listens on port 80 and forwards traffic to the same port on the selected
  pods.
- The Service selects pods with the label 'name: user'.

sock-shop-2/manifests/27-user-db-dep.yaml
- This manifest defines a Deployment in Kubernetes.
- The Deployment is named 'user-db' and is located in the 'sock-shop' namespace.
- It specifies that there should be 2 replicas of the 'user-db' pod running.
- The pods are selected based on the label 'name: user-db'.
- Each pod runs a single container using the image 'weaveworksdemos/user-db:0.3.0'.
- The container exposes port 27017, labeled as 'mongo', which is typically used for
  MongoDB.
- Security settings are applied to drop all capabilities and only add CHOWN, SETGID,
  and SETUID, with a read-only root filesystem.
- A temporary volume is mounted at '/tmp' using an in-memory emptyDir volume.
- The pods are scheduled to run on nodes with the operating system labeled as 'linux'.

sock-shop-2/manifests/28-user-db-svc.yaml
- This manifest defines a Kubernetes Service.
- The Service is named 'user-db'.
- It is labeled with 'name: user-db'.
- The Service is created in the 'sock-shop' namespace.
- It exposes port 27017 and directs traffic to the same port on the target pods.
- The Service selects pods with the label 'name: user-db' to route traffic to them.

<<Resiliency issues/weaknesses in the manifests>>
Issue #0: Missing Resource Requests
- details: Pods may not get scheduled if the cluster is under resource pressure, leading to potential downtime.
- manifests having the issues: [
  'sock-shop-2/manifests/03-carts-db-dep
   .yaml',
  'sock-shop-2/manifests/07-catalogue-db-dep
   .yaml',
  'sock-shop-2/manifests/13-orders-db-dep
   .yaml',
  'sock-shop-2/manifests/19-rabbitmq-dep
   .yaml',
  'sock-shop-2/manifests/21-session-db-dep
   .yaml',
  'sock-shop-2/manifests/27-user-db-dep.yaml'
  ]
- problematic config: The deployments for carts-db, catalogue-db, orders-db, rabbitmq, 
  session-db, and user-db do not specify resource requests.

Issue #1: Single Replica Deployment
- details: The front-end deployment has only one replica, which can lead to downtime if the pod fails.
- manifests having the issues: ['sock-shop-2/manifests/09-front-end-dep.yaml']
- problematic config: spec.replicas: 1

Issue #2: Missing Liveness and Readiness Probes
- details: Without liveness and readiness probes, Kubernetes cannot determine the health of the containers, which may lead to serving traffic to unhealthy pods.
- manifests having the issues: [
  'sock-shop-2/manifests/01-carts-dep.yaml',
  'sock-shop-2/manifests/03-carts-db-dep
   .yaml',
  'sock-shop-2/manifests/05-catalogue-dep
   .yaml',
  'sock-shop-2/manifests/07-catalogue-db-dep
   .yaml',
  'sock-shop-2/manifests/11-orders-dep.yaml',
  'sock-shop-2/manifests/13-orders-db-dep
   .yaml',
  'sock-shop-2/manifests/15-payment-dep.yaml',
  'sock-shop-2/manifests/17-queue-master-dep
   .yaml',
  'sock-shop-2/manifests/19-rabbitmq-dep
   .yaml',
  'sock-shop-2/manifests/21-session-db-dep
   .yaml',
  'sock-shop-2/manifests/23-shipping-dep
   .yaml',
  'sock-shop-2/manifests/25-user-dep.yaml',
  'sock-shop-2/manifests/27-user-db-dep.yaml'
  ]
- problematic config: The deployments for carts, carts-db, catalogue, catalogue-db,
  orders, orders-db, payment, queue-master, rabbitmq, session-db, shipping, user, and user-db do not have liveness or readiness probes configured.

Issue #3: Hardcoded Environment Variables
- details: Hardcoded environment variables, such as passwords, can lead to security
  vulnerabilities.
- manifests having the issues: [
  'sock-shop-2/manifests/07-catalogue-db-dep
   .yaml']
- problematic config: env: MYSQL_ROOT_PASSWORD is hardcoded with value: fake_password.

<<Application of the manifests>>
Thoughts:
The manifests provided are part of a well-known microservices demo application called 'Sock Shop'. This application is designed to demonstrate the use of microservices architecture and Kubernetes orchestration. The application simulates an e-commerce website for selling socks, with various services such as carts, catalogue, orders, payment, and user management. The file names, labels, and Docker images used in the manifests (e.g., 'weaveworksdemos') are indicative of the Sock Shop application, which is a common reference application in the Kubernetes community.

Assumed application:
Sock Shop is an e-commerce application that simulates an online sock store. It offers users the ability to browse a catalogue of socks, add items to a cart, place orders, and manage user accounts. The application is composed of multiple microservices, each responsible for a specific function, such as managing the product catalogue, handling user sessions, processing payments, and managing orders.

<<Summary of user instructions>>
- The Chaos-Engineering experiment must be completed within 1 minute.
- When using k6 in steady-state definition, always select a request URL from the
  following options (other requests are invalid):
  1. http://front-end.sock-shop.svc.cluster.local/
  2. http://front-end.sock-shop.svc.cluster.local/
   catalogue?size=10
  3. http://front-end.sock-shop.svc.cluster.local/
   detail.html?id=<ID> 
   Replace <ID> with an available ID: [03fef6ac-1896-4ce8-bd69-b798f85c6e0b , 3395a43e-2d88-40de-b95fe00e1502085b , 510a0d7e-8e83-4193-b483-e27e09ddc34d , 808a2de1-1aaa-4c25-a9b9-6612e8f29a38 , 819e1fbf-8b7e-4f6d-811f-693534916a8b , 837ab141-399e-4c1f-9abc-bace40296bac , a0a4f044-b040-410d-8ead-4de0446aec7e , d3588630-ad8e-49df-bbd7-3167f7efb246 , zzz4f044-b040-410d-8ead-4de0446aec7e ]
  4. http://front-end.sock-shop.svc.cluster.local/
   category/
  5. http://front-end.sock-shop.svc.cluster.local/
   category?tags=<TAG>
   Replace <TAG> with an available tag: [ magic , action , blue , brown , black , sport , formal , red , green , skin , geek]
  6. http://front-end.sock-shop.svc.cluster.local/
   basket.html
\end{customdialogue}

\begin{customdialogue}[Steady state \#1: carts-db-replicas]
<<Description>>
The first issue to address is the 'Missing Resource Requests' for the 'carts-db' deployment. This is a critical issue because without resource requests, the pods may not get scheduled if the cluster is under resource pressure, leading to potential downtime. To ensure the system's steady state, we need to verify that the 'carts-db' deployment maintains the expected number of replicas, which is 2, even under resource constraints. This can be measured by checking the number of running pods for the 'carts-db' deployment.

<<Current state inspection>>
To inspect the 'front-end-replica' state, I will use the K8s API to check the number of ready replicas for the 'front-end' deployment in the 'sock-shop' namespace. This is because the task involves verifying the number of running pods, which is a Kubernetes resource state check. The K8s API is suitable for this purpose as it allows us to directly query the deployment status and ensure that the single replica is always ready. The script will run for a duration of 5 seconds, checking the status every second, to ensure the deployment maintains at least one ready replica 100% of the time.
tool: k8s duration: 5s
Inspection script (Python) k8s_front-end-replica.py :
```
import os
import time
import argparse
from kubernetes import client, config

def check_carts_db_replicas(
  namespace='sock-shop', deployment_name='carts-db', duration=5):
  # Load Kubernetes configuration based on the environment
  if os.getenv('KUBERNETES_SERVICE_HOST'):
  config.load_incluster_config()
  else:
  config.load_kube_config()

  v1 = client.AppsV1Api()

  for _ in range(duration):
  try:
    deployment = v1.read_namespaced_deployment(
    deployment_name, namespace)
    replicas = deployment.status.replicas
    ready_replicas = deployment.status.ready_replicas
    print(f"Total replicas: {replicas}, Ready replicas: {ready_replicas}")
  except client.exceptions.ApiException as e:
    print(f"Exception when calling AppsV1Api->read_namespaced_deployment: {e}")
  time.sleep(1)

if __name__ == '__main__':
  parser = argparse.ArgumentParser(
  description='Check carts-db replicas')
  parser.add_argument('--duration', type=int, default=5, help='Duration to check the replicas')
  args = parser.parse_args()
  check_carts_db_replicas(
  duration=args.duration)
```

The inspection script's results (current states) are as follows:
```
Total replicas: 2, Ready replicas: 2
Total replicas: 2, Ready replicas: 2
Total replicas: 2, Ready replicas: 2
Total replicas: 2, Ready replicas: 2
Total replicas: 2, Ready replicas: 2
```

<<Threshold>>
The steady state we are considering is the number of replicas for the 'carts-db' deployment. The current state shows that there are 2 total replicas and 2 ready replicas consistently over a period of 5 seconds. This indicates that the system is currently stable with all replicas being ready. Given that the deployment is configured to have 2 replicas, the threshold should ensure that both replicas are ready to maintain the steady state. To account for some fluctuations, we can set a tolerance that allows for a brief period where not all replicas are ready, but the system should recover quickly. A reasonable threshold would be to have at least 1 ready replica 100% of the time and both replicas ready at least 80% of the time during the monitoring period. This ensures that the system remains operational even if one replica temporarily becomes unavailable. 

At least 1 ready replica 100% of the time and 2 ready replicas at least 80% of the time during the monitoring period.


<<Unit test to validate the steady state>>
To verify the steady state of the 'carts-db' deployment, I will write a unit test that checks the number of ready replicas over a specified duration. The test will ensure that there is at least 1 ready replica 100% of the time and 2 ready replicas at least 80% of the time. I will implement a loop that checks the status every second for the specified duration and use assertions to validate the threshold conditions. The test will be run from the command line with a '--duration' argument to specify the monitoring period.
```
import os
import time
import argparse
from kubernetes import client, config
from unittest_base import K8sAPIBase

class TestCartsDBReplicas(K8sAPIBase):
  def __init__(self, namespace='sock-shop', deployment_name='carts-db', duration=5):
    super().__init__()
    self.namespace = namespace
    self.deployment_name = deployment_name
    self.duration = duration

  def test_replicas_threshold(self):
    # Initialize counters for ready replicas
    total_checks = 0
    ready_replicas_count = 0
    fully_ready_replicas_count = 0

    for _ in range(self.duration):
      try:
        # Read the deployment status
        deployment = self.v1.read_namespaced_deployment(
          self.deployment_name, self.namespace)
        replicas = deployment.status.replicas
        ready_replicas = deployment.status.ready_replicas
        print(f"Total replicas: {replicas}, Ready replicas: {ready_replicas}")

        # Increment the total checks
        total_checks += 1

        # Check if at least 1 replica is ready
        if ready_replicas >= 1:
          ready_replicas_count += 1

        # Check if both replicas are ready
        if ready_replicas == 2:
          fully_ready_replicas_count += 1

      except client.exceptions.ApiException as e:
        print(f"Exception when calling AppsV1Api->read_namespaced_deployment: {e}")

      # Wait for 1 second before the next check
      time.sleep(1)

    # Calculate the percentage of time conditions are met
    one_ready_percentage = (ready_replicas_count / total_checks) * 100
    two_ready_percentage = (fully_ready_replicas_count / total_checks) * 100

    # Assert the threshold conditions
    assert one_ready_percentage == 100, "At least 1 ready replica was not available 100% of the time."
    assert two_ready_percentage >= 80, "2 ready replicas were not available at least 80% of the time."

    print("Test passed: Steady state conditions are satisfied.")


def main():
  parser = argparse.ArgumentParser(description='Test carts-db replicas threshold')
  parser.add_argument('--duration', type=int, default=5, help='Duration to check the replicas')
  args = parser.parse_args()

  # Create an instance of the test class with the specified duration
  test = TestCartsDBReplicas(duration=args.duration)
  # Run the test
  test.test_replicas_threshold()


if __name__ == '__main__':
  main()
```

Pod unittestcarts-db-replicasmod0-pod has failed.
```
Traceback (most recent call last):
  File "/chaos-eater/sandbox/cycle_20241127_043136/
        hypothesis/unittest_carts-db-replicas
        _mod0.py", line 68, in <module>
  main()
  File "/chaos-eater/sandbox/cycle_20241127_043136/
        hypothesis/unittest_carts-db-replicas
        _mod0.py", line 64, in main
  test.test_replicas_threshold()
  File "/chaos-eater/sandbox/cycle_20241127_043136/
        hypothesis/unittest_carts-db-replicas
        _mod0.py", line 23, in test_replicas_threshold
  deployment = self.v1.read_namespaced_deployment(
    self.deployment_name, self.namespace)
AttributeError: 'CoreV1Api' object has no attribute 'read_namespaced_deployment'. Did you mean: 'read_namespaced_event'?
```

The error occurs because the 'CoreV1Api' object does not have a method called 'read_namespaced_deployment'. The correct API to use for reading deployments is 'AppsV1Api'. I will fix the error by changing the API client from 'CoreV1Api' to 'AppsV1Api' in the 'K8sAPIBase' class initialization. This will allow the test to correctly access the deployment information.
```
import os
import time
import argparse
from kubernetes import client, config
from unittest_base import K8sAPIBase

class TestCartsDBReplicas(K8sAPIBase):
  def __init__(self, namespace='sock-shop', deployment_name='carts-db', duration=5):
    super().__init__()
    self.namespace = namespace
    self.deployment_name = deployment_name
    self.duration = duration
    # Use the correct API client for deployments
    self.apps_v1 = client.AppsV1Api()

  def test_replicas_threshold(self):
    # Initialize counters for ready replicas
    total_checks = 0
    ready_replicas_count = 0
    fully_ready_replicas_count = 0

    for _ in range(self.duration):
      try:
        # Read the deployment status using the correct API client
        deployment = self.apps_v1.read_namespaced_deployment(
          self.deployment_name, self.namespace)
        replicas = deployment.status.replicas
        ready_replicas = deployment.status.ready_replicas
        print(f"Total replicas: {replicas}, Ready replicas: {ready_replicas}")

        # Increment the total checks
        total_checks += 1

        # Check if at least 1 replica is ready
        if ready_replicas >= 1:
          ready_replicas_count += 1

        # Check if both replicas are ready
        if ready_replicas == 2:
          fully_ready_replicas_count += 1

      except client.exceptions.ApiException as e:
        print(f"Exception when calling AppsV1Api->read_namespaced_deployment: {e}")

      # Wait for 1 second before the next check
      time.sleep(1)

    # Calculate the percentage of time conditions are met
    one_ready_percentage = (ready_replicas_count / total_checks) * 100
    two_ready_percentage = (fully_ready_replicas_count / total_checks) * 100

    # Assert the threshold conditions
    assert one_ready_percentage == 100, "At least 1 ready replica was not available 100% of the time."
    assert two_ready_percentage >= 80, "2 ready replicas were not available at least 80% of the time."

    print("Test passed: Steady state conditions are satisfied.")


def main():
  parser = argparse.ArgumentParser(description='Test carts-db replicas threshold')
  parser.add_argument('--duration', type=int, default=5, help='Duration to check the replicas')
  args = parser.parse_args()

  # Create an instance of the test class with the specified duration
  test = TestCartsDBReplicas(duration=args.duration)
  # Run the test
  test.test_replicas_threshold()


if __name__ == '__main__':
  main()
```

Pod unittestcarts-db-replicasmod1-pod has completed sucessfully.
The inspection script's results (current states) are as follows:
```
Total replicas: 2, Ready replicas: 2
Total replicas: 2, Ready replicas: 2
Total replicas: 2, Ready replicas: 2
Total replicas: 2, Ready replicas: 2
Total replicas: 2, Ready replicas: 2
Test passed: Steady state conditions are satisfied.
```
\end{customdialogue}

\begin{customdialogue}[Steady state \#2: front-end-replica]
<<Description>>
The next issue to address is the 'Single Replica Deployment' for the 'front-end' deployment. This is a critical issue because having only one replica can lead to downtime if the pod fails. To ensure the system's steady state, we need to verify that the 'front-end' deployment maintains at least one running pod at all times. This can be measured by checking the number of running pods for the 'front-end' deployment. Given that the deployment is configured to have only 1 replica, the threshold should ensure that this single replica is always ready to maintain the steady state. A reasonable threshold would be to have at least 1 ready replica 100% of the time during the monitoring period. This ensures that the system remains operational even if there are
temporary issues with the pod.

<<Current state inspection>>
To inspect the 'front-end-replica' state, I will use the K8s API to check the number of ready replicas for the 'front-end' deployment in the 'sock-shop' namespace. This is because the task involves verifying the number of running pods, which is a Kubernetes resource state check. The K8s API is suitable for this purpose as it allows us to directly query the deployment status and ensure that the single replica is always ready. The script will run for a duration of 5 seconds, checking the status every second, to ensure the deployment maintains at least one ready replica 100% of the time.
tool: k8s duration: 5s
Inspection script (Python) k8s_front-end-replica.py :
```
import os
import time
import argparse
from kubernetes import client, config

def check_front_end_replicas(namespace, deployment_name, duration):
  # Load Kubernetes configuration based on the environment
  if os.getenv('KUBERNETES_SERVICE_HOST'):
    config.load_incluster_config()
  else:
    config.load_kube_config()

  v1 = client.AppsV1Api()
  ready_replicas_count = 0

  for _ in range(duration):
    resp = v1.read_namespaced_deployment_status(
      deployment_name, namespace)
    ready_replicas = resp.status.ready_replicas or 0
    print(f"Ready replicas for {deployment_name}: {ready_replicas}")
    if ready_replicas >= 1:
      ready_replicas_count += 1
    time.sleep(1)

  print(f"{deployment_name} was ready {ready_replicas_count}/{duration} times.")

if __name__ == '__main__':
  parser = argparse.ArgumentParser(description='Check front-end replicas')
  parser.add_argument('--duration', type=int, default=5, help='Duration to check the replicas')
  args = parser.parse_args()
  check_front_end_replicas('sock-shop', 'front-end', args.duration)
```

Pod k8sfront-end-replica-pod has been completed successfully.
The inspection script's results (current states) are as follows:
```
Ready replicas for front-end: 1
Ready replicas for front-end: 1
Ready replicas for front-end: 1
Ready replicas for front-end: 1
Ready replicas for front-end: 1
front-end was ready 5/5 times.
```

<<Threshold>>
The steady state we are considering is the availability of the 'front-end' deployment, which currently has only one replica. The current state shows that the 'front-end' deployment has 1 ready replica consistently over a monitoring period of 5 seconds. To ensure the system remains operational, the threshold should be set such that the 'front-end' deployment maintains at least 1 ready replica 100% of the time during the monitoring period. This accounts for the fact that there is only one replica, and any downtime would mean the service is unavailable. Given the current state satisfies this condition, the threshold should be set to ensure that at least 1 ready replica is present 100% of the time during the monitoring period.

At least 1 ready replica must be present 100% of the time during the monitoring period.

<<Unit test to validate the steady state>>
To verify the steady state of the 'front-end' deployment, I will write a unit test that checks if there is at least 1 ready replica 100% of the time during the specified monitoring period. The test will use a for-loop to check the status every second for the given duration. The threshold is set to ensure that the 'front-end' deployment maintains at least 1 ready replica 100% of the time, which means the number of times the ready replicas are 1 or more should equal the duration. I will use the Kubernetes Python API to interact with the cluster and check the deployment status. The test will include an assertion to verify that the steady state satisfies the threshold, and it will print a summary of the results at the end.
```
import os
import time
import argparse
from kubernetes import client, config
from unittest_base import K8sAPIBase

class TestFrontEndReplica(K8sAPIBase):
  def __init__(self, namespace, deployment_name, duration):
    super().__init__()
    self.namespace = namespace
    self.deployment_name = deployment_name
    self.duration = duration

  def test_steady_state(self):
    ready_replicas_count = 0

    # Loop for the specified duration
    for _ in range(self.duration):
      # Get the deployment status
      resp = self.v1.read_namespaced_deployment_status(
        self.deployment_name, self.namespace)
      ready_replicas = resp.status.ready_replicas or 0
      print(f"Ready replicas for {self.deployment_name}: {ready_replicas}")

      # Check if the number of ready replicas is at least 1
      if ready_replicas >= 1:
        ready_replicas_count += 1

      # Wait for 1 second before the next check
      time.sleep(1)

    # Calculate the percentage of time the deployment was ready
    readiness_percentage = (ready_replicas_count / self.duration) * 100
    print(f"{self.deployment_name} was ready {ready_replicas_count}/{self.duration} times.")

    # Assert that the deployment was ready 100% of the time
    assert readiness_percentage == 100, f"{self.deployment_name} readiness was {readiness_percentage}%, expected 100%."


def main():
  parser = argparse.ArgumentParser(description='Test front-end replica readiness')
  parser.add_argument('--duration', type=int, default=5, help='Duration to check the replicas')
  args = parser.parse_args()

  # Create a test instance and run the test
  test = TestFrontEndReplica('sock-shop', 'front-end', args.duration)
  test.test_steady_state()


if __name__ == '__main__':
  main()
```

Pod unittestfront-end-replicamod0-pod has failed.
```
Traceback (most recent call last):
  File "/chaos-eater/sandbox/cycle_20241127_043136/
        hypothesis/unittest_front-end-replica
        _mod0.py", line 50, in <module>
  main()
  File "/chaos-eater/sandbox/cycle_20241127_043136/
       hypothesis/unittest_front-end-replica
       _mod0.py\", line 46, in main
  test.test_steady_state()
  File \"/chaos-eater/sandbox/cycle_20241127_043136/
         hypothesis/unittest_front-end-replica
         _mod0.py\", line 20, in test_steady_state
  resp = self.v1.read_namespaced_deployment_status(
    self.deployment_name, self.namespace)
AttributeError: 'CoreV1Api' object has no attribute 'read_namespaced_deployment_status'. Did you mean: 'read_namespaced_pod_status'?
```

The error occurs because the CoreV1Api class does not have a method called read_namespaced_deployment_status . Instead, the correct class to use for reading deployment status is AppsV1Api . I will fix the error by changing the API client from CoreV1Api to AppsV1Api in the K8sAPIBase class. This will allow the test to correctly
call the read_namespaced_deployment_status method and check the deployment status. The rest of the test logic remains the same, ensuring that the deployment maintains at least 1 ready replica 100% of the time during the specified duration.
```
import os
import time
import argparse
from kubernetes import client, config
from unittest_base import K8sAPIBase

class TestFrontEndReplica(K8sAPIBase):
  def __init__(self, namespace, deployment_name, duration):
    super().__init__()
    self.namespace = namespace
    self.deployment_name = deployment_name
    self.duration = duration
    # Use AppsV1Api to interact with deployments
    self.apps_v1 = client.AppsV1Api()

  def test_steady_state(self):
    ready_replicas_count = 0

    # Loop for the specified duration
    for _ in range(self.duration):
      # Get the deployment status
      resp = self.apps_v1
             .read_namespaced_deployment_status(
        self.deployment_name, self.namespace)
      ready_replicas = resp.status.ready_replicas or 0
      print(f"Ready replicas for {self.deployment_name}: {ready_replicas}")

      # Check if the number of ready replicas is at least 1
      if ready_replicas >= 1:
        ready_replicas_count += 1

      # Wait for 1 second before the next check
      time.sleep(1)

    # Calculate the percentage of time the deployment was ready
    readiness_percentage = (ready_replicas_count / self.duration) * 100
    print(f"{self.deployment_name} was ready {ready_replicas_count}/{self.duration} times.")

    # Assert that the deployment was ready 100% of the time
    assert readiness_percentage == 100, f"{self.deployment_name} readiness was {readiness_percentage}%, expected 100%."


def main():
  parser = argparse.ArgumentParser(description='Test front-end replica readiness')
  parser.add_argument('--duration', type=int, default=5, help='Duration to check the replicas')
  args = parser.parse_args()

  # Create a test instance and run the test
  test = TestFrontEndReplica('sock-shop', 'front-end', args.duration)
  test.test_steady_state()


if __name__ == '__main__':
  main()
```

Pod unittestfront-end-replicamod1-pod has completed sucessfully.
The inspection script's results (current states) are as follows:
```
Ready replicas for front-end: 1
Ready replicas for front-end: 1
Ready replicas for front-end: 1
Ready replicas for front-end: 1
Ready replicas for front-end: 1
front-end was ready 5/5 times.
```
\end{customdialogue}

\begin{customdialogue}[Failure scenario: Black Friday Sale]
<<Description>>
During a Black Friday sale, the system is expected to handle a significant increase in traffic, which can stress the system's resources and reveal weaknesses. The 'carts-db' deployment lacks resource requests, which can lead to scheduling issues under resource pressure. The 'front-end' deployment has only one replica, making it a single point of failure. To simulate the Black Friday event, we will first inject a 'StressChaos' fault to simulate high CPU usage on the 'carts-db' pods, testing if the system can maintain the steady state of having at least 1 ready replica 100% of the time and 2 ready replicas 80% of the time. Next, we will inject a 'PodChaos' fault to kill the 'front-end' pod, testing if the system can maintain the steady state of having at least 1 ready replica 100% of the time. This sequence simulates the increased load and potential failures
during a high-traffic event like Black Friday.

<<Failure-injection sequence>>
StressChaos ({'namespace': 'sock-shop', 'label': 'name=carts-db'}) -> PodChaos
({'namespace': 'sock-shop', 'label': 'name=front-end'})

<<Detailed failure parameters>>
Detailed parameters of StressChaos ({'namespace': 'sock-shop', 'label': 'name=carts-db'})
```
{
  "mode": "all",
  "selector": {
  "namespaces": [
    "sock-shop"
  ],
  "labelSelectors": {
    "name": "carts-db"
  },
  "stressors": {
    "cpu": {
    "workers": 2,
    "load": 80
    }
  },
  "containerNames": [
    "carts-db"
  ]
  }
}
```

Detailed parameters of PodChaos ({'namespace': 'sock-shop', 'label': 'name=front-end'})
```
{
  "action": "pod-kill",
  "mode": "one",   
  "selector":{
  "namespaces": ["sock-shop"],
  "labelSelectors": {
    "name": "front-end"
  }
  },
  "value": "1"
}
```
\end{customdialogue}

\begin{customdialogue}[Chaos experiment (Planning)]
<<Time Schedule>>
The chaos engineering experiment is designed to test the system's resilience under stress conditions and potential failures. Given the constraints, the experiment must be completed within 1 minute. The experiment is divided into three phases: pre-validation, fault-injection, and post-validation.

1. Pre-validation Phase: This phase ensures that the system is in a steady state before any faults are injected. We will allocate 20 seconds for this phase. During this time, we will verify that the 'carts-db' deployment maintains at least 1 ready replica 100% of the time and 2 ready replicas 80% of the time, and that the 'front-end' deployment maintains at least 1 ready replica 100% of the time.

2. Fault-injection Phase: This phase involves injecting faults to simulate the Black Friday event. We will allocate 20 seconds for this phase. The faults include a 'StressChaos' fault to simulate high CPU usage on the 'carts-db' pods and a 'PodChaos' fault to kill the 'front-end' pod. These faults will test the system's ability
to maintain the defined steady states under stress and failure conditions.

3. Post-validation Phase: This phase ensures that the system returns to its steady state after the faults are removed. We will allocate 20 seconds for this phase. During this time, we will again verify the steady states for the 'carts-db' and 'front-end' deployments to ensure they meet the defined thresholds.

The total time for the experiment is 60 seconds, with each phase receiving an equal allocation of 20 seconds. This allocation allows for a balanced approach to validating the system's steady state, injecting faults, and confirming recovery.
Total experiment time: 60s
Pre-validation Phase: 20s
Fault-injection Phase: 20s
Post-validation Phase: 20s

<<Pre-validation Phase>>
In the pre-validation phase, we need to ensure that the system is in a steady state before we introduce any faults. This involves verifying that the current state of the system meets the defined steady state thresholds. Given the constraints of a 20-second total time for this phase, we will execute the unit tests for both steady states simultaneously to maximize efficiency. The first steady state, 'carts-db-replicas requires checking that at least 1 replica is ready 100% of the time and 2 replicas are ready at least 80% of the time. The second steady state, 'front-end-replica', requires ensuring that the single replica is ready 100% of the time. Both tests will run for the full 20 seconds, with checks occurring every second. This approach ensures that we have
sufficient data to confirm the system's readiness before proceeding to the fault injection phase.

- Verified Steady State #0: carts-db-replicas
  - Workflow Name: pre-unittest-carts-db-replicas
  - Grace Period: 0s
  - Duration: 20s

- Verified Steady State #1: front-end-replica
  - Workflow Name: pre-unittest-front-end-replica
  - Grace Period: 0s
  - Duration: 20s

<<Failure-injection Phase>>
In this fault-injection phase, we aim to simulate a Black Friday sale scenario by introducing two types of faults: 'StressChaos' and 'PodChaos'. The goal is to observe how the system behaves under stress and potential failure conditions. The phase is limited to 20 seconds, so we need to carefully time the injections and tests to fit within this window.

First, we will inject a 'StressChaos' fault on the 'carts-db' deployment to simulate high CPU usage. This will help us test the system's ability to maintain the steady state of having at least 1 ready replica 100% of the time and 2 ready replicas 80% of the time. We will start this fault injection immediately at the beginning of the phase and let it run for 10 seconds.

Simultaneously, we will run the unit test for the 'carts-db-replicas' steady state to verify if the system can maintain the required number of ready replicas during the stress period. This test will also start at the beginning of the phase and run for 10 seconds.

Next, we will inject a 'PodChaos' fault to kill the 'front-end' pod. This will test the system's ability to maintain the steady state of having at least 1 ready replica 100% of the time for the 'front-end' deployment. We will start this fault injection at the 10-second mark and let it run for 5 seconds.

Finally, we will run the unit test for the 'front-end-replica' steady state to verify if the system can maintain the required readiness during the pod failure. This test will start at the 10-second mark and run for 5 seconds.

By staggering the fault injections and unit tests in this manner, we can effectively simulate the Black Friday scenario and assess the system's resilience under stress and failure conditions.

- Verified Steady State #0: carts-db-replicas
  - Workflow Name: fault-unittest-carts-db-replicas
  - Grace Period: 0s
  - Duration: 10s

- Verified Steady State #1: front-end-replica
  - Workflow Name: fault-unittest-front-end-replica
  - Grace Period: 10s
  - Duration: 5s

- Injected Faults #0: StressChaos
  - Workflow Name: fault-stresschaos
  - Grace Period: 0s
  - Duration: 10s

- Injected Faults #1: PodChaos
  - Workflow Name: fault-podchaos
  - Grace Period: 10s
  - Duration: 5s

<<Post-validation Phase>>
In the post-validation phase, we need to ensure that the system has returned to its steady states after the fault injection. The steady states we are concerned with are the 'carts-db-replicas' and 'front-end-replica'. The post-validation phase will involve running unit tests to verify these steady states. Given the total time of 20 seconds for this phase, we will stagger the unit tests to ensure both steady states are verified within
this timeframe.

The 'carts-db-replicas' steady state requires that at least 1 ready replica is available 100% of the time and 2 ready replicas are available at least 80% of the time. We will allocate 10 seconds for this test, with a grace period of 0 seconds, as we want to immediately verify the recovery of the 'carts-db' deployment after the stress test.

The 'front-end-replica' steady state requires that at least 1 ready replica is available 100% of the time. We will allocate the remaining 10 seconds for this test, also with a grace period of 0 seconds, to ensure the 'front-end' deployment has recovered from the pod kill fault.

By staggering the tests in this manner, we ensure that both steady states are verified within the 20-second post-validation phase, allowing us to confirm that the system has returned to its expected operational state.

- Verified Steady State #0: carts-db-replica
  - Workflow Name: post-unittest-carts-db-replicas
  - Grace Period: 0s
  - Duration: 10s

- Verified Steady State #1: front-end-replica
  - Workflow Name: post-unittest-front-end-replica
  - Grace Period: 0s
  - Duration: 10s

<<Summary>>
The chaos engineering experiment is structured into three phases, each lasting 20 seconds, to test the system's resilience. In the pre-validation phase, both unit tests for verifying steady states are executed simultaneously for the entire 20 seconds. The 'carts-db-replicas' test checks that at least one replica is ready 100% of the time and two replicas are ready 80% of the time, while the 'front-end-replica' test ensures one replica is ready 100% of the time. In the fault-injection phase, the 'StressChaos' fault is injected on the 'carts-db' deployment to simulate high CPU usage, starting immediately and lasting for 10 seconds. Concurrently, the 'carts-db-replicas' unit test runs for 10 seconds to verify readiness under stress. At the 10-second mark, the 'PodChaos' fault is injected to kill the 'front-end' pod, lasting for 5 seconds, while the 'front-end-replica' unit test runs for 5 seconds to check readiness during the fault. In the post-validation
phase, the 'carts-db-replicas' unit test runs for the first 10 seconds to verify recovery, followed by the 'front-end-replica' unit test for the next 10 seconds, ensuring both deployments return to their steady states.

The generated Chaos Mesh workflow manifest:
```
apiVersion: chaos-mesh.org/v1alpha1
kind: Workflow
metadata:
  name: chaos-experiment-20241127-045539
spec:
  entry: the-entry
  templates:
  #-------------------------------
  # entry point of whole workflow
  #-------------------------------
  - name: the-entry
    templateType: Serial
    deadline: 30m45s
    children:
    - pre-validation-phase
    - fault-injection-phase
    - post-validation-phase

  #-------------------------------------
  # Entry point of pre-validation-phase
  #-------------------------------------
  - name: pre-validation-phase
    templateType: Serial
    deadline: 10m20s
    children:
    - pre-validation-parallel-workflows
  
  - name: pre-validation-parallel-workflows
    templateType: Parallel
    deadline: 5m20s
    children:
    - pre-unittest-carts-db-replicas
    - pre-unittest-front-end-replica

  # Definitions of children of pre-validation-phase
  - name: pre-unittest-carts-db-replicas
    templateType: Task
    deadline: 5m20s
    task:
    container:
      name: pre-unittest-carts-db-replicas-container
      image: chaos-eater/k8sapi:1.0
      imagePullPolicy: IfNotPresent
      command: ["/bin/bash", "-c"]
      args: ["python
        /chaos-eater/sandbox/cycle_20241127
        _043136/unittest_carts-db-replicas
        _mod0.py --duration 20"]
      volumeMounts:
      - name: pvc-volume
        mountPath: /chaos-eater
    volumes:
      - name: pvc-volume
      persistentVolumeClaim:
        claimName: pvc

  - name: pre-unittest-front-end-replica
    templateType: Task
    deadline: 5m20s
    task:
    container:
      name: pre-unittest-front-end-replica-container
      image: chaos-eater/k8sapi:1.0
      imagePullPolicy: IfNotPresent
      command: ["/bin/bash", "-c"]
      args: ["python
        /chaos-eater/sandbox/cycle_20241127
        _043136/unittest_front-end-replica
        _mod0.py --duration 20"]
      volumeMounts:
      - name: pvc-volume
        mountPath: /chaos-eater
    volumes:
      - name: pvc-volume
      persistentVolumeClaim:
        claimName: pvc
  
  #--------------------------------------
  # Entry point of fault-injection-phase
  #--------------------------------------
  - name: fault-injection-phase
    templateType: Serial
    deadline: 10m15s
    children:
    - fault-injection-overlapped-workflows

  - name: fault-injection-parallel-workflow
    templateType: Parallel
    deadline: 5m10s
    children:
    - fault-unittest-carts-db-replicas
    - fault-stresschaos

  - name: fault-injection-suspend-workflow
    templateType: Serial
    deadline: 5m15s
    children:
    - fault-injection-suspend
    - fault-injection-parallel-workflows

  - name: fault-injection-suspend
    templateType: Suspend
    deadline: 10s

  - name: fault-injection-parallel-workflows
    templateType: Parallel
    deadline: 5m5s
    children:
    - fault-unittest-front-end-replica
    - fault-podchaos

  - name: fault-injection-overlapped-workflows
    templateType: Parallel
    deadline: 5m15s
    children:
    - fault-injection-parallel-workflow
    - fault-injection-suspend-workflow

  # Definitions of children of pre-validation-phase
  # unit tests
  - name: fault-unittest-carts-db-replicas
    templateType: Task
    deadline: 5m10s
    task:
    container:
      name: fault-unittest-carts-db-replicas-container
      image: chaos-eater/k8sapi:1.0
      imagePullPolicy: IfNotPresent
      command: ["/bin/bash", "-c"]
      args: ["python
        /chaos-eater/sandbox/cycle_20241127
        _043136/unittest_carts-db-replicas
        _mod0.py --duration 10"]
      volumeMounts:
      - name: pvc-volume
        mountPath: /chaos-eater
    volumes:
      - name: pvc-volume
      persistentVolumeClaim:
        claimName: pvc

  - name: fault-unittest-front-end-replica
    templateType: Task
    deadline: 5m5s
    task:
    container:
      name: fault-unittest-front-end-replica-container
      image: chaos-eater/k8sapi:1.0
      imagePullPolicy: IfNotPresent
      command: ["/bin/bash", "-c"]
      args: ["python
        /chaos-eater/sandbox/cycle_20241127
        _043136/unittest_front-end-replica
        _mod0.py --duration 5"]
      volumeMounts:
      - name: pvc-volume
        mountPath: /chaos-eater
    volumes:
      - name: pvc-volume
      persistentVolumeClaim:
        claimName: pvc

  # fault_injections
  - name: fault-stresschaos
    templateType: StressChaos
    deadline: 10s
    stressChaos:
    containerNames:
      - carts-db
    mode: all
    selector:
      labelSelectors:
      name: carts-db
      namespaces:
      - sock-shop
    stressors:
      cpu:
      load: 80
      workers: 2


  - name: fault-podchaos
    templateType: PodChaos
    deadline: 5s
    podChaos:
    action: pod-kill
    mode: one
    selector:
      labelSelectors:
      name: front-end
      namespaces:
      - sock-shop
    value: '1'

  
  #--------------------------------------
  # Entry point of post-validation phase
  #--------------------------------------
  - name: post-validation-phase
    templateType: Serial
    deadline: 10m10s
    children:
    - post-validation-parallel-workflows

  - name: post-validation-parallel-workflows
    templateType: Parallel
    deadline: 5m10s
    children:
    - post-unittest-carts-db-replicas
    - post-unittest-front-end-replica

  # Definitions of children of pre-validation-phase
  - name: post-unittest-carts-db-replicas
    templateType: Task
    deadline: 5m10s
    task:
    container:
      name: post-unittest-carts-db-replicas-container
      image: chaos-eater/k8sapi:1.0
      imagePullPolicy: IfNotPresent
      command: ["/bin/bash", "-c"]
      args: ["python
        /chaos-eater/sandbox/cycle_20241127
        _043136/unittest_carts-db-replicas
        _mod0.py --duration 10"]
      volumeMounts:
      - name: pvc-volume
        mountPath: /chaos-eater
    volumes:
      - name: pvc-volume
      persistentVolumeClaim:
        claimName: pvc

  - name: post-unittest-front-end-replica
    templateType: Task
    deadline: 5m10s
    task:
    container:
      name: post-unittest-front-end-replica-container
      image: chaos-eater/k8sapi:1.0
      imagePullPolicy: IfNotPresent
      command: ["/bin/bash", "-c"]
      args: ["python
        /chaos-eater/sandbox/cycle_20241127
        _043136/unittest_front-end-replica
        _mod0.py --duration 10"]
      volumeMounts:
      - name: pvc-volume
        mountPath: /chaos-eater
    volumes:
      - name: pvc-volume
      persistentVolumeClaim:
        claimName: pvc
```
\end{customdialogue}

\begin{customdialogue}[Analysis]
The chaos engineering experiment aimed to test the resilience of the 'sock-shop' system under simulated fault conditions, specifically focusing on the 'carts-db' and 'front-end' deployments. The experiment was structured into three phases: pre-validation, fault-injection, and post-validation. The results indicate that while the 'carts-db' deployment maintained its steady state throughout the experiment, the 'front-end' deployment failed to meet the expected steady state during both the fault-injection and post-validation phases.

1. Pre-Validation Phase:
  - Both the 'carts-db-replicas' and 'front-end-replica' tests passed, indicating that the system was in a stable state before any faults were injected. This confirms that the initial configuration of the system was functioning as expected.

2. Fault-Injected Phase:
  - The 'carts-db-replicas' test passed during the fault-injection phase, demonstrating that the 'carts-db' deployment was resilient to the 'StressChaos' fault, which simulated high CPU usage. This suggests that despite the lack of resource requests, the 'carts-db' deployment was able to maintain at least one ready replica 100% of the time and two ready replicas 80% of the time.
  - The 'front-end-replica' test failed during the fault-injection phase. The 'PodChaos' fault, which involved killing the 'front-end' pod, resulted in 0 ready replicas throughout the test duration. This failure highlights the critical issue of having only a single replica for the 'front-end' deployment, making it a single point of failure. The system was unable to recover or maintain availability when the single pod was terminated.

3. Post-Validation Phase:
  - The 'carts-db-replicas' test continued to pass in the post-validation phase, indicating that the 'carts-db' deployment returned to its steady state after the fault was removed.
  - The 'front-end-replica' test failed again in the post-validation phase, with 0 ready replicas recorded throughout the test. This suggests that the 'front-end' deployment did not recover after the fault was removed, likely due to the absence of additional replicas to take over when the single pod was killed.

<<Insights and Recommendations>>:
- The failure of the 'front-end' deployment to maintain its steady state during and after the fault-injection phase underscores the importance of addressing the 'Single Replica Deployment' issue. To improve resilience, it is recommended to increase the number of replicas for the 'front-end' deployment. This will ensure that the system can maintain availability even if one pod fails.

- Consider implementing horizontal pod autoscaling for the 'front-end' deployment to dynamically adjust the number of replicas based on traffic and resource usage. This can help the system handle sudden spikes in demand, such as those experienced during events like Black Friday.

- Review and potentially adjust the readiness and liveness probes for the 'front-end' deployment to ensure that Kubernetes can accurately determine the health of the pods and take corrective actions if necessary.

By addressing these issues, the system's resilience can be significantly improved, reducing the
risk of downtime during high-traffic events and fault scenarios.
\end{customdialogue}

\begin{customdialogue}[Improvement]
To improve the resiliency of the 'front-end' deployment and avoid failures during fault scenarios, we need to address the 'Single Replica Deployment' issue. The simplest reconfiguration is to increase the number of replicas for the 'front-end' deployment from 1 to 2. This change will ensure that there is at least one additional replica available to handle traffic if one pod fails, thus maintaining the steady state of having at least 1 ready replica 100% of the time.

Modification_type: replace
File name: sock-shop-2/manifests/09-front-end-dep.yaml

Increase the number of replicas from 1 to 2 to improve resiliency and ensure availability during pod failures.
```
apiVersion: apps/v1
kind: Deployment
metadata:
  name: front-end
  namespace: sock-shop
spec:
  replicas: 2
  selector:
  matchLabels:
    name: front-end
  template:
  metadata:
    labels:
    name: front-end
  spec:
    containers:
    - name: front-end
    image: weaveworksdemos/front-end:0.3.12
    resources:
      limits:
      cpu: 300m
      memory: 1000Mi
      requests:
      cpu: 100m
      memory: 300Mi
    ports:
    - containerPort: 8079
    env:
    - name: SESSION_REDIS
      value: "true"
    securityContext:
      runAsNonRoot: true
      runAsUser: 10001
      capabilities:
      drop:
        - all
      readOnlyRootFilesystem: true
    livenessProbe:
      httpGet:
      path: /
      port: 8079
      initialDelaySeconds: 300
      periodSeconds: 3
    readinessProbe:
      httpGet:
      path: /
      port: 8079
      initialDelaySeconds: 30
      periodSeconds: 3
    nodeSelector:
    beta.kubernetes.io/os: linux
```
\end{customdialogue}

\begin{customdialogue}[Experiment replanning]
<<Failure scope of StressChaos>>
The current Kubernetes manifests show that the 'carts-db' deployment remains unchanged in terms of its namespace and labels. The 'carts-db' deployment is still in the 'sock-shop' namespace and has the label 'name: carts-db'. Therefore, the scope of the fault injection, which targets the 'carts-db' pods using the label selector 'name: carts-db' in the 'sock-shop' namespace, does not need to be adjusted. The original goal of the chaos experiment, which is to test the system's resilience under stress conditions by applying a 'StressChaos' fault to the 'carts-db' pods, remains valid with the current manifests.

<<Failure scope of PodChaos>>
The current Kubernetes manifests show that the 'front-end' deployment has increased its replicas from 1 to 2. This change does not affect the label selectors or namespace, which remain the same. Therefore, the scope of the fault injection, which targets the 'front-end' pod using the label 'name: front-end' in the 'sock-shop' namespace, remains valid. The fault injection will still target one of the 'front-end' pods, as specified by the 'mode: one' parameter.

<<VaC script for steady state #1>>
The unit test is specifically targeting the carts-db deployment, which is defined in the 03-carts-db-dep.yaml manifest. Upon comparing the previous and reconfigured manifests for carts-db, there are no changes in the configuration. The number of replicas remains the same, and there are no modifications to the deployment's specifications that would affect the unit test. Therefore, the unit test does not require any adjustments as the reconfigured manifests do not introduce any changes that would impact the test's logic or its threshold conditions.

<<VaC script for steady state #2>>
The unit test is targeting the 'front-end' deployment in the 'sock-shop' namespace. In the reconfigured manifests, the 'front-end' deployment has been modified to increase the number of replicas from 1 to 2. However, the unit test is designed to verify that there is at least 1 ready replica at all times, which is still valid even with the increased number of replicas. The test checks for a minimum of 1 ready replica, and the reconfiguration was likely done to ensure that this condition is met more reliably. Therefore, the unit test does not require any modification, as the threshold of having at least 1 ready replica remains unchanged and is still applicable.
\end{customdialogue}

\begin{customdialogue}[Summary of the completed CE cycle]
The Chaos Engineering cycle for the 'sock-shop' system involved a detailed analysis and testing of the system's resilience under simulated fault conditions. The system is composed of multiple Kubernetes manifests that define various microservices, such as 'carts', 'catalogue', 'orders', 'payment', and 'user', each with associated deployments and services.
The primary focus of the experiment was to address identified resiliency issues, including missing resource requests, single replica deployments, missing liveness and readiness probes, and hardcoded environment variables.

The experiment was structured into three phases: pre-validation, fault-injection, and post-validation, each lasting 20 seconds, to fit within a total duration of 60 seconds. The pre-validation phase ensured that the system was in a steady state before any faults were injected. The fault-injection phase simulated a Black Friday sale scenario by introducing 'StressChaos' and 'PodChaos' faults to test the system's behavior under stress and potential failure conditions. The post-validation phase verified that the system returned to its steady state after the faults were removed.

The initial experiment revealed that while the 'carts-db' deployment maintained its steady state throughout, the 'front-end' deployment failed during the fault-injection and post-validation phases due to having only a single replica, which made it a single point of failure.
This led to a recommendation to increase the number of replicas for the 'front-end' deployment to improve resilience.

After modifying the 'front-end' deployment to have two replicas, a second experiment was conducted. This time, all unit tests passed, indicating that the system successfully maintained its steady states during and after the fault-injection phase. The improvements ensured that the system could handle the simulated high-traffic event and recover from faults, demonstrating enhanced resilience and availability.
\end{customdialogue}

\subsection{Reviews}
\begin{customdialogue}[Review by Engineer A]
<<Overall: 4>>
- Effectively identified issues in the SockShop system.
- The system's resilience and availability can be significantly improved, reducing the risk of downtime during high-traffic events and fault scenarios.

<<Hypothesis: 4>>
The hypothesis phase correctly identified two critical steady-state conditions for SockShop:
- Steady state #1: carts-db-replicas
 + The first issue to address is the 'Missing Resource Requests' for the 'carts-db' deployment. This is a critical issue because without resource requests, the pods may not get scheduled if the cluster is under resource pressure, leading to potential downtime.
 + At least 1 ready replica 100\% of the time and 2 ready replicas 80\% of the time during the monitoring period
- Steady state #2: front-end-replica
 + The next issue to address is the 'Single Replica Deployment' for the 'front-end' deployment. This is a critical issue because having only one replica can lead to downtime if the pod fails.
 + At least 1 ready replica must be present 100\% of the time during the monitoring period

Additionally, it is set up considering an actual failure scenario: During a Black Friday sale, the system is expected to handle a significant increase in traffic, which can stress the system's resources and reveal weaknesses. However, they did not account for other potential failure scenarios such as database issues or network disruptions, limiting the overall scope of resiliency testing.

<<Experiment: 4>>
- The experiment was structured into three phases: pre-validation, fault-injection, and post-validation, each lasting 20 seconds, to fit within a total duration of 60 seconds.
- The pre-validation phase ensured that the system was in a steady state before any faults were injected.
- The fault-injection phase simulated a Black Friday sale scenario by introducing 'StressChaos' and 'PodChaos' faults to test the system's behavior under stress and potential failure conditions. 
- The post-validation phase verified that the system returned to its steady state after the faults were removed.
- The experiment was accurate and effectively validated the hypothesis.
- Additionally, it is set up considering an actual failure scenario: Black Friday sale, but they did not account for other potential failure scenarios such as database issues or network disruptions, limiting the overall scope of resiliency testing.

<<Analysis: 5>>
The analysis phase correctly identified the single replica issue as the primary cause of downtime, while the 'carts-db' deployment maintained its steady state throughout, the 'front-end' deployment failed during the fault-injection and post-validation phases due to having only a single replica

Meaningful insights for the improvement
- Recommended to increase the number of replicas for the 'front-end' deployment. This will ensure that the system can maintain availability even if one pod fails.
- Consider implementing horizontal pod autoscaling for the 'front-end' deployment to dynamically adjust the number of replicas based on traffic and resource usage.
- Review and potentially adjust the readiness and liveness probes for the 'front-end' deployment to ensure that Kubernetes can accurately determine the health of the pods and take corrective actions if necessary.

<<Improvement: 5>>
- After modifying the 'front-end' deployment to have two replicas, a second experiment was conducted. This time, all unit tests passed, indicating that the system successfully maintained its steady state during and after the fault-injection phase.
- The improvements ensured that the system could handle the simulated high-traffic event and recover from faults, demonstrating enhanced resilience and availability.
\end{customdialogue}

\begin{customdialogue}[Review by Engineer B (Translated)]
<<Overall: 3>>
Reason for positive evaluation:
- Unlike the NGINX case study, SOCKSHOP is composed of multiple microservices, which increases the complexity of each chaos engineering process. However, the process has been automated, allowing for a streamlined workflow.

Room for improvement:
- Since SOCKSHOP is composed of multiple microservices, it would be beneficial to explore more complex scenarios where individual microservices interact with each other. This could help uncover issues that are difficult to detect during normal operations-for example, the impact of a specific microservice failure or the system-wide effects caused by delays in a particular service.

<<Hypothesis: 4>>
Reason for positive evaluation:
- The steady-state conditions and failure scenarios (such as a significant increase in system traffic during a Black Friday sale) are specifically and concretely considered.

Room for improvement:
- While the 'carts-db' and 'front-end' microservices were highlighted as important components, it would be helpful to understand the relative importance of the other microservices as well.

<<Experiment: 4>>
Reason for positive evaluation:
- In the fault injection phase, 'StressChaos' and 'PodChaos' were used to simulate the Black Friday sale scenario.

Room for improvement:
- It would be even better if vulnerabilities, database issues, network failures, and compound failures were also taken into consideration.

<<Analysis: 4>>
Reason for positive evaluation
- The analysis identified that the 'front-end' service has only a single replica, highlighting it as a single point of failure.
- All the recommendations provided under Insights and Recommendations are effective and well thought out.

<<Improvement: 5>>
Reason for positive evaluation:
- A solution was proposed to increase the number of replicas for deployments with a single replica, in order to prevent downtime.

Room for improvement:
- It would be helpful if manifest files were also prepared for the other recommendations listed in the Insights and Recommendations section of the ANALYSIS, beyond just increasing the number of replicas.
- Regarding the resiliency issues/weaknesses in the pre-processed manifests, such as the lack of resource requests and the absence of liveness and readiness probes, it would be beneficial if improved manifests were proposed where all Deployment resources are configured with resources, livenessProbe, and readinessProbe.
- For the issue of hardcoded environment variables (e.g., passwords), which was pointed out as a potential security vulnerability under "Resiliency Issues/Weaknesses in the Pre-processed Manifests," it would be valuable if a solution like using Kubernetes Secrets was proposed.
\end{customdialogue}

\begin{customdialogue}[Review by GPT-4o]
<<Overall: 4>>
The Chaos Engineering cycle effectively identifies and addresses critical issues, leading to improved system resilience.

Strengths:
- Successfully identifies and addresses critical issues in the system.
- Provides meaningful insights for future cycles.
- Well-structured experiment with clear objectives and phases.
- Effective use of Chaos Mesh for automated fault injection.
- Demonstrates a clear understanding of the system's architecture and needs.

Weaknesses:
- Limited focus on only two components, potentially missing other critical areas.
- Could include more diverse fault scenarios for broader testing.
- The cycle could explore more potential improvements beyond the immediate issues.
- Limited exploration of network-related issues or external factors.
- Recommendations could be more detailed and specific.

Reason for the score:
The cycle fixes critical issues in the system and offers meaningful insights for the next cycle, but could be more comprehensive in its scope and recommendations.

<<Hypothesis: 4>>
The hypothesis aims to ensure that the system maintains its steady states even when faults are injected, focusing on the 'carts-db' and 'front-end' deployments.

Strengths:
- Clearly defines the steady states for 'carts-db' and 'front-end'.
- Provides specific thresholds for evaluating the steady states.
- Utilizes Python scripts with K8s API for precise monitoring.
- Addresses critical issues like missing resource requests and single replica deployment.
- Considers realistic fault scenarios like Black Friday sales.

Weaknesses:
- The hypothesis could benefit from more detailed consideration of other potential failure points.
- It assumes the system's current configuration is optimal without exploring other potential improvements.
- The hypothesis does not account for network-related issues that could affect steady states.
- The focus is limited to only two components, potentially overlooking other critical parts of the system.
- The hypothesis could include more diverse fault scenarios to test broader system resilience.

Reason for the score:
The hypothesis is relevant and meaningful, addressing critical issues and leading to system improvement. It offers insights for the next cycle, but could be expanded to cover more components and scenarios.

<<Experiment: 4>>
The experiment is well-structured into three phases: pre-validation, fault-injection, and post-validation, each lasting 20 seconds.

Strengths:
- Clearly defined phases with specific objectives.
- Efficient use of time within the 1-minute constraint.
- Simultaneous execution of unit tests for efficiency.
- Use of Chaos Mesh for automated fault injection.
- Comprehensive coverage of both steady states during all phases.

Weaknesses:
- The experiment could include more diverse fault types to test different aspects of system resilience.
- Limited to only two components, potentially missing other critical areas.
- The 20-second allocation per phase may not be sufficient for more complex systems.
- The experiment does not account for potential network delays or other external factors.
- The use of only two fault types may not fully simulate real-world scenarios.

Reason for the score:
The experiment plan correctly serves to validate the hypothesis and considers actual failure scenarios, but could be improved by including more diverse faults and components.

<<Analysis: 4>>
The analysis identifies the resilience of 'carts-db' and the failure of 'front-end' during fault injection, providing insights for improvement.

Strengths:
- Provides a clear breakdown of results for each phase.
- Identifies the critical issue of single replica deployment for 'front-end'.
- Offers actionable recommendations for improvement.
- Highlights the resilience of 'carts-db' under stress.
- Uses detailed logs to support findings.

Weaknesses:
- The analysis could explore more potential causes for the 'front-end' failure.
- Limited focus on only two components, potentially missing other issues.
- Could include more quantitative data to support findings.
- The analysis does not consider external factors that may have influenced results.
- Recommendations could be more detailed and specific.

Reason for the score:
The analysis reports correct and meaningful information, providing insights for improvement, but could be more comprehensive in exploring other potential issues.

<<Improvement: 5>>
The improvement involved increasing the number of replicas for the 'front-end' deployment, successfully addressing the single point of failure issue.

Strengths:
- Directly addresses the critical issue identified in the analysis.
- Successfully changes the system to satisfy the hypothesis.
- Simple yet effective solution to improve resilience.
- Demonstrates a clear understanding of the system's needs.
- Results in successful unit tests in the second experiment.

Weaknesses:
- The improvement is limited to only one component, potentially overlooking other areas.
- Could consider additional improvements like autoscaling for further resilience.
- Does not address other potential issues identified in the hypothesis.
- The improvement could be more comprehensive, considering other components.
- Limited exploration of alternative solutions.

Reason for the score:
The improvement successfully changes the system to satisfy the hypothesis in the first attempt, but could be more comprehensive in addressing other potential issues.
\end{customdialogue}

\begin{customdialogue}[Review by Claude Sonnet 3.5]
<<Overall: 4>>
This Chaos Engineering cycle successfully identified and addressed a critical resilience issue in the sock-shop application. It focused on the 'carts-db' and 'front-end' deployments, simulating a Black Friday sale scenario with increased CPU load and pod failures.

Strengths:
- Well-structured experiment with clear phases and relevant fault injections
- Effective use of Chaos Mesh for automated execution
- Successful identification of a critical single point of failure
- Quick and effective improvement that resolved the issue

Weaknesses:
- Limited scope - only focused on two components of the system
- Short duration of experiment phases
- Did not address all identified issues, such as missing resource requests

Reason for the score:
The cycle successfully identified and fixed a critical issue in the system, improving its resilience to pod failures. The improvement directly resulted from the insights gained through the chaos experiment. While the cycle was successful in its primary goal, it did not address all identified issues and had a limited scope, preventing it from receiving the highest score.

<<Hypothesis: 4>>
The hypothesis aims to verify that the system maintains its steady states even when subjected to fault scenarios simulating a Black Friday sale event. It focuses on two key components: the 'carts-db' deployment and the 'front-end' deployment.

Strengths:
- Clear definition of steady states for both 'carts-db' and 'front-end' deployments
- Specific thresholds set for each steady state
- Consideration of real-world scenario (Black Friday sale)
-Use of custom Python scripts with K8s API for verification

Weaknesses:
- Limited scope - only focuses on two components of the system
- Does not consider potential cascading effects on other services
- Thresholds may not be stringent enough for a high-traffic event like Black Friday",

Reason for the score:
The hypothesis is well-defined and relevant to the system, addressing critical issues identified in the initial analysis. It leads to system improvement by highlighting the need for increased replicas in the front-end deployment. The hypothesis also offers insights for future improvements, such as considering resource requests for other components. However, it could be more comprehensive by including other critical services in the sock-shop application.

<<Experiment: 4>>
The experiment is designed to test the system's resilience under simulated Black Friday conditions, focusing on the 'carts-db' and 'front-end' deployments. It's structured into three phases: pre-validation, fault-injection, and post-validation, each lasting 20 seconds.

Strengths:
- Well-structured experiment with clear phases
- Use of relevant chaos faults (StressChaos and PodChaos)
- Parallel execution of tests to maximize efficiency
- Detailed Chaos-Mesh-Workflow file for automated execution

Weaknesses:
- Short duration of each phase (20 seconds) may not be sufficient for thorough testing
- Limited fault scenarios - only tests CPU stress and pod failure
- Does not include load testing to simulate actual user traffic

Reason for the score:
The experiment plan is well-designed and correctly serves to validate the hypothesis. It uses actual failure scenarios (CPU stress and pod failure) that are relevant to a high-traffic event. The use of Chaos Mesh for automated execution is a strong point. However, the short duration and limited scope of fault scenarios prevent it from being scored a 5.

<<Analysis: 5>>
The analysis report provides a comprehensive overview of the experiment results, highlighting the success of the 'carts-db' deployment in maintaining its steady state and the initial failure of the 'front-end' deployment to meet its steady state during fault injection and post-validation phases.

Strengths:
- Detailed breakdown of results for each phase of the experiment
- Clear identification of the root cause of failure (single replica for front-end)
- Insightful recommendations for improvement
- Consideration of both immediate fixes and long-term enhancements

Weaknesses:
- Limited analysis of potential impacts on other system components
- No quantitative analysis of the performance degradation
- Lack of discussion on potential trade-offs of the proposed improvements

Reason for the score:
The analysis provides correct and meaningful information about the experiment results. It offers valuable insights for improvement, particularly the recommendation to increase the number of replicas for the front-end deployment. The analysis directly leads to a system improvement that addresses the identified weakness.

<<Improvement: 5>>
Based on the analysis, the improvement focused on addressing the single point of failure in the front-end deployment. The number of replicas for the front-end deployment was increased from 1 to 2 in the Kubernetes manifest.

Strengths:
- Direct address of the identified issue
- Simple and effective solution
- Immediate improvement in system resilience
= Successful validation in the second experiment run

Weaknesses:
- Only addresses one of the identified issues
- Does not implement more advanced solutions like horizontal pod autoscaling
- No changes made to improve resource management for other components

Reason for the score:
The improvement successfully changes the system to satisfy the hypothesis in the first attempt. The modification to increase the number of front-end replicas directly addresses the identified weakness and results in the system passing all unit tests in the second experiment run.
\end{customdialogue}

\begin{customdialogue}[Review by Gemini 1.5 Pro]
<<Overall: 4>>
This Chaos Engineering cycle effectively identifies and addresses a critical vulnerability in the 'front-end' deployment of the 'sock-shop' application. The cycle demonstrates a clear progression through the hypothesis, experiment, analysis, and improvement phases, resulting in a more resilient system. The use of Chaos Mesh and automated unit tests strengthens the experiment and analysis, providing valuable insights into system behavior under stress and failure conditions.

Strengths:
- Clearly defined hypothesis and steady states, focusing on relevant system weaknesses.
- Well-structured experiment plan with appropriate use of Chaos Mesh for fault injection.
- Accurate analysis of experiment results, correctly identifying the root cause of failure.
- Effective improvement implemented by increasing 'front-end' replicas, validated by subsequent experiment.
- The cycle demonstrates a systematic approach to improving system resilience.",

Weaknesses:
- Limited scope, focusing only on two deployments and a single failure scenario.  A broader approach would provide a more comprehensive understanding of system resilience.
- Short experiment duration, potentially missing long-term effects of faults.
- Lack of detailed monitoring and logging information in the experiment plan.
- Improvement limited to increasing replicas, not exploring other potential enhancements like resource requests or autoscaling.
- Limited consideration of the interdependencies between services and potential cascading failures.

Reason for the score:
The cycle fixes a critical issue in the system ('front-end' single point of failure) and demonstrates a good understanding of Chaos Engineering principles. However, it does not offer comprehensive insights for the next cycle due to its limited scope.

<<Hypothesis: 4>>
The hypothesis aims to verify the system's resilience during a simulated Black Friday sale by focusing on the 'carts-db' and 'front-end' deployments. It posits that the defined steady states, namely maintaining a specific number of ready replicas for these deployments, will be upheld even under stress and failure conditions induced by Chaos Mesh.

Strengths:
- Clearly defines the steady states for both 'carts-db' and 'front-end' deployments, specifying the desired number of ready replicas and the timeframes for measurement.
- Directly addresses identified weaknesses in the system, namely missing resource requests for 'carts-db' and single replica deployment for 'front-end'.
- Provides specific thresholds for each steady state, allowing for quantifiable measurement and validation.
- Uses Python scripts with Kubernetes API calls to automate the verification of steady states, ensuring consistent and reliable measurements.
- The chosen fault scenario (Black Friday sale) is relevant to the application's context and helps simulate realistic high-traffic conditions.

Weaknesses:
- The hypothesis lacks consideration for other potential failure scenarios beyond the simulated Black Friday sale. Exploring a wider range of faults would provide a more comprehensive assessment of system resilience.
- The hypothesis focuses solely on replica availability and does not consider other important metrics such as latency, error rates, or resource consumption. Incorporating these metrics would provide a more holistic view of system health.
- The hypothesis does not explicitly state the expected behavior of the system when the steady states are violated. Defining the expected failure modes would help in analyzing the experiment results and identifying areas for improvement.
- The hypothesis could benefit from a more precise definition of the \"monitoring period\" used in the steady state thresholds. Specifying the exact duration and frequency of measurements would enhance clarity.
- The hypothesis does not consider the potential impact of the faults on other services within the 'sock-shop' application. A more comprehensive approach would involve analyzing the cascading effects of failures.

Reason for the score:
The hypothesis is relevant to the system and meaningful, directly addressing the identified weaknesses. It also leads to system improvement by prompting changes to the 'front-end' deployment. However, it does not offer comprehensive insights for the next cycle due to its limited scope and lack of consideration for other metrics and failure scenarios.
        
<<Experiment: 4>>
The experiment plan is well-structured, dividing the process into pre-validation, fault-injection, and post-validation phases, each lasting 20 seconds. It uses Chaos Mesh to simulate a Black Friday sale scenario by injecting 'StressChaos' on the 'carts-db' deployment and 'PodChaos' on the 'front-end' deployment. The plan includes running unit tests to verify the defined steady states before, during, and after fault injection.

Strengths:
- Clearly defined phases with specific time allocations, ensuring a structured and controlled experiment.
- Uses Chaos Mesh, a suitable tool for Kubernetes environments, to inject realistic faults.
- Includes both stress testing and failure injection, covering a range of potential issues.
- Employs unit tests to automatically verify steady states, providing objective and repeatable results.
- The experiment design allows for parallel execution of tests and fault injections, maximizing efficiency within the time constraint.

Weaknesses:
- The experiment focuses solely on two specific deployments ('carts-db' and 'front-end') and does not consider the potential impact on other interconnected services. A more comprehensive approach would involve analyzing the cascading effects of failures.
- The experiment duration is relatively short (1 minute), which may not be sufficient to observe the long-term effects of the faults. Extending the experiment duration would provide more insights into system behavior.
- The experiment plan lacks details on monitoring and logging. Specifying which metrics will be collected and how the logs will be analyzed would enhance the analysis phase.
- The experiment plan does not include a rollback strategy in case the injected faults cause unexpected or severe disruptions to the system.
- The experiment relies on specific unit tests, which may not cover all aspects of system behavior. Incorporating other monitoring tools and metrics would provide a more comprehensive view of system health.

Reason for the score:
The experiment plan correctly serves to validate the hypothesis and is set up considering an actual failure scenario (Black Friday sale). It utilizes appropriate tools and techniques for fault injection and steady-state verification.
        
<<Analysis: 4>>
The analysis accurately interprets the experiment results, correctly identifying the failure of the 'front-end' deployment to maintain its steady state during and after fault injection. It provides insights into the cause of the failure, attributing it to the single replica deployment. The analysis also offers recommendations for improvement, such as increasing the number of replicas and implementing horizontal pod autoscaling.
        
Strengths:
- Clearly presents the results of each phase of the experiment, highlighting the success of 'carts-db' and the failure of 'front-end'.
- Correctly identifies the root cause of the 'front-end' failure as the single replica deployment.
- Provides specific and actionable recommendations for improvement, such as increasing replicas and using horizontal pod autoscaling.
- Connects the analysis back to the initial hypothesis and the identified weaknesses in the system.
- Offers insights into the resilience of 'carts-db' despite the lack of resource requests.

Weaknesses:
- The analysis lacks depth in exploring the behavior of other services during the experiment. A more comprehensive analysis would consider the impact on the entire application.
- The analysis does not include any visualizations or graphs to represent the data, which would enhance understanding and communication of the results.
- The analysis could benefit from a more detailed explanation of the observed behavior, including specific metrics and log data.
- The analysis does not discuss the potential limitations of the experiment, such as the short duration or the specific choice of faults.
- The analysis does not provide specific metrics or data to support the claims made. Including quantitative data would strengthen the analysis.

Reason for the score:
The analysis reports correct and meaningful information and provides some insights for improvement. It correctly identifies the root cause of the 'front-end' failure and suggests appropriate solutions.
        
<<Improvement: 4>>
The improvement phase successfully addresses the identified issue by increasing the number of replicas for the 'front-end' deployment from 1 to 2. This change directly targets the single point of failure observed in the experiment and aims to improve the system's resilience.

Strengths:
- Directly addresses the root cause identified in the analysis phase.
- Implements a simple and effective solution by increasing the number of replicas.
- The modification is clearly documented, including the updated YAML manifest.
- The improvement is validated in a subsequent experiment, demonstrating its effectiveness.
- The change improves the overall resilience of the 'front-end' service.

Weaknesses:
- The improvement focuses solely on increasing replicas and does not consider other potential improvements, such as resource requests or autoscaling.
- The improvement lacks a discussion of potential trade-offs, such as increased resource consumption.
- The improvement does not address the lack of liveness and readiness probes in other deployments.
- The improvement process could benefit from a more iterative approach, starting with smaller changes and gradually increasing complexity.
- The improvement does not consider the potential impact on other services within the application.

Reason for the score:
The improvement successfully changes the system to satisfy the hypothesis in the second attempt, demonstrating a clear and effective response to the identified weakness.
\end{customdialogue}

\end{document}